\documentclass[11pt]{article}
\usepackage{graphicx}
\usepackage{epsfig}
\usepackage{hyperref}
\usepackage[usenames,dvipsnames]{color}%mjt-this uses familiar color names
\usepackage{amsmath}  %%%get rid of this
\usepackage{xspace} 
\newcommand{\sqs}{\mbox{$\sqrt{s}$}\xspace}
\newcommand{\sqsn}{\mbox{$\sqrt{s_{_{NN}}}$}\xspace}

\def\lsim{\raise0.3ex\hbox{$<$\kern-0.75em\raise-1.1ex\hbox{$\sim$}}}
\def\gsim{\raise0.3ex\hbox{$>$\kern-0.75em\raise-1.1ex\hbox{$\sim$}}}

\def\mean#1{\left<#1\right>}
\def\Journal#1#2#3#4{{#1}{\bf #2} (#4) #3}
\def\IJMPA{{Int. J. Mod. Phys. A}}

\def\EPJC{{Eur. Phys. J. C}}
\def\JPG{{J. Phys. G}}
\def\JPCS{{J. Phys: Conf. Series\ }}
\def\NCA{Nuovo Cimento\ }

\def\NIMA{{Nucl. Instrum. Methods A}}
\def\NPA{{Nucl. Phys. A}}
\def\NPB{{Nucl. Phys. B}}
\def\PLB{{Phys. Lett. B}}
\def\PLC{Phys. Repts.\ }

\def\PRL{Phys. Rev. Lett.\ }
\def\PRD{{Phys. Rev. D}}
\def\PRC{{Phys. Rev. C}}
\def\PR{Phys. Rev.\ }

\def\ZPC{{Z. Phys. C}}
\def\ARNPS{{Ann. Rev. Nucl. Part. Sci.\ }}

\def\RPP{Rep. Prog. Phys.\ }
\def\QGP{{\color{Red} Q}{\color{Blue} G}{\color{Green} P}} 
\def\QCD{{\color{Red} Q}{\color{Green} C}{\color{Blue} D}} 
\begin{document}
\markboth{M.~J.~Tannenbaum}{Highlights from BNL-RHIC 2011-2013}

\title{Highlights from BNL-RHIC 2011-2013}
\author{M.~J.~Tannenbaum
\thanks{Research supported by U.~S.~Department of Energy, DE-AC02-98CH10886.}
\\Physics Department, 510c,\\
Brookhaven National Laboratory,\\
Upton, NY 11973-5000, USA\\
mjt@bnl.gov} \maketitle

\begin{abstract}
Highlights from Brookhaven National Laboratory (BNL) and experiments at the BNL Relativistic Heavy Ion Collider (RHIC) are presented for the years 2011--2013. This review is a combination of lectures which discussed the latest results each year at a three year celebration of the 50th anniversary of the International School of Subnuclear Physics in Erice, Sicily, Italy. Since the first collisions  in the year 2000, RHIC has provided nucleus-nucleus and polarized proton-proton collisions over a range of nucleon-nucleon c.m. energies (\sqsn ) from 7.7 to 510 GeV with nuclei from deuterium to uranium, most often gold. The objective was the discovery of the Quark Gluon Plasma, which was achieved, and the measurement of its properties, which were much different than expected, namely a `perfect  fluid' of quarks and gluons with their color charges exposed rather than a gas. Topics including quenching of light and heavy quarks at large transverse momentum, thermal photons, search for a \QCD\ critical point as well as measurements of collective flow, two-particle correlations and $J/\Psi$ suppression are presented. During this period, results from the first and subsequent heavy ion measurements at the Large Hadron Collider (LHC) at CERN became available. These confirmed and extended the RHIC discoveries and have led to ideas for new and improved measurements. \\
 
\noindent{\it Keywords:} {RHIC; LHC; s\QGP; heavy ion collisions; jet quenching; thermal radiation; collective flow; \QCD\ critical point; Cumulants; $J/\Psi$ suppression; deconfinement .} \\

\noindent{PACS numbers:25.75.-q}
\end{abstract}

\tableofcontents
	
\section{Introduction}\label{sec:introduction}
High energy nucleus-nucleus collisions provide the means of creating nuclear matter in conditions of extreme temperature and density~\cite{BearMountain,seeMJTROP,MJTROP}.  
 The kinetic energy of the incident projectiles would be dissipated in the large 
volume of nuclear matter involved in the reaction.  At large energy or baryon densities, a phase transition is expected from a state of nucleons containing confined quarks and gluons to a state of ``deconfined'' (from their individual nucleons) quarks and gluons, in chemical and thermal equilibrium, covering a volume that is many units of the confining length scale. This state of nuclear matter was originally given the name Quark Gluon Plasma (\QGP)~\cite{Shuryak80}, a plasma being an ionized gas. 

\begin{figure}[!htb]
\begin{center}
%\begin{minipage}{1.0\textwidth}
%\begin{center}
\includegraphics[width=0.6\textwidth]{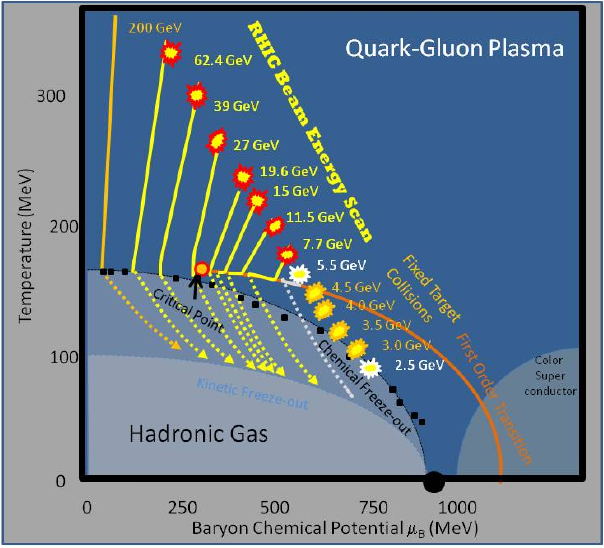}
\end{center}%\vspace*{-1.5pc}
\caption[]{(left) A proposed phase diagram for nuclear matter from STAR Beam Use Request for 2014-2015: Temperature,  $T$,  vs Baryon Chemical Potential, $\mu_B$. \label{fig:phase_boundary}}%\vspace*{-1.0pc}
%\end{minipage}
%\end{center}
\end{figure}

 A typical proposed phase diagram of nuclear matter is shown in Fig.~\ref{fig:phase_boundary} together with 
the idealized trajectories of the evolution of the medium for Au+Au collisions at c.m. energies proposed for the Beam Energy Scan at RHIC in search for a \QCD\ critical point. The bursts represent the hottest and densest stage of the medium when thermal equilibrium is  reached shortly after the collision. The axes are the temperature $T$ vs. the baryon chemical potential $\mu_B$. The temperature for the transition from the Quark Gluon Plasma (\QGP) to a hadron gas is taken as 170 MeV for $\mu_B=0$ and the phase boundary is predicted to be a smooth crossover down to a critical point below which the phase boundary becomes a first order phase transition.

\subsection{Discovery of the \QGP}
The \QGP\ was discovered at RHIC, and announced on April 19, 2005. 
However the results at RHIC~\cite{seeMJTROP} indicated that instead of behaving like a gas of free quarks and gluons, the matter created in heavy ion collisions at nucleon-nucleon c.m. energy $\sqrt{s_{NN}}=200$ GeV appears to be more like a {\em liquid}. This matter interacts much more strongly than originally expected, as elaborated in peer reviewed articles by the 4 RHIC experiments~\cite{BRWP,PHWP,STWP,PXWP}, which inspired the theorists~\cite{THWPS} to give it the new name ``s\QGP" (strongly interacting \QGP). These properties were quite different from the ``new state of matter'' claimed in a press-conference~\cite{CERNBaloney} by the CERN fixed target heavy ion program on February 10, 2000, which was neither peer-reviewed nor published. 

In spite of not being published, the CERN press-release had a major effect on the press in the United States resulting in an article on the front page of the New York Times~\cite{NYT02102000}. Ironically, on this very same front page was an article announcing that the true version of the famous Italian sausage, Mortadella, would, for the first time, be allowed to be imported into the United States. A photograph of the iconic Bologna sausages appeared right next to the article about the CERN ``qgp''. Unfortunately, the first European Baloney to arrive in the U.~S. was the CERN announcement.~\footnote{It is important for the reader of these proceedings to be aware that a high official of CERN was in the audience during this talk and made no objection to this comment. Furthermore, the author has a long, positive and productive relationship with this great laboratory and has praised its many successes. Thus he feels justified in commenting on one of their rare misjudgments. See Ref.~\cite {MJTROP} for a detailed scientific discussion. } 

However the situation at CERN improved dramatically in November 2010 with the startup of Pb+Pb collisions in the CERN-LHC at $\sqrt{s_{NN}}=2760$ GeV (2.76 TeV) where the real \QGP\ was observed. The LHC Pb+Pb measurements confirm the RHIC discoveries~\cite{EriceProcPR,MJTIJMPA2011,RATCUP} and add some new information---notably with fully reconstructed jets~\cite{ATLASdijet,CMSdijet}.

\section{Progress at BNL and the RHIC machine 2011-2013.}
\subsection{RHIC}
\begin{figure}[!t]
\begin{center}
\includegraphics[width=0.90\textwidth]{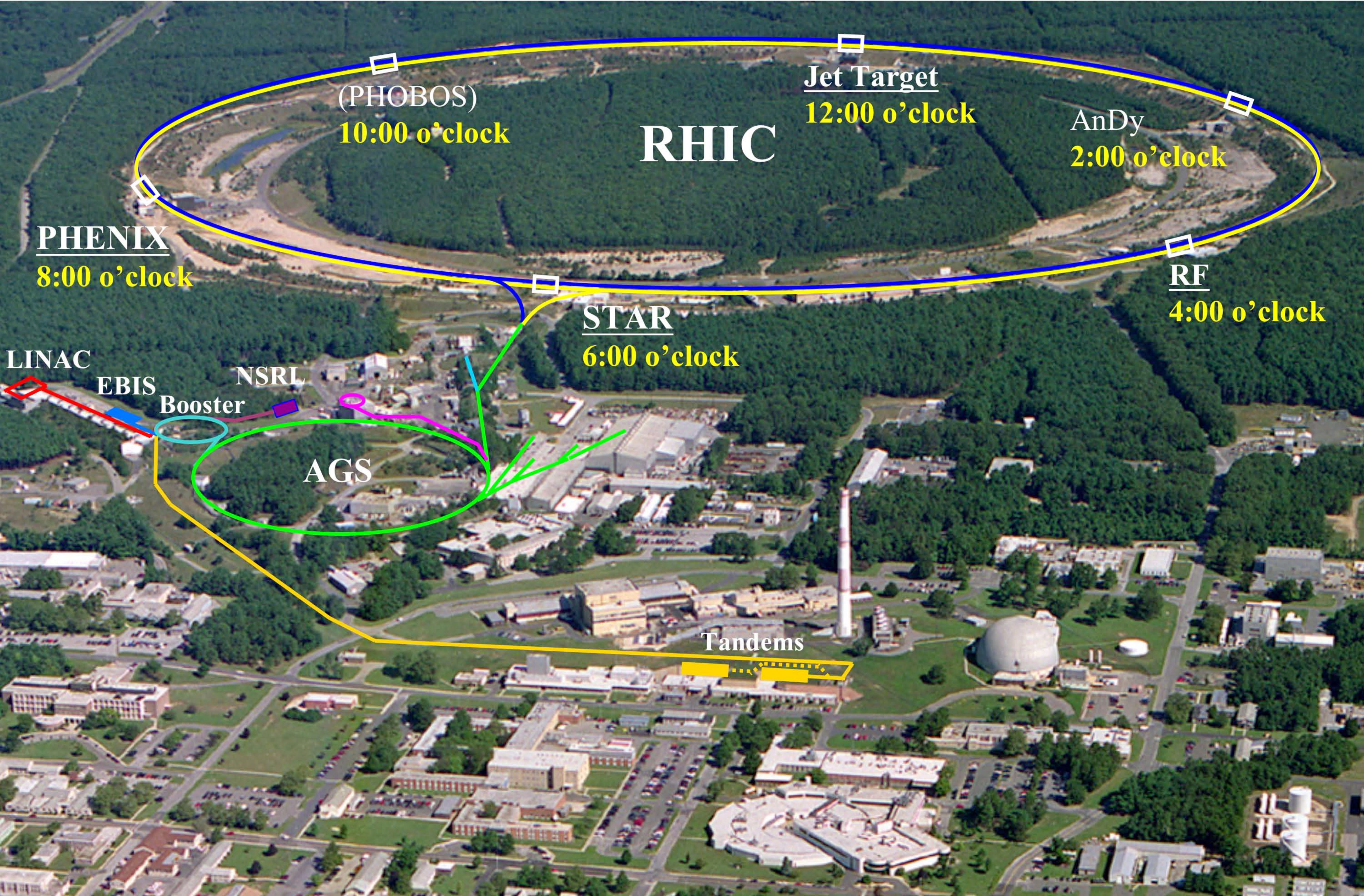}
\end{center}\vspace*{-0.8pc}
\caption[]{ Aerial view of the RHIC facility~\cite{RHICNIM}. The six crossing points are labelled as on a clock. The two principal experiments still running are PHENIX and STAR. Two smaller experiments PHOBOS~\cite{PHWP} and BRAHMS~\cite{BRWP} have been completed; a test run, AnDy, occupies the former location of BRAHMS. The LINAC is the injector for polarized protons into the Booster/AGS/RHIC chain; with the Jet Target used for precision beam polarization measurements. The TANDEM injector for Ions has been replaced by the Electron Beam Ion Source (EBIS) starting with the 2012 run. }
\label{fig:RHICoverview}\vspace*{-1.0pc}
\end{figure}
\begin{figure}[!h]%\vspace*{-1.0pc}
\begin{center}
\includegraphics[width=0.49\textwidth]{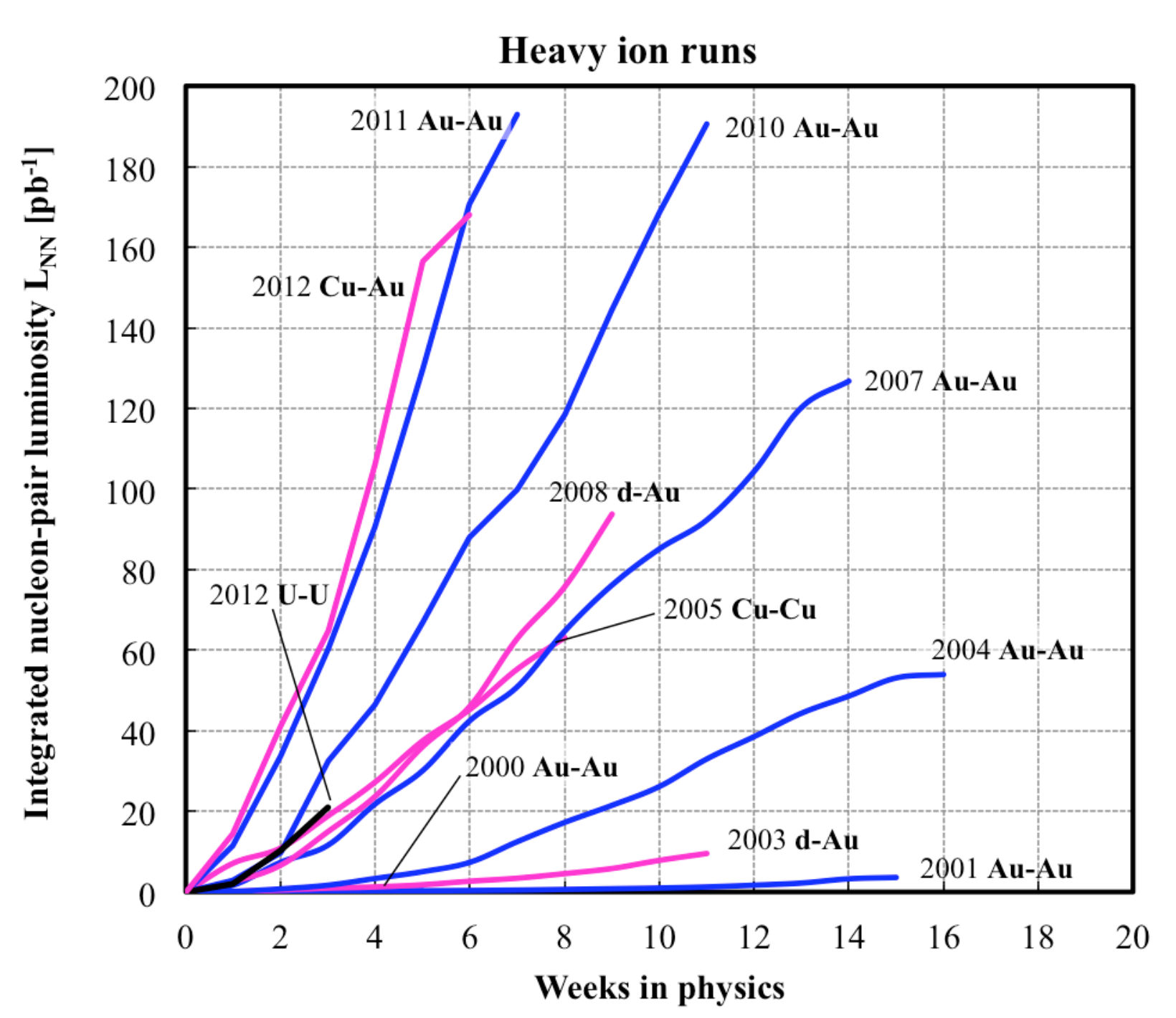}%\hspace*{1pc}
\includegraphics[width=0.49\textwidth]{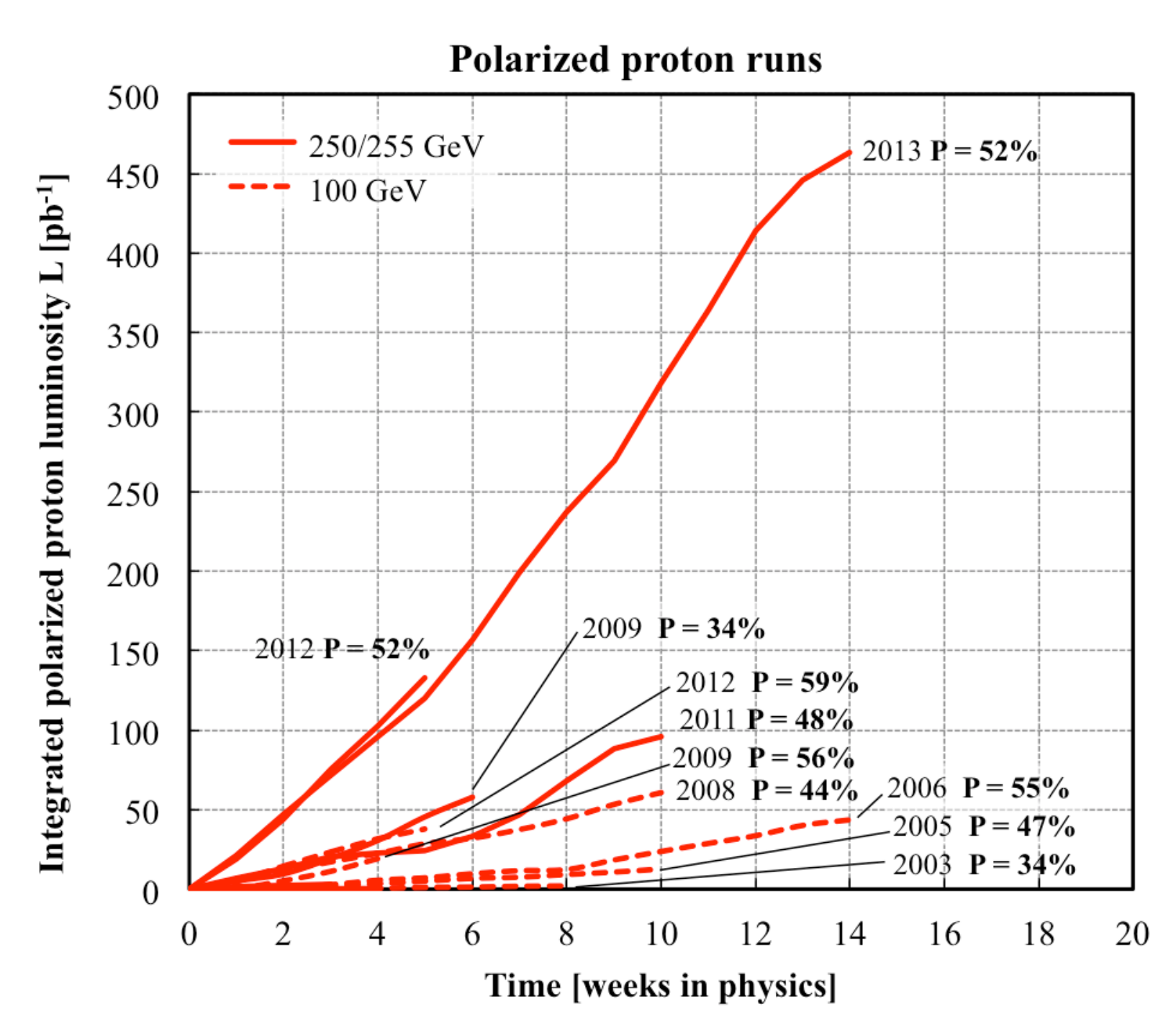}
\end{center}\vspace*{-1.0pc}
\caption[]{a)(left) Au+Au performance, where the nucleon-pair luminosity is defined as $L_{\rm NN}=A\times B\times L$, where $L$ is the luminosity and $A$, $B$ are the number of nucleons in the colliding species. b) (right) Polarized p-p performance. Courtesy Wolfram Fischer.}
\label{fig:RHICperf}%\vspace*{-1.0pc}
\end{figure}

With the shutdown of the Tevatron at FERMILAB, on September 30, 2011 after 28 years of operation, the Relativistic Heavy Ion Collider (RHIC)  at Brookhaven National Laboratory (Fig.~\ref{fig:RHICoverview}) is the only hadron collider in the U.S. and one of only two hadron-colliders in the world, 
the other being the CERN-LHC. RHIC is also the world's first and only polarized proton collider.
RHIC is composed of two independent rings, of circumference 3.8 km, containing a total of 1,740 superconducting magnets(see Fig.~\ref{fig:RHICring}a, below). RHIC can collide any species with any other species and since beginning operation in the year 2000 has provided collisions at 13 different values of nucleon-nucleon c.m. energy, $\sqrt{s_{NN}}$, and nine different species combinations including Au+Au, d+Au, Cu+Cu, Cu+Au, U+U, if differently polarized protons are counted as different species.  For the runs in 2010-2011, an Au+Au energy scan was performed with $\sqrt{s_{NN}}=$ 7.7, 11.5, 19.6, 27, 39, 62.4, 200 GeV. The performance history of RHIC with A+A and polarized p-p collisions is shown in Fig.~\ref{fig:RHICperf}. 

At present RHIC operates at 15 times design luminosity for Au+Au and has shown a factor 2 progress in integrated luminosity $\cal{L}_{\rm int}$ per week from Run-4 (2004) to Run-7 to Run-10 to Run-11. In 2011, 3 dimensional stochastic cooling (with only 2-d transverse cooling active) was introduced for the Au+Au collisions to improve the storage lifetime. Further improvements in the longitudinal profile, i.e. smaller diamond size without bucket migration, will be made with increased longitudinal focusing from new 56 MHz radio frequency storage cavity.

A significant improvement to the A+A program starting with Run-12 in 2012 was the replacement of the 40 year old Tandem Van de Graaff injector  with an Electron Beam Ion Source (EBIS). A 10 A electron beam creates the desired charge state(s) in a trap within a 5 T superconducting solenoid. This is then accelerated through the Radio Frequency Quadrupole (RFQ)  and linac and injected into the AGS Booster (Fig.~\ref{fig:RHICoverview}). All ion species including noble gases, uranium and polarized $^3$He are available.  Commissioning of the EBIS started during early 2011 and supplied He$^+$, He$^{2+}$, Ne$^{5+}$, Ne$^{8+}$, Ar$^{11+}$, Ti$^{18+}$ and Fe$^{20+}$ for the NASA Space Radiation Research Laboratory (NSRL) at BNL~\cite{NSRL}. %{\color{Red}steady note UU run at RHIC with 3d cooling while AuAu only had 2d}
For the first time in a collider, Cu+Au and U+U collisions at $\sqrt{s_{NN}}=200$ GeV were studied.  Also, the polarized p-p runs in 2012 included both $\sqrt{s}=200$ and 510 GeV with  improved polarization and luminosity for the purposes of comparison data for the new silicon vertex detectors introduced (200 GeV) and for measurements of flavor-identified parton spin distribution functions using the parity violating single spin asymmetry in $W^{\pm}$ production (510 GeV)~\cite{Bunce-ARNPS50}. The performance in Run-12 was even more outstanding than usual thanks to the new EBIS source as well as 3-dimensional stochastic cooling~\cite{3DSC-CC1012} of which the third dimension (horizontal) was active only for the U+U segment at the end of the run.

The 2013 run (Run-13) was devoted entirely to polarized proton-proton collisions at \sqsn=510 GeV for the measurement of flavor-identified spin structure functions using parity-violating production of $W^{\pm}$-bosons.  Machine improvements included an upgraded Optically Pumped Polarized H$^-$ source (OPPIS) with an order of magnitude increase in beam current to 10 mA and an increase in polarization from 85\% to 90\%. The components of the upgrade are an atomic hydrogen injector (in collaboration with BINP, Novosibirsk), a 3 Tesla superconducting solenoid and improved beam diagnostics and polarization measurement. A further improvement which, together with the improved OPPIS, is expected to yield a doubling of the luminosity (which is now limited by the head-on beam-beam effect) is the addition of electron lenses for partial compensation of head-on beam-beam tune shift. The idea is that in addition to the two regions of beam-beam collisions with positively charged beam, add another another collision point with a negatively charged beam with the same amplitude dependence to have the effect from p-e collisions partially cancel the p-p effect. The commissionning of the electron lenses with a new lattice proved to be challenging, so the latter part of the run was done with the lattice tested in 2012 and provided record luminosity and polarizations near 60\%. 
\subsection{BNL's Superconducting Magnet Division}
All of the upgrades to the RHIC machine mentioned above involved superconducting magnets built or developed in BNL's Superconducting Magnet Division. The Magnet Division also developed the RHIC machine superconducting magnets (Palmer magnet~\cite{CBAmagnetsNIMA235}) which are the basis for the other post-Tevatron machines such as HERA and the LHC. 
\begin{figure}[!b]
\begin{center}
\includegraphics[width=0.47\textwidth]{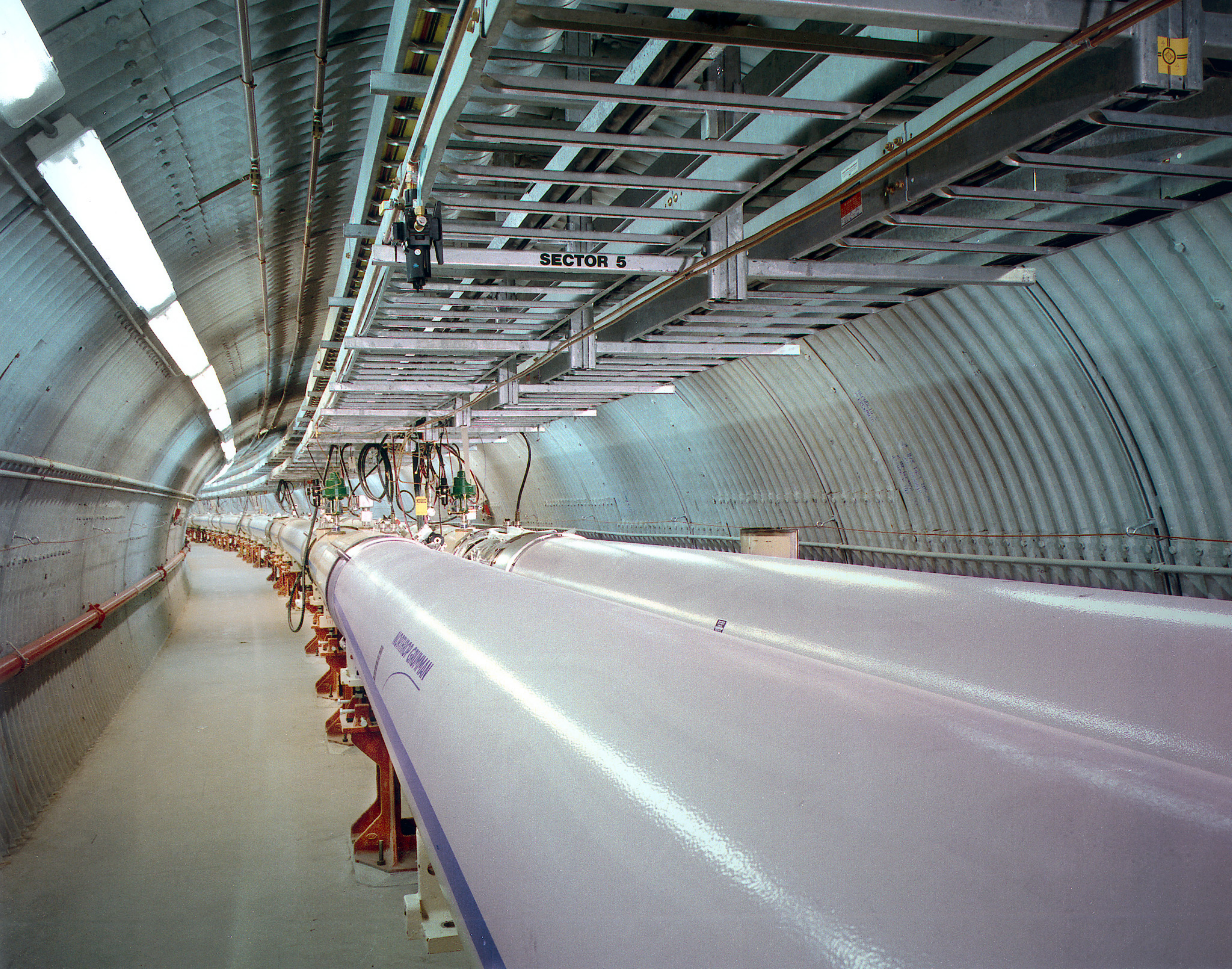}\hspace*{0.1pc}
\includegraphics[width=0.51\textwidth]{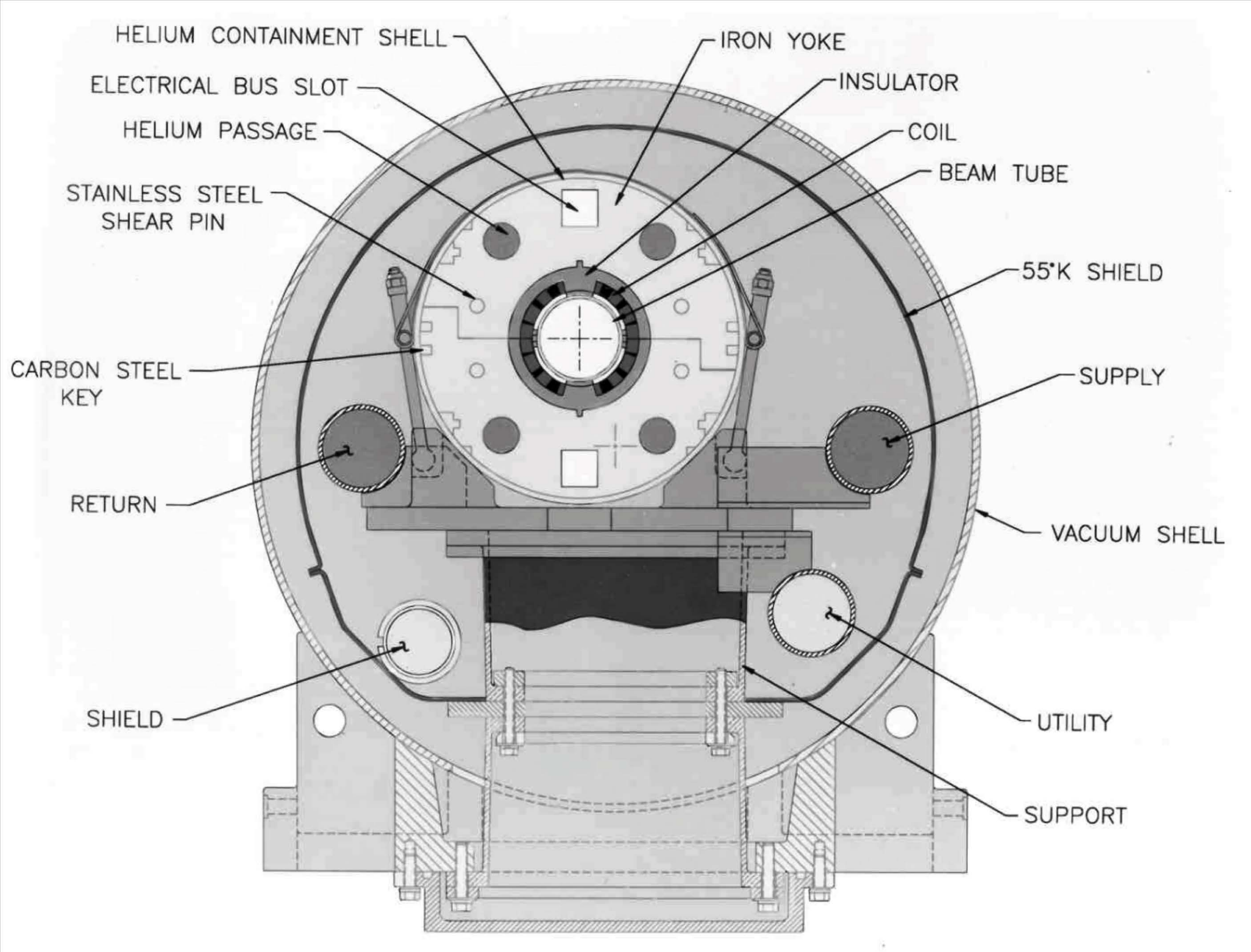}
\end{center}
\caption[]{a)(left) Photo of the RHIC machine composed of two independent rings, with a total of 1740 superconducting dipole, quadrupole and corrector magnets.  b) (right) Cross section of RHIC dipole.}
\label{fig:RHICring}\vspace*{-1.0pc}
\end{figure}
The RHIC dipole design (Fig.~\ref{fig:RHICring}b) is based on a relatively large bore (80 mm inner diameter), single-layer ``cosine theta" coil, wound from a (partially) keystoned, kapton-insulated, 30-strand Rutherford-type cable, arranged in coil blocks with intervening copper wedges, in order to meet the stringent field quality specifications, and mechanically supported by a laminated, cold steel yoke encased in a stainless steel shell. The shell contains the helium and is also a load bearing part of the assembly. 

\begin{figure}[!hbt]
\begin{center}
\includegraphics[width=0.55\textwidth,angle=1.0]{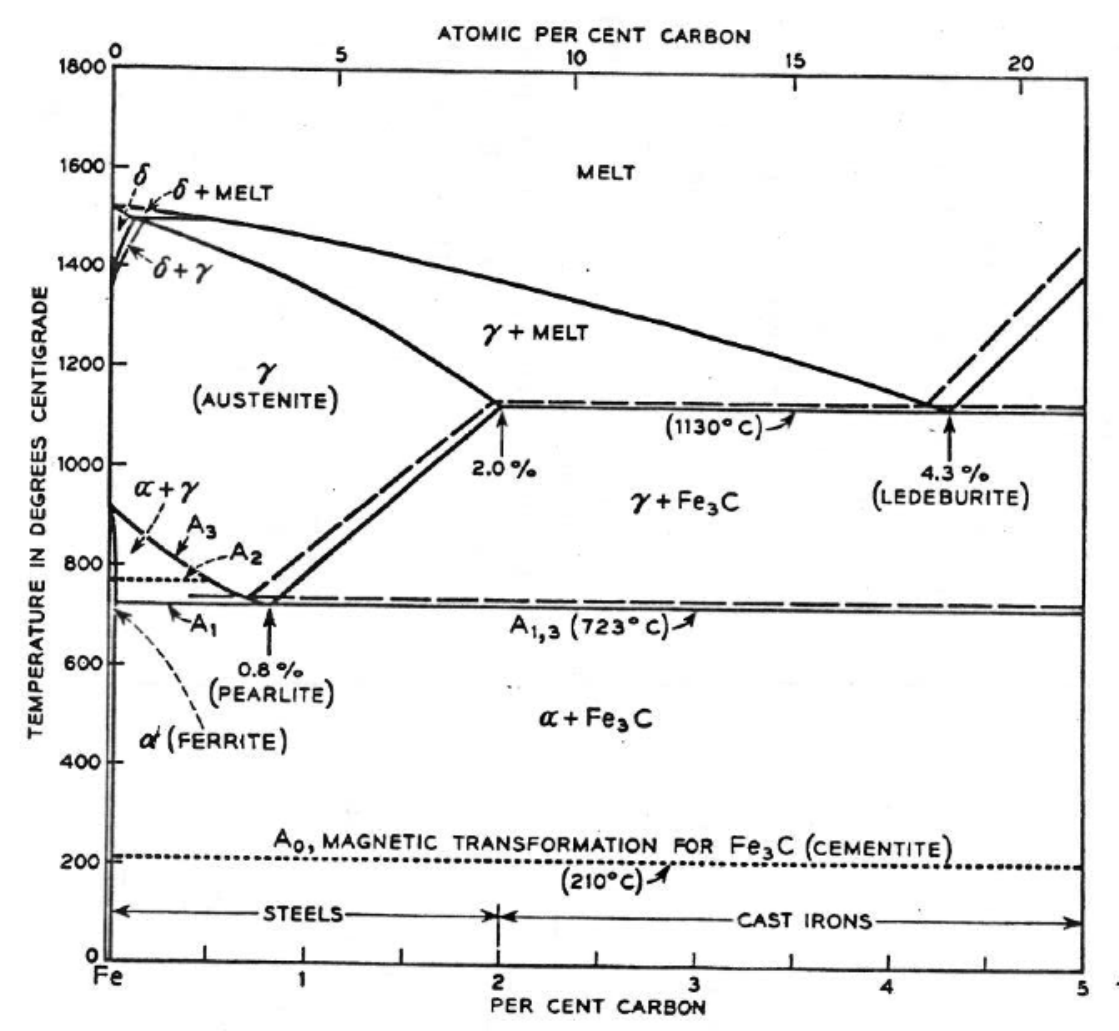}
\end{center}\vspace*{-1.0pc}
\caption[]{Phase diagram of iron-carbon alloys~\cite{Bozorth}. \label{fig:FeCphasediagram}}
\end{figure}

It is amusing to note that in order to build the RHIC magnets for the purpose of making the \QGP\ and studying its phase diagram, it is important to understand another phase-diagram, that of Fe+C, i.e. magnet steel (Fig.~\ref{fig:FeCphasediagram}).
The Fe+C phase diagram is quite complicated with many phases, but is well known; while the proposed phase diagram of nuclear matter (Fig.~\ref{fig:phase_boundary}) seems much simpler, probably because it is largely unknown. 
%\subsection{Latest results from the Magnet Division}

\begin{figure}[!t]
\begin{center}
\includegraphics[width=0.30\textwidth]{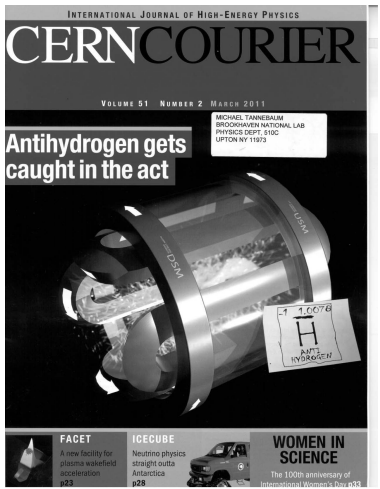}\hspace*{1pc}
\includegraphics[width=0.67\textwidth]{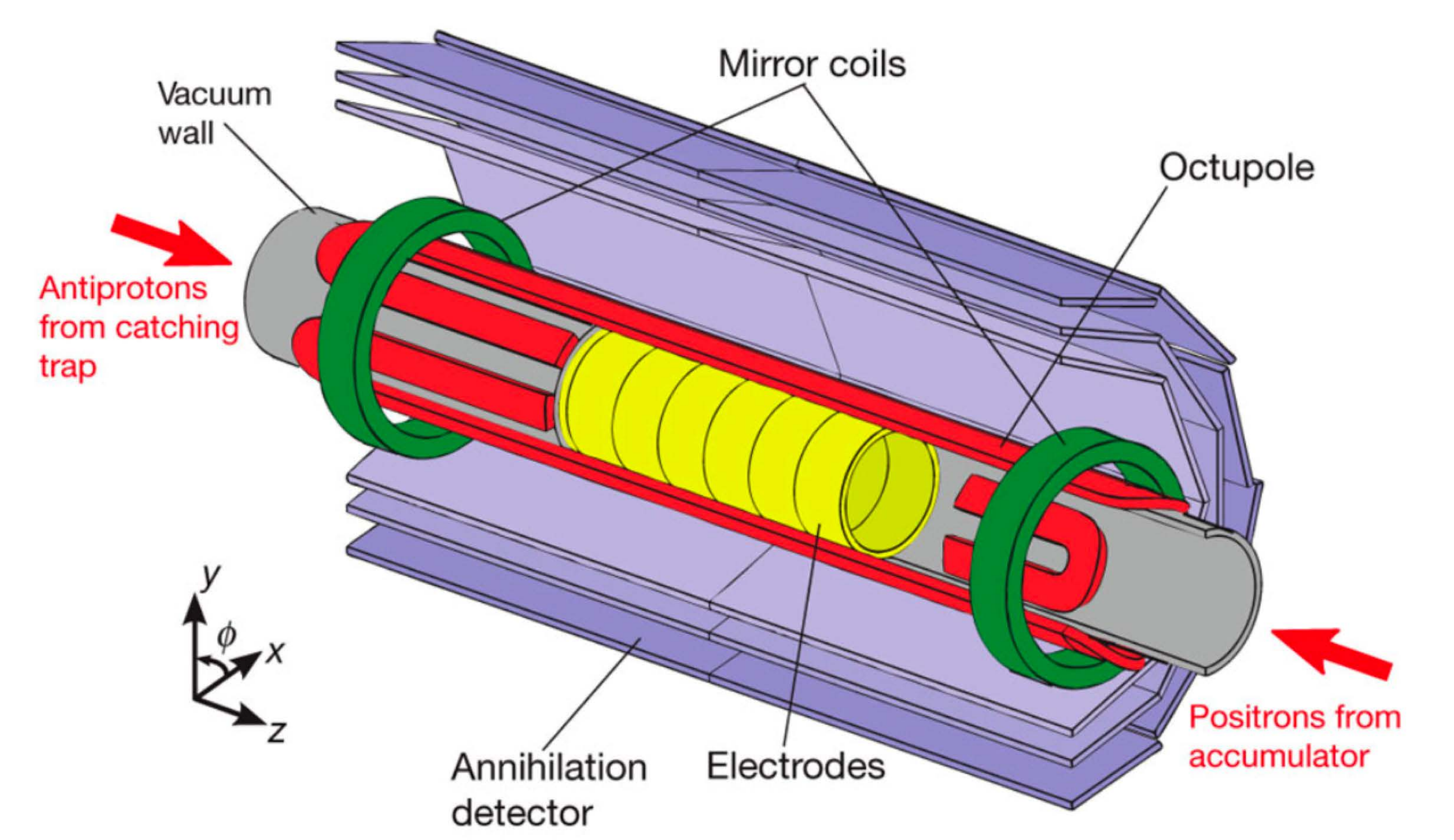}
\end{center}
\caption[]{a)(left) CERN Courier cover March 2011. b) (right) Antihydrogen synthesis and trapping region of the ALPHA apparatus~\cite{NaturePhysics7}. The atom-trap magnets, the modular annihilation detector and some of the Penning trap electrodes are shown (not to scale).  }
\label{fig:ALPHAexpt}
\end{figure}

BNL's superconducting magnet division is an international resource and actively participates in many projects, several in the news recently. The CERN Courier of March 2011 featured on its cover (Fig.~\ref{fig:ALPHAexpt}a) the beautiful magnet used to trap anti-hydrogen for 1000 seconds in the ALPHA experiment~\cite{NaturePhysics7} in the Antiproton Decelerator at CERN.
The octupole magnet (Fig.~\ref{fig:ALPHAexpt}b) was built by the BNL Magnet Division and provides a very pure octupole field which is zero on the axis and rises sharply near the vacuum wall to keep the anti-hydrogen confined radially. 

Another recent press release featuring BNL magnets was the ``Indication of Electron Neutrino Appearance'' at the T2K (Tokai to Kamioka) Experiment in Japan on June 15, 2011~\cite{T2KPR}. BNL provided 5 superconducting dipole corrector magnets in the proton beam at the Japan Proton Accelerator Research Complex (J-PARC) in Tokai which produced the $\mu$-neutrinos that were detected in the Super-Kamiokande detector after transforming to $e$-neutrinos over the 295 km flight path. 

Other work of the Magnet Division includes: NbTi magnets for the RHIC Electron Lens upgrade; NbTi final focus quadrupole for the ILC; Nb$_3$Sn 11.5 Tesla strand-test-barrel magnet and two coils for the LHC luminosity upgrade; High Temperature Superconductor (HTS) Quadrupole for the Facility for Rare Isotope Beams at Michigan State University; Spare NbTi dipole for LHC; HTS solenoid R\&D for muon collider and Energy Storage; Nb$_3$Sn Open Midplane Dipole for the Muon Collider.

\subsection{National Synchrotron Light Source-II}
In addition to accelerators for high energy particle and nuclear physics, Brookhaven has been innovative in synchrotron radiation light sources. The National Synchrotron Light Source (NSLS) which started operations in 1982 was the first to use the Chasman-Green double-bend achromat lattice~\cite{IEEENS22}, which is now the standard lattice at the major synchrotron light sources worldwide. A new third generation light source, NSLS-II, 4.66 times larger in circumference than NSLS, is now under construction at BNL, with unique design features of high brightness, small source size and long beam lines which will replace NSLS in 2014. 
%%%I cut out the figure to save space.
NSLS-II is designed to deliver photons with high average spectral brightness in the 2 keV to 10 keV energy range exceeding $10^{21}$ ph/s/0.1\%BW/mm$^2$/mrad$^2$. The spectral flux density should exceed $10^{15}$ ph/s/0.1\%BW in all spectral ranges. This cutting-edge performance requires the storage ring to support a very high-current electron beam (I = 500 mA) with sub-nm-rad horizontal emittance (down to 0.5 nm-rad) and diffraction-limited vertical emittance at a wavelength of 1 \AA (vertical emittance of 8 pm-rad)~\cite{NSLSIIsource}. 

\subsection{BNL g-2 Magnet moves to FERMILAB}
In June 2013, the 15 m diameter precision storage ring from the BNL muon \mbox{$g-2$} experiment~\cite{BNL-g-2} began a circuitous (5000 km) very delicate cross-contry trip to Fermilab (only 1440 km in a straight line) involving custom built trucks and a specially prepared barge which brought the magnet down the East Coast, around the tip of Florida and up the Mississippi River to Illinois. This new muon $g-2$ experiment at Fermilab would be the fifth such experiment, which was pioneered at CERN. 

By some incredible coincidence, a few weeks before the ISSP2013 school, the June 2013 CERN Courier reprinted an article from 1970 with the title ``Preparing for a third `$g-2$'.'' 
This brought back good memories to me because I had worked on the second $g-2$ experiment when I was a post-doc at CERN in 1965-66~\cite{g-2-2-1966}; but, of course, Prof. Zichichi worked on the ground-breaking original $g-2$ experiment at CERN in 1959-1961~\cite{CERN-first-g-2-final}. 
My thesis research, done at the BNL-AGS from 1961-64, was muon-proton elastic scattering~\cite{MuP1968} to find out ``Why does the muon weigh heavy?'' Even in 2013, with Prof. Higgs in the audience, we still don't know! Other experiments at that time did better: the ``two-neutrino experiment''~\cite{Danby1962} (Nobel Prize) was in the beam to the left of ``my'' muon beam; while on the right, over the AGS machine, in ``inner Mongolia'', CP violation~\cite{Fitch1963} was discovered (Nobel Prize). Those were the days; but even more excitement lay ahead. 

\section{ICHEP1972: Hard-Scattering, Quarks, and QCD} At the International Conference on High Energy Physics (ICHEP) in 1972, there were three momentous developments that inform our work today: 
\begin{itemize}
\item The discovery in p-p collisions at the CERN ISR of production of particles with large transverse momentum ($p_T$) which proved that the partons of Deeply Inelastic Scattering (DIS) interacted with each other much more strongly than electromagnetically. 
\item Measurements of DIS in neutrino scattering presented by Don Perkins who proclaimed that ``In terms of constituent models, the fractionally charged (Gell-Mann/Zweig) quark model is the only one which fits both the electron and neutrino data.''
\item The origin of \QCD\ in the presentation by Harald Fritzsch and Murray Gell-Mann with the title ``Current Algebra: Quarks and What Else?''
\end{itemize}

\begin{figure}[!h]
   \begin{center}
\includegraphics[width=0.48\textwidth]{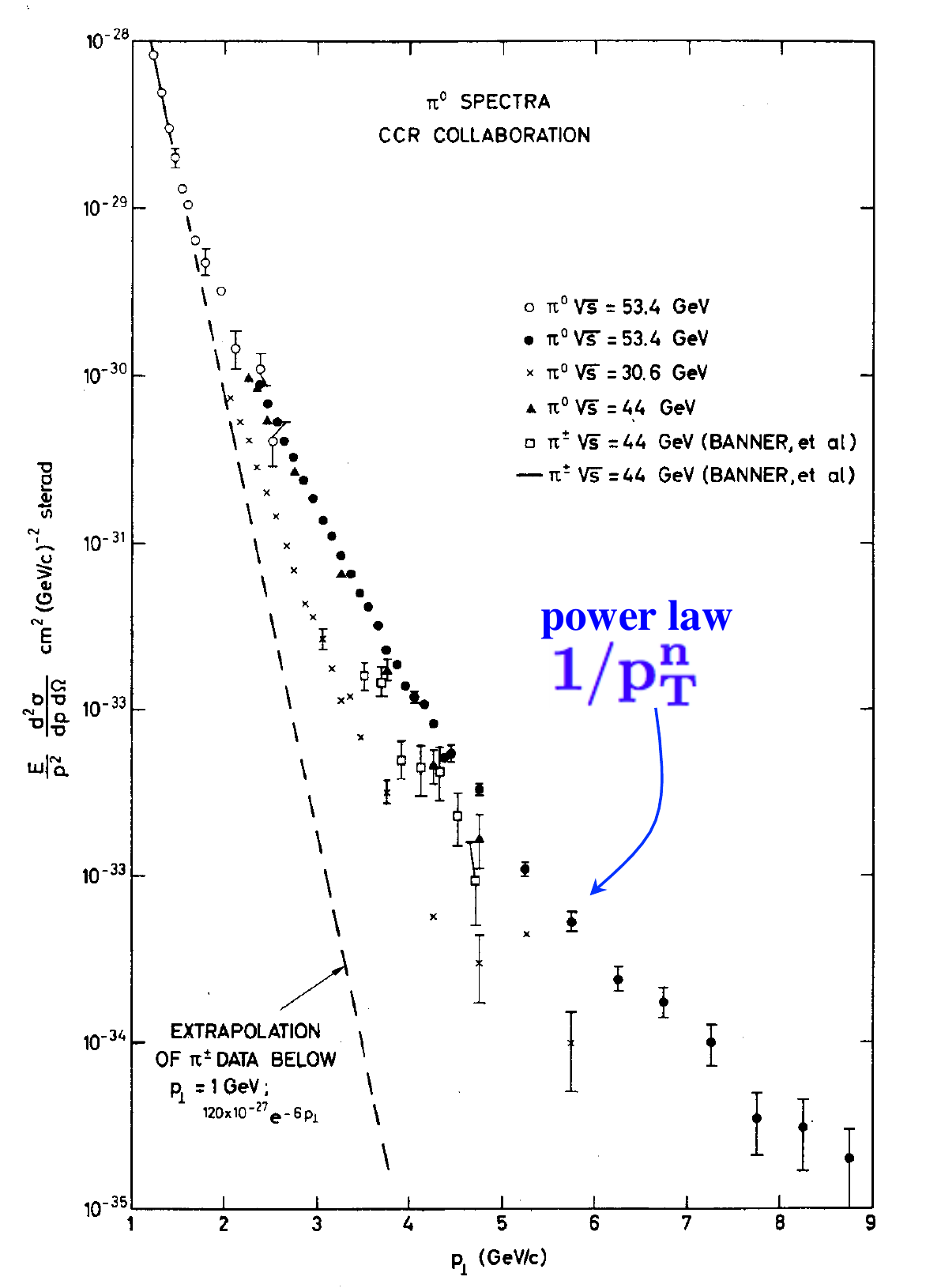}
\raisebox{0.2pc}{\includegraphics[width=0.51\textwidth]{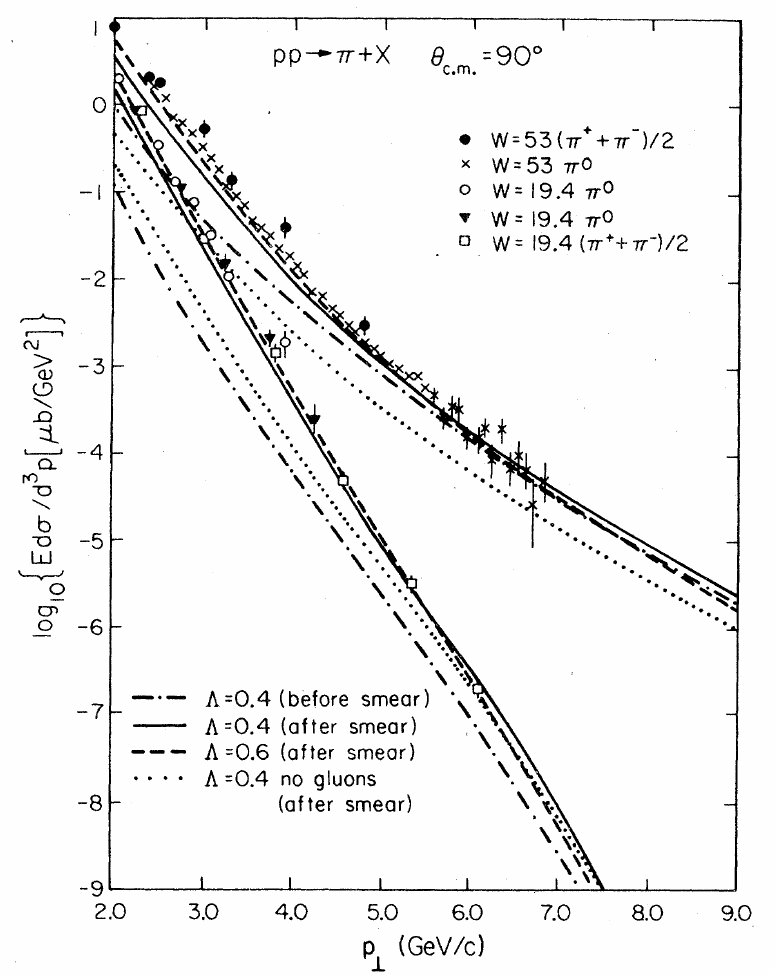}}
\end{center}\vspace*{-1.0pc}
\caption[]{a)(left) Plot of invariant single $\pi^0$ cross section vs. $p_T$ for several $\sqrt{s}$ from CCR at the CERN-ISR~\cite{CoolICHEP72}. b)(right) Feynman, Field, and Fox~\cite{FFF} \QCD\ calculation of mid-rapidity high-$p_T$ $\pi$ spectra at $\sqrt{s}=19.4$ and 53 GeV, with and without $k_T$ smearing, for two values of $\Lambda_{\rm QCD}$.} 
\label{fig:QCDworks}%\vspace*{-0.8pc}
\end{figure}

Figure~\ref{fig:QCDworks}a shows the first observation of scattering at large $p_T$~\cite{CoolICHEP72}. The $\pi^0$ spectrum breaks away from the $e^{-6p_T}$ dependence known since cosmic ray measurements, with a power-law spectrum that flattens as the c.m energy, $\sqrt{s}$, is increased. Excellent cooperation of experimentalists and theorists showed in 1978 that these data could be explained by QCD~\cite{OwensKimel,FFF} if the quarks in a nucleon had ``intrinsic'' transverse momentum, $k_T\approx 1$ GeV/c (Fig.~\ref{fig:QCDworks}b). 
Although these \QCD\ calculations in agreement with the high $p_T$ single particle spectra were published in 1978, most experimentalists in the U.~S., notably at the first Snowmass conference in July 1982, were skeptical because of evidence against jets presented at the ICHEP1980 by a CERN experiment, NA5~\cite{NA5PLB112}. 

\begin{figure}[!t]
   \begin{center}
\includegraphics[width=0.48\textwidth,height=0.5\textwidth]{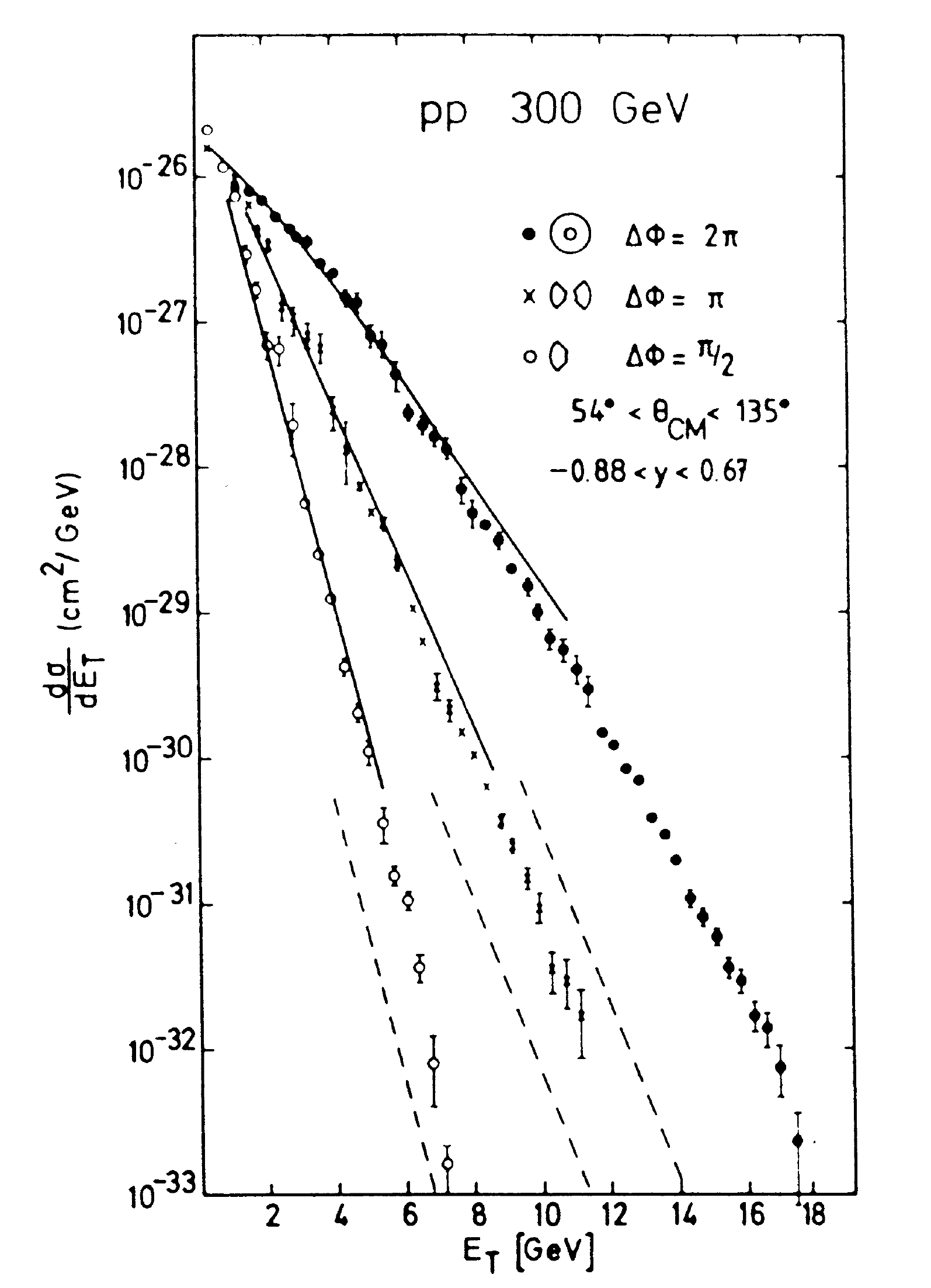}\hspace*{-1pc}
\raisebox{0.2pc}{\includegraphics[width=0.53\textwidth]{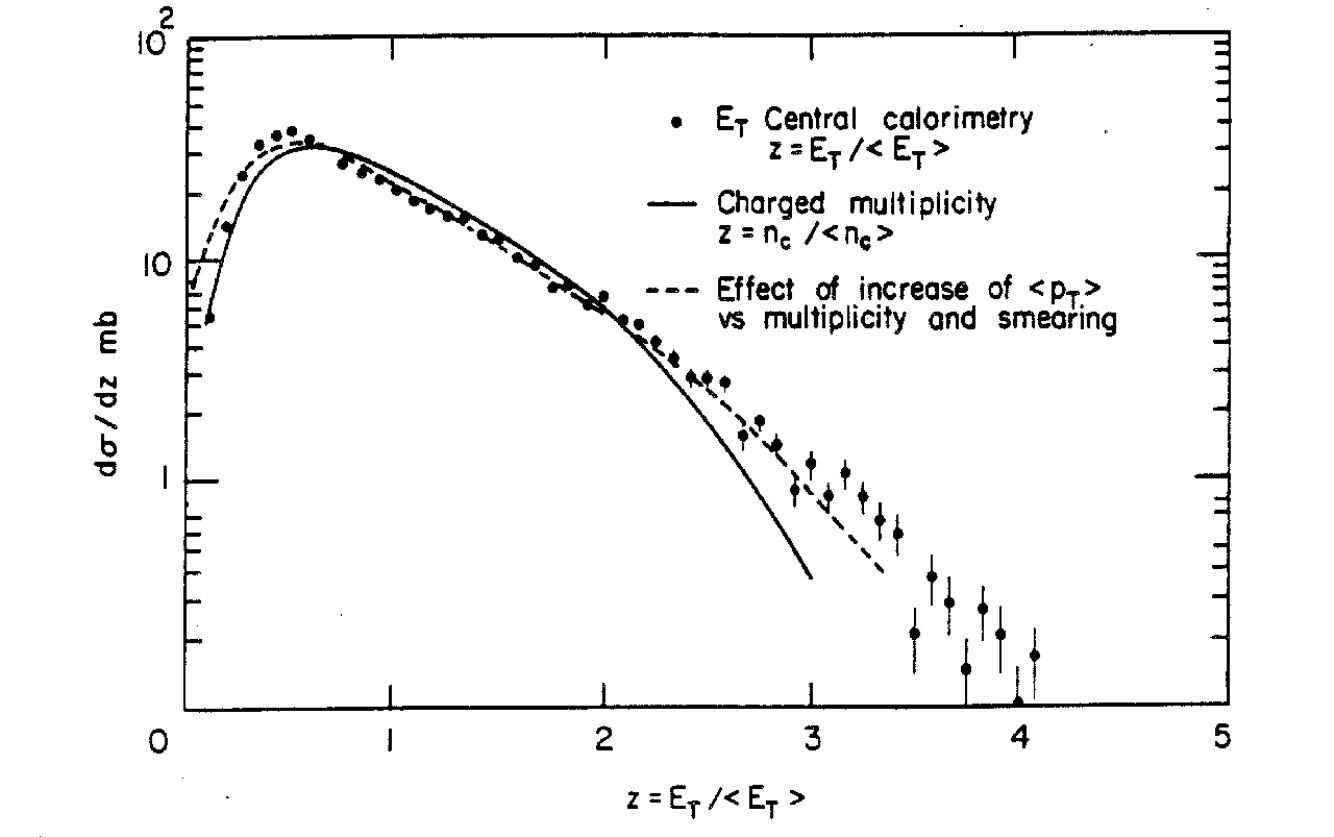}}
\end{center}\vspace*{-1.0pc}
\caption[]{a)(left) $E_T$ distributions~\cite{NA5PLB112} in the solid angles indicated. Predictions from soft (low $p_T$) multiparticle production and \QCD\ hard-scattering are shown by solid and dashed curves respectively. b) (right) $E_T$ distribution~\cite{UA1Paris82} for $|\eta|\leq3$ from $\bar{\rm p}$-p collisions at $\sqrt{s}=540$ GeV in the KNO type variable, $z=E_T/\mean{E_T}$, used for multiplicity~\cite{UA1PLB107}. } 
\label{fig:firstET}\vspace*{-0.1pc}
\end{figure}

Bjorken had proposed in 1973~\cite{BjPRD8} that jets from the fragmentation of high $p_T$ scattered partons should be observed using ``$4\pi$'' hadron calorimeters.  The first large aperture measurement was by NA5~\cite{NA5PLB112} at the CERN-SpS (Fig.~\ref{fig:firstET}a) who showed a transverse energy ($E_T$) spectrum at ICHEP1980, where the sum:\vspace*{-0.2pc}
      \begin{equation}
          E_T=\sum_i E_i\ \sin\theta_i
          \label{eq:ETdef}\vspace*{-0.3pc}
          \end{equation}
is taken over all particles emitted into a fixed solid angle for each event. In Fig.~\ref{fig:firstET}a~\cite{NA5PLB112}, 
the solid angle varies from full azimuthal acceptance, $\Delta\phi=2\pi$, in the c.m. rapidity range $-0.88<y<0.76$, to smaller azimuthal regions as shown on the figure. The striking results, which contradicted a previous claim from Fermilab~\cite{E260NPB134}, were: i) no jets were seen in the full azimuth data; ii) all the data were far above the \QCD\ predictions; iii) the large $E_T$ observed was the result of ``a large number of particles with a rather small transverse momentum''. 

As we shall see below, $E_T$ distributions are very important in Relativistic Heavy Ion (RHI) physics since they can be used to characterize and study the nuclear geometry of an A+B reaction on an event-by-event basis. The strong relation between $E_T$ and multiplicity distributions and the absence of jets in these distributions was emphasized in a talk by UA1 at ICHEP 1982 (Fig.~\ref{fig:firstET}b)~\cite{UA1Paris82}. Ironically, this talk immediately followed a talk by UA2~\cite{UA2JetICHEP82} which provided the first evidence for a di-jet from hard-scattering at a level 5-6 orders of magnitude down in the $E_T$ distribution from $\bar{\rm p}$-p collisions at $\sqrt{s}=540$ GeV.

There are many additional important results in high-energy physics from this period that are relevant to both \QCD\ and RHI physics, which must be skipped in this brief introduction. I have covered some of these results in previous ISSP lectures and proceedings~\cite{EriceProcPR,MJTIJMPA2011}; but I wrote a book with Jan Rak~\cite{RATCUP}, which was published in mid-2013, that covers this information in detail and which is the real introduction to what follows.

\section{Introduction to \QGP\ Physics}

Given that I already said that the \QGP\ was discovered at RHIC, what further studies are important? 
   The \QGP\  is the only place in the universe where we can in principle and in practice study Quantum Chromo-Dynamics (\QCD) for color-charged quarks in a color-charged medium. 
\begin{figure}[!h]
\begin{center}
\includegraphics[width=0.55\linewidth]{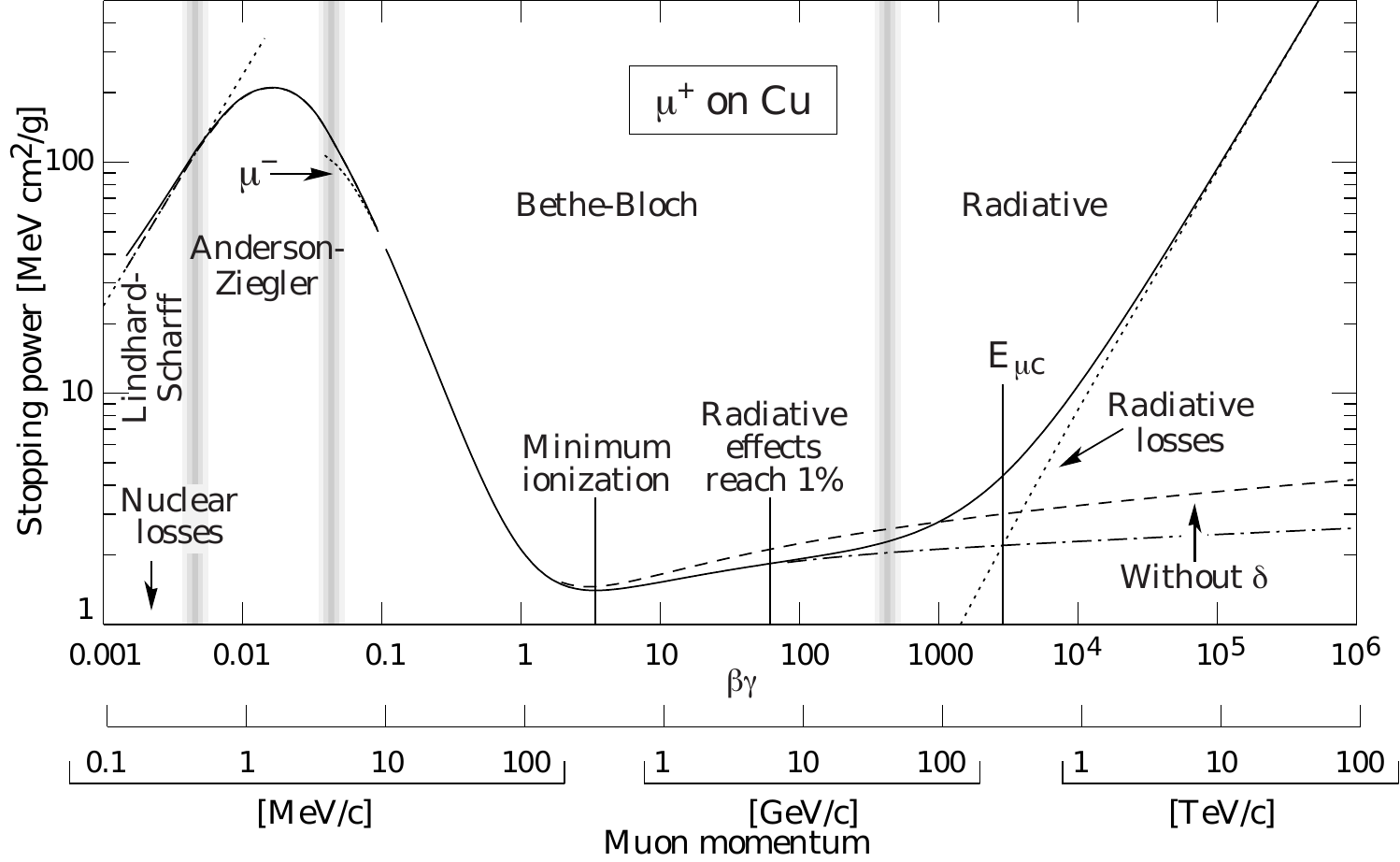}
\end{center}\vspace*{-1pc}
\caption[]{$dE/dx$ of a $\mu^+$ in Copper as a function of muon momentum~\cite{PDG}.}
\label{fig:muCu}\vspace*{-1pc}
\end{figure}
For instance, how long will it take before we understand the passage of a quark through a \QGP\ medium as well as we understand the passage of a muon through Copper in QED (Fig.~\ref{fig:muCu})~\cite{PDG}. 

Of course, in addition to understanding the behavior of \QCD\ in a medium, the central goal of our field is a quantitative study of the phases of nuclear matter. This requires a broad, quantitative study of the fundamental properties of the \QGP\  including the extraction of the transport coefficients of the medium such as critical temperature, $T_c$, speed of sound, $c_s$, the ratio of shear viscosity to entropy density, $\eta/s$, etc. To help understand how we shall proceed to address these issues, it is important to understand how we got to this point.  

The year 2013 was the 13th year of RHIC operation, and the two major detectors PHENIX and STAR which study the \QGP\ at RHIC are basically first round detectors with a few incremental upgrades. Thus, the design of these detectors was heavily influenced by the c. 1990 view of the signatures of the \QGP, which as noted above were quite different from what was discovered.  

\subsection{$J/\Psi$ suppression---the original ``gold-plated" \QGP\ signature}

   Since 1986, the `gold-plated' signature of deconfinement was thought to be $J/\Psi$ suppression. Matsui and Satz~\cite{MatsuiSatz86} proposed that $J/\Psi$ production in A+A collisions would be suppressed by Debye screening of the quark
color charge in the \QGP. The $J/\Psi$ is produced when two gluons
interact to produce a $c, \bar c$ pair which then resonates to form the
$J/\Psi$. In the plasma the $c, \bar c$ interaction is screened so that the 
$c, \bar c$ go their separate ways and eventually pick up other quarks at
the periphery to become {\it open charm}. 

\begin{figure}[!tb]
\begin{center}
\includegraphics[width=0.44\linewidth]{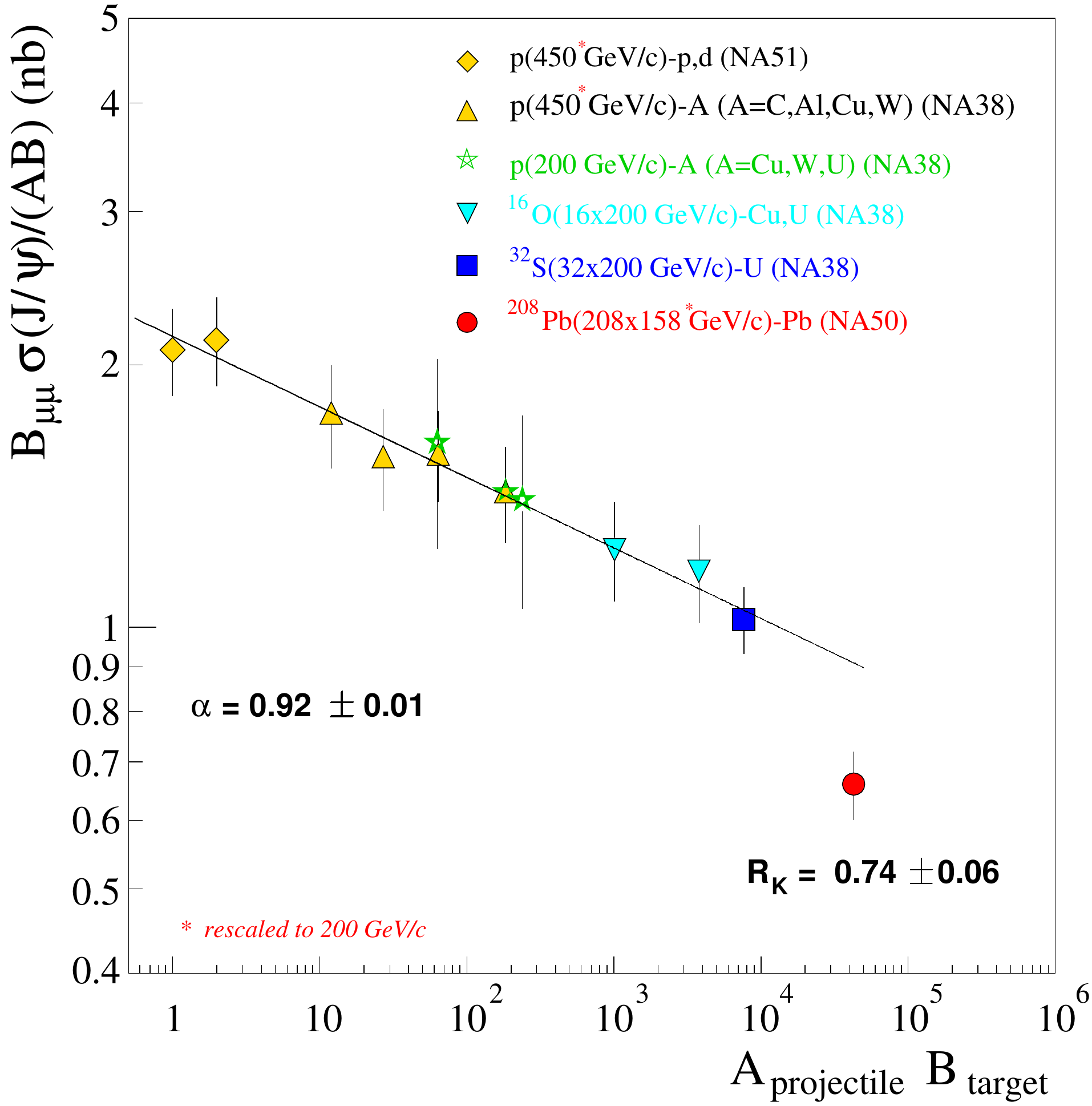}\hspace*{1pc}
\includegraphics[width=0.53\linewidth]{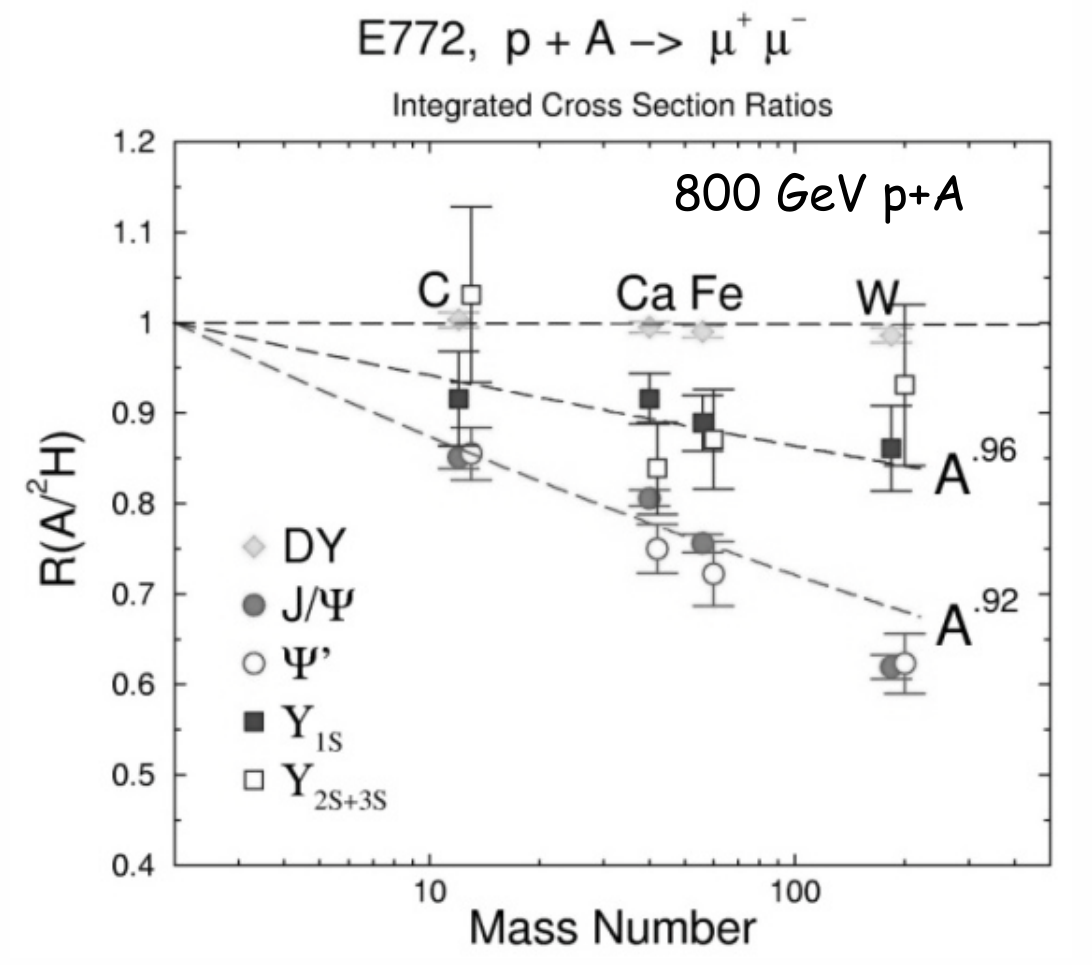}
\end{center}\vspace*{-1.0pc}
\caption[]{\small a) (left) Total cross section for $J/\Psi$ production divided by $AB$ in A+B collisions at 158--200$A$ GeV~\cite{NA50EPJC39} (note the {\color{Red}*}). b) (right) $A$ dependence of charmonium and Drell-Yan pair production in 800 GeV p+A collisions~\cite{E772} expressed as the ratio of heavy nucleus to deuterium cross sections per nucleon. The dashed lines are fits to $A^\alpha$ for the CNM effect, with the values of $\alpha$ indicated. \label{fig:JPsiAB}}
\end{figure}

``Anomalous suppression'' of $J/\Psi$ was found in $\sqrt{s_{NN}}=17.2$ GeV Pb+Pb collisions at the CERN-SpS~\cite{NA50EPJC39} (Fig.~\ref{fig:JPsiAB}a). This is the CERN fixed target heavy ion program's main claim to fame: but the situation has always been complicated because the $J/\Psi$ is suppressed in p+A collisions. For example, in $\sqrt{s_{NN}}=38.8$ GeV p+A collisions~\cite{E772} (Fig.~\ref{fig:JPsiAB}b) the Drell-Yan $\bar{q}q\rightarrow\mu^+ \mu^-$ cross-section per nucleon is constant as a function of mass number, A, which indicates the expected absence of shadowing in a nucleus for point-like production processes; while the $J/\Psi$ and $\Upsilon$ cross sections per nucleon are suppressed by an amount $A^\alpha$ with $\alpha=0.920\pm0.008$ for both $J/\Psi$ and $\Psi^{'}$ and $\alpha=0.96\pm0.01$ for both the $\Upsilon_{1s}$ and $\Upsilon_{2s+3s}$.  This is called a Cold Nuclear Matter or CNM effect and is shown as the line with $\alpha=0.92$ on Fig.~\ref{fig:JPsiAB}a. The ``Anomalous suppression'' is the difference between the data point at $AB=208\times 208$ and the line, provided that the CNM effect is the same at $\sqrt{s_{NN}}=17.2$ and 38.8 GeV.  

The later development of $J/\Psi$ measurements, after it was shown that $J/\Psi$ suppression was the same at RHIC at $\sqrt{s_{NN}}=200$ GeV as at the CERN-SpS (e.g. see Ref.~\cite {MJTIJMPA2011}),  has not been concerned with $J/\Psi$ suppression as a signature of deconfinement, but rather with the strong c.m. energy dependence of the CNM effect and the possibility of regeneration of $J/\Psi$ from recombination of the large number of $c$ and $\bar c$ quarks produced in the \QGP\ .  
Nevertheless, the search for $J/\Psi$ suppression (as well as thermal photon/dilepton radiation from the \QGP ) drove the design of the RHIC experiments~\cite{RHICNIM} and the ALICE experiment at the LHC~\cite{LHCJINST}. Only recently have results from the ALICE experiment, to be discussed below, reopened the issue of whether recombination implies deconfinement. 

	\subsection{Detector issues in A+A compared to p-p collisions} 
\begin{figure}[!h]
\begin{center}
\begin{tabular}{cc}
\psfig{file=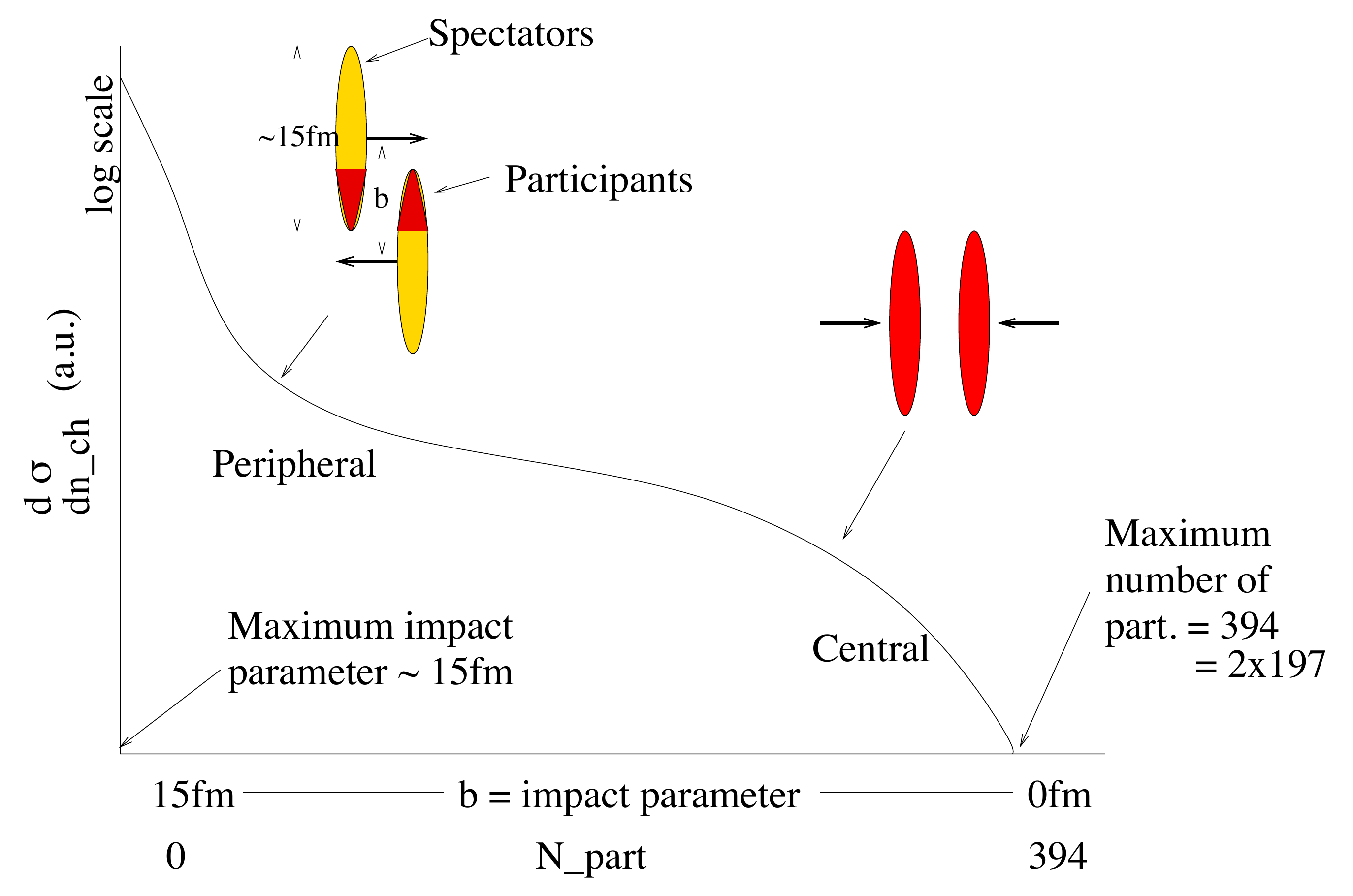,width=0.50\linewidth}\hspace*{-1pc}
\psfig{file=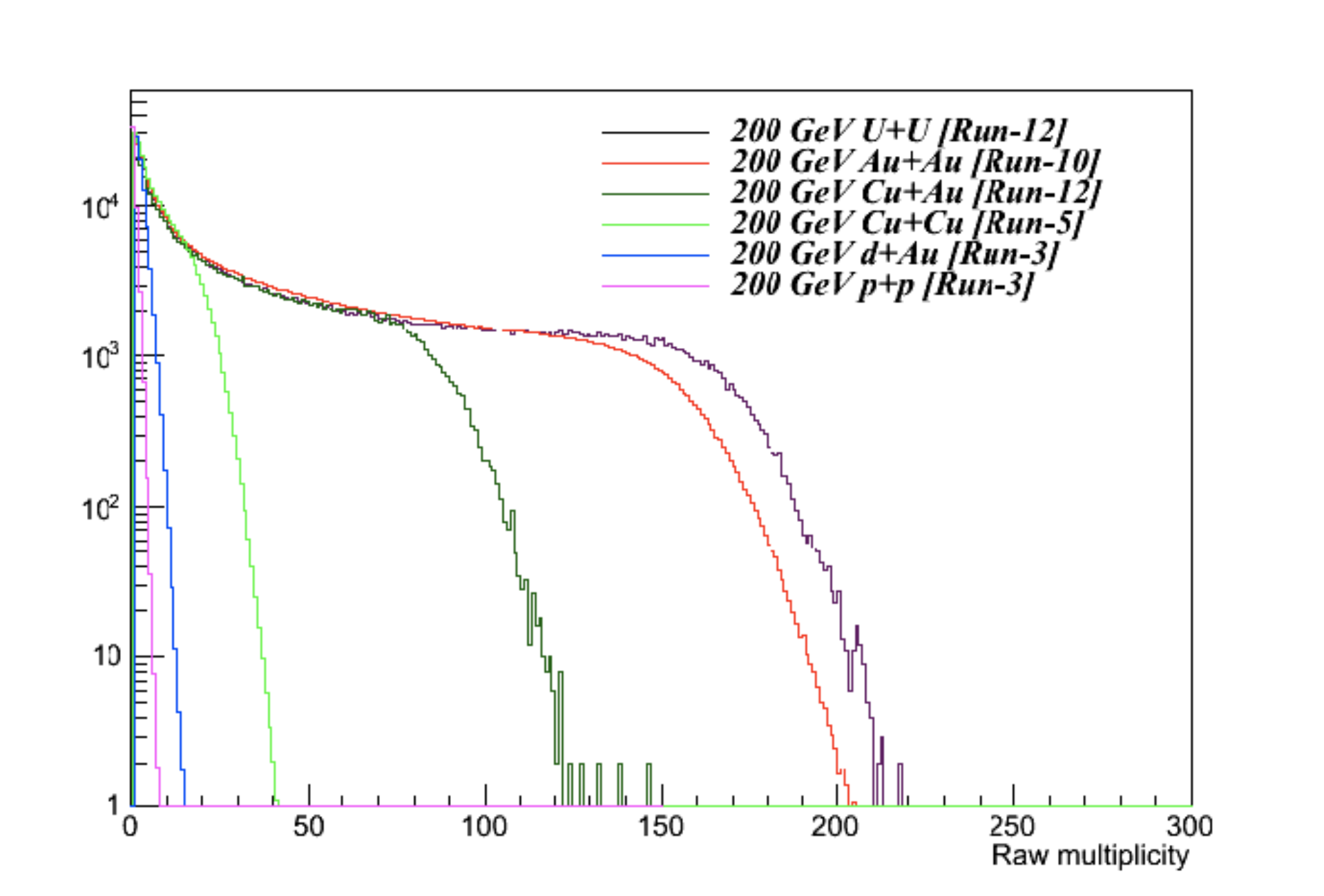,width=0.52\linewidth,angle=0}\end{tabular}
\end{center}\vspace*{-1.0pc}
%-90-->0 height-->width
\caption[]{a) (left) Schematic of collision in the $N$-$N$ c.m. system of two Lorentz contracted nuclei with radius $R$ and impact parameter $b$. The curve with the ordinate labeled $d\sigma/d n_{\rm ch}$ represents the relative probability of charged particle  multiplicity $n_{\rm ch}$ which is directly proportional to the number of participating nucleons, $N_{\rm part}$. b)(right) raw $n_{\rm ch}$ distributions in p-p to U-U collisions at $\sqrt{s_{NN}}=200$ GeV from PHENIX~\cite{JTMQM12}.  
\label{fig:nuclcoll}}%\vspace*{-2pc}
\end{figure}

 	A main concern of experimental design in RHI collisions is the huge multiplicity in A+A central collisions compared to  p-p collisions. 
A schematic drawing of a collision of two relativistic Au nuclei is shown in Fig.~\ref{fig:nuclcoll}a. 
In the center of mass system of the nucleus-nucleus collision, the two Lorentz-contracted nuclei of radius $R$ approach each other with impact parameter $b$. In the region of overlap, the ``participating" nucleons interact with each other, while in the non-overlap region, the ``spectator" nucleons simply continue on their original trajectories and can be measured in Zero Degree Calorimeters (ZDC), so that the number of participants can be determined. The degree of overlap is called the centrality of the collision, with $b\sim 0$, being the most central and $b\sim 2R$, the most peripheral. The maximum time of overlap is $\tau_\circ=2R/\gamma\,c$ where $\gamma$ is the Lorentz factor and $c$ is the speed of light in vacuum.  

\begin{figure}[!t]
\begin{center}
\begin{tabular}{cc}
%\hspace*{-4cm}
\psfig{file=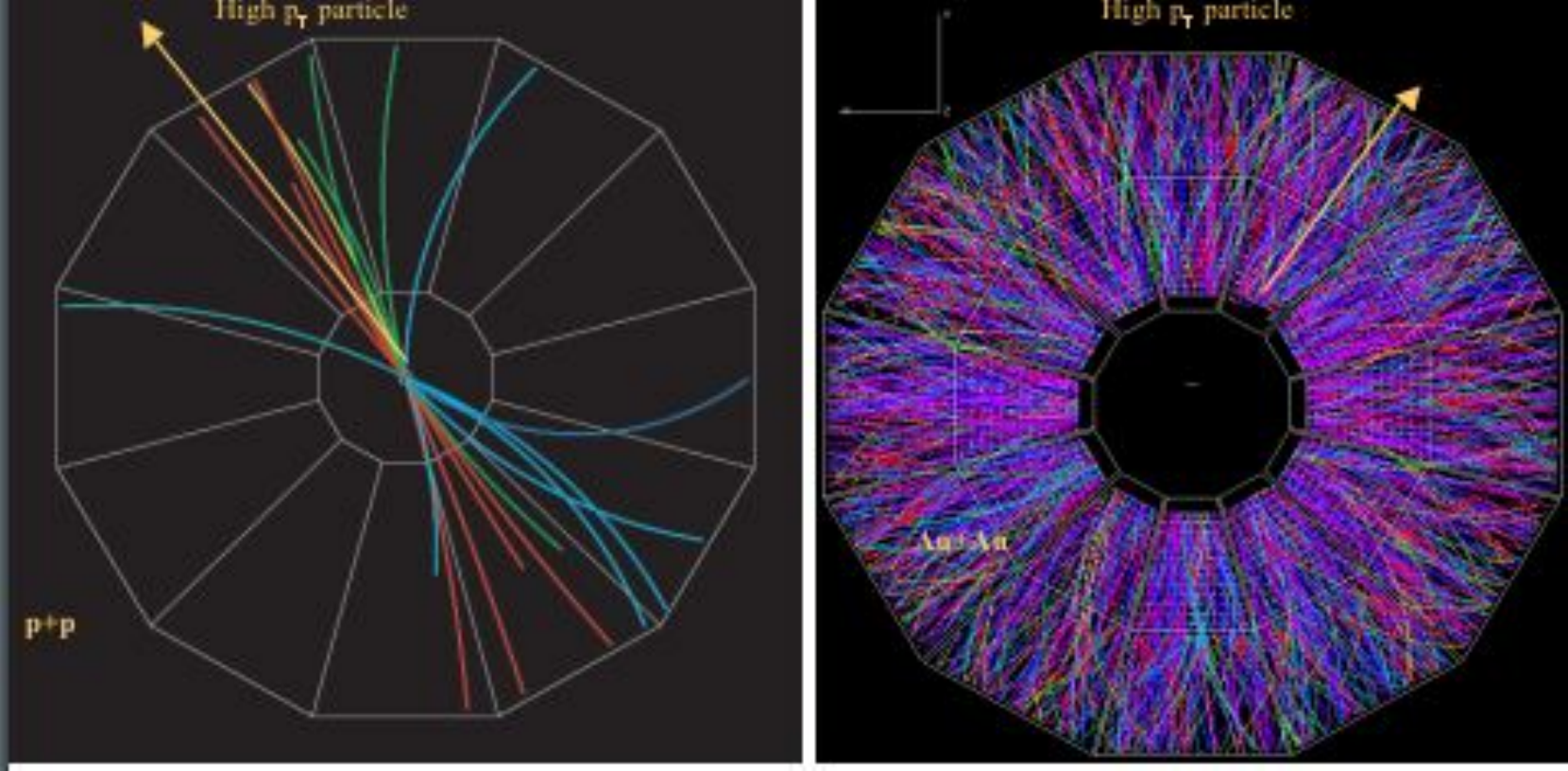,width=0.64\linewidth}&\hspace*{-0.025\linewidth}
\psfig{file=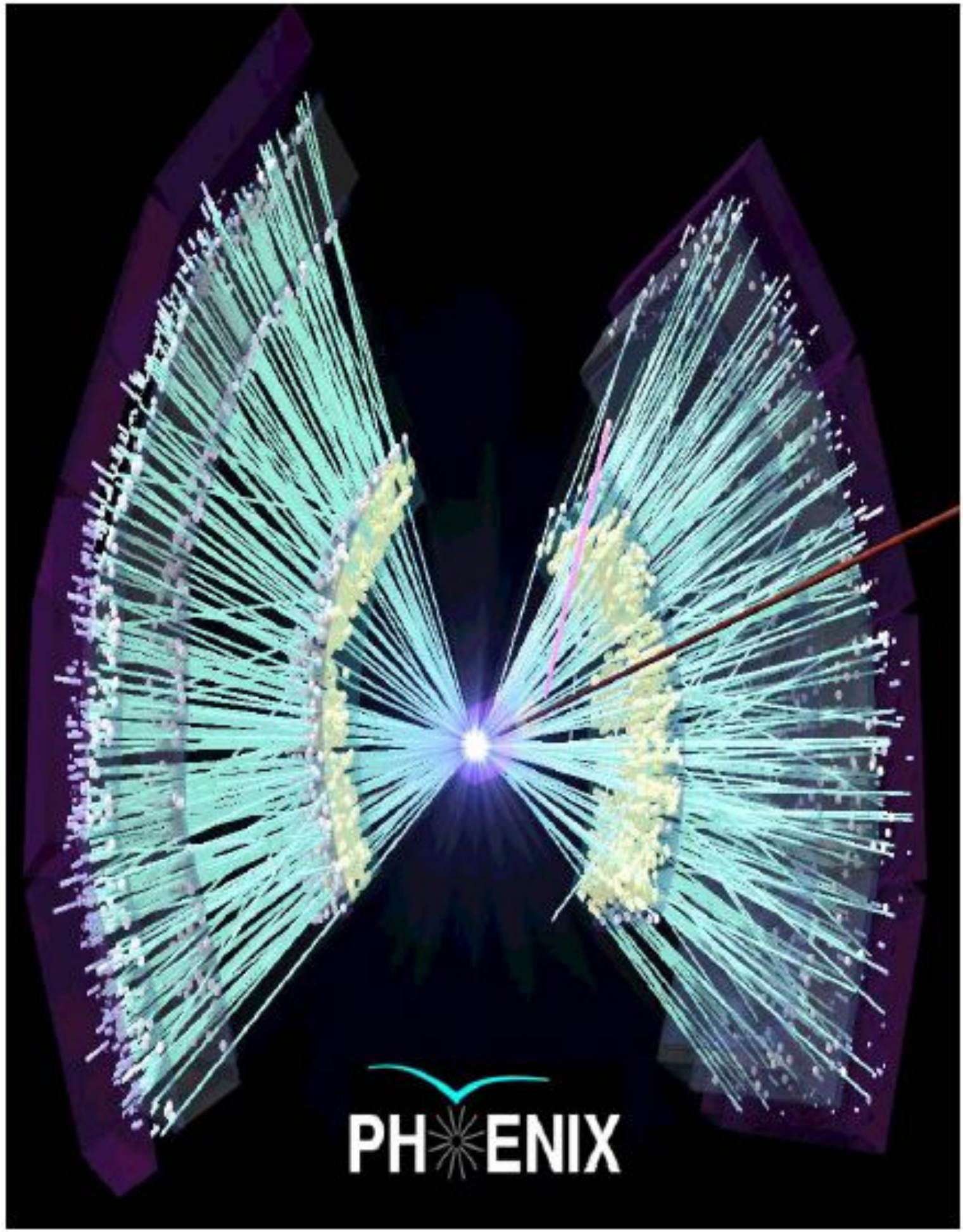,width=0.315\linewidth,height=0.315\linewidth}
\end{tabular}
\end{center}%\vspace*{-2pc}%\vspace*{-0.35in}
\caption[]{ a) (left) A p-p collision in the STAR detector viewed along the collision axis; b) (center) Au+Au central collision at $\sqrt{s_{NN}}=200$ GeV in STAR;  c) (right) Au+Au central collision at $\sqrt{s_{NN}}=200$ GeV in PHENIX.  
\label{fig:collstar}}\vspace*{-0.5pc}
\end{figure}

The energy of the inelastic collision is predominantly dissipated by multiple particle production, where $N_{\rm ch}$, the number of charged particles produced, is directly proportional~\cite{PXWP} to the number of participating nucleons ($N_{\rm part}$) as sketched on Fig.~\ref{fig:nuclcoll}a. Thus, $N_{\rm ch}$ or the total transverse energy $E_T$ in central Au+Au collisions is roughly $A$ times larger than in a p-p collision, as shown in actual events from the STAR and PHENIX detectors at RHIC (Fig.~\ref{fig:collstar}). 
Figure~\ref{fig:nuclcoll}b shows the measured distributions~\cite{JTMQM12} of the charged particle multiplicity, $n_{\rm ch}$, at mid-rapidity at $\sqrt{s_{NN}}=200$ GeV for all the combinations of A+B collisions measured at RHIC. The increase of $n_{\rm ch}$ with both $A$ and $B$ is evident. The impact parameter $b$ can not be measured directly, so the centrality of a collision is defined in terms of the upper percentile of $n_{\rm ch}$ or $E_T$ distributions, e.g. top 10\%-ile, upper $10-20$\%-ile. Unfortunately the ``upper'' and ``-ile'' are usually not mentioned which sometimes confuses the uninitiated. 

\begin{figure}[!t]
\begin{center}
\includegraphics[width=0.55\textwidth]{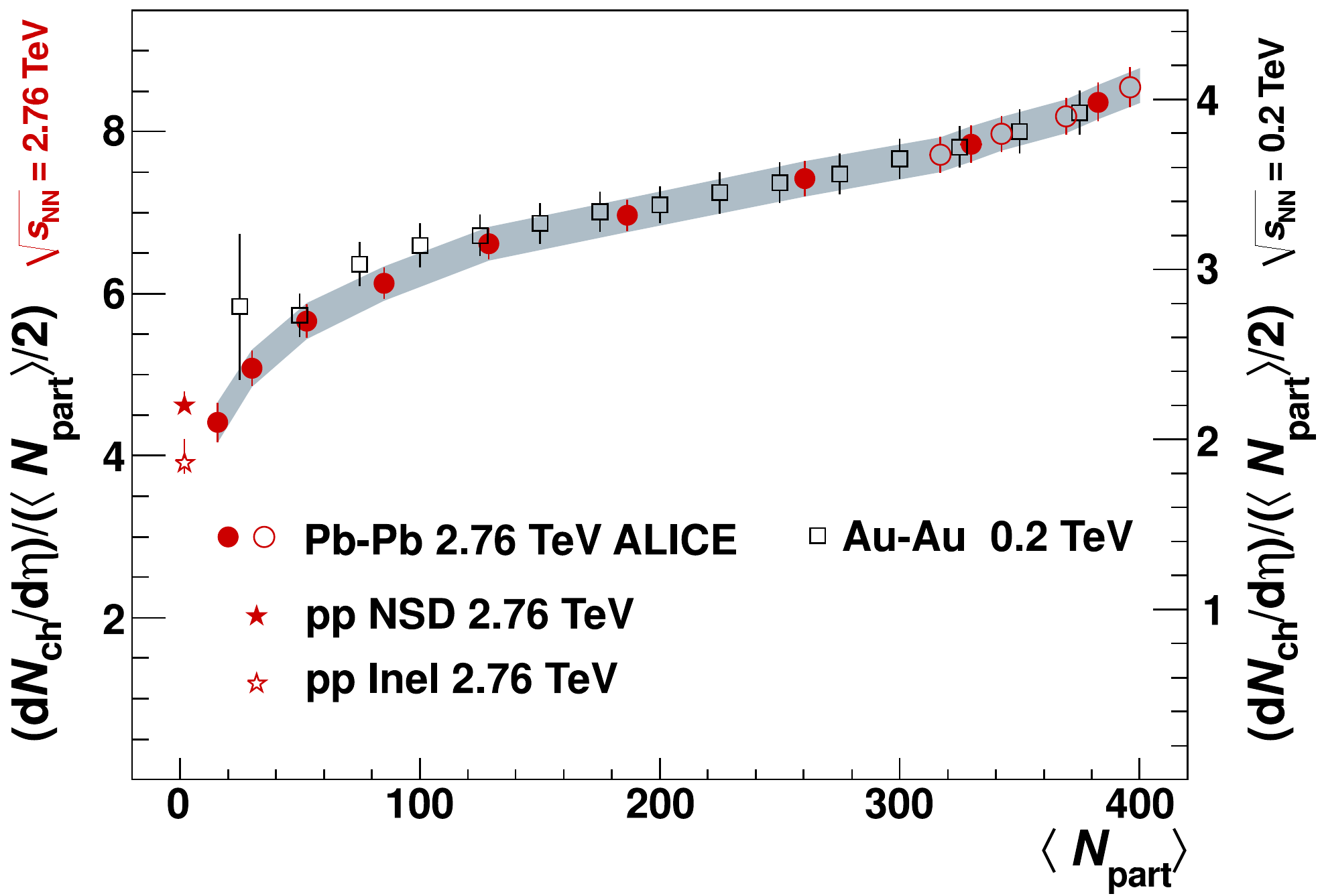}
\end{center}\vspace*{-1.0pc}
\caption[]{Dependence of $(dN_{\rm ch}/d\eta)/(N_{\rm part}/2)$ on the average number of participants $\mean{N_{\rm part}}$   in bins of centrality, for Pb+Pb collisions at $\sqrt{s_{NN}}=2.76$ TeV~\cite{ALICEmult} and Au+Au collisions at $\sqrt{s_{NN}}=0.200$ TeV~\cite{ppg019}. The scale for the lower-energy data (right side) differs by a factor of 2.1 from the scale for the higher-energy data (left side).          \label{fig:dNdeta}}\vspace*{-1.0pc}
\end{figure}

In Fig.~\ref{fig:dNdeta}, measurements of the charged particle multiplicity density $dN_{\rm ch}/d\eta$ at mid-rapidity, $|\eta|<0.5$, relative to the number of participating nucleons, $N_{\rm part}$, are shown as a function of centrality for $\sqrt{s_{NN}}=200$ GeV Au+Au collisions at RHIC~\cite{ppg019} together with new results from \mbox{ALICE} in $\sqrt{s_{NN}}=2.76$ TeV Pb+Pb collisions at LHC~\cite{ALICEmult}. The results are expressed as $(dN_{\rm ch}/d\eta)/(N_{\rm part}/2)$ for easy comparison to p-p collisions. 

The LHC data show the effect well known from RHIC that $dN_{\rm ch}/d\eta$ does not depend linearly on $N_{\rm part}$, since $(dN_{\rm ch}/d\eta)/(N_{\rm part}/2)$ is not a constant for all $N_{\rm part}$. However the data also show the amazing effect that the ratio of $(dN_{\rm ch}/d\eta)/(N_{\rm part}/2)$ from LHC to RHIC is simply a factor of 2.1 in every  centrality bin.  Thus the LHC and RHIC data lie one on top of each other by simple scaling of the RHIC measurements by a factor of 2.1. This is an incredibly beautiful result which shows that in going from p-p to A+A collisions, the charged particle production is totally dominated by the nuclear geometry of the A+A collisions represented by the number of participating nucleons, $N_{\rm part}$, independently of the nucleon-nucleon c.m. energy, $\sqrt{s_{NN}}$.  

	Since it is a huge task to reconstruct the momenta and identity of all the particles produced in these events, the initial detectors at RHIC~\cite{RHICNIM} concentrated on the measurement of single-particle or multi-particle inclusive variables to analyze RHI collisions, with inspiration from the CERN ISR which emphasized those techniques before the era of jet reconstruction~\cite{egseeMJTISSP2009}. There are two major detectors in operation at RHIC, STAR and PHENIX, and there were also two smaller detectors, BRAHMS and PHOBOS, which have completed their program. As may be surmised from Fig.~\ref{fig:collstar}, STAR, which emphasizes hadron physics, is most like a conventional general purpose collider detector, a TPC to detect all charged particles over the full azimuth ($\Delta\phi=2\pi$) and  $\pm 1$ units of pseudo-rapidity ($\eta$); while PHENIX is a very high granularity high resolution special purpose detector covering a smaller solid angle at mid-rapidity, together with a muon-detector at forward rapidity~\cite{egseePT}. 
	
\begin{figure}[!h]\vspace*{-1.0pc}
\begin{center}
\includegraphics[width=0.40\textwidth]{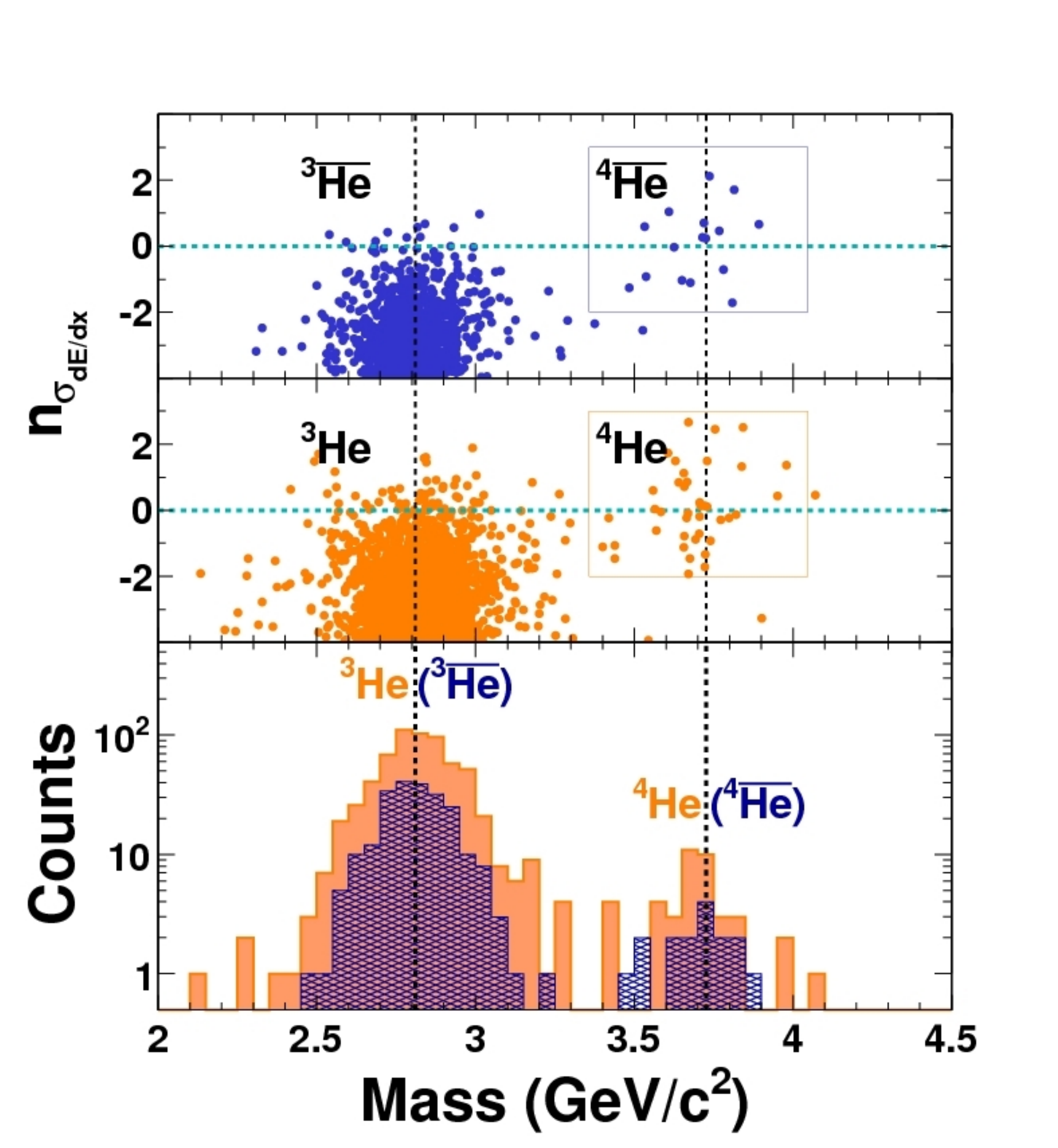}\hspace*{1pc}
\includegraphics[width=0.50\textwidth]{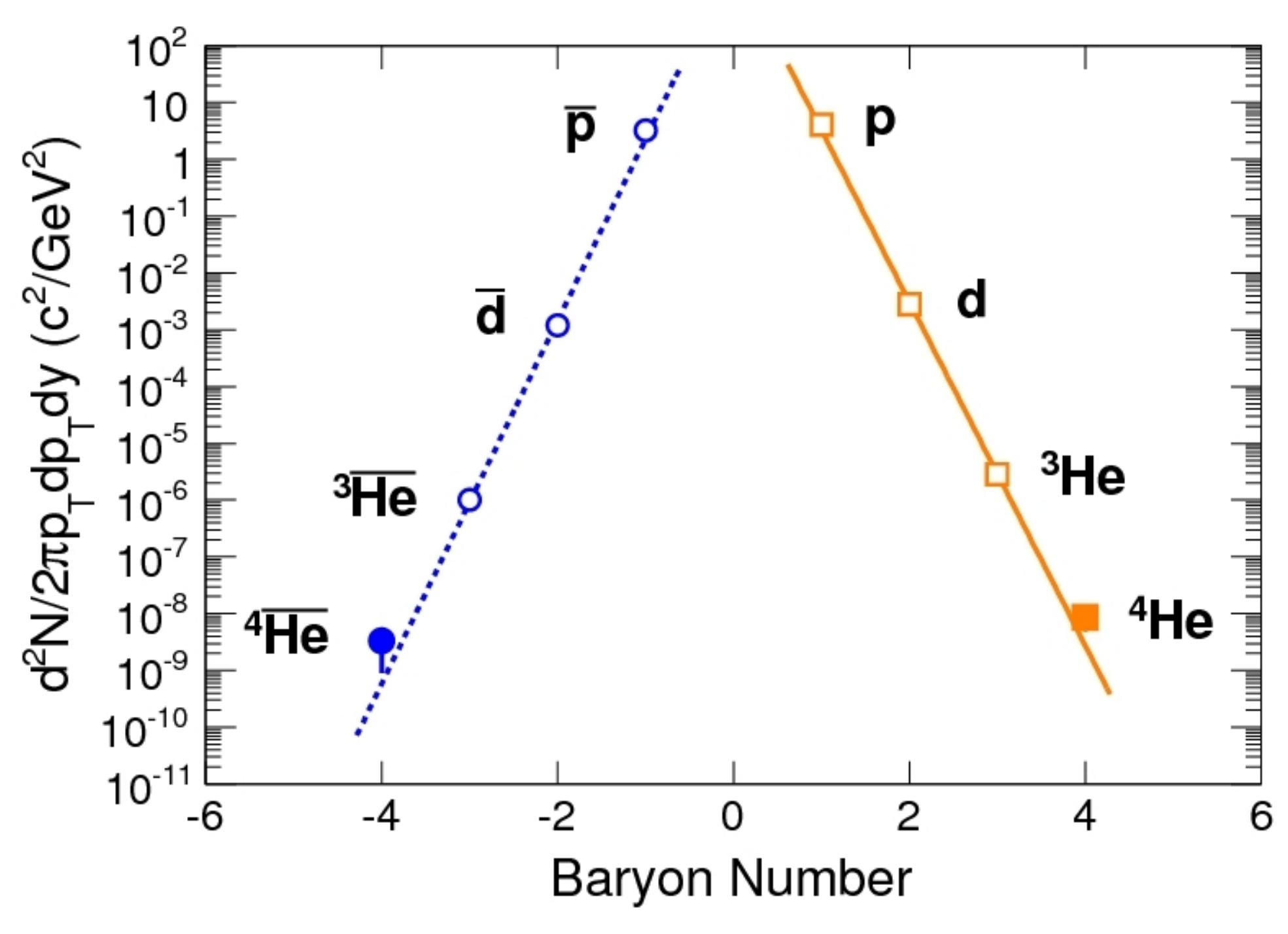}
\end{center}\vspace*{-1.0pc}
\caption[]{a)(left) Number of standard deviations, $n_{\sigma_{dE/dx}}$, of $dE/dx$ resolution from the expected value for $^4$He, for negative and positive particles as a function of calculated mass, and projected counts for $-2<n_{\sigma_{dE/dx}}<3$. b) Differential invariant yields as a function of baryon number, B.}
\label{fig:STARanti}\vspace*{-0.1pc}
\end{figure}

	One nice feature of the STAR detector is the ability to measure the mass/charge of a particle from its momentum/charge  and time of flight, and then use $dE/dx$ measured in the TPC to determine the charge. In this way STAR has observed many anti-nuclei, notably in 2011 the ``Observation of the antimatter helium-4 nucleus'' in Au+Au collisions~\cite{STARNature473} (Fig.~\ref{fig:STARanti}a). 
The differential invariant yields $d^2N/(2\pi p_T dp_T dy)$ per central Au+Au collision at $\sqrt{s_{NN}}=200$ GeV as a function of baryon number B, evaluated at $p_T/|B|=0.875$ GeV/c, are shown in Fig.~\ref{fig:STARanti}b and show a steady exponential decrease with increasing $B$~\cite{0909.0566}. The anti-nuclei are made by coalescence of the large number of $\bar{n}$ and $\bar{p}$ produced, an advantage of the high multiplicity. 	
	
	PHENIX is designed to measure and trigger on rare processes involving leptons, photons and identified hadrons at the highest luminosities with the special features: i) a minimum of material (0.4\% $X_\circ$) in the aperture to avoid photon conversions; ii) possibility of zero magnetic field on axis to prevent de-correlation of $e^+ e^-$ pairs from photon conversions; iii) Electro-Magnetic Calorimeter (EMCal) and Ring Imaging Cherenkov Counter (RICH) for $e^{\pm}$ identification and level-1 $e^{\pm}$ trigger; iv) a finely segmented EMCal ($\delta\eta$, $\delta\phi=0.01 \times$ 0.01) to avoid overlapping showers due to the high multiplicity and for separation of single-$\gamma$ and $\pi^0$ up to $p_T\sim 25$ GeV/c; v) EMCal and precision Time of Flight measurement for particle identification. Some results uniquely possible with this detector such as measurements of direct photons via internal conversion to $e^+ e^-$ pairs will be discussed below.   	

	In addition to the large multiplicity, there are two other issues in RHI physics which are different from p-p physics: i) space-time issues, both in momentum space and coordinate space---for instance what is the spatial extent of fragmentation? is there a formation time/distance?; ii) huge azimuthal anisotropies of particle production in non-central collisions (colloquially collective flow) which are very interesting in their own right and provide much richer features than originally envisaged. 
	\section{Collective Flow} 
   A distinguishing feature of A+A collisions compared to either p-p or p+A collisions has been the collective flow observed. This effect is seen over the full range of energies studied in heavy ion collisions, from incident kinetic energy of $100A$ MeV to c.m. energy of $\sqrt{s_{NN}}=200$ GeV~\cite{LaceyQM05}. Collective flow, or simply flow, is a collective effect which can not be obtained from a superposition of independent N-N collisions.  
      \begin{figure}[!thb]
   \begin{center}
\includegraphics[width=0.45\linewidth]{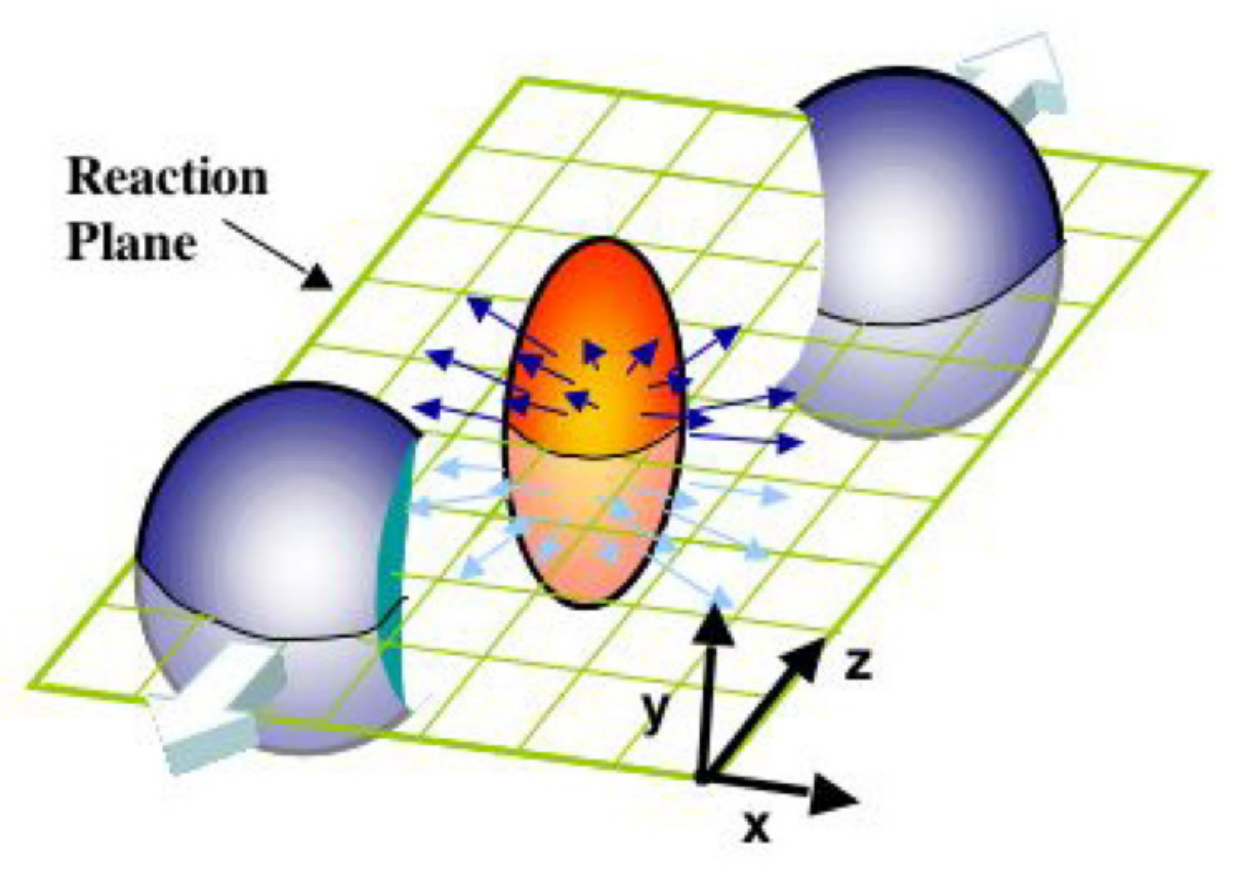}
\includegraphics[width=0.54\linewidth]{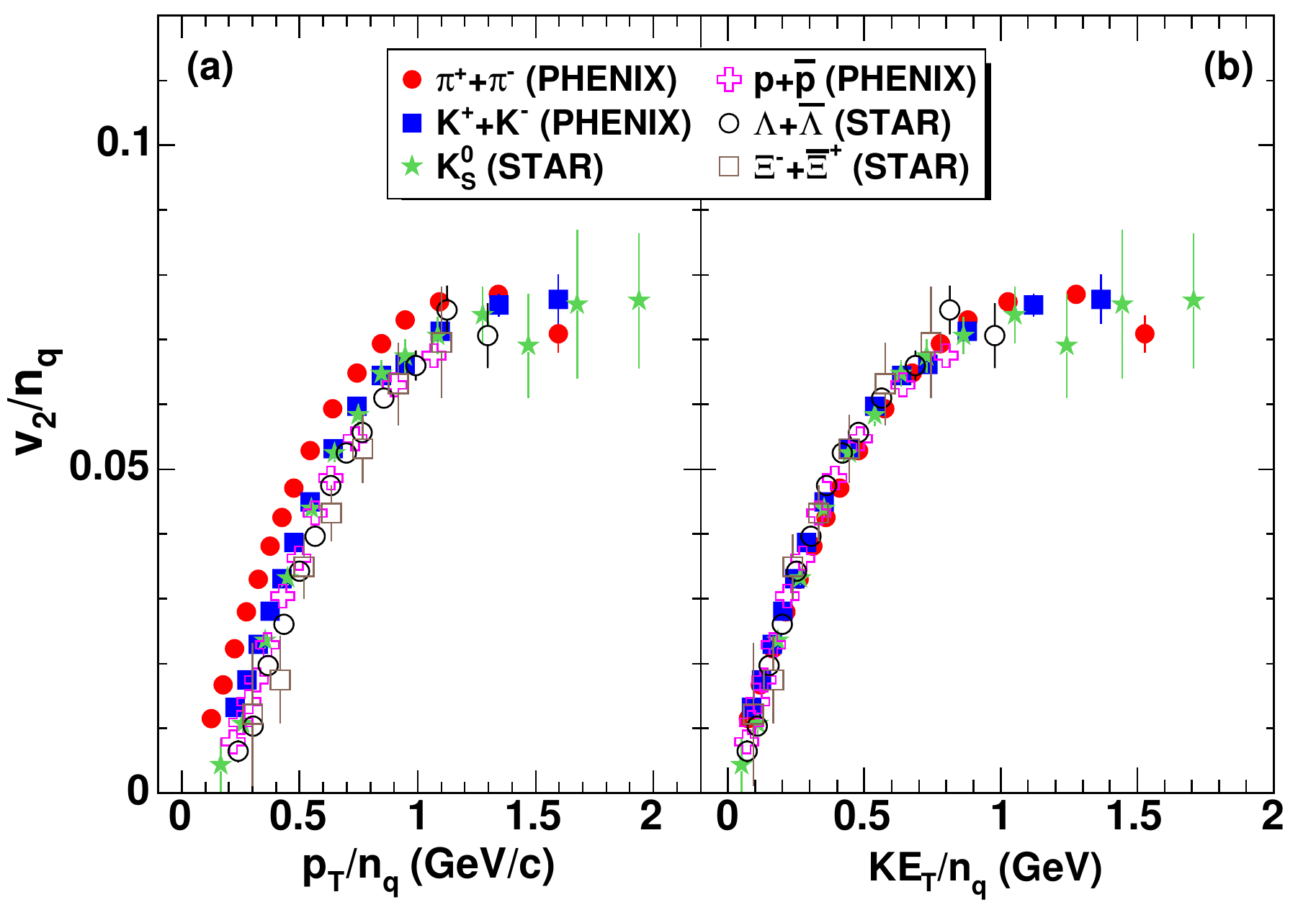}
\end{center}\vspace*{-1.0pc}
\caption[]{(left) Almond shaped overlap zone generated just after an A+A collision where the incident nuclei are moving along the $\pm z$ axis. The reaction plane by definition contains the impact parameter vector (along the $x$ axis)~\cite{KanetaQM04}. (right) Measurements of elliptical-flow ($v_2$)  for identified hadrons plotted as $v_2$ divided by the number of constituent quarks $n_q$ in the hadron as a function of (a) $p_T/n_q$, (b) $KE_T/n_q$~\cite{PXArkadyQM06}.   
\label{fig:MasashiFlow}}
\end{figure}
Immediately after an A+A collision, the overlap region defined by the nuclear geometry is almond shaped (see Fig~\ref{fig:MasashiFlow}) with the shortest axis along the impact parameter vector. Due to the reaction plane breaking the $\phi$ symmetry of the problem, the semi-inclusive single particle spectrum is modified by an expansion in harmonics~\cite{Ollitrault} of the azimuthal angle of the particle with respect to the reaction plane, $\phi-\Phi_R$~\cite{HeiselbergLevy}, where the angle of the reaction plane $\Phi_R$ is defined to be along the impact parameter vector, the $x$ axis in Fig.~\ref{fig:MasashiFlow}: 
  \begin{equation}
\frac{Ed^3 N}{dp^3}=\frac{d^3 N}{p_T dp_T dy d\phi}
=\frac{d^3 N}{2\pi\, p_T dp_T dy} \left[ 1+\sum_n 2 v_n \cos n(\phi-\Phi_R)\right] .
\label{eq:siginv2} 
\end{equation} 
The expansion parameter $v_2$, called elliptical flow, is predominant at mid-rapidity. In general, the fact that flow is observed in final state hadrons  shows that thermalization is rapid, so that hydrodynamics comes into play at a time, $\tau_0$, which is before the spatial anisotropy of the overlap almond dissipates. At this early stage hadrons have not formed and it has been proposed that the constituent quarks flow~\cite{VoloshinQM02}, so that the flow should be proportional to the number of constituent quarks $n_q$, in which case $v_2/n_q$ as a function of $p_T/n_q$ would represent the constituent quark flow as a function of constituent quark transverse momentum and would be universal. However, in relativistic hydrodynamics, at mid-rapidity, the transverse kinetic energy, $m_T-m_0=(\gamma_T-1) m_0\equiv KE_T$, rather than $p_T$, is the relevant variable; and in fact $v_2/n_q$ as a function of $KE_T/n_q$ seems to exhibit nearly perfect scaling~\cite{PXArkadyQM06} (Fig.~\ref{fig:MasashiFlow}b). 

\begin{figure}[!thb]
   \begin{center}
\includegraphics[width=0.50\linewidth]{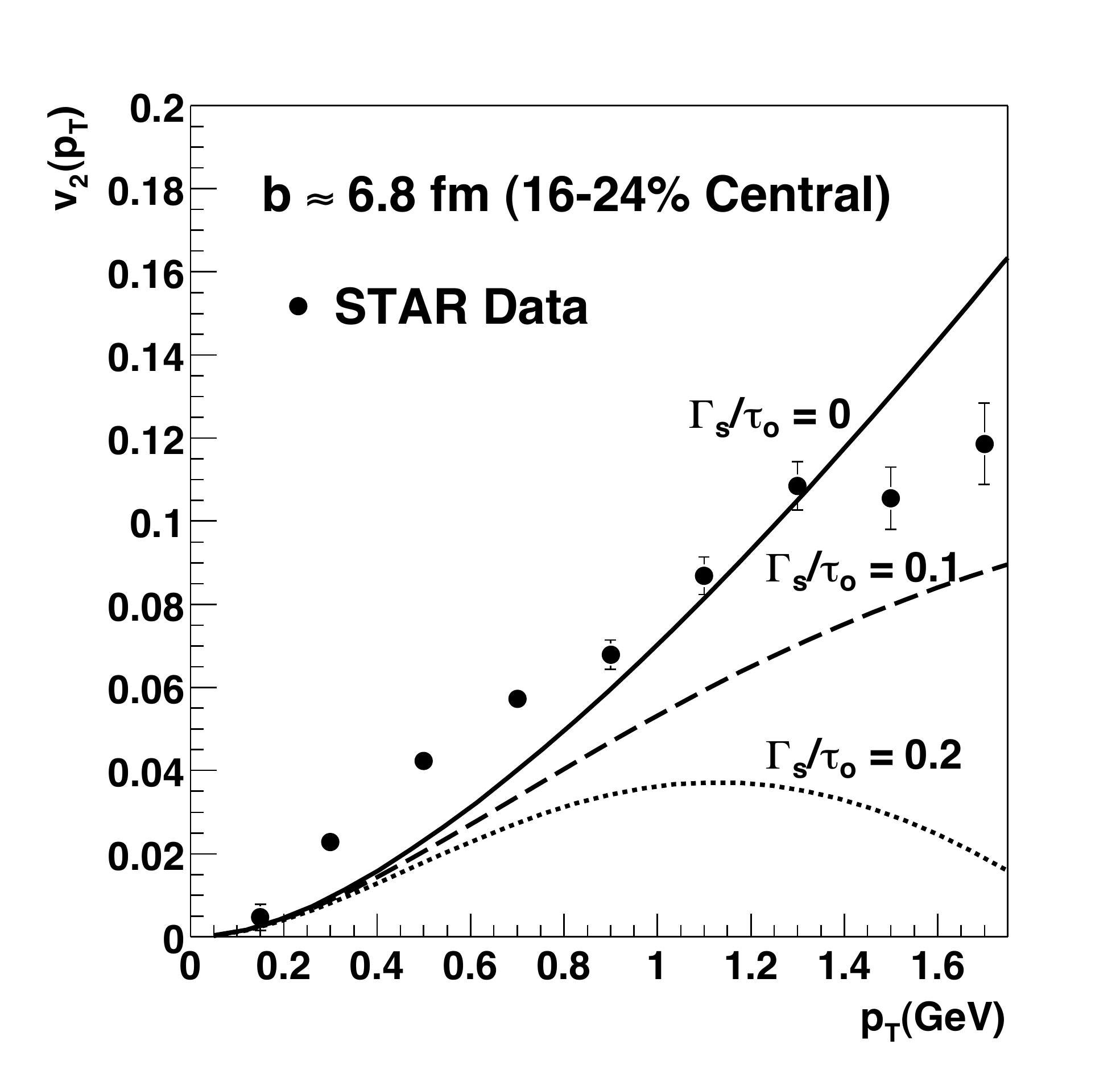}
\includegraphics[width=0.44\linewidth]{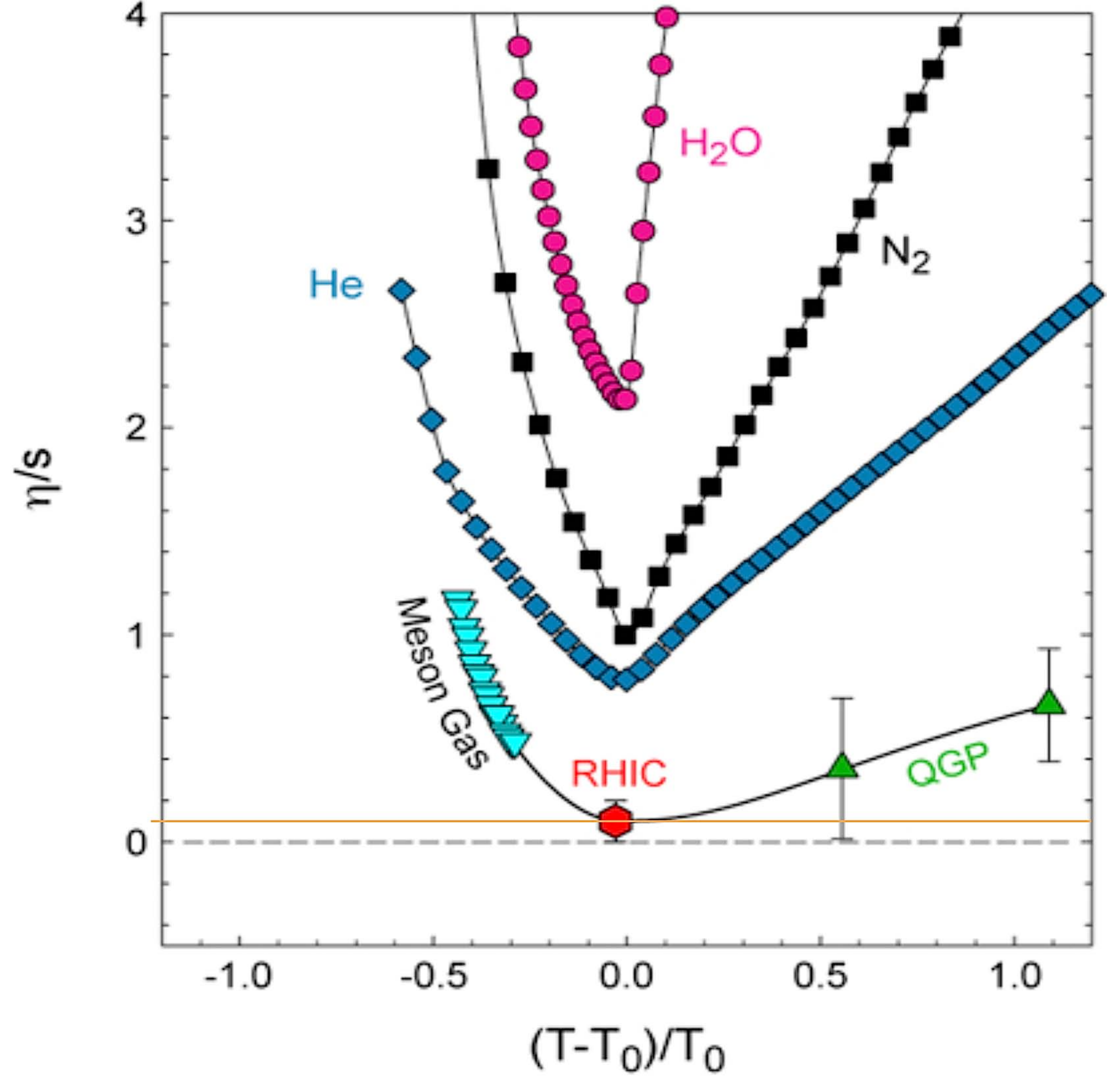}
\end{center}\vspace*{-1.0pc}
\caption[]{a) (left) Teaney's~\cite{TeaneyPRC68} predictions for $v_2(p_T)$ for ideal ($\Gamma_s/\tau_0=0$) and viscous hydrodynamics, where $\Gamma_s=\frac{4}{3}\frac{\eta}{sT}$ is the sound attenuation length, and $\tau_0$ is the thermalization time. b) $\eta/s$ for various fluids at fixed pressure as a function of temperature $T$, where $T_0$ is the temperature of the critical point of the liquid-gas phase transition~\cite{LaceyPRL98,CKMPRL97}.  }
\label{fig:TeaneyFlow}
\end{figure}

    The fact that the flow persists for $p_T>1$ GeV/c (Fig.~\ref{fig:TeaneyFlow}a) implies that the viscosity is small~\cite{TeaneyPRC68}, perhaps as small as a quantum viscosity bound from string theory~\cite{Kovtun05}, $\eta/s=1/(4\pi)$ where $\eta$ is the shear viscosity and $s$ the entropy density per unit volume.  This has led to the description of the ``s\QGP'' produced at RHIC as ``the perfect fluid''~\cite{THWPS}. 
An estimate~\cite{LaceyPRL98} of $\eta/s$ for nuclear matter and for several common fluids, as a function of the fractional difference of the temperature from the critical temperature, at fixed pressure, is shown in Fig.~\ref{fig:TeaneyFlow}b. This particular estimate~\cite{LaceyPRL98} for the \QGP\ at RHIC is quite close to the quantum bound (solid line).  Also, empirically, for all common fluids $\eta/s$ is a minimum at or near the critical point~\cite{CKMPRL97} which might suggest that the conditions at RHIC energies are near the \QCD\ critical point.   
\subsection{Two-Particle Correlations and Flow}
In addition to measuring flow by the correlation of individual particles to the reaction plane, it is also possible to measure flow by the correlation of two particles to each other. The advantage of this method is that one does not have to determine the reaction plane. Thus if two particles $A$ and $B$ are correlated to the reaction plane, but not otherwise correlated to each other, 
\[ \frac{dN^A}{d\phi^A}\propto 1+\sum_n 2 v_n^A \cos(n(\phi^A-\Psi_n)) ,\quad \frac{dN^B}{d\phi^B}\propto 1+\sum_n 2 v_n^B \cos(n(\phi^B-\Psi_n)) \]
then the correlation to the reaction plane induces a correlation of these two particles to each other which can be measured without knowledge of the reaction plane,
\begin{equation} \frac{dN^{AB}}{d\phi^A d\phi^B}\propto \left[1+2 v_2^A v_2^B \cos 2(\phi^A-\phi^B) + 2 v_3^A v_3^B \cos 3(\phi^A-\phi^B) +\ldots\right] \ .\label{eq:2partdphi}\end{equation}

In p-p collisions there is no collective flow but there are strong two-particle azimuthal correlations due to di-jet production in hard-scattering (see Fig.~\ref{fig:collstar}a), which also exist in A+A collisions but are obscured by the large multiplicity [e.g. can you find a jet in Fig.~\ref{fig:collstar}b]. Before the discovery of jets, two-particle correlations were used extensively at the CERN-ISR to study hard-scattering (the production of particles with large transverse momentum) in p-p collisions which was discovered there~\cite{egseeMJTISSP2009}. Due to the huge multiplicities in Au+Au collisions at RHIC, where for central Au+Au collisions at $\sqrt{s_{NN}}=200$ GeV, there is an estimated $\pi\Delta r^2\times{1\over {2\pi}} {dE_T\over{d\eta}}\sim 375$ GeV of energy 
in one unit of the nominal jet-finding cone,  $\Delta r=\sqrt{(\Delta\eta)^2 + (\Delta\phi)^2}$, two-particle correlations  were used exclusively for the first 10 years to study hard scattering at RHIC.

\begin{figure}[!thb]
\begin{center}
\begin{tabular}{cc}
\begin{tabular}[b]{c}
\hspace*{-0.10in}\includegraphics[width=0.52\linewidth]{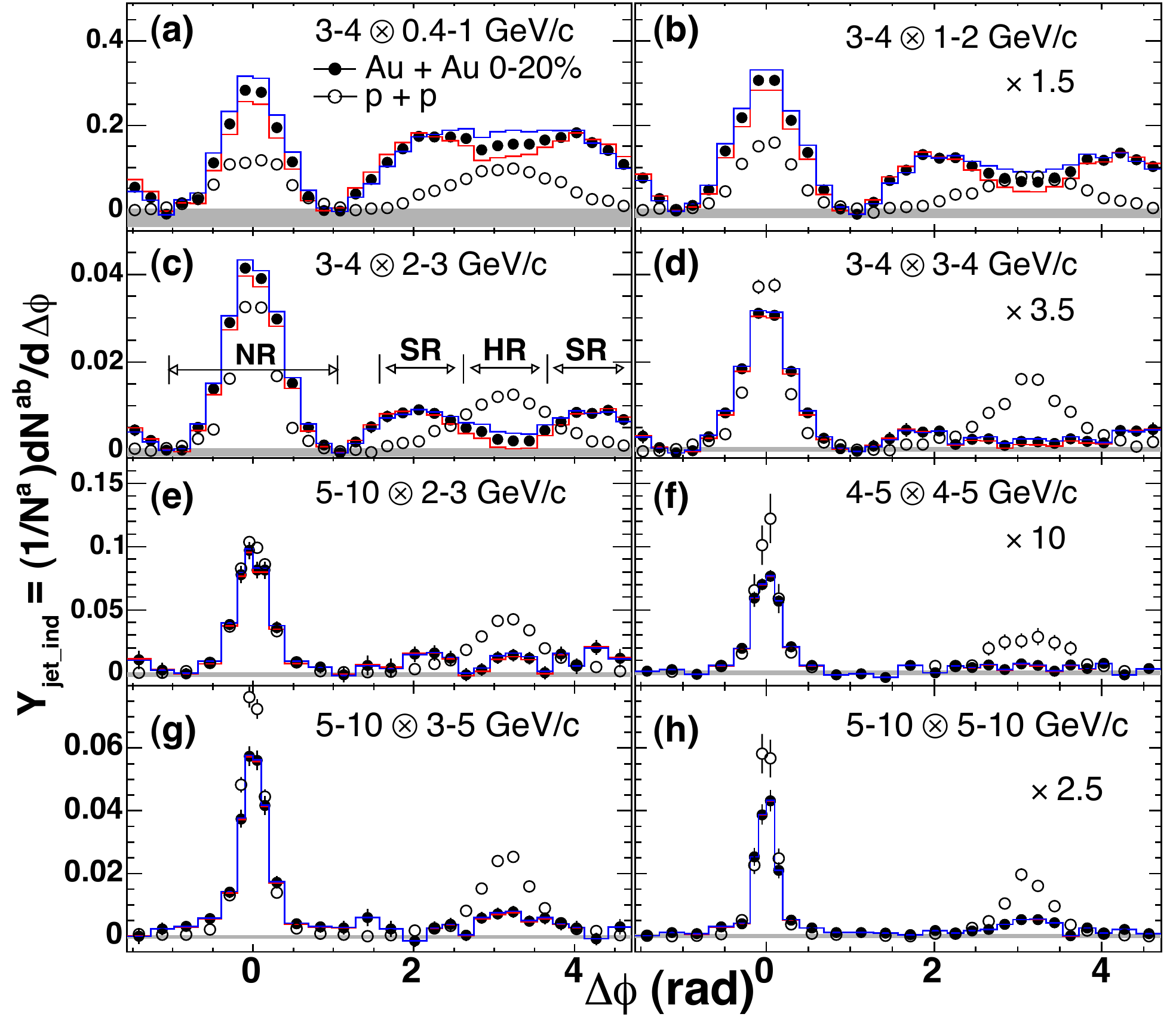}
\end{tabular}\hspace*{-0.2pc}
\begin{tabular}[b]{c}
\includegraphics[width=0.43\linewidth]{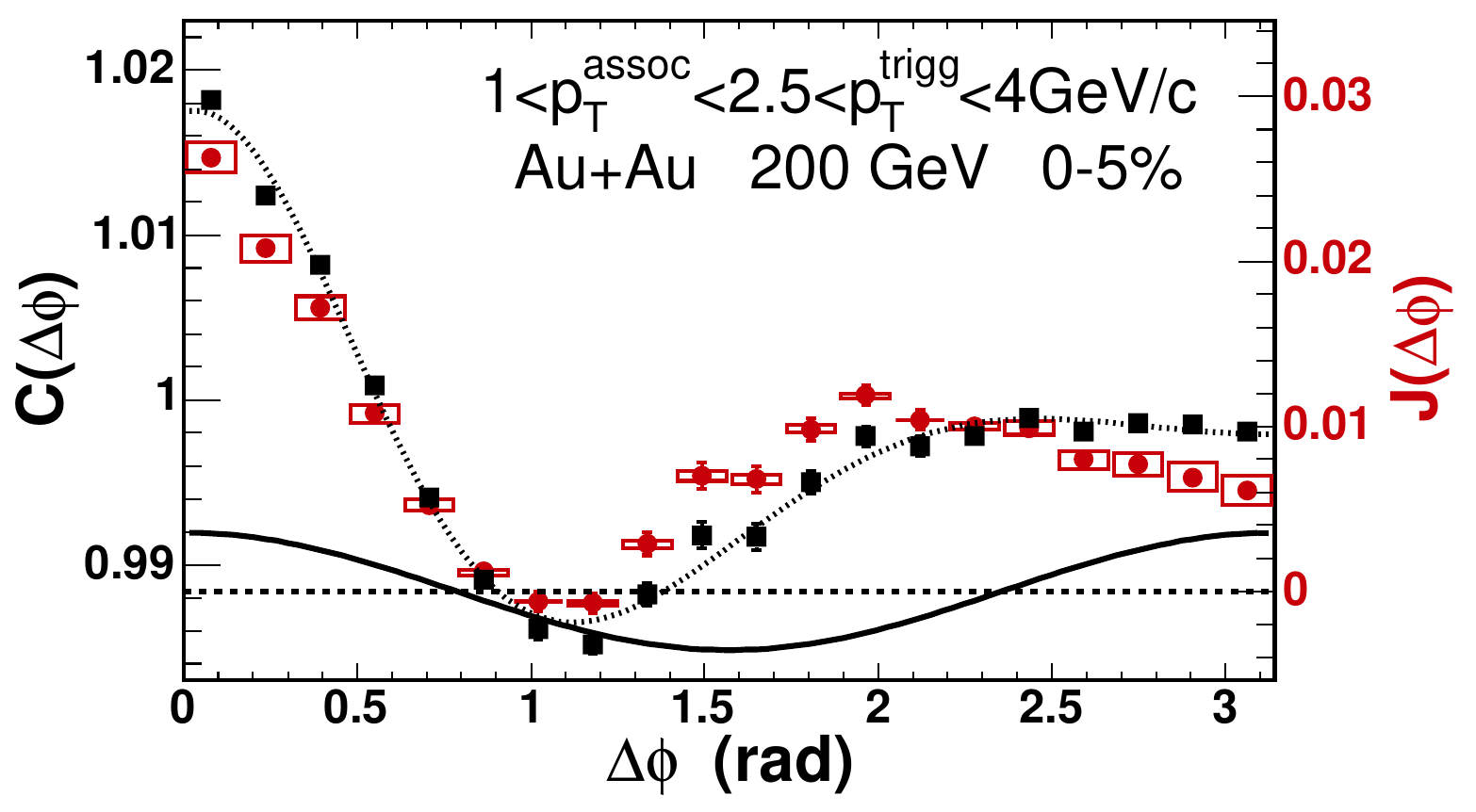}\\[0.3pc]
%\psfig{file=figs-global/ccorxt,width=0.30\linewidth}\cr
%\vspace*{3pc}
\includegraphics[width=0.42\linewidth]{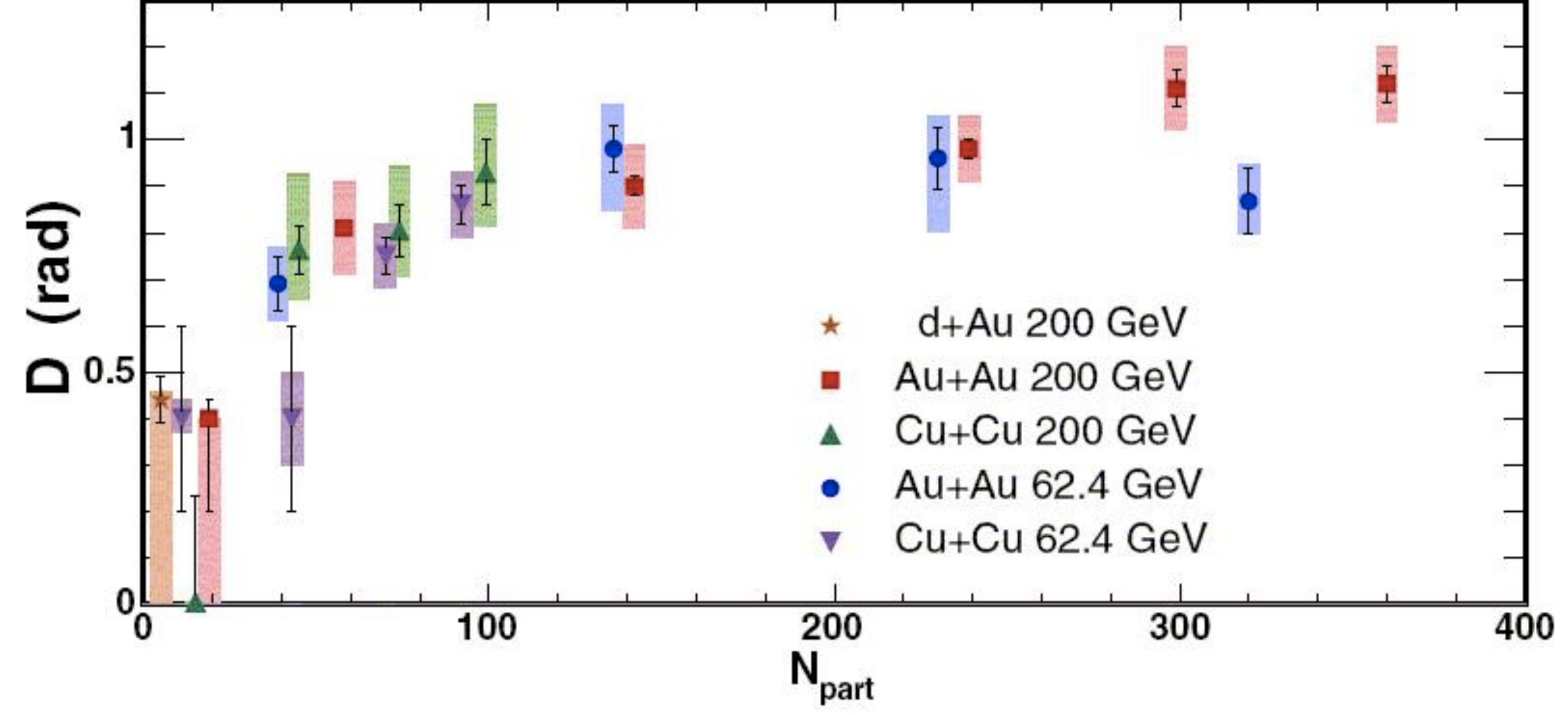}
\end{tabular}
\end{tabular}
\end{center}
\vspace*{-1.0pc}
\caption[]
{a-h) (left) PHENIX~\cite{ppg083} azimuthal correlation conditional yield of associated $h^{\pm}$ particles with $p_{T_a}$ for trigger $h^{\pm}$ with $p_{T_t}$ for the various $p_{T_t} \otimes p_{T_a}$ combinations shown. i) (right)-(top)  PHENIX~\cite{ppg067} azimuthal correlation function $C(\Delta\phi)$ of $h^{\pm}$ with $1\leq p_{T_a}\leq 2.5$ GeV/c with respect to a trigger $h^{\pm}$ with $2.5\leq p_{T_t} \leq 4$ GeV/c in Au+Au central collisions, where the line with data points indicates $C(\Delta\phi)$ before correction for the azimuthally modulated ($v_2$) background, and the other line is the $v_2$ correction which is subtracted to give the jet correlation function $J(\Delta\phi)$ (data points). j) (right)-(bottom) PHENIX $D$ parameters~\cite{ppg067}, the angular distance of the apparently displaced peak of the $J(\Delta\phi)$ distribution from the angle $\Delta\phi=\pi$ as a function of centrality, represented as the number of participants $N_{\rm part}$, for the systems and c.m. energies indicated.
\label{fig:HSD} }\vspace*{-1.0pc}
\end{figure}

Typical examples of the di-hadron measurements in p-p and Au+Au central (0--20\%) collisions at $\sqrt{s_{NN}}=200$ GeV are shown in Fig.~\ref{fig:HSD}~\cite{ppg083,ppg067} which are presented as azimuthal distributions of the conditional yields of associated particles, with $p_{T_a}$, with respect to trigger particles with $3\leq p_{T_t}\leq10$ GeV/c. The di-jet structure in p-p collisions 
is clearly indicated by the gaussian-like strong azimuthal correlation peaks on the same side ($\Delta\phi=\phi_a-\phi_t\sim0$) and away side ($\Delta\phi\sim\pi$ rad.) relative to the trigger particle for all ranges of $p_{T_t}$ and $p_{T_a}$ measured.      
However, one of the many interesting features in Au+Au collisions is that the away side azimuthal jet-like correlations (Fig.~\ref{fig:HSD}c) are much wider than in p-p collisions and show a two-lobed structure (``the shoulder'' (SR)) at lower $p_{T_t}$ with a dip at 180$^\circ$,  reverting to the more conventional structure of a peak at 180$^\circ$ (``the head'' (HR)) for larger $p_{T_t}$. 

\begin{figure}[!hbt]
\begin{center}
\includegraphics[width=0.44\linewidth]{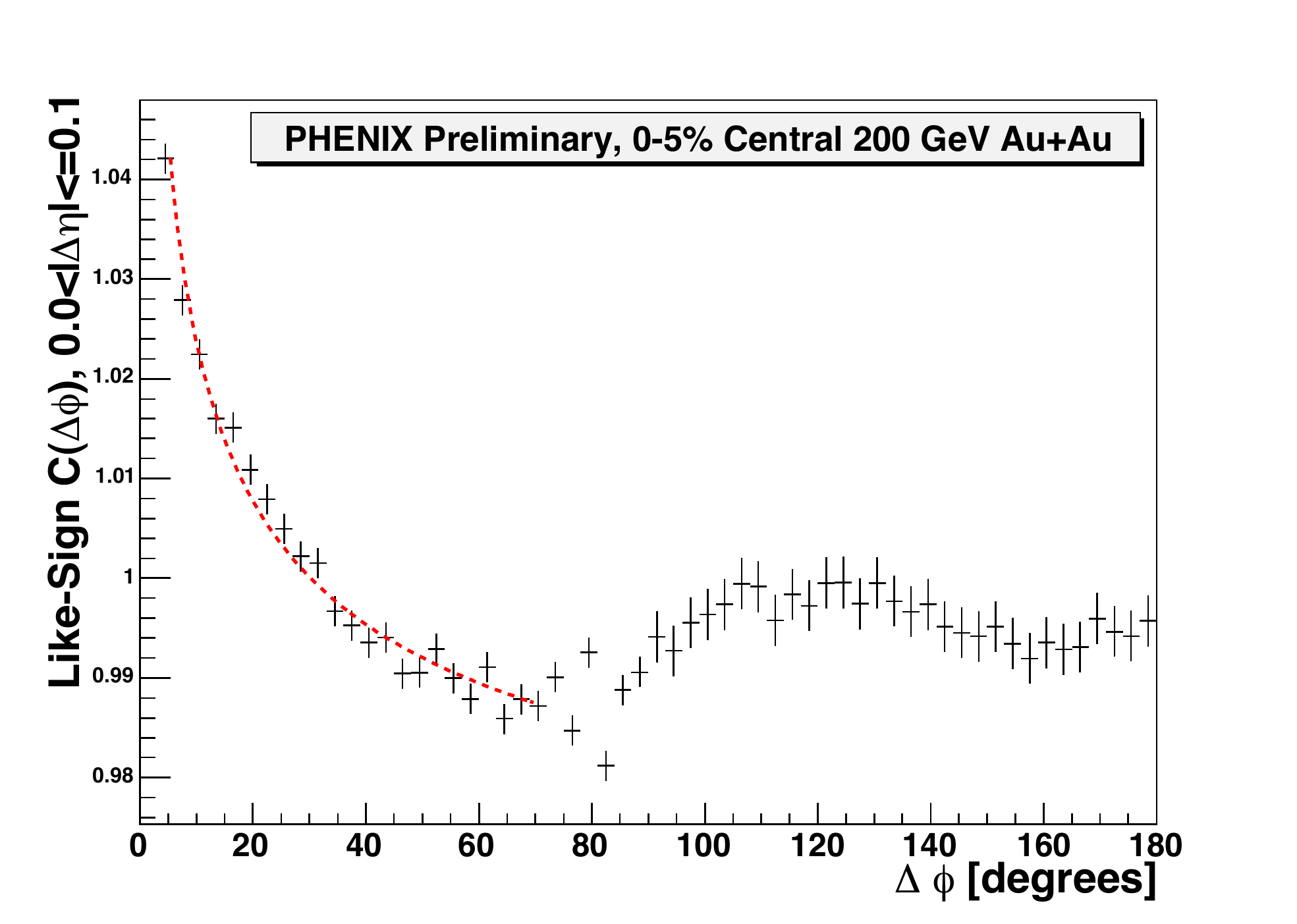}\hspace*{-0.5pc}
\raisebox{-0.06\linewidth}{\includegraphics[width=0.56\linewidth]{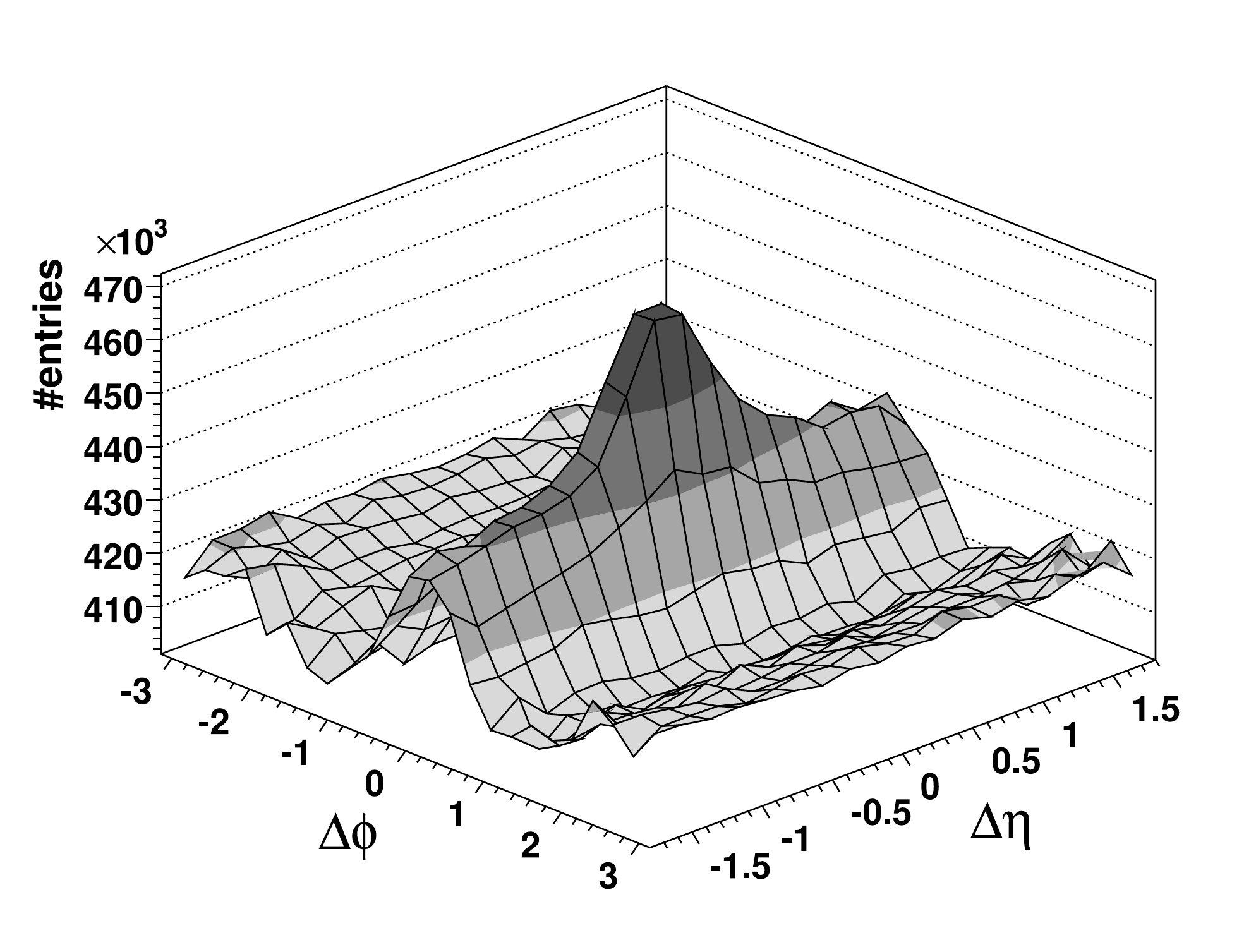}}
\end{center}
\vspace*{-0.28in}
\caption[]
{a) (left) Low $p_T$ like-sign pair azimuthal correlation function for 0-5\% central Au+Au collisions at $\sqrt{s_{NN}}=200$ GeV from charged hadrons with $0.2\leq p_{T_1}, p_{T_2}\leq 0.4$ GeV/c~\cite{JTMQM06}. b) (right) ``The Ridge''~\cite{PutschkeHP06,JacobsHP04}.\label{fig:moreartifacts} }
\end{figure}

The wide away-side correlation in central Au+Au collisions is significantly obscured by the large multiparticle background which is modulated in azimuth by the $v_2$ collective flow of a comparable width to the jet correlation (Fig.~\ref{fig:HSD}i). After the $v_2$ correction, the double peak structure $\sim \pm 1$ radian from $\pi$, with a dip at $\pi$ radians, becomes evident. The double-peak structure may indicate a reaction of the medium to a passing parton in analogy to a ``sonic boom'' or the wake of a boat, which was given the name ``Mach Cone''~\cite{CSST-Coney05}, and has been under active study both theoretically~\cite{egseeppg083} and experimentally. PHENIX characterizes this effect by the half-width $D$ ($\sim 1.1$ radian) of the Jet function, $J(\Delta\phi)$, the angular distance of the displaced peak of the distribution from the angle $\Delta\phi=\pi$. One of the striking features of the wide away side correlation is that the width $D$ (Fig.~\ref{fig:HSD}j)  does not depend on centrality, angle to the reaction plane,  $p_{T_a}$ and $\sqrt{s_{NN}}$, which always seemed problematic to me if the effect were due to a reaction to the medium. Another suspicious issue is that the same effect occurs even for auto-correlations of particles with very low $p_T$ between 0.2 and 0.4 GeV/c where any effect of hard-scattered partons should be submerged by the predominant soft physics (Fig.~\ref{fig:moreartifacts}a)~\cite{JTMQM06}.    

In addition to the Head/Shoulder or Mach Cone effect in two-particle correlations on the away-side, same-side correlations also show a new effect in A+A collisions called ``the Ridge''~\cite{STARridgePRC80}. This is seen in two-dimensional correlations in $\Delta\eta, \Delta\phi$ (Fig.~\ref{fig:moreartifacts}b)~\cite{PutschkeHP06,JacobsHP04} where the associated yield distribution can be decomposed into a narrow jet-like peak at small angular separation which has a similar shape to that found in p-p collisions, and a component that is narrow in $\Delta\phi$ but depends only weakly on $\Delta\eta$, the ``ridge.'' 
However, new results in 2010--2011 dramatically changed this picture.

\subsection{Triangular flow, odd harmonics}
For the first 10 years of RHIC running, and dating back to the Bevalac, all the experts thought that the odd harmonics in Eq.~\ref{eq:siginv2} would vanish by the symmetry $\phi\rightarrow \phi+\pi$ of the almond shaped overlap region~\cite{AlverOllitrault} (Fig.~\ref{fig:MasashiFlow}). However, in the year 2010, an MIT graduate student an his Professor in experimental physics, seeking (at least since 2006) how to measure the fluctuations of $v_2$ in the PHOBOS experiment at RHIC,  realized that fluctuations in the collision geometry on an event-by-event basis, i.e. the distribution of participants from event-to-event, did not respect the average symmetry. This resulted in what they called ``participant triangularity'' and ``triangular flow'', or $v_3$ in Eq.~\ref{eq:siginv2}, which they measured using both PHOBOS and STAR data~\cite{AlverRoland}.\footnote{It was pointed out by Leticia Palhares in the discussion that a Brazilian group showed in 2009 that  
that the ridge and the cone, i.e. $v_3$, does appear in an event-by-event hydrodynamics calculation without jets~\cite{BrazilNuXuv3}, but the MIT group~\cite{AlverRoland} was the first to show it with real data.}  

\begin{figure}[!h]
\begin{center}
\includegraphics[width=0.75\linewidth]{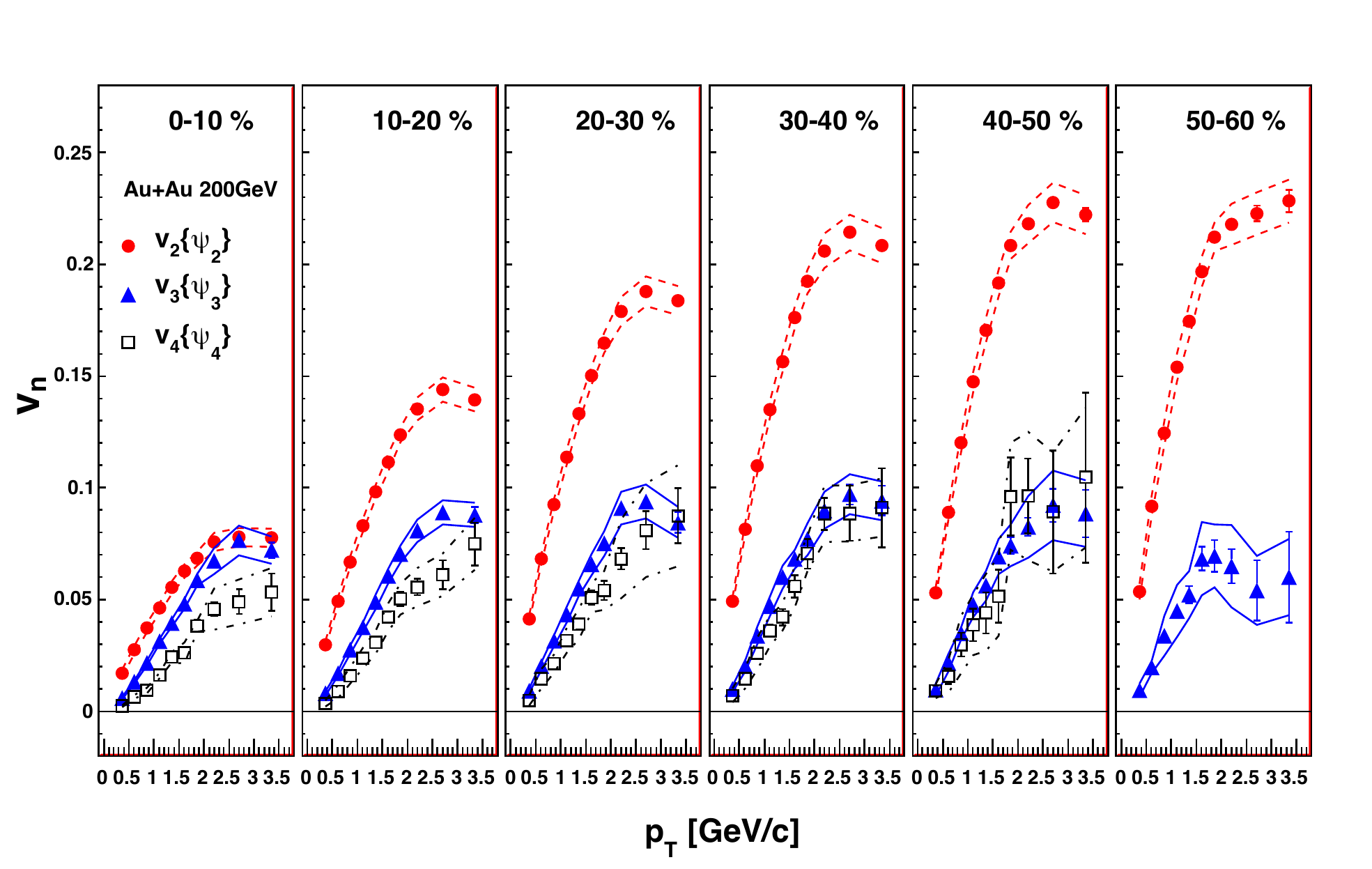}
\end{center}
%\vspace*{-0.28in}
\caption[]
{PHENIX~\cite{EsumiQM11} measurements of the $v_n$ parameters using Eq.~\ref{eq:siginv2} (with the appropriate reaction plane) as a function of $p_T$ for different centrality slices in $\sqrt{s_{NN}}=200$ GeV Au+Au collisions. \label{fig:PXv3} }
\end{figure}

Many experiments presented measurements of $v_3$ at Quark Matter 2011, e.g. Fig.~\ref{fig:PXv3}~\cite{EsumiQM11},   and it was one of the most exciting results of that year. 
There are two striking observations from Fig.~\ref{fig:PXv3} which indicate that fluctuations of the initial collision geometry are driving the observed $v_3$: i) the centrality dependence of $v_3(p_T)$ is weak as one would expect from fluctuations, but $v_2(p_T)$ which is most sensitive to the geometry of the ``almond''-shaped overlap region tracks the change in eccentricity with centrality; ii) for the most central collisions (0-10\%), where the overlap region is nearly circular so that all the $v_n$ are driven by fluctuations, $v_2(p_T)$, $v_3(p_T)$, $v_4(p_T)$ are comparable.  The fact that the observed collective flow of final state particles follows the fluctuations in the initial state geometry points to real hydrodynamic flow of a nearly perfect fluid (and convinces this author of the validity of hydrodynamics in RHI collisions, of which he was quite skeptical previously~\cite{MJTIJMPA2011}). It is evident that $v_3$, a $\cos 3(\Delta\phi)$ term with lobes at $\Delta\phi=0, 2\pi/3$ and $4\pi/3 \approx 0, 2, 4$ radians, would explain the peaks at $\pi\pm D$ radian in the two-particle correlations (Fig.~\ref{fig:HSD}i,j) and also why $D\approx 1$ radian independent of centrality and kinematic variables; while the lobe at $\Delta\phi=0$ explains the ridge (Fig.~\ref{fig:moreartifacts}b). There is presently lots of activity to confirm in detail whether taking account of the odd harmonics in addition to $v_2$ and $v_4$ in the background of Fig.~\ref{fig:HSD}i will result in gaussian-like away-jet peaks in A+A collisions and the disappearance of the same-side ridge.

\section{RHIC beam energy scan---In search of the critical point}  
\begin{figure}[!thb]
   \begin{center}
\includegraphics[width=0.32\linewidth]{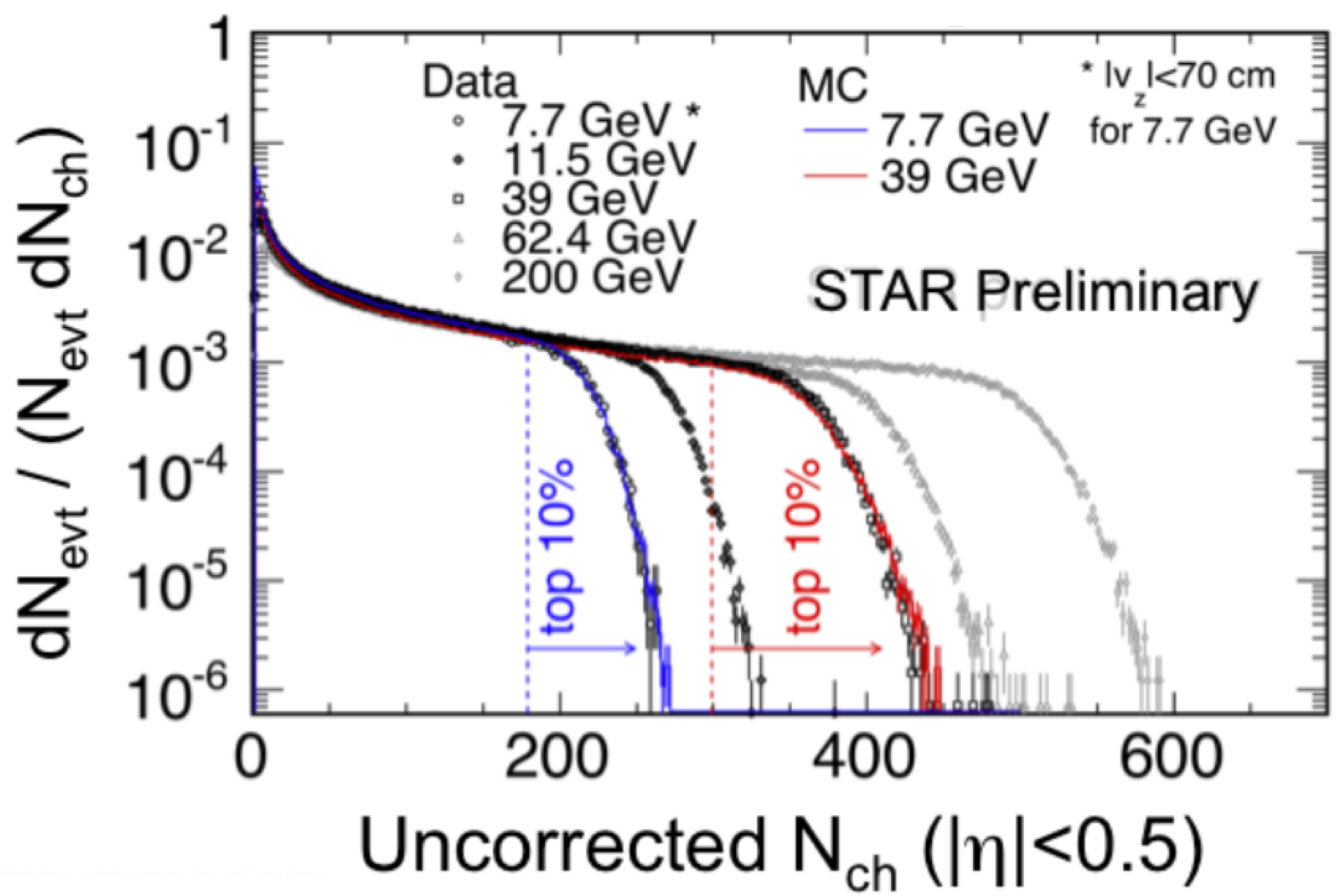}
\includegraphics[width=0.32\linewidth]{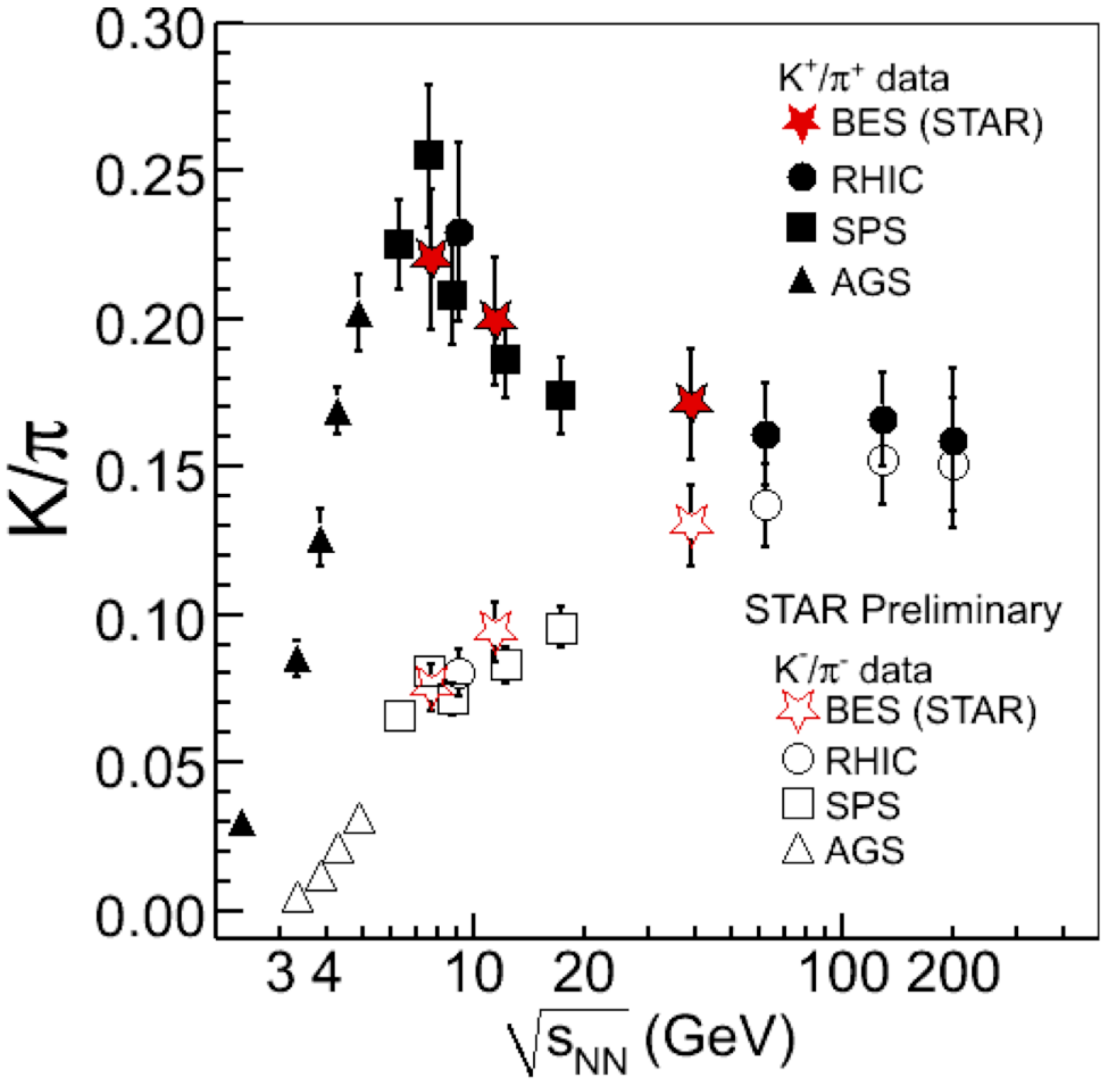}
\includegraphics[width=0.32\linewidth]{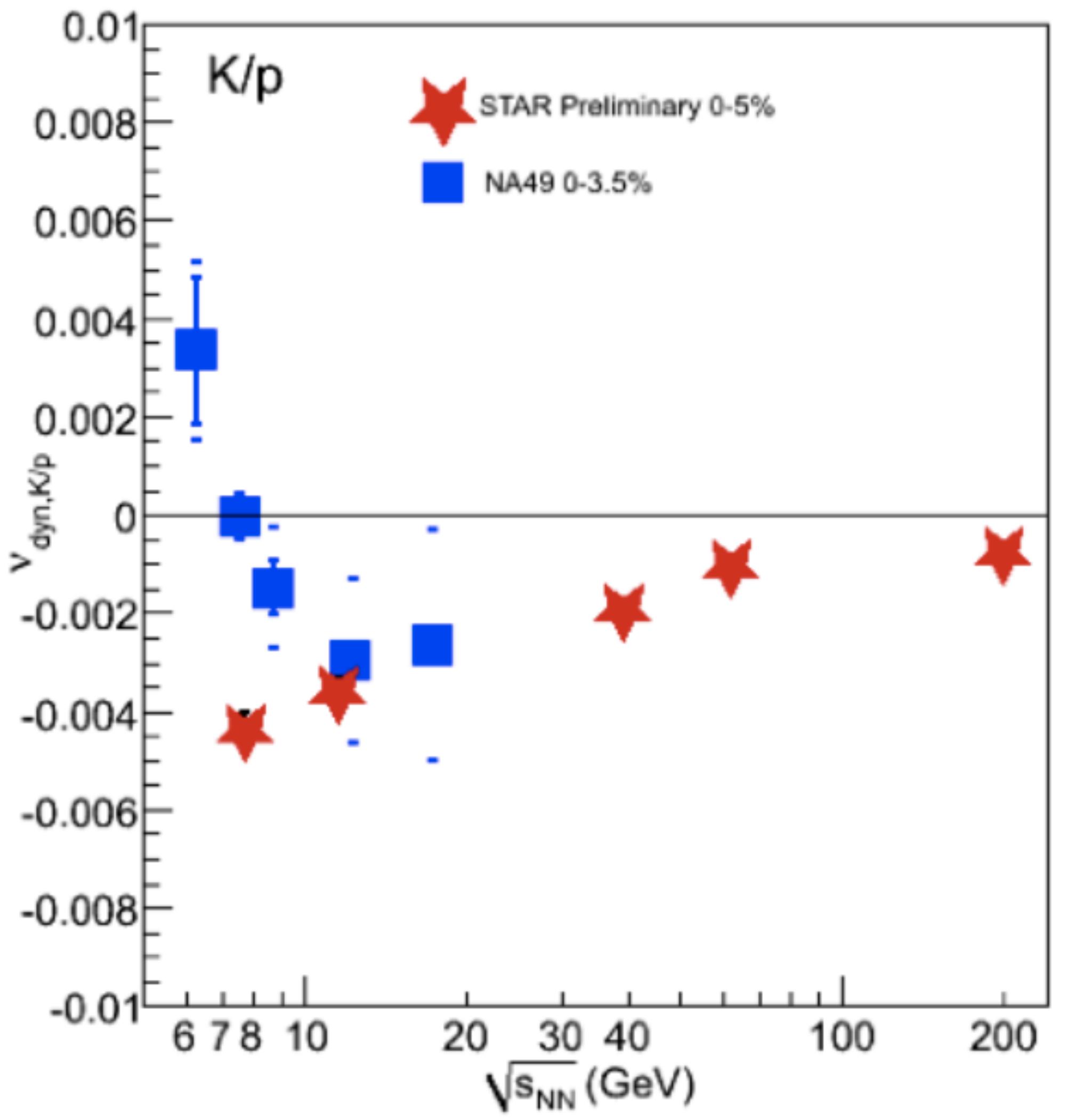}
\end{center}\vspace*{-1.0pc}
\caption[]{a) (left) $N_{\rm ch}$ distribution in the STAR detector for 5 values of $\sqrt{s_{NN}}$~\cite{LKBMQM11}; b) (center) $K/\pi$ ratio vs $\sqrt{s_{NN}}$~\cite{LKBMQM11}; c) (right) Event-by-event fluctuation of $K/p$ ratio~~\cite{LKBMQM11}. }
\label{fig:byebyeMarek}
\end{figure}

In the years 2010--2011, RHIC made runs with Au+Au collisions at c.m. energies $\sqrt{s_{NN}}=7.7, 11.5$, $39$, 19.6, 62.4 GeV, in addition to previous runs at 130 and 200 GeV, to search for the onset of large fluctuations which should occur near a critical point in the phase diagram (Fig.~\ref{fig:phase_boundary}). Such fluctuations in the $K/\pi$ and $K/p$ ratio had been claimed at the CERN SPS fixed target program near $\sqrt{s_{NN}}=8$ GeV and were presented~\cite{MarekQM11} as ``evidence of the onset of the deconfinement phase transition''. At QM2011, STAR~\cite{LKBMQM11} presented many excellent results on this subject of which I show a small selection in Fig.~\ref{fig:byebyeMarek}. 

Fig.~\ref{fig:byebyeMarek}a shows the multiplicity distribution $dN_{\rm ch}/d\eta$ which maintains the characteristic Nuclear Geometry dominated shape (as in Fig.~\ref{fig:nuclcoll}), stretching to higher multiplicity, $dN_{\rm ch}/d\eta$, with increasing $\sqrt{s_{NN}}$. Fig.~\ref{fig:byebyeMarek}b shows the $K^+/\pi^+$ and $K^-/\pi^-$ ratios over the entire range of $\sqrt{s_{NN}}$ measurements. The maximum of the $K^+/\pi^+$ ratio near $\sqrt{s_{NN}}=8$ GeV is naturally explained~\cite{CleymansOeschlerPLB615} by the change in dominant $K^+$ production from $K^+ \Lambda$ to $K^+ K^-$ whose smooth increase with $\sqrt{s_{NN}}$ can be seen from the $K^-/\pi^-$ ratio. The famous ``horn'', or apparent discontinuity, at $\sqrt{s_{NN}}=8$ GeV from the SPS data~\cite{MarekQM11} is greatly smoothed when the new STAR data are added. Fig.~\ref{fig:byebyeMarek}c shows a smooth variation of the the STAR measurements of the fluctuations of the event-by-event $K/p$ ratio as a function of $\sqrt{s_{NN}}$, which differs dramatically from the huge effect claimed by the SPS Fixed Target measurements below 12 GeV, notably the change from negative to positive~\cite{MarekQM11}. There is no doubt in this author's mind that one must prefer the collider measurements, where the detector position at mid-rapidity in the c.m. system is constant for all values of $\sqrt{s_{NN}}$, to the fixed target measurements, where the rapidity of the c.m. system moves dramatically with respect to the the detector as $\sqrt{s_{NN}}$ varies. This is a major strength of the RHIC Beam Energy Scan program.   
\subsection{A press release during the 2011 school}
On June 23, 2011, shortly before I was to present my 2011 lectures, a press release from LBL arrived claiming that ``By comparing theory with data from STAR, Berkeley Lab scientists and their colleagues map phase changes in the QGP''~\cite{LBLJune23}. Since I criticized ``physics by press-release'' concerning the discovery of the \QGP\ (above), I felt that I was obliged to review the physics behind this latest example, presumably a ``Highlight from RHIC''.  

The subject is ``Fluctuations of conserved quantities'', in this case the net baryon distribution taken as $p-\bar{p}$. Since there can be no fluctuations of conserved quantities such as net charge or net baryon number in the full phase space, one has go to small intervals~\cite{AsakawaHMPRL85} to detect a small fraction of the protons and anti-protons which then fluctuates, i.e. varies from event to event. The argument is that, e.g. the fluctuation of one charged particle in or out of the considered interval produces a larger mean square fluctuation of the net electric charge if the system is in the hadron gas phase with integral charges than for the \QGP\ phase with fractional charges. 

However, while there are excellent statistical mechanical arguments about the utility of fluctuations of conserved quantities   such as net baryon number as a probe of a critical point~\cite{KochCFRNC06}, there are, so far, no adequate treatments of the mathematical statistics of the experimental measurements. Theoretical analyses tend to be made in terms of a Taylor expansion of the free energy $F=-T\ln Z$ around the critical temperature $T_c$ where $Z$ is the partition function, or sum over states, which is of the form $Z\propto e^{-(E-\sum_i \mu_i Q_i)/kT}$ and $\mu_i$ are chemical potentials associated with conserved charges $Q_i$~\cite{KochCFRNC06}.  The terms of the Taylor expansion, which are obtained by differentiation, are called susceptibilities, denoted $\chi$. The only connection of this method to mathematical statistics is that the Cumulant generating function in mathematical statistics is also a Taylor expansion of the $\ln$ of an exponential:
\begin{equation}
g_x (t)=\ln\mean{e^{tx}}=\sum_{n=1}^\infty \kappa_n \frac{t^n}{n!} \qquad \kappa_m =\left.\frac{d^m g_x (t)}{dt^m}\right|_{t=0} \qquad .
\label{eq:cumgenfn}
\end{equation}
Thus, the susceptibilities are Cumulants in mathematical statistics terms, where, in general, the Cumulant $\kappa_m$ represents the $m^{\rm th}$ central moment with all $m$-fold combinations of the lower order moments subtracted.\footnote{Note that factorial Cumulants, also known as ``Mueller moments''~\cite{AMuellerPRD4}, have been used previously for multiplicity distributions.} For instance, 
$\kappa_2=\mean{(x-\mu)^2}\equiv \sigma^2$, $\kappa_3=\mean{(x-\mu)^3}$, $\kappa_4=\mean{(x-\mu)^4}-3\kappa_2^2$, $\kappa_5=\mean{(x-\mu)^5}-10\kappa_3 \kappa_2$, where $\mu=\mean{x}$. Two so-called normalized or standardized Cumulants are common in this field, the skewness, $S=\kappa_3/\sigma^3$ and the kurtosis, $\kappa=\kappa_4/\sigma^4=\mean{(x-\mu)^4}/\sigma^4-3$.  

\begin{figure}[!t]
   \begin{center}
\includegraphics[width=0.46\linewidth]{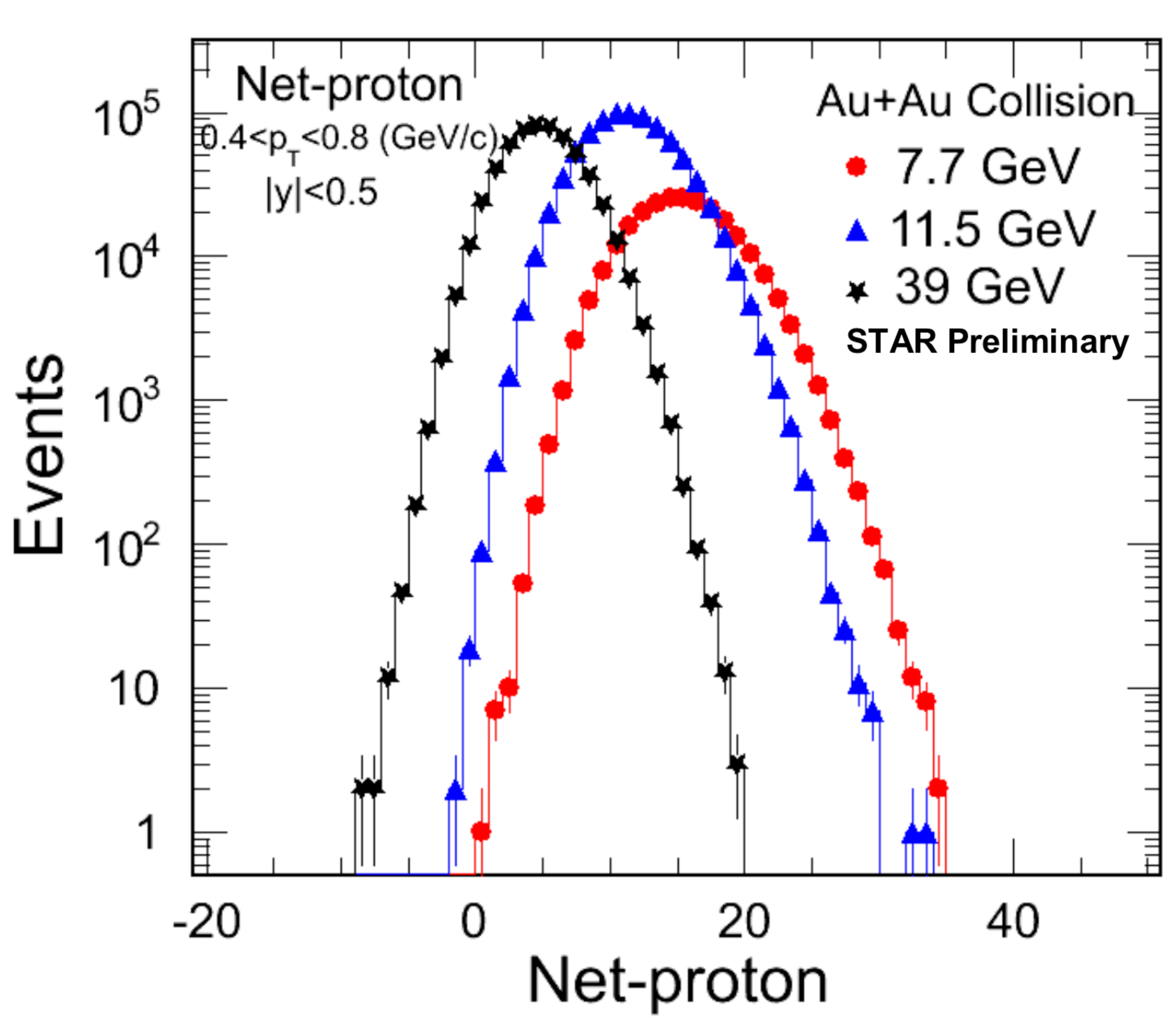}
\includegraphics[width=0.53\linewidth]{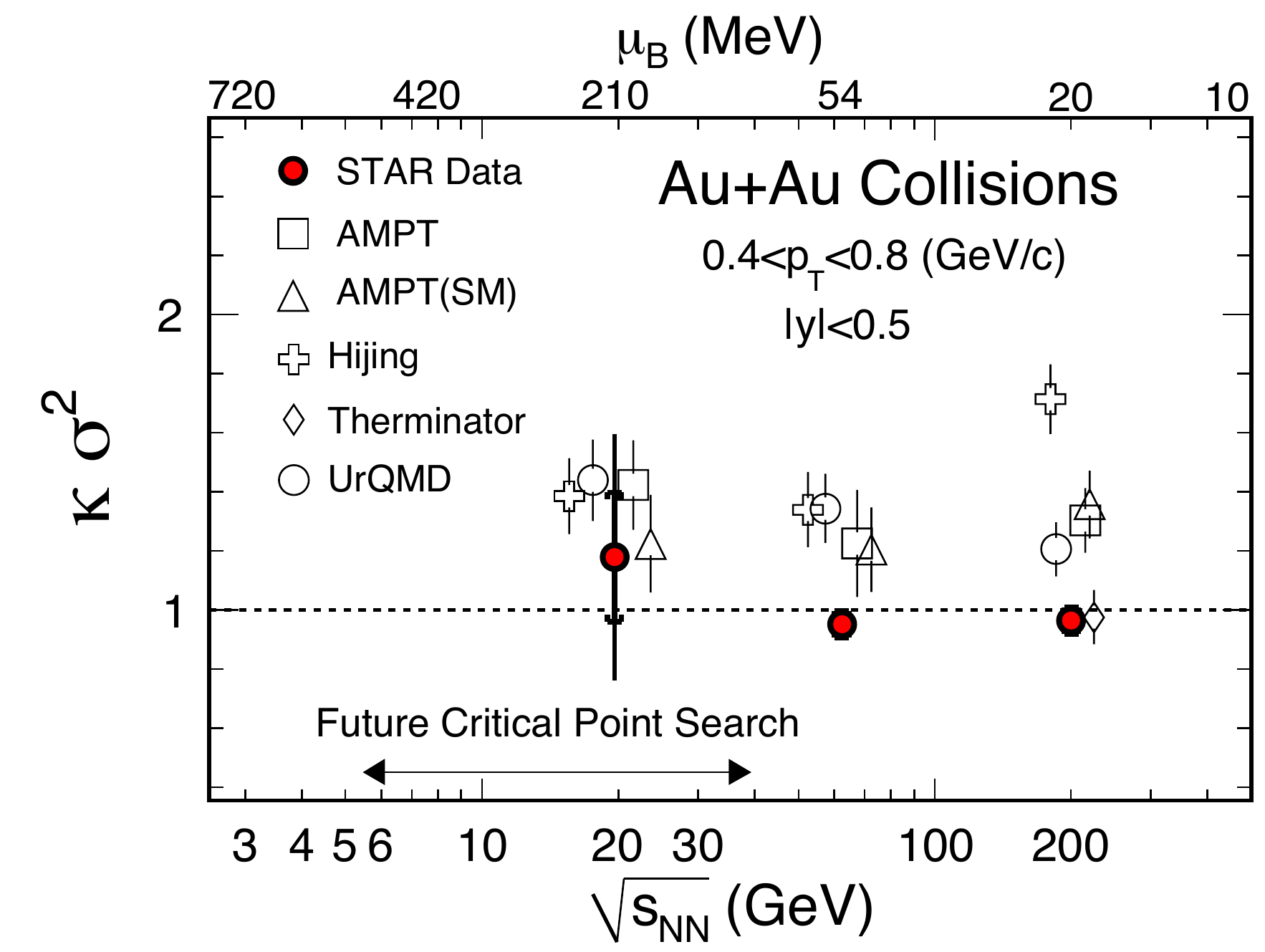}
\includegraphics[width=0.48\linewidth]{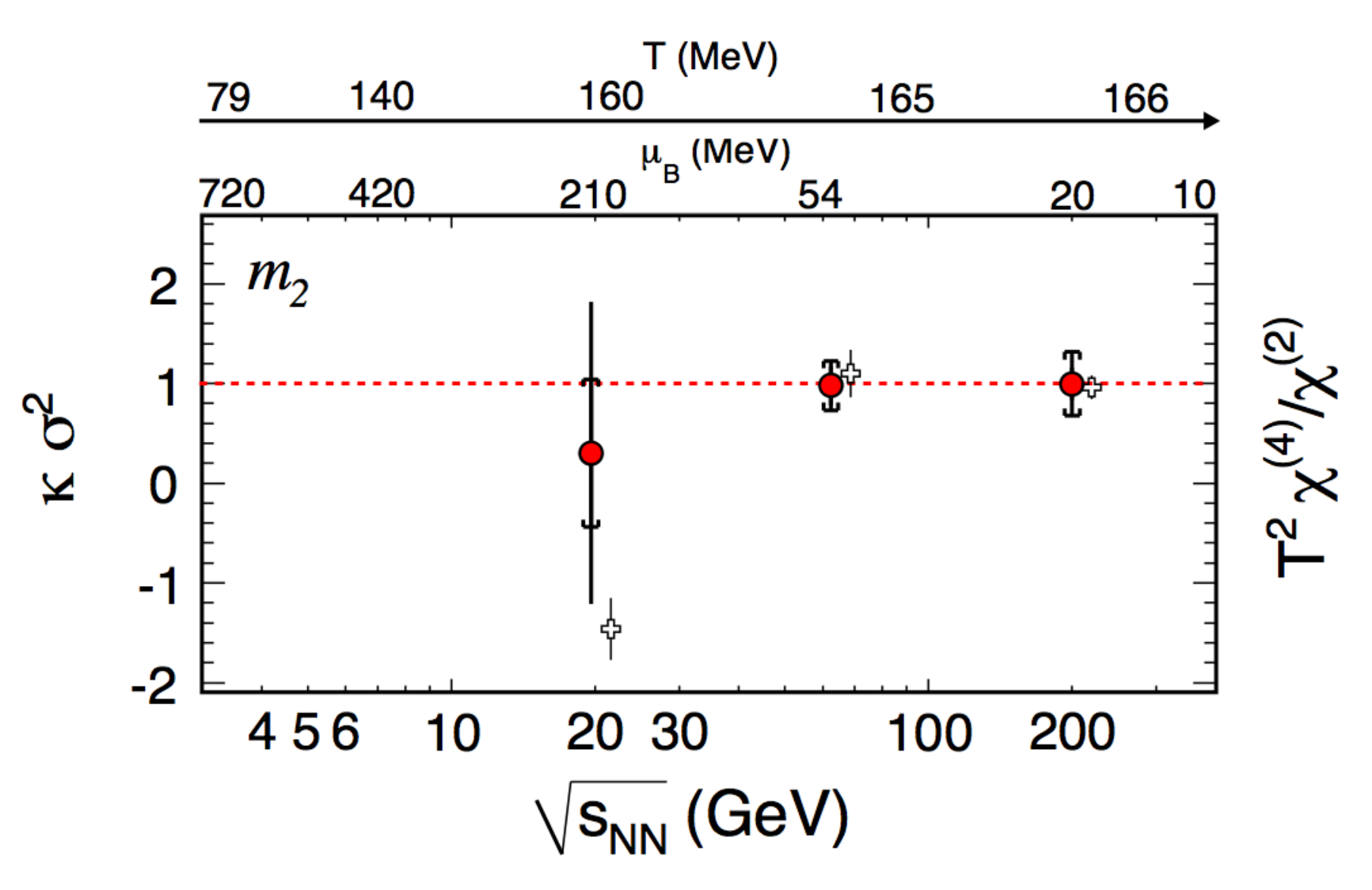}
\includegraphics[width=0.50\linewidth]{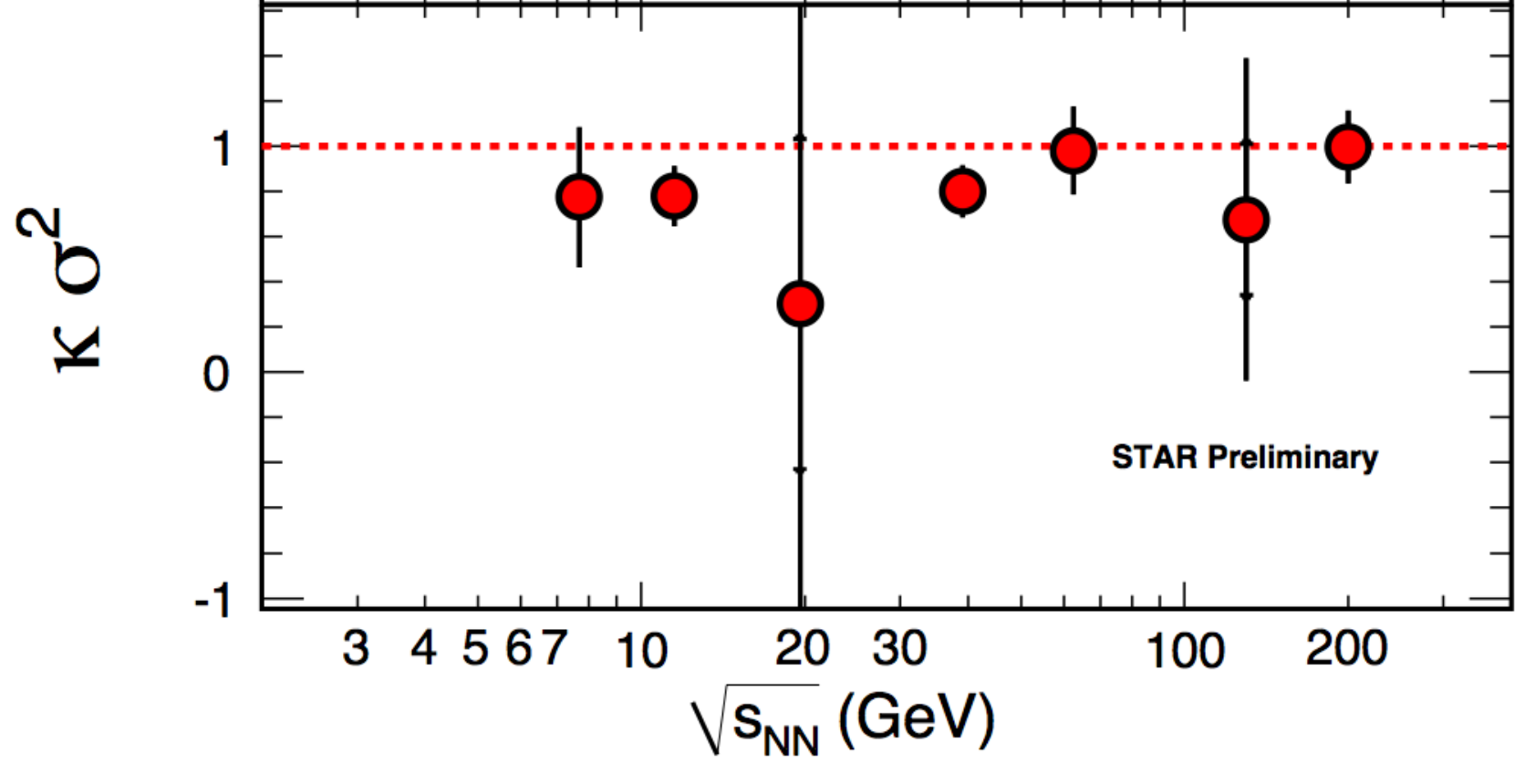}
\end{center}\vspace*{-1.0pc}
\caption[]{a) (top-left) STAR~\cite{TarnowskyQM2011} distribution of event-by-event $p-\bar{p}$ at 3 values of $\sqrt{s_{NN}}$; b) (top-right) STAR published~\cite{STARnetPPRL105} measurements of $\kappa\sigma^2$; c) (bottom-left) Measurements from (b) as shown in Ref.~\cite {Science332} compared to the predicted ratio of susceptibilities (open crosses); d) (bottom-right) compilation~\cite{TarnowskyQM2011} of STAR measurements of $\kappa\sigma^2$.   }
\label{fig:STAR-NetP}
\end{figure}

A sample~\cite{TarnowskyQM2011} of STAR measurements of the distribution of net-protons in Au+Au collisions in the small interval $0.4\leq p_T\leq 0.8$ GeV/c, $|y|<0.5$ for different $\sqrt{s_{NN}}$ is shown in Fig.~\ref{fig:STAR-NetP}a. 

The moments in the form $\kappa\sigma^2=\kappa_4/\kappa_2$ are shown from a previous STAR publication~\cite{STARnetPPRL105} in  Fig.~\ref{fig:STAR-NetP}b while a plot, alleged to be of this same data, presented in the Lattice \QCD\ theory publication that generated the press-release, is shown in Fig.~\ref{fig:STAR-NetP}c~\cite{Science332}; and a plot of the $\kappa\sigma^2$ from the data of Fig.~\ref{fig:STAR-NetP}c, combined with the results from Fig.~\ref{fig:STAR-NetP}b, is shown in Fig.~\ref{fig:STAR-NetP}d~\cite{TarnowskyQM2011}. 
There are many interesting issues to be gleaned from Fig.~\ref{fig:STAR-NetP}. 

The data point at 20 GeV in Fig.~\ref{fig:STAR-NetP}c is not the published one from (b), as stated in the caption~\cite{Science332}, but the one from (d), which is different and with a much larger error. This, in my opinion, makes the data point look better compared to the predicted discontinuous value of $\kappa\sigma^2=-1.5$ for the critical point at 20 GeV (open crosses) in contrast to the predictions of 1.0 for both 62.4 and 200 GeV. The published measurements in (b) together with the new measurements in (d) are all consistent with $\kappa\sigma^2=1$; but clearly indicate the need for a better measurement at $\sqrt{s_{NN}}=20$ GeV. Apart from these issues, the main problem of comparing Lattice \QCD\ ``data'' to experimental measurements is that it is like comparing peaches to fish, since the prediction is the result of derivatives of the log of the calculated partition function of an idealized system, which may have little bearing on what is measured using finite sized nuclei in an experiment with severe kinematic cuts. Maybe this is too harsh a judgement; but since this is the first such comparison (hence the press release), perhaps the situation will improve in the future. 

When I first saw the measured distributions in Fig.~\ref{fig:STAR-NetP}a in 2011, my immediate reaction was that STAR should fit them to Negative Binomial distributions (NBD) so that they would know all the Cumulants. However, I subsequently realized that my favorite 3 distributions for integer random variables, namely, Poisson, Binomial, and Negative Binomial, are all defined only for positive integers, while the number of net-protons on an event can be negative as well as positive, especially at higher c.m. energies. This is why somebody should work out the mathematical statistics of the net proton distribution as we did in PHENIX for the distribution of the difference in foreground (opposite charge) and background (like charge) di-lepton events when both are Poisson distributed~\cite{ppg104}. Until then, it is instructive to compare the values of $\kappa\sigma^2$ in Fig.~\ref{fig:STAR-NetP} to those from the well-known distributions: Poisson, $\kappa\sigma^2=1$; Binomial, $\kappa\sigma^2=1-6p+6p^2<1$; Negative Binomial, $\kappa\sigma^2=1+6\mu/k+6\mu^2/k^2>1$; Gaussian, all Cumulants=0 for $k>2$, so $\kappa\sigma^2=0$. The data favor Poisson (i.e. no correlation) everywhere, with some hint of Binomial. Nevertheless, if a future measurement would show a significant huge discontinuity of $\kappa\sigma^2$ similar to the theoretical prediction at $\sqrt{s_{NN}}=20$ GeV, then even I would admit that such a discovery would deserve a press release, maybe more!  

\subsection{The future measurement has appeared without a press release.}
In the intervening period since 2011, the STAR collaboration has improved the preliminary measurements to a publication~\cite{STAR14021558} and has improved the analysis by comparing to both Poisson and Negative Binomial distributions which involved finding the formula for the Cumulants of the difference between two distributions~\cite{Westfall2013}. The first four Cumulants of the Poisson, Binomial and Negative Binomial distributions  are given in Table~\ref{tab:Cumulant}. These three distributions fall into the class of ``integer valued L\'evy processes~\cite{Barndorff2013}'' for which the Cumulants $\kappa_j$ for the difference between two such distributions, $P(m)=P^{+}(m) - P^{-}(m)$ with Cumulants $\kappa_j^+$ and $\kappa_j^-$, respectively, are~\cite{Westfall2013,Barndorff2013}:
\begin{equation}
\kappa_j=\kappa_j^+ +(-1)^j \kappa_j^- \qquad. \label{eq:difCum}
\end{equation} 

\begin{table}[!h]
\centering
\caption[]{Cumulants for Poisson, Binomial and Negative Binomial Distributions}
{\begin{tabular}{llll} 
\hline
Cumulant & Poisson & Binomial & Negative Binomial \\
\hline
$\kappa_1=\mu$ & $\mu$ & $np$ &$\mu$\\
$\kappa_2=\mu_2=\sigma^2$ &$\mu$& $\mu (1-p)$ & $\mu (1+{\mu}/{k})$\\[0.2pc]
$\kappa_3=\mu_3$ &$\mu$& $\sigma^2 (1-2p)$ & $\sigma^2 (1+2{\mu}/{k})$\\[0.2pc] 
$\kappa_4=\mu_4-3\kappa_2^2$\hspace*{1pc}&$\mu$& $\sigma^2 (1-6p+6p^2)$ & $\sigma^2 (1+6{\mu}/{k}+6{\mu^2}/{k^2})$\\[0.2pc]
\hline\\[-0.6pc]
$S\equiv{\kappa_3}/{\sigma^3}$& $ {1}/{\sqrt{\mu}}$ &${(1-2p)}/{\sigma}$ & $(1+2{\mu}/{k})/{\sigma}$\\[0.2pc]
$\kappa\equiv{\kappa_4}/{\kappa_2^2}$ & ${1}/{\mu}$ & ${(1-6p+6p^2)}/{\sigma^2}$\hspace*{1pc}& $(1+6{\mu}/{k}+6{\mu^2}/{k^2})/{\sigma^2}$\\[0.2pc]
$S\sigma$ & $1$ & $(1-2p)$ & $(1+2{\mu}/{k})$\\[0.2pc]
$\kappa\sigma^2$ & $1$ & $(1-6p+6p^2)$ & $(1+6{\mu}/{k}+6{\mu^2}/{k^2})$\\[0.2pc] 
\hline
\end{tabular}} \label{tab:Cumulant}
\end{table}

\begin{figure}[!h]
   \begin{center}
\includegraphics[width=0.46\linewidth]{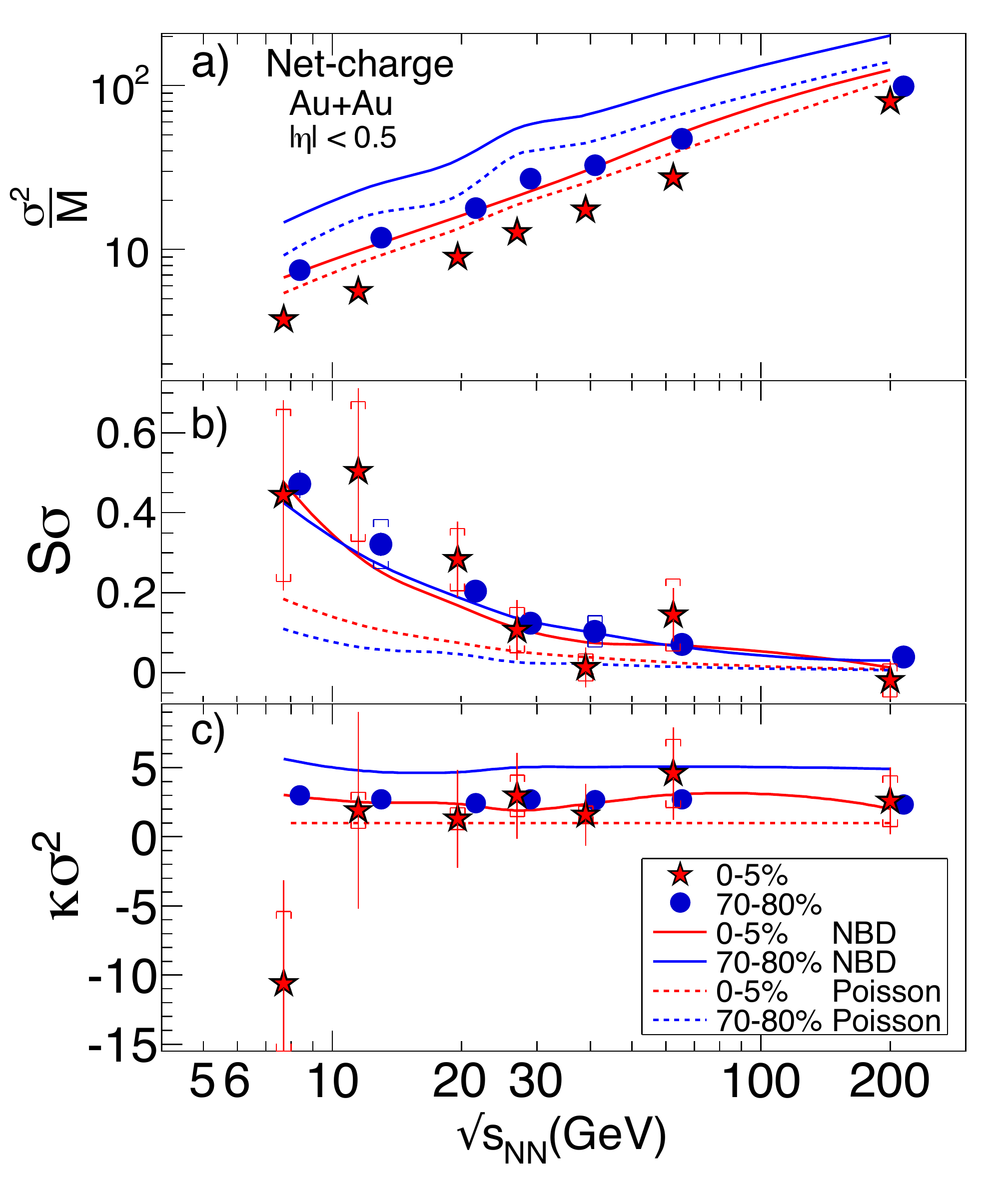}
\end{center}\vspace*{-0.25in}
\caption[]{ $\sqrt{s_{NN}}$ dependence of combinations of net-charge Cumulants in Au+Au from STAR~\cite{STAR14021558}, where in (a) $M$ is used to represent the mean, $\mu$.   }
\label{fig:STARCumpub}
\end{figure}

Figure~\ref{fig:STARCumpub}~\cite{STAR14021558} shows the STAR measurements of Cumulants of the net charge $(N^+ -N^-)$ distributions        
from the ``number of positive ($N^+$) and negative ($N^-$) charged particles within $|\eta|<0.5$ and $0.2< p_T< 2.0$ GeV/c on each event (after removing protons and antiprotons with $p_T<400$ MeV/c)~\cite{STAR14021558}''. The corresponding Poisson and NBD Cumulants were calculated from the measured mean, $\mu$, and variance, $\sigma^2$, of the $N^+$ and $N^-$ distributions, respectively, and then calculated using Eq.~\ref{eq:difCum}.  
In contrast to Fig.~\ref{fig:STAR-NetP}, no non-monotonic behavior with $\sqrt{s_{NN}}$ is observed (or claimed) and the measurements of $S\sigma$ and $\kappa\sigma^2$ are all above the Poisson baseline. The $S\sigma$ measurements clearly favor the NBD. The new results indicate that the search for a \QCD\  critical point may not be as easy as originally assumed.

\section{Hard Scattering as an in-situ probe}
One of the most important innovations at RHIC was the use of hard scattering as an in-situ probe of the medium produced in A+A collisions by the effect of the medium on outgoing hard-scattered partons produced by the 
\mbox{initial}  A+A collision. 
%One of the best probes found at RHIC to study the \QGP\ is the hard-scattering of quarks and gluons (the constituents of the nucleon) by the effect of the medium on outgoing hard-scattered partons produced which were 
This was observed primarily via inclusive single particle production at large transverse momentum ($p_T$) or by two-particle correlations with a high $p_T$ trigger.   The use of hard-scattering to probe the thermal or ``soft'' medium produced in RHI collisions was stimulated by p\QCD\ studies~\cite{BDMPS} of the energy loss of partons produced by hard scattering, ``with their color charge fully exposed'', in traversing a medium ``with a large density of similarly exposed color charges''. The conclusion was that ``Numerical estimates of the loss suggest that it may be significantly greater in hot matter than in cold. {\em This makes the magnitude of the radiative energy loss a remarkable signal for \QGP\  formation}''~\cite{BDMPS}. In addition to being a probe of the \QGP\ the fully exposed color charges allow the study of parton-scattering with $Q^2 \ll 1-5$ (GeV/c)$^2$ in the medium where new collective \QCD\ effects may possibly be observed.

\begin{figure}[h]
 \centering 
 \includegraphics[width=0.9\textwidth]{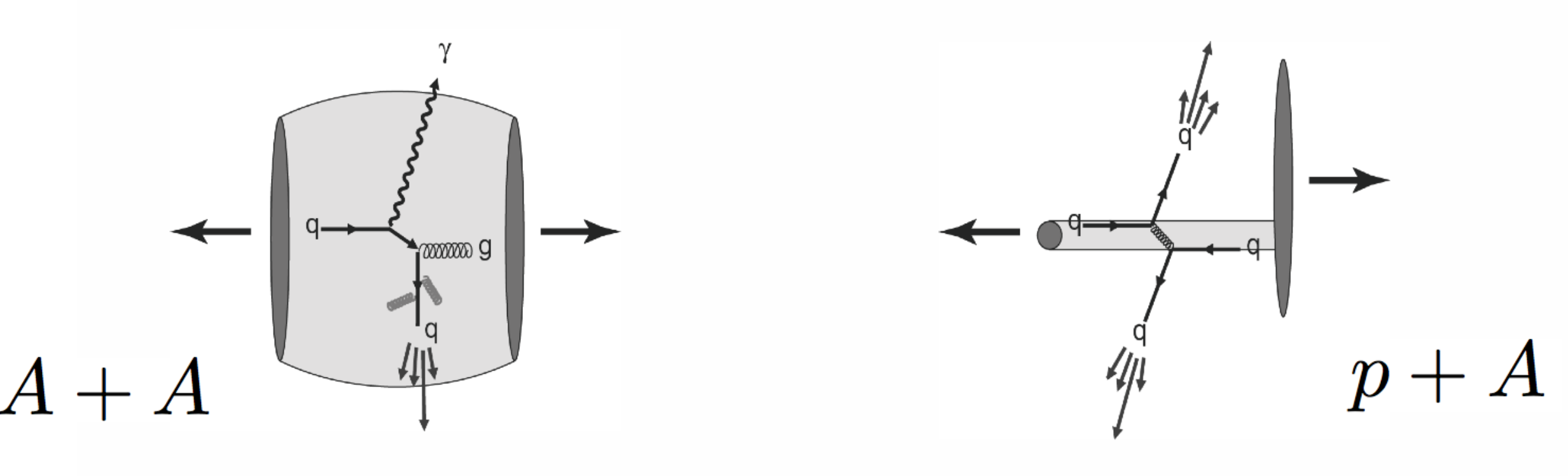}
\caption[]{Schematic drawing of hard-scatterings in relativistic A+A and p+A collisions.}
\label{fig:colldwg}
\end{figure}

The hard-scattering takes place in the initial collision of the highly Lorentz contracted nuclei.  The scattered partons which emerge near 90$^\circ$ to the collision axis (the sweet-spot for such observations) pass through the medium formed and then fragment to jets of particles which are detected (Fig.~\ref{fig:colldwg}). 
The most likely  
constituent reactions are: $g+g\rightarrow g+g$, $g+q\rightarrow g+q$, $q+q\rightarrow q+q$, and $g+q\rightarrow \gamma+q$ (shown). This last reaction~\cite{QCDCompton}, called direct-$\gamma$ production or the inverse \QCD\ Compton effect, is the most beautiful because the $\gamma$-ray participates directly in the hard scattering, then emerges from the medium without interacting and can be measured with high precision. 
No fragmentation is involved and, in fact, these direct-$\gamma$ are distinguished from e.g. $\gamma$ rays from fragmenting partons because they are isolated, with no accompanying fragments. Triggering on a direct-$\gamma$ of a given $\hat{p}_{T_t}$ provides a `beam' of partons with exactly opposite (thus well-known) initial $\hat{p}_T=-\hat{p}_{T_t}$, so that the effect of the medium can be determined by measuring, for instance, the ratio of the transverse momentum $\hat{p}_{T_a}$ of the jet from the away-parton to that of the direct-$\gamma$ trigger, denoted $\hat{x}_h=\hat{p}_{T_a}/\hat{p}_{T_t}$, or equivalently, the fractional jet imbalance, $1-\hat{x}_h$, as used by CMS at LHC~\cite{CMSdijet}.

Since hard-scattering at high $p_T >2$ GeV/c is point-like, with distance scale $1/p_T < 0.1$ fm, the cross section in p+A (A+A) collisions, compared to p-p, should be larger by the relative number of possible point-like encounters, a factor of $A$ ($A^2$) for p+A (A+A) minimum bias collisions. When the impact parameter or centrality of the collision is defined, the proportionality factor becomes $\mean{T_{AA}}$, the average overlap integral of the nuclear thickness functions. Measurements in p+A (or d+A) collisions, where no (or negligible) medium is produced, allow correction for any modification of the nuclear structure function from an incoherent superposition of proton and neutron structure functions.

\subsection{Jet quenching---suppression of high $p_T$ particles}
\begin{figure}[!t]
\includegraphics[height=0.25\textheight]{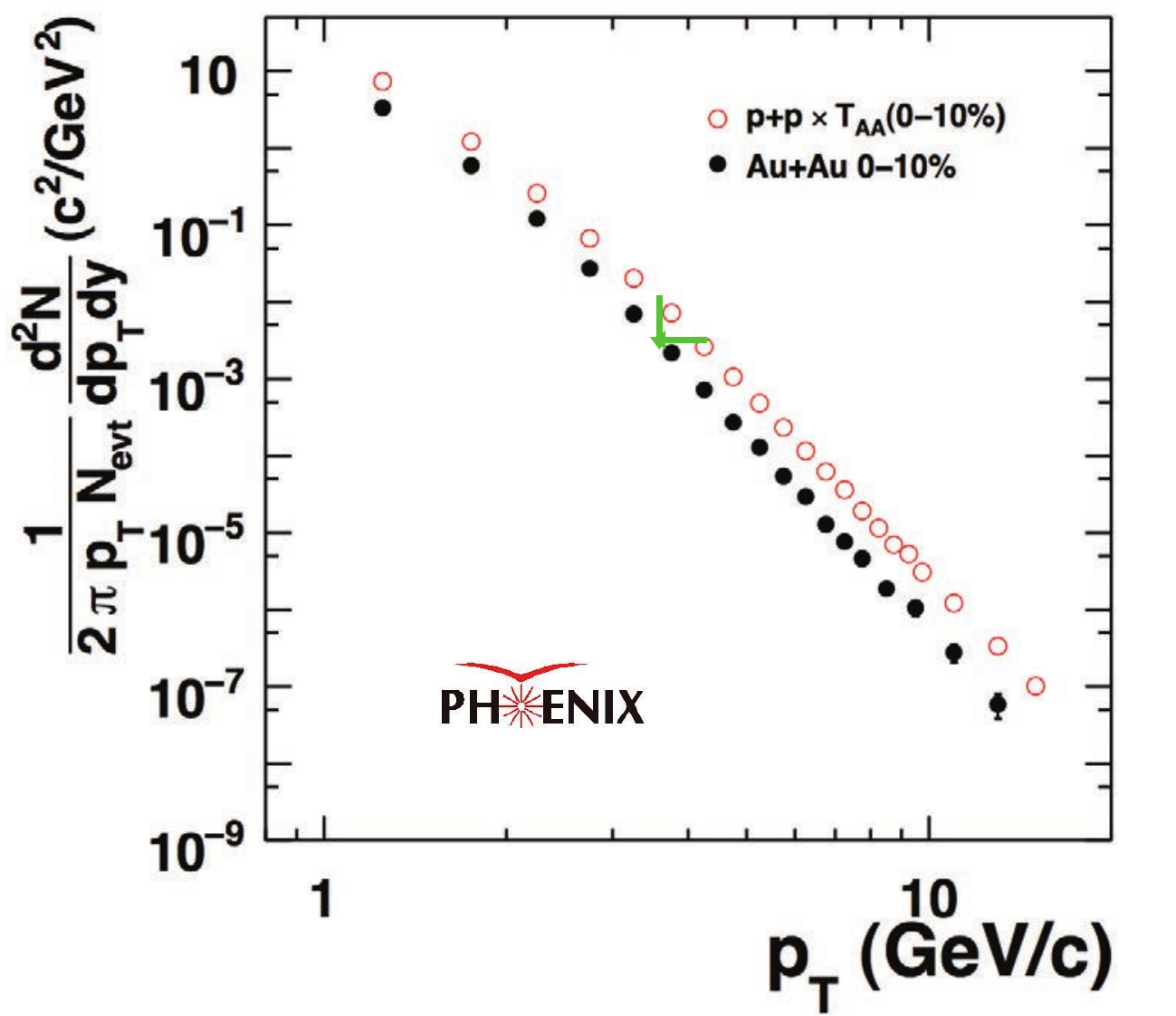}
\hspace*{-0.011\textwidth} \includegraphics[height=0.25\textheight]{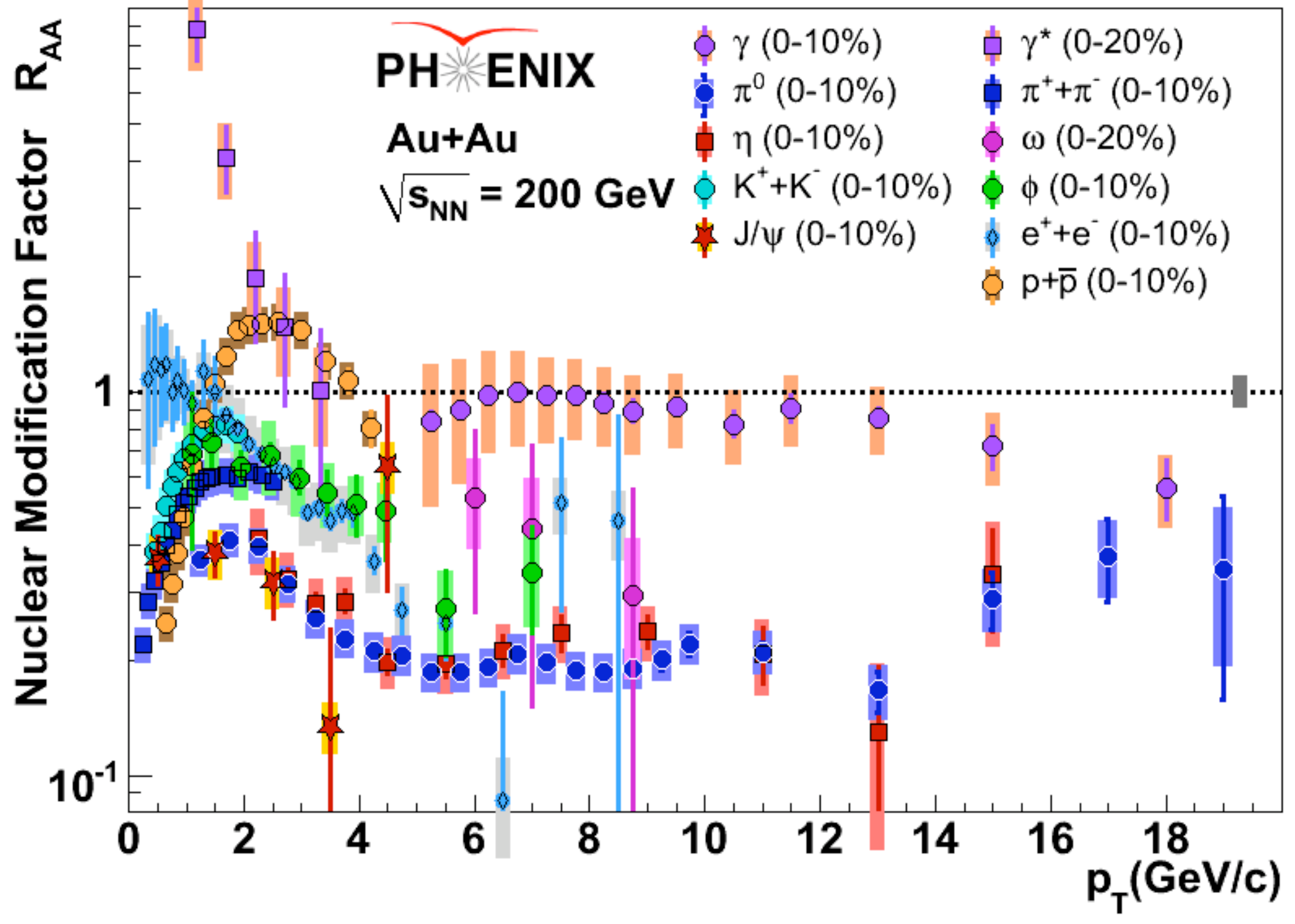}\vspace*{-0.5pc}
\caption{a) (left) Log-log plot of invariant yield of $\pi^0$ at $\sqrt{s_{NN}}=200$ GeV as a function of transverse momentum $p_T$ in p-p collisions multiplied by $\mean{T_{AA}}$ for Au+Au central (0--10\%) collisions compared to the Au+Au measurement~\cite{ppg054}. b) (right) $R_{AA}(p_T)$ for all identified particles so far measured by PHENIX in Au+Au central collisions at $\sqrt{s_{NN}}=200$ GeV.}
\label{fig:Tshirt}\vspace*{-0.5pc}
\end{figure}

   The discovery, at RHIC~\cite{ppg003}, that $\pi^0$'s produced at large transverse momenta are suppressed in central Au+Au collisions by a factor of $\sim5$ compared to point-like scaling from p-p collisions is arguably {\em the}  major discovery in Relativistic Heavy Ion Physics. For $\pi^0$ (Fig.~\ref{fig:Tshirt}a)~\cite{ppg054} the hard-scattering in p-p collisions is indicated by the power law behavior $p_T^{-n}$ for the invariant cross section, $E d^3\sigma/dp^3$, with $n=8.1\pm 0.05$ for $p_T\geq 3$ GeV/c.  The Au+Au data at a given $p_T$ can be characterized either as shifted lower in $p_T$ by $\delta p_T$ from the point-like scaled p-p data at $p'_T=p_T+\delta p_T$, or shifted down in magnitude, i.e. suppressed. In Fig.~\ref{fig:Tshirt}b, the suppression of the many identified particles measured by PHENIX at RHIC is presented as the Nuclear Modification Factor, 
$R_{AA}(p_T)$, the ratio of the yield of e.g. $\pi$ per central Au+Au collision (upper 10\%-ile of observed multiplicity)  to the point-like-scaled p-p cross section, where $\mean{T_{AA}}$ is the average overlap integral of the nuclear thickness functions: 
   \begin{equation}
  R_{AA}(p_T)=\frac{{d^2N^{\pi}_{AA}/dp_T dy N_{AA}}} { \mean{T_{AA}} d^2\sigma^{\pi}_{pp}/dp_T dy} \quad . 
  \label{eq:RAA}
  \end{equation}

The striking differences of $R_{AA}(p_T)$ in central Au+Au collisions for the many particles measured by PHENIX  (Fig.~\ref{fig:Tshirt}b) illustrates the importance of particle identification for understanding the physics of the medium produced at RHIC. Most notable are: the equal suppression of $\pi^0$ and $\eta$ mesons by a constant factor of 5 ($R_{AA}=0.2$) for $4\leq p_T \leq 15$ GeV/c, with suggestion of an increase in $R_{AA}$ for $p_T > 15$ GeV/c; the equality of suppression of direct-single $e^{\pm}$ (from heavy quark ($c$, $b$) decay) and $\pi^0$ at $p_T\gsim 5$ GeV/c; the non-suppression of direct-$\gamma$ for $p_T\geq 4$ GeV/c; the exponential rise of $R_{AA}$ of direct-$\gamma$ for $p_T<2$ GeV/c~\cite{ppg086}, which is totally and dramatically different from all other particles and attributed to thermal photon production by many authors (e.g. see citations in Ref.~\cite {ppg086}). For $p_T\gsim 4$ GeV/c, the hard-scattering region,  the fact that all hadrons are suppressed, but direct-$\gamma$ are not suppressed, indicates that suppression is a medium effect on outgoing color-charged partons likely due to energy loss by coherent Landau-Pomeranchuk-Migdal radiation of gluons, predicted in p\QCD~\cite{BDMPS}, which is sensitive to properties of the medium. Measurements of two-particle correlations~\cite{EriceProcPR} confirm the loss of energy of the away-jet relative to the trigger jet in Au+Au central collisions compared to p-p collisions. However, there are still many details which remain to be understood, such as the apparent suppression of direct-$\gamma$ for $p_T\gsim\ 18$ GeV/c, approaching that of the $\pi^0$. Interesting new results have extended and clarified these observations. 

\begin{figure}[!h]
   \begin{center}
\includegraphics[width=0.49\textwidth,height=0.38\textwidth]{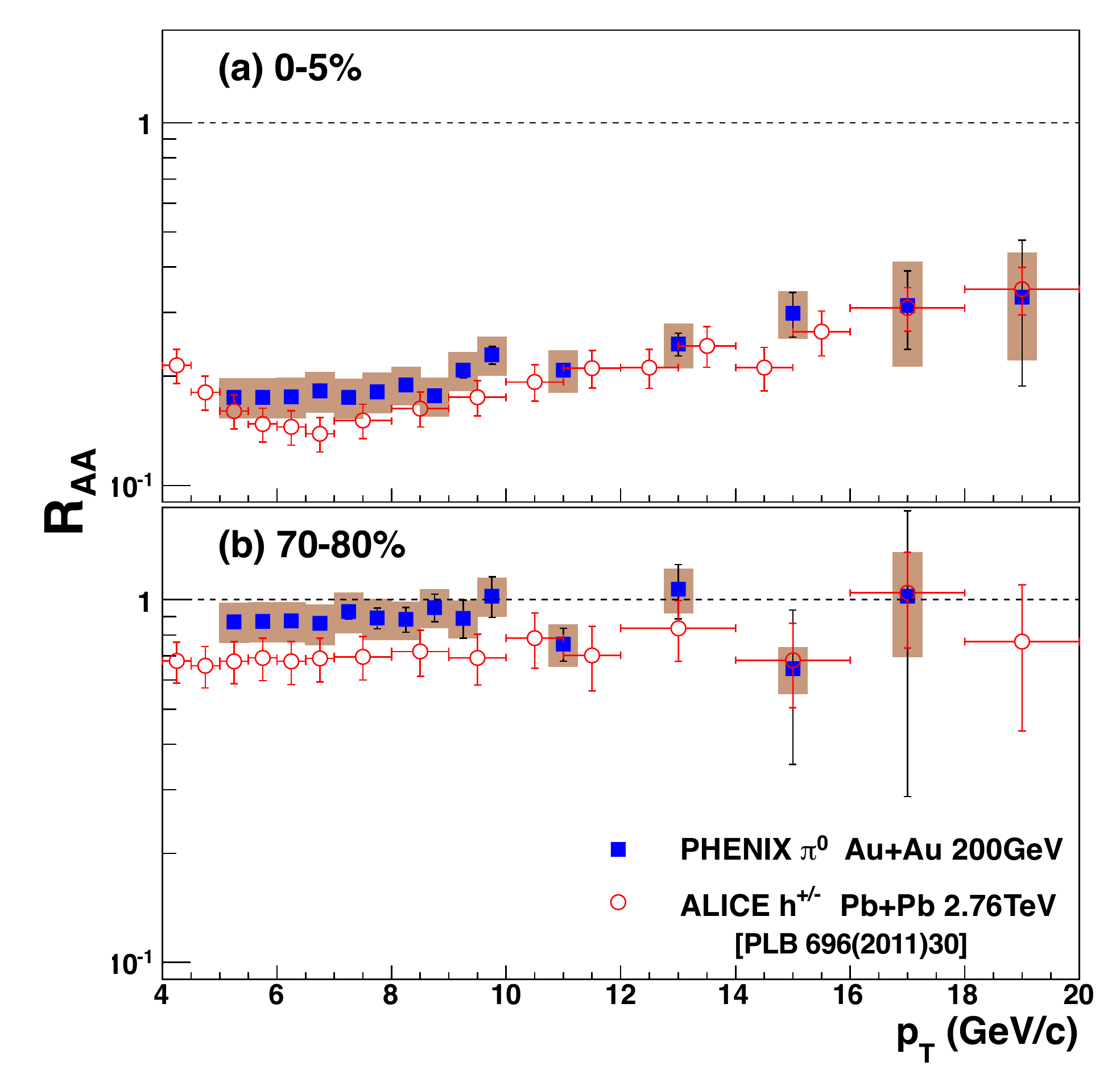}
\includegraphics[width=0.49\textwidth]{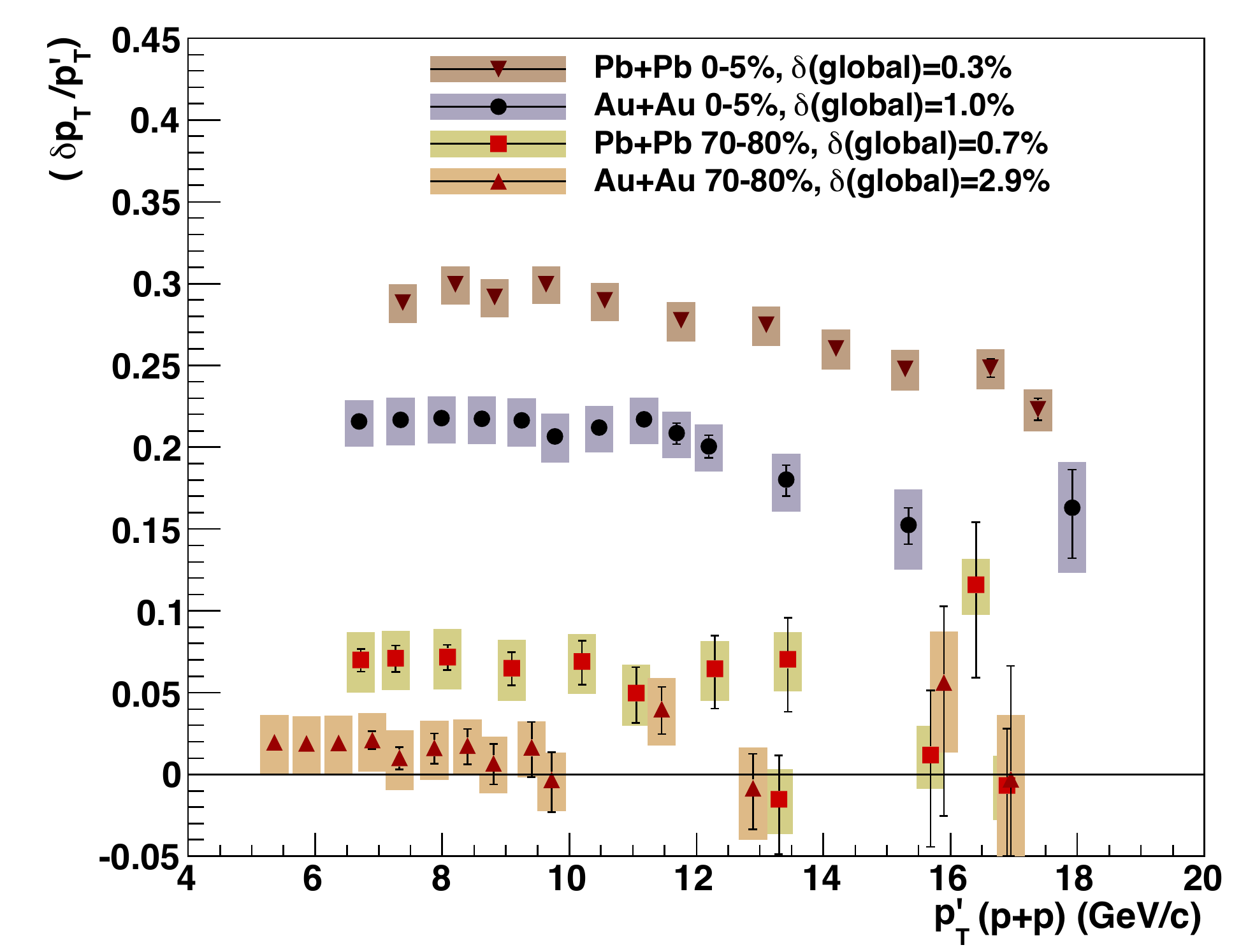}
\end{center}\vspace*{-1.0pc}
\caption[]{a) (left) $R_{AA}$ of $\pi^0$ in $\sqrt{s_{NN}}=200$ GeV central (0-5\%) and peripheral (70-80\%) Au+Au collisions~\cite{ppg133} at RHIC compared to non-identified charged hadron (${\rm h}^{\pm}$) $R_{AA}$ in $\sqrt{s_{NN}}=2.76$ TeV Pb+Pb collisions at LHC.~\cite{ALICEPLB696} b) (right) Fractional shift of $p_T$ spectrum $\delta p_T/p'_T$ vs. $p'_T$ (p-p) calculated by PHENIX~\cite{ppg133} for RHIC and LHC. 
\label{fig:ppg133}}\vspace*{-0.8pc}
\end{figure}

An improved measurement of $\pi^0$ production in Au+Au and p-p collisions by PHENIX~\cite{ppg133} now clearly shows a significant increase of $R_{AA}$ (decrease in suppression) with increasing $p_T$ over the range $7<p_T<20$ GeV/c  for 0-5\% central Au+Au collisions at $\sqrt{s_{NN}}=200$ GeV (Fig.~\ref{fig:ppg133}a).  
Comparisons of the suppression of non-identified hadrons in $\sqrt{s_{NN}}=2.76$ TeV Pb+Pb collisions at LHC to the RHIC Au+Au $\pi^0$ data are also very interesting. This is shown both in terms of the suppression, $R_{AA}(p_T)$ (Fig.~\ref{fig:ppg133}a), and the fractional shift in the $p_T$ spectrum $\delta{p_T}/p'_T$ (Fig.~\ref{fig:ppg133}b).
Interestingly, despite more than a factor of 20 higher c.m. energy, the ALICE $R_{AA}$ data from LHC~\cite{ALICEPLB696} are nearly identical to the RHIC measurement~\cite{ppg133} for $5< p_T <20$ GeV/c. Since the exponent of the power-law at LHC ($n\approx 6$) is flatter than at RHIC ($n\approx 8$), a $\sim 40$\% larger  shift $\delta p_T/p'_T$ in the spectrum from p-p to A+A is required at LHC (Fig.~\ref{fig:ppg133}b) to get the same $R_{AA}$, which likely indicates $\sim 40$\% larger fractional energy loss at LHC in this $p_T$ range due to the probably hotter and denser medium. 

\begin{figure}[!b]
   \begin{center}
\includegraphics[width=0.99\textwidth]{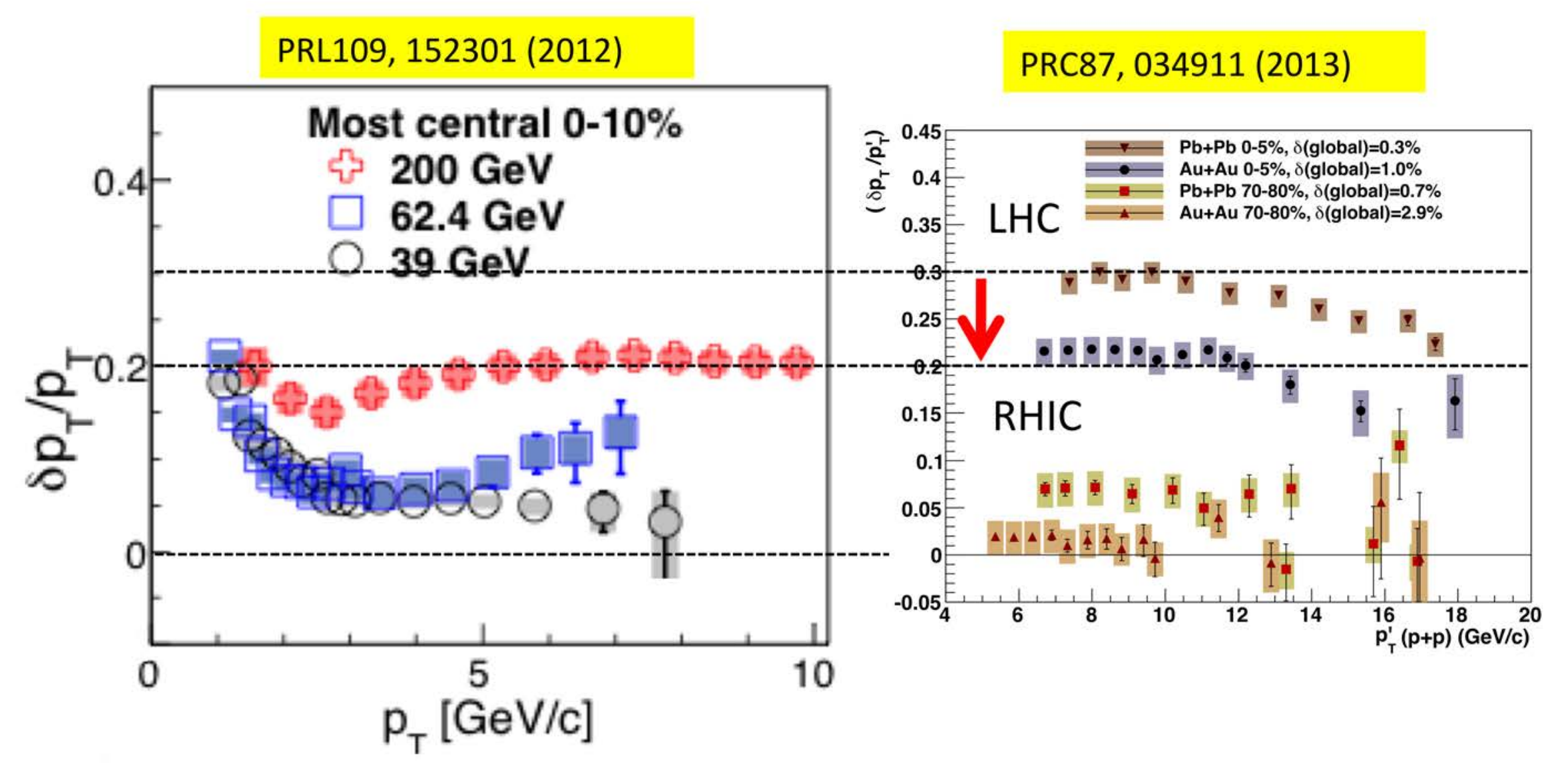}
\end{center}\vspace*{-1.0pc}
\caption[]{Fractional shift of $p_T$ spectrum $\delta p_T/p'_T$ in central A+A collisions from $\sqrt{s_{NN}}=39$~GeV to 2.76 TeV 
\label{fig:shiftallRHICLHC}}\vspace*{-0.08pc}
\end{figure}
\begin{figure}[!t]
   \begin{center}
\includegraphics[width=0.49\textwidth]{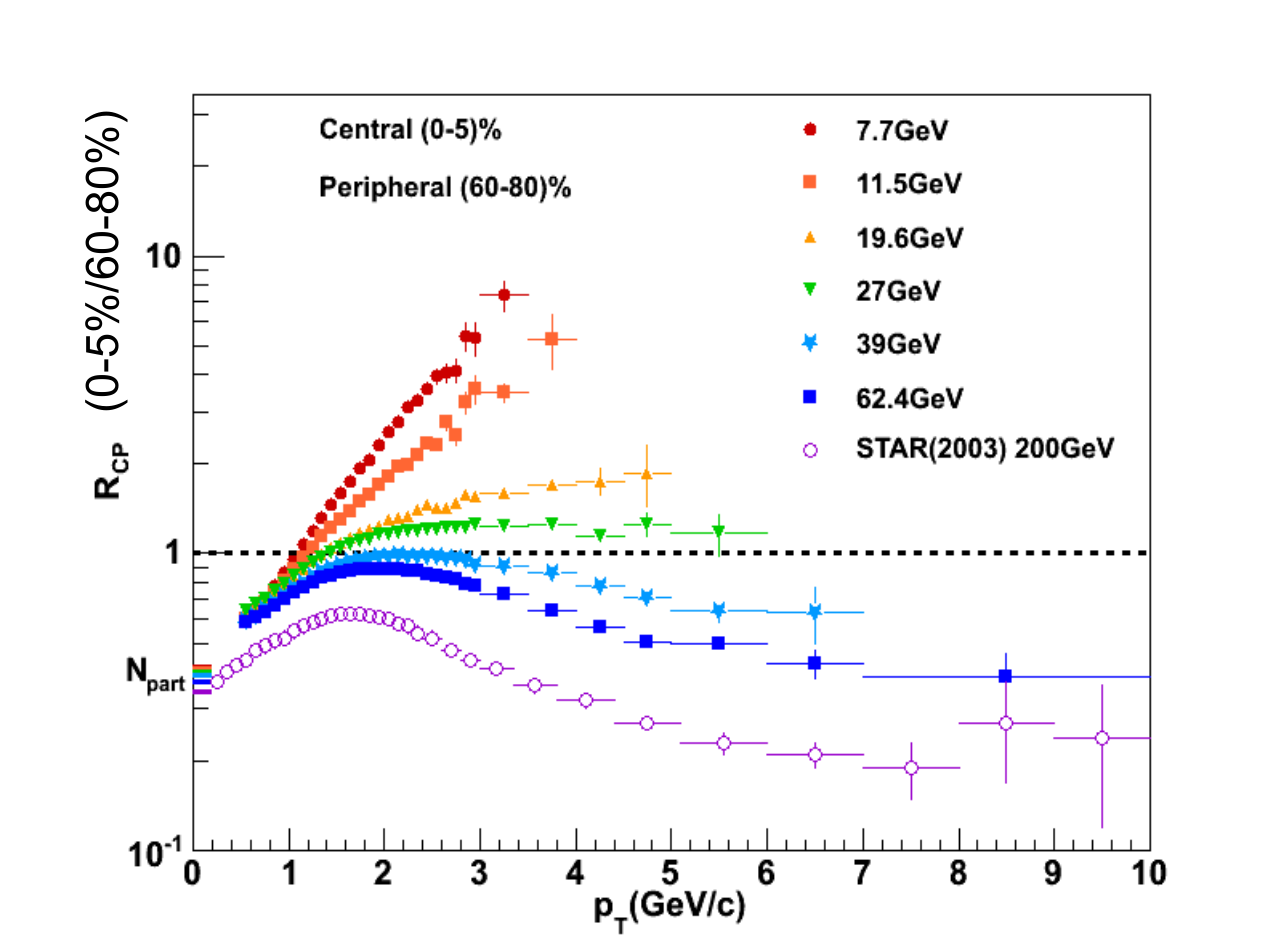}
\includegraphics[width=0.49\textwidth,height=0.30\textwidth]{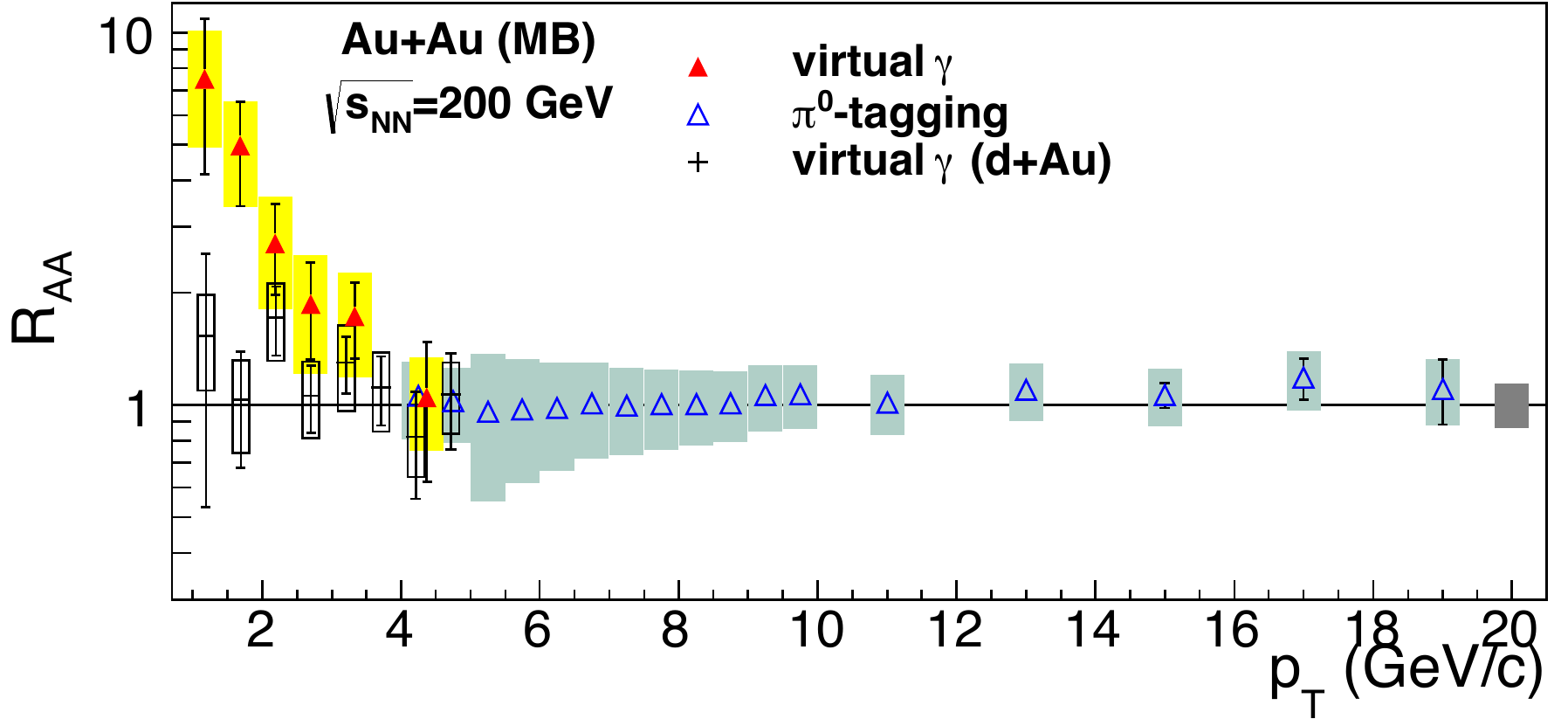}
\end{center}\vspace*{-1.0pc}
\caption[]{a) (left) STAR $R_{\rm CP} (p_T)$ for $h^{\pm}$ as a function of $\sqrt{s_{NN}}$ in Au+Au collisions~\cite{STARhiQM2012}. b)(right) PHENIX $R_{AA} (p_T)$ of direct-$\gamma$ in d+Au and Au+Au minimum bias collisions at $\sqrt{s_{NN}}=200$ GeV~\cite{ppg140}.   
\label{fig:twoRAA}}\vspace*{-1pc}
\end{figure}

In Fig.~\ref{fig:shiftallRHICLHC}, these measurements are combined with the previous measurements at RHIC for $\sqrt{s_{NN}}=39$ and 62.4 GeV~\cite{ppg138} to reveal a systematic increase of $\delta p_T/p'_T$ in central A+A collisions at $p'_T=7$ GeV/c, going from 5\% to 30\% over the c.m. energy range $\sqrt{s_{NN}}=39$~GeV to 2.76 TeV.   
Measurements by STAR (Fig.~\ref{fig:twoRAA}a)~\cite{STARhiQM2012} of the evolution of charged hadron suppression with $\sqrt{s_{NN}}$ in Au+Au collisions, using the variable $R_{CP}=(R_{AA}^{0-5\%}/R_{AA}^{60-80\%})$, which does not require the p-p cross section (see Eq.~\ref{eq:RAA}) but is usually smaller than $R_{AA}$~\cite{BRWP}, show the transition from suppression ($R_{\rm CP}<1$) to enhancement ($R_{\rm CP}>1$) for $\sqrt{s_{NN}}\ \lsim\ 27$ GeV.          

Improved measurements of direct-$\gamma$ production in p-p, dAu and Au+Au collisions by PHENIX~\cite{ppg140} show several interesting results. In Fig.~\ref{fig:twoRAA}b, new measurements of $R_{AA}\approx 1$ for d+Au using internal conversions in the thermal region, $p_T<4$ GeV,  reinforce the uniqueness of the exponential rise of the Au+Au minimum bias photon spectrum, thus confirming that the exponential for $p_T<4$ GeV/c in Au+Au is a hot matter effect, i.e. thermal photon production.   Also in Fig.~\ref{fig:twoRAA}b, improved measurements of real direct-$\gamma$ in Au+Au collisions, by eliminating background from $\gamma$-rays associated with a second $\gamma$ in the $\pi^0$ mass range ($\pi^0$ tagging), no longer show an ``apparent suppression'' but are consistent with $R_{AA}=1$ out to 20 GeV/c.

\begin{figure}[!t]
   \begin{center}
\includegraphics[width=0.49\textwidth]{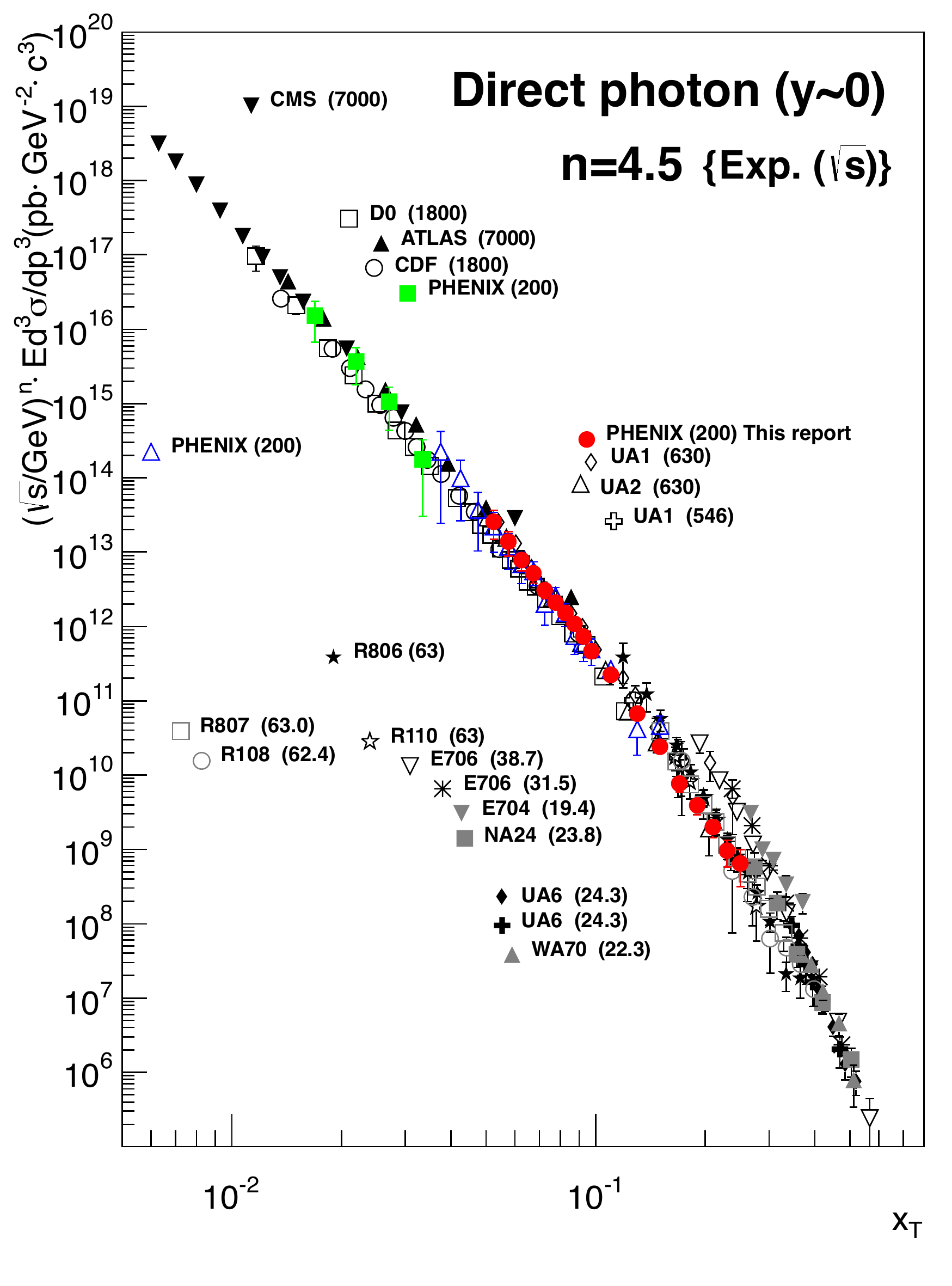}
\raisebox{0.2pc}{\includegraphics[width=0.492\textwidth]{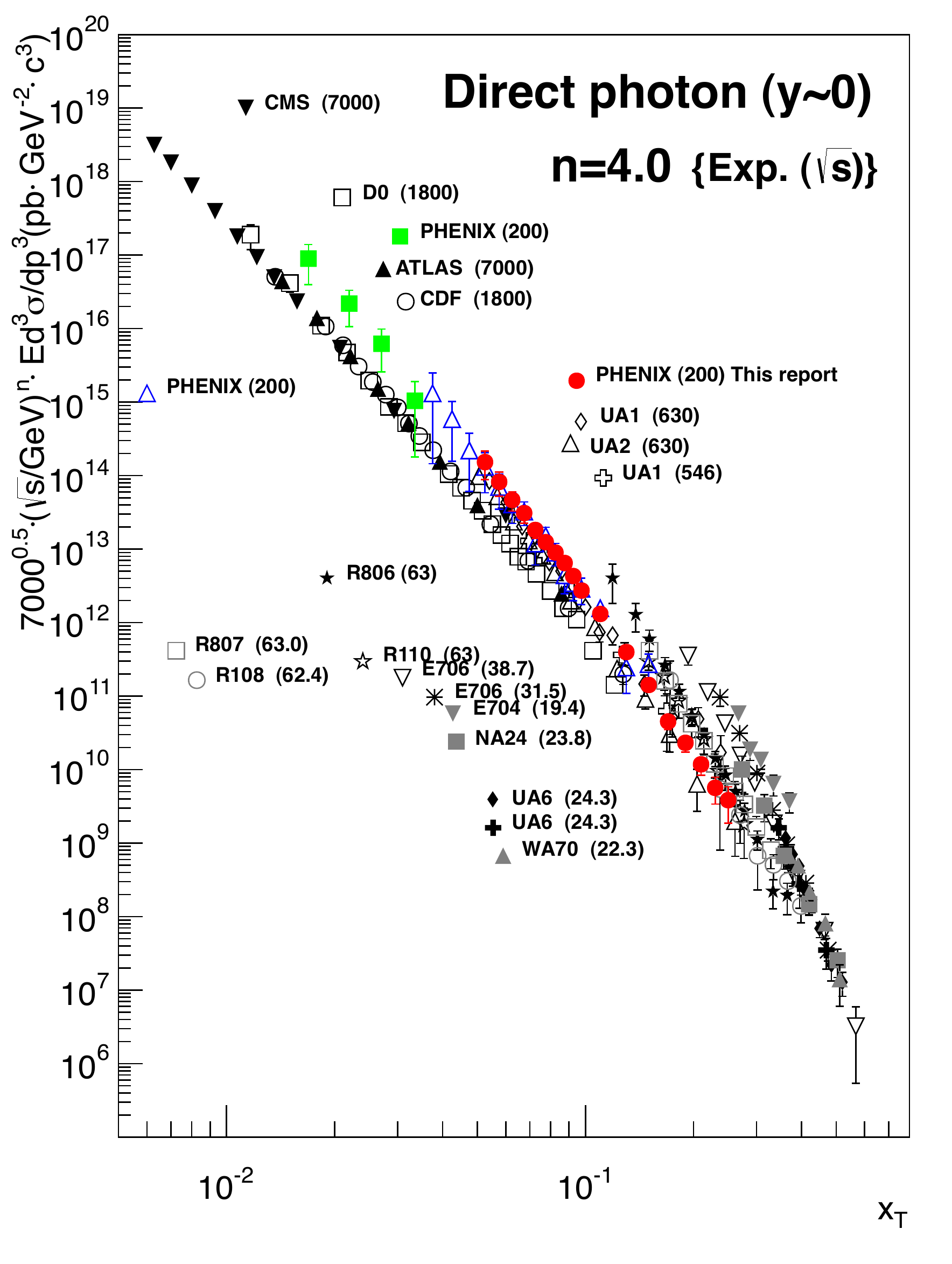}}
\end{center}\vspace*{-2pc}
\caption[]{a)(left) $\sqrt{s}^{\,n_{\rm eff}} \times E d^3\sigma/dp^3$, as a function of $x_T=2p_T/\sqrt{s}$, with $n_{\rm eff}=4.5$,  for direct-$\gamma$ measurements in p-p and $\bar{\rm p}$-p experiments at the ($\sqrt{s}$ GeV) indicated.~\cite{ppg136}. b)(right) same as (a) with $n_{\rm eff}=4.0$  
\label{fig:ggg}}\vspace*{-1pc}
\end{figure}

My favorite direct-$\gamma$ result in 2012 was the improved PHENIX measurement in p-p collisions at $\sqrt{s}=200$ GeV out to $p_T=25$ GeV/c~\cite{ppg136}, in excellent agreement with p\QCD. A more direct way to show this without a detailed theory calculation is to use $x_T$ scaling. Figure~\ref{fig:ggg}a shows $x_T$ scaling for all presently existing direct-$\gamma$ data\footnote{This includes the PHENIX p-p direct-virtual-$\gamma$ measurement down to $p_T\approx1$ GeV/c, further confirming the absence of a soft production mechanism for direct-$\gamma$ in p-p collisions~\cite{EriceProcPR}.}, with $n_{\rm eff}=4.5$, very close to the pure-scaling parton-parton Rutherford scattering result of $n_{\rm eff}=4.0$ (Fig.~\ref{fig:ggg}b). The deviation of the data points in Fig.~\ref{fig:ggg}b from the universal curve for $\sqrt{s}>38.7$ GeV is an illustration of the non-scaling of the coupling constant, structure and fragmentation functions  in \QCD---what I like to call ``\QCD\ in action". For $\sqrt{s}\leq 38.7$ GeV, the deviation of the data from the universal curve in Fig.~\ref{fig:ggg}a (and from p\QCD) is claimed to be due to the $k_T$ effect (transverse momentum of the quarks in a nucleon).
%\subsection{Two-particle correlations}

\subsection{Two-Particle Correlations and Fragmentation Functions} 
The key to measuring the fragmentation function of the jet of particles from a hard-scattered parton  is to know the energy of the original parton which fragments, as pioneered at LEP~\cite{Tingpizff}. Since the $p^{\gamma}_{T}$ of a direct-$\gamma$ can be measured very precisely, the fragmentation function of the jet from the away quark in the reaction $g+q\rightarrow \gamma +q$ can be measured by the direct-$\gamma-h$ correlations (where $h$ represents charged hadrons opposite in azimuth to the direct-$\gamma$) because the  $p_T$ of the away-quark at production is equal and opposite to $p^{\gamma}_{T}$, thus known to high precision (modulo a small $k_T$-smearing effect). 
A further advantage is that the identity of the away-quark fragmenting to the jet is also known to reasonable precision:  8/1 $u$-quark, in p+p collisions (maybe 8/2 if the $\bar{q}+q\rightarrow \gamma+g$ channel is included) and $\sim 3/1$ in A+A collisions (not counting the $\bar{q}+q\rightarrow \gamma+g$ channel). The main disadvantage is the low rate since the $\gamma-q$ vertex is electromagnetic.     
     	%Following the terminology of the pioneering hadron-collider measurements~\cite{DarriulatNPB107},
		Two-particle correlations are analyzed in terms of the two variables~\cite{ppg029}: $p_{\rm out}=p_T \sin(\Delta\phi)$, the out-of-plane transverse momentum of a track;  
 and $x_E$, where:\\ 
 \begin{minipage}[c]{0.48\textwidth}
\vspace*{-0.30in}
\[%\begin{equation}
x_E=\frac{-\vec{p}_T\cdot \vec{p}_{Tt}}{|p_{Tt}|^2}=\frac{-p_T \cos(\Delta\phi)}{p_{Tt}}\simeq \frac {z}{z_{\rm trig}}  
\]%\end{equation}
\vspace*{0.06in}
\end{minipage}
\hspace*{0.01\textwidth}
\begin{minipage}[b]{0.48\textwidth}
%\begin{figure}[!h] %%was 0.55--.0.6
\vspace*{0.06in}
\includegraphics[scale=0.5]{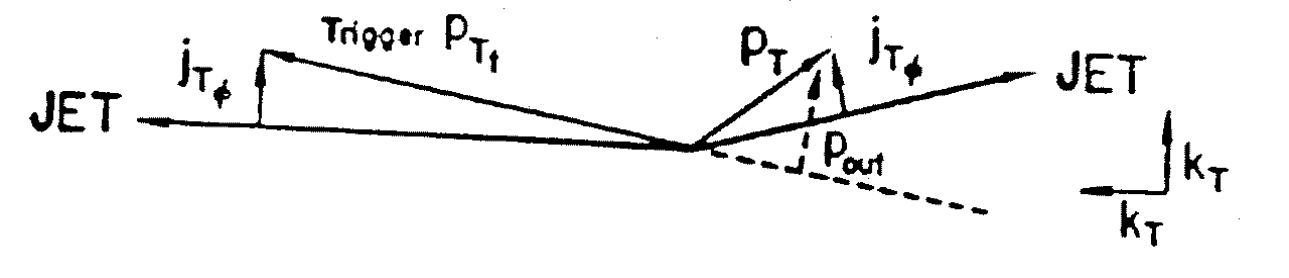}
\vspace*{-0.12in}
\label{fig:mjt-poutxe} \nonumber
%\end{figure}
%\addtocounter{figure}{-1}
\end{minipage}%\vspace*{-0.12in}

\noindent $z_{\rm trig}\simeq p_{Tt}/p_{T{\rm jet}}$ is the fragmentation variable of the trigger jet, and $z$ is the fragmentation variable of the away jet. Note that $x_E$ would equal the fragmenation fraction $z$ of the away jet, for $z_{\rm trig}\rightarrow 1$, if the trigger and away jets balanced transverse momentum. The beauty of direct-$\gamma$ for this purpose is that $z_{\rm trig}\equiv 1$. 

Following the approach of
Borghini and Wiedemann~\cite{BW06} who predicted the medium modification of fragmentation functions in the hump-backed or $\xi=\ln(1/z)$ representation, PHENIX measured $x_E$ distributions in p-p collisions~\cite{ppg095} and converted them to the $\xi=-\ln\, x_E$       \begin{figure}[!h] 
      \centering
\raisebox{0pc}{\begin{minipage}[b]{0.45\linewidth} \includegraphics[width=\linewidth]{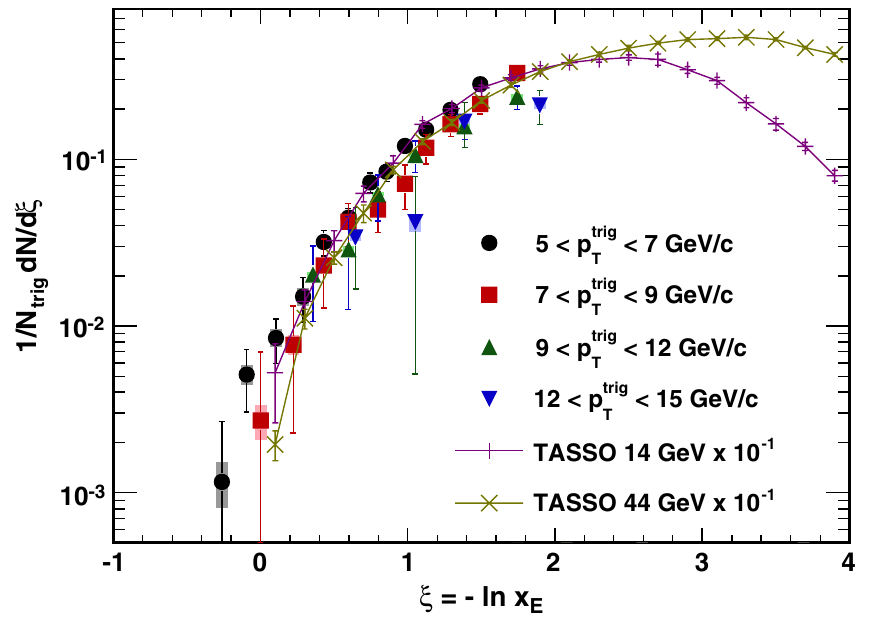} 
\caption[] {a) (left) PHENIX measurement~\cite{ppg095}  of direct-$\gamma$-h correlations in p-p collisions at \sqs=200 GeV in the variable \mbox{$\xi=-\ln x_E\approx-\ln(p_T^{h}/p_T^{\gamma})$} compared to fragmentation functions measured in $e^+ e^-$ collisions at \sqs=14 and 44 GeV by TASSO~\cite{TASSOZPC47}. b)(right)-(top) $\xi$ distributions of direct-$\gamma$-h correlations in Au+Au and p-p collisions at \sqs=200 GeV. c) (right)-(bottom) Ratio of the Au+Au/p-p distributions, $I_{AA}(\xi)$ when the away side azimuthal range is restricted as indicated~\cite{PXPRL111gamh}}\hspace*{2pc}
       \label{fig:PXgamma-h} \end{minipage}}
\raisebox{0.0pc}{\begin{minipage}[b]{0.49\linewidth} \includegraphics[width=\linewidth]{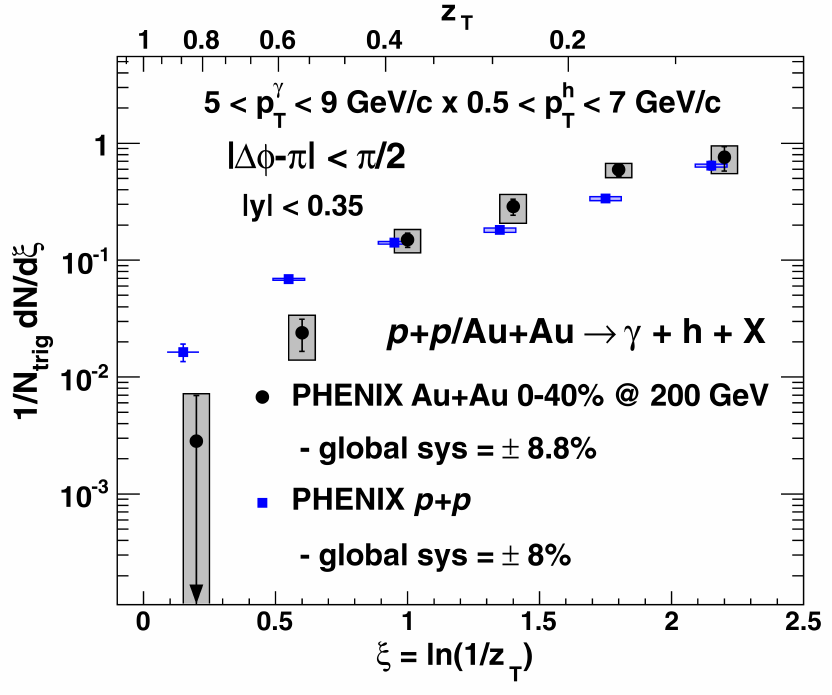} 
\includegraphics[width=\linewidth]{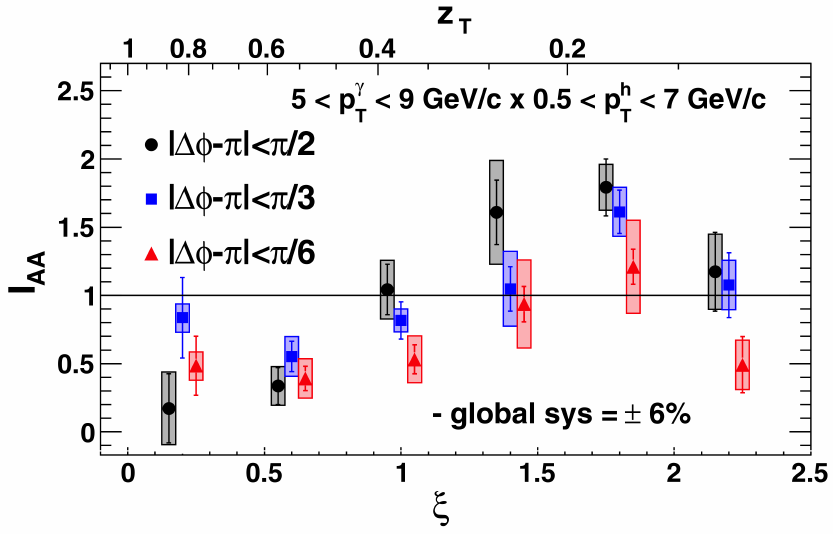} \end{minipage}}\vspace*{-1.0pc}
   \end{figure}
   representation (Fig.~\ref{fig:PXgamma-h}a) which are in quite excellent agreement with the dominant $u$-quark fragmentation functions measured in $e^+ e^-$ collisions at $\sqrt{s}/2=7$ and 22 GeV~\cite{TASSOZPC47}, which cover a comparable range in jet energy. 
   In 2013, improved measurements by PHENIX~\cite{PXPRL111gamh} in both p-p and Au+Au collisions (Fig.~\ref{fig:PXgamma-h}b) now indicate a significant modification of the fragmentation function in Au+Au (0-40\%) central collisions compared to p-p , with an enhancement at low $z_T=p^{h}_{T}/p^{\gamma}_{T}$ (large $\xi=-\ln z_T$)  and a suppression at large $z_T$ (small $\xi$)  which is more clearly seen as $I_{AA}(\xi)$, the ratio of the fragmentation functions in Au+Au/pp (Fig.~\ref{fig:PXgamma-h}c). As shown in Fig.~\ref{fig:PXgamma-h}c, restricting the away-side azimuthal range reduces the large $\xi>0.9$ ($p^{h}_{T}\lsim 3$ GeV/c) enhancement but leaves the suppression at small $\xi<0.9$  relatively unchanged, which shows that the large $\xi$ enhancement is predominantly at large angles, similar to the effect observed by CMS with actual jets.~\cite{CMSQM2012}. 
Fragmentation functions from full jet reconstruction in A+A collisions are not yet available at RHIC. 

\subsection{Two-Particle Correlations and Jet Imbalance}
\begin{figure}[!t]
\includegraphics[height=0.45\textwidth]{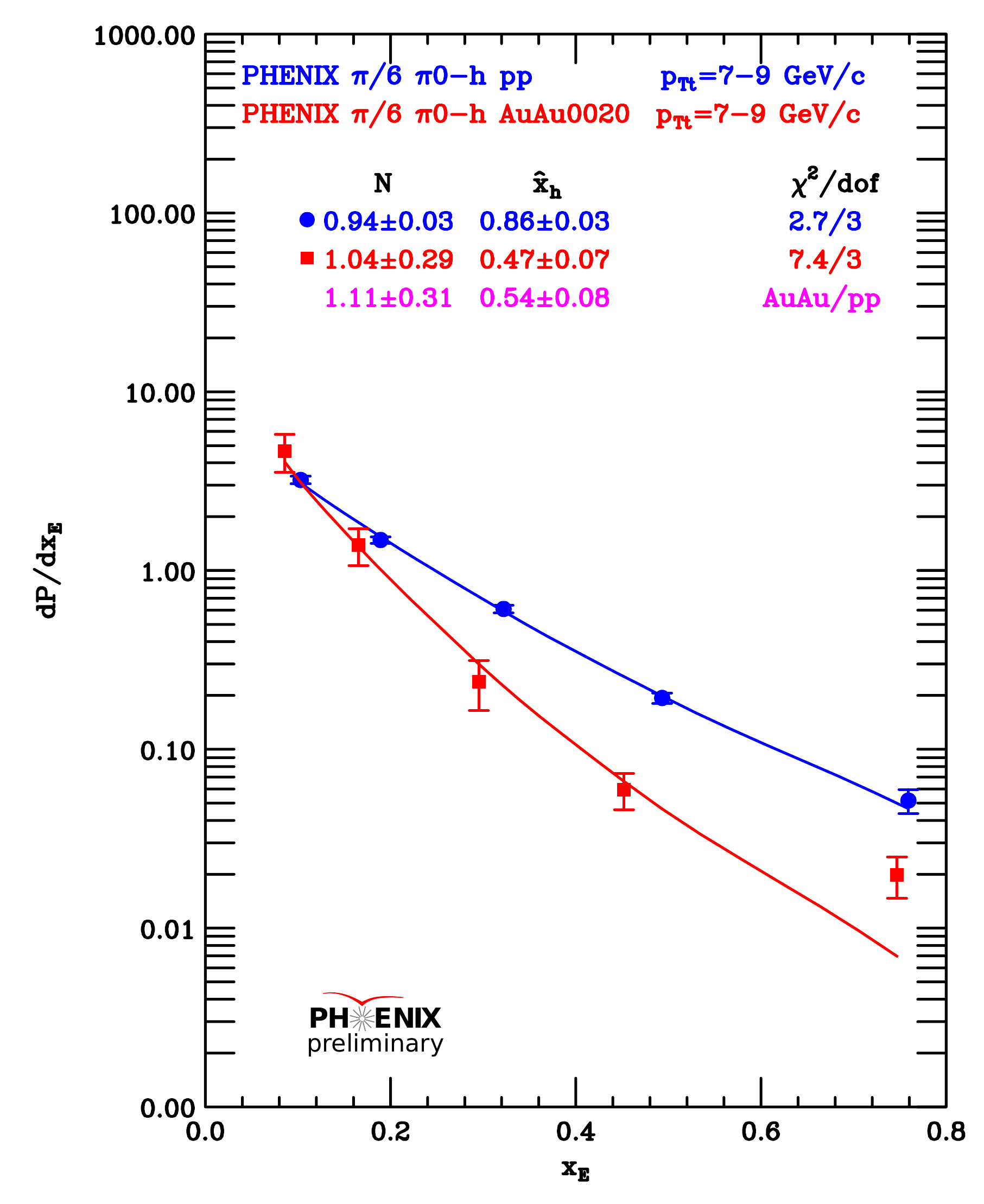}
\includegraphics[height=0.45\textwidth]{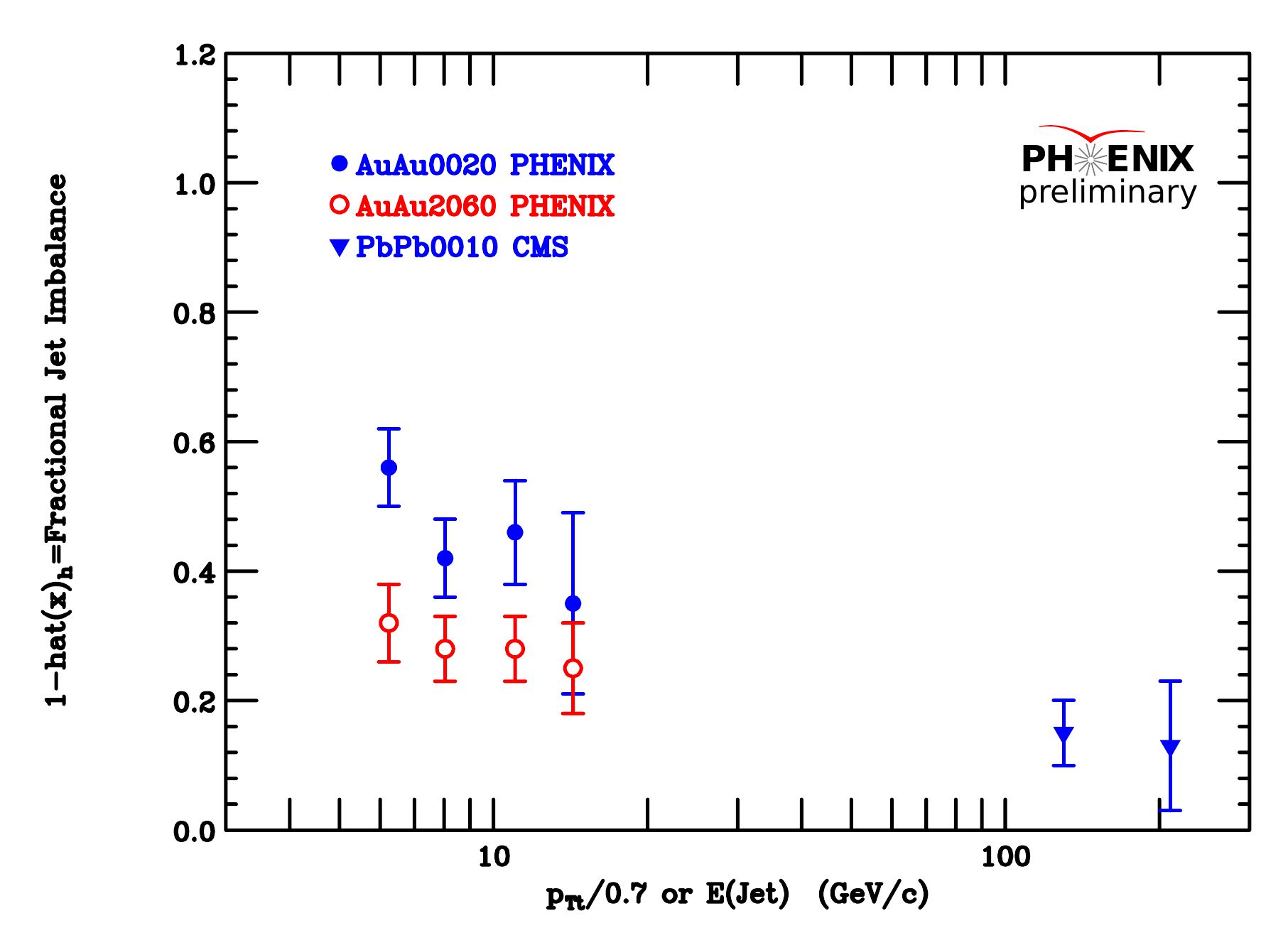}
\caption{(left) a) $x_E$ distributions~\cite{MJT-Utrecht} from p-p (circles) and AuAu 0-20\% centrality (squares) for $p_{T_t}=7-9$ GeV/c, together with fits to Eq.~\ref{eq:condxeN2} (solid lines) with parameters indicated.  The ratios of the fitted parameters for AuAu/pp are also given. b) (right) Fractional jet imbalance~\cite{MJT-Utrecht}, \hbox{$1-\hat{x}_h^{AA}/\hat{x}_h^{pp}$}, for RHIC and CMS data. }
\label{fig:AuAupp79}%\vspace*{-1.0pc}
\end{figure}

One of the important lessons learned at RHIC~\cite{ppg029} about fragmentation functions is that the away-side $x_E$ distribution of particles opposite to a trigger particle (e.g. a $\pi^0$), which is itself the fragment of a jet, does not measure the fragmentation function, but, instead, measures the ratio of $\hat{p}_{T_a}$ of the away-parton to $\hat{p}_{T_t}$ of the trigger-parton and depends only on the same power $n$ as the invariant single particle spectrum:  
		     \begin{equation}
\left.{dP \over dx_E}\right|_{p_{T_t}}\approx {N\,(n-1)}{1\over\hat{x}_h} {1\over
{(1+ {x_E \over{\hat{x}_h}})^{n}}} \, \qquad . 
\label{eq:condxeN2}
\end{equation}
This equation gives a simple relationship between the ratio, $x_E\approx p_{T_a}/p_{T_t}\equiv z_T$, of the transverse momenta of the away-side particle to the trigger particle, and the ratio of the transverse momenta of the away-jet to the trigger-jet, $\hat{x}_{h}=\hat{p}_{T_a}/\hat{p}_{T_t}$. PHENIX measurements~\cite{ppg106} of the $x_E$ distributions of $\pi^0$-h correlations in p-p and Au+Au collisions at $\sqrt{s_{NN}}=200$ GeV were fit to Eq.~\ref{eq:condxeN2} (Fig.~\ref{fig:AuAupp79}a)~\cite{MJT-Utrecht}. 
The steeper distribution in Au+Au shows that the away parton has lost energy. 
The results for the fitted parameters are shown on the figure. 

\begin{figure}[!t]
\begin{center}
\includegraphics[width=0.66\textwidth]{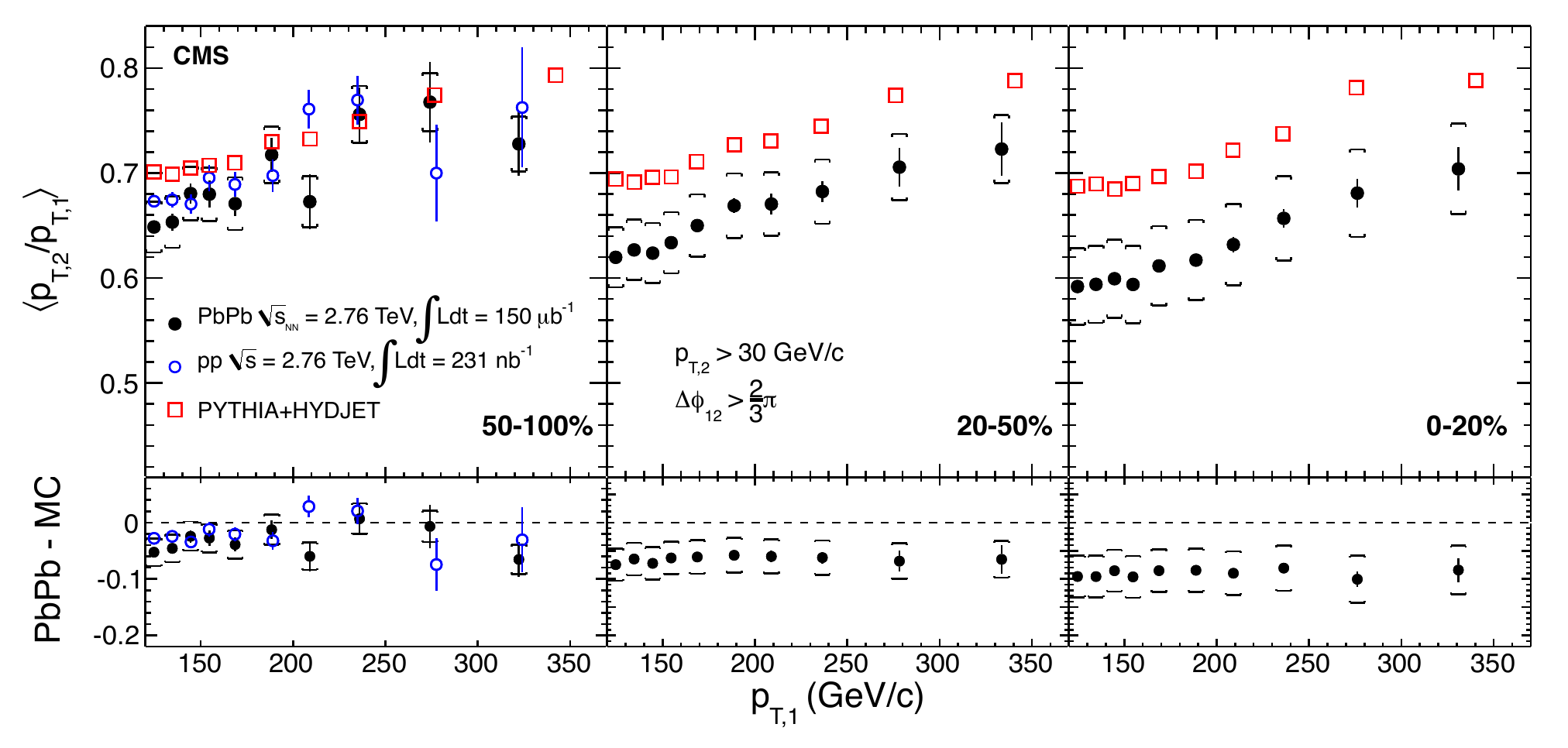}\hspace*{0.1pc}
\raisebox{1pc}{\includegraphics[height=0.24\textwidth]{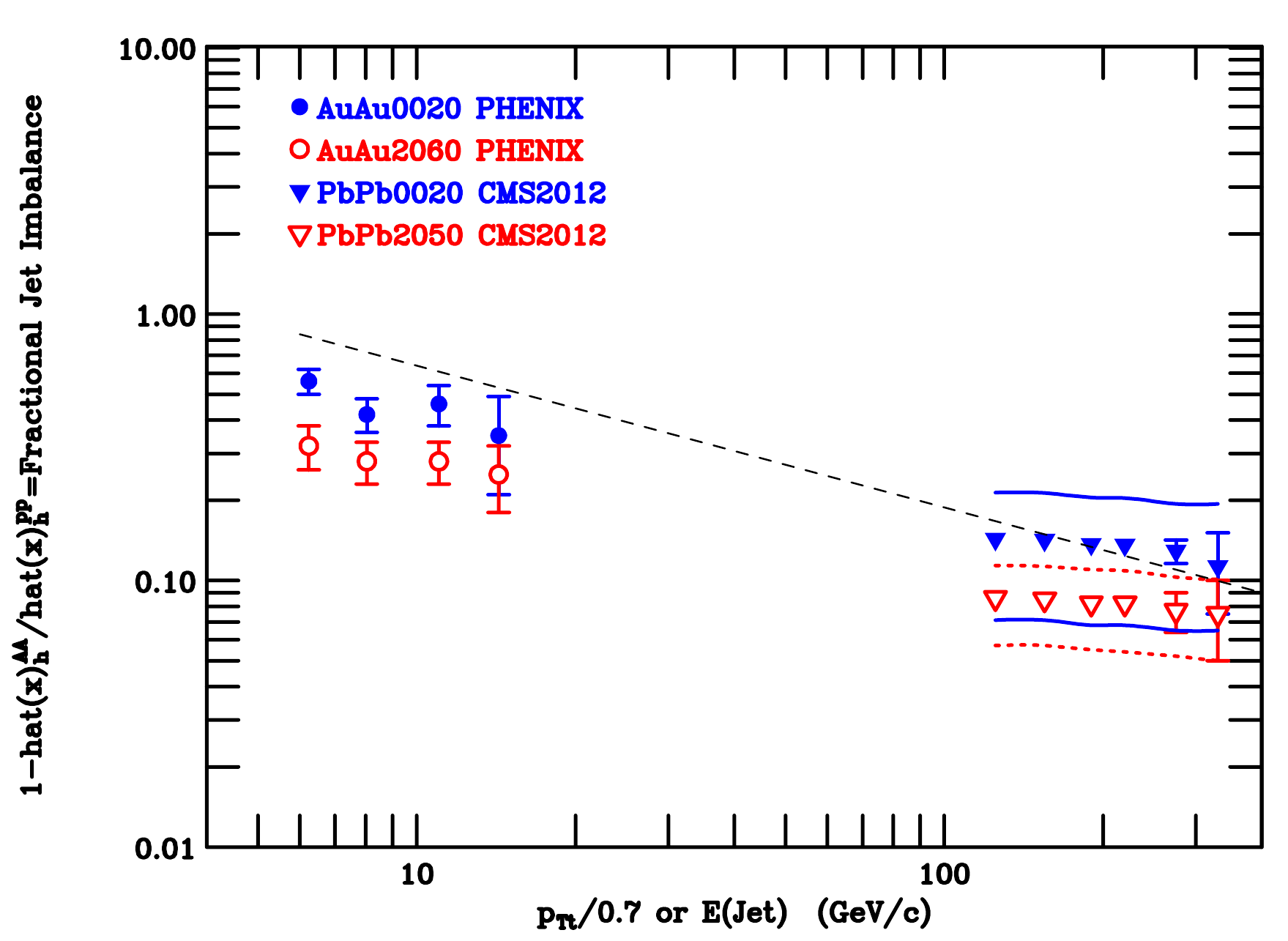}}
\end{center}\vspace*{-0.2pc}
\caption[]{a) (left 3 panels) CMS~\cite{CMS-dijet-PLB712} measurements of average di-jet transverse momentum ratio,  $\hat{x}_h=\hat{p}_{T,2}/\hat{p}_{T,1}$, as a function of leading jet $\hat{p}_{T,1}$ at $\sqrt{s_{NN}}=2.76$ TeV in p-p collisions and for 3 centralities in Pb+Pb collisions, as well as simulated p-p di-jets embedded in heavy ion events. b) (right) Fractional jet imbalance~\cite{MJT-Utrecht}, $1-\hat{x}_h^{AA}/\hat{x}_h^{pp}$, for the RHIC data from Fig.~\ref{fig:AuAupp79} with CMS measurement from (a). The solid (dotted) lines represent the systematic uncertainty of the CMS 0-20\% (20-50\%) results. The dashed line is my estimate of the Fractional Jet Imbalance vs. E(Jet) at \sqsn=2.76 TeV made the previous year from Fig.~\ref{fig:AuAupp79}b. }   
\label{fig:CMSdijet2012}
\end{figure}

In general the values of $\hat{x}^{pp}_h$ do not equal 1 but range between $0.8<\hat{x}^{pp}_h<1.0$ due to $k_T$ smearing and the range of $x_E$ covered. In order to take account of the imbalance ($\hat{x}^{pp}_h <1$) observed in the p-p data, the ratio $\hat{x}_h^{AA}/\hat{x}_h^{pp}$ is taken as the measure of the energy of the away jet relative to the trigger jet in A+A compared to p-p collisions. The fractional jet imbalance was also measured directly with reconstructed di-jets by the CMS collaboration at the LHC in Pb+Pb central collisions at $\sqrt{s_{\rm NN}}=2.76$ TeV~\cite{CMSdijet}. There was also a large imbalance  in p-p collisions due to the cuts on jet $\hat{p}_T$ that were used. I calculated $\hat{x}_h$ from their p-p and Pb+Pb results and compared these LHC values of $1-\hat{x}_h^{AA}/\hat{x}_h^{pp}$ to those from PHENIX (Fig.~\ref{fig:AuAupp79}b).  
Newer results in 2012 by CMS (Fig.~\ref{fig:CMSdijet2012})~\cite{CMS-dijet-PLB712} significantly extended and improved their previous measurement and confirmed my correction~\cite{MJT-Utrecht}. 

In Figs.~\ref{fig:AuAupp79}b and~\ref{fig:CMSdijet2012}b, there is a large difference in fractional jet imbalance in the different $\hat{p}_T$ ranges covered by the RHIC and LHC measurements. This could be due to the difference in jet $\hat{p}_{T_t}$ between RHIC ($\sim 20$ GeV/c) and LHC ($\sim 200$ GeV/c), a difference in the properties of the medium, the difference in $n$ for the different $\sqrt{s_{NN}}$, or a problem with Eq.~\ref{eq:condxeN2} which has not been verified by direct comparison to di-jets. In any case the strong $\hat{p}_T$ dependence of the fractional jet imbalance (apparent energy loss of a parton) also seems to disfavor purely radiative energy-loss in the \QGP~\cite{BDMPS} and indicates that the details of energy loss in a \QGP\ remain to be understood.  Future measurements at both RHIC and LHC will need to sort out these issues by extending di-jet and two-particle correlation measurements to overlapping regions of $\hat{p}_T$.   

\section{Anisotropic flow ({$v_2$}) of direct-{$\gamma$} }
Although direct-$\gamma$ production~\cite{QCDCompton} is the most beautiful \QCD\  subprocess, it has a very serious problem: an overwhelming background of photons from high $p_T$ $\pi^0\rightarrow \gamma+\gamma$ and $\eta\rightarrow \gamma+\gamma$  decays  makes it a very difficult experiment.  One must measure all the background sources: $\pi^0$, $\eta$, \ldots, and calculate their contributions to the inclusive $\gamma$-ray spectrum. In principle, the background can be calculated whatever the $p_T$ distribution of the $\pi^0$ and $\eta$. However nature has been kind in that the invariant cross section for hard-scattering is a power law, $d\sigma/p_T dp_T\propto 1/p_T^n$, with $n=8.1\pm0.05$ at $\sqrt{s_{NN}}=200$ GeV, for $\pi^0$ with $p_T\geq 3$ GeV/c (Fig.~\ref{fig:Tshirt}a)~\cite{ppg054}; also for $p_T\geq 3$ GeV/c, $\eta/\pi^0=0.48\pm 0.03$ is a constant. This implies that for $\pi^0\rightarrow \gamma +\gamma$, the spectrum of decay photons has the same power as the parent $\pi^0$ so that the ratio at any $p_T$ is a constant: 
\begin{equation}
\frac{\gamma}{\pi^0} \biggm |_{\pi^0}=\frac{2}{n-1} \qquad.
\label{eq:2nm1}
\end{equation}
The resulting background inclusive $\gamma$ spectrum from $\pi^0$ and $\eta$ decays at $\sqrt{s_{NN}}=200$ GeV is: 
\begin{equation} \gamma_{\rm background}/\pi^0 \approx (1+0.48\times0.39)\times 2/7.1 = 1.19\times2/7.1 = 0.334 \label{eq:gbkg} 
\end{equation}
where 0.39 is the branching ratio for $\eta\rightarrow \gamma +\gamma$. 
In PHENIX we plot what we call the double ratio: 
 \[R_{\gamma}=(\gamma_{\rm inclusive}/\pi^0)/(\gamma_{\rm background}/\pi^0)=\gamma_{\rm inclusive}/\gamma_{\rm background} \]
where it is important to see the calculated $\gamma_{\rm background}/\pi^0$ ratio, which usually comes from some opaque Monte Carlo program, to understand whether it makes sense according to Eqs.~\ref{eq:2nm1} and \ref{eq:gbkg}. 

\begin{figure}[!h]
\begin{center}
\includegraphics[width=0.95\textwidth]{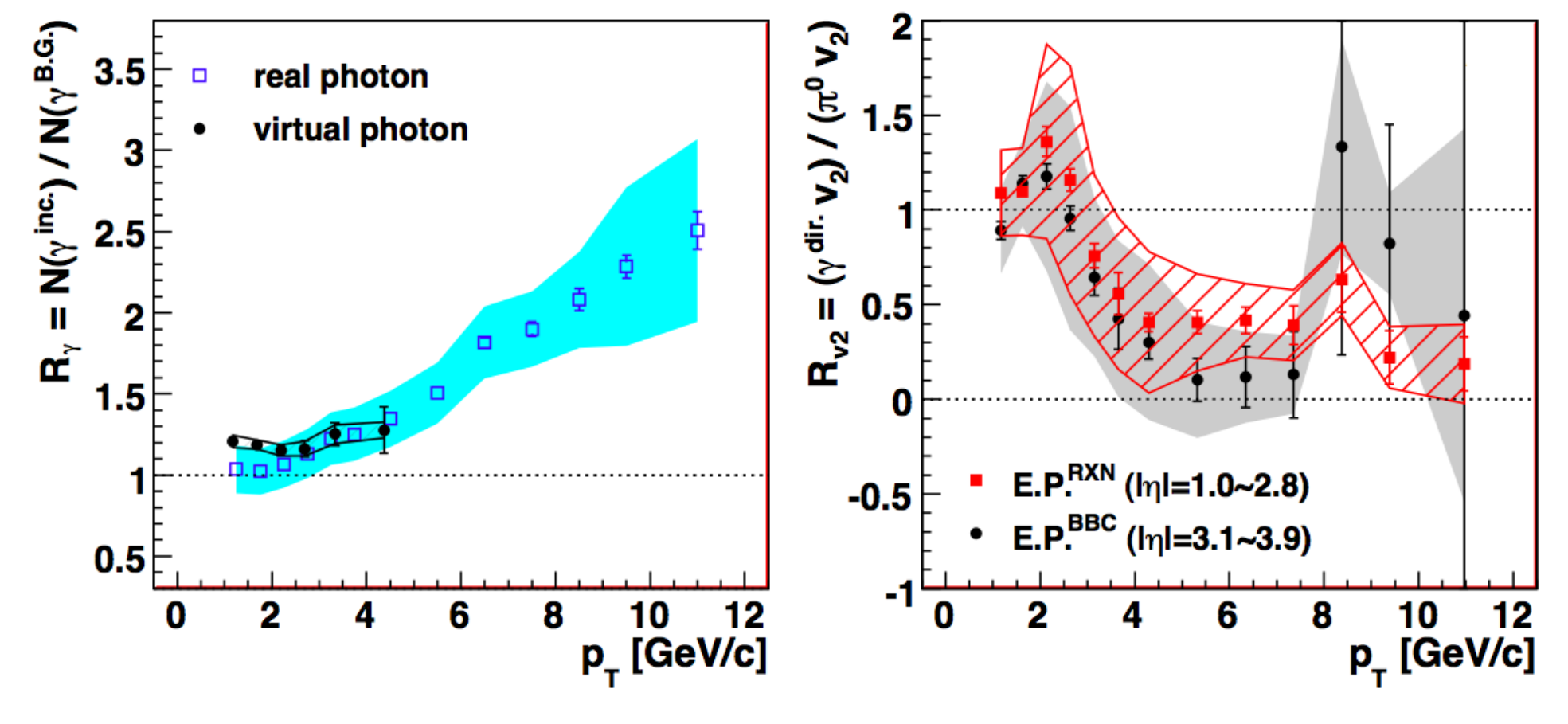}
\end{center}\vspace*{-2.0pc}
\caption[]{a) (left) $R_{\gamma}$ vs $p_T$~\cite{ppg126} for 
virtual photons (solid circles)~\cite{ppg086} and  
real photons (open squares)~\cite{ppg042} for minimum bias Au+Au collisions at $\sqrt{s_{NN}}=200$ GeV. b) (right) Ratio of direct-$\gamma$ $v_2$ to $\pi^0$ $v_2$ for two reaction plane detectors~\cite{ppg126}. }
\label{fig:dirgv2}
\end{figure}

Fig.~\ref{fig:dirgv2}a~\cite{ppg126}  
shows $R_{\gamma}$ for real photons, measured in an EM calorimeter, and virtual photons, which are $e^+ e^-$ pairs from internal conversion of the direct-$\gamma$, with $0.12< m_{ee}<0.30$ GeV/c$^2$ where there is no background from $\pi^0$ Dalitz decay. This reduces the background by a factor of $\geq1.19/0.19\approx 6$~\cite{egseeMJTISSP2009}, and allows the precision of $R_{\gamma}$ to be greatly improved as shown. 
Then, using the precise virtual photon $R_{\gamma}$     
with the much higher statistics inclusive real-$\gamma$ data, one can derive $v_2$ for direct-$\gamma$ from the measured $v_2$ of inclusive real-$\gamma$ compared to the measured $v_2$ of $\gamma$'s from $\pi^0$ and $\eta$ decay. 

The result (Fig.~\ref{fig:dirgv2}b)~\cite{ppg126} is that the $v_2$ of direct-$\gamma$ is large in the range $1\leq p_T\leq 3$ GeV/c but drops to zero for $p_T \geq 5$ GeV/c where the photons are produced from initial hard-scattering and do not interact with the medium so that they do not flow. Since thermal radiation is produced in the flowing medium, the observed large $v_2$ in the range $1\leq p_T\leq 3$ GeV/c, where the direct-$\gamma$ $p_T$ spectrum is exponential (Fig.~\ref{fig:Tshirt}b), confirms that these $\gamma$ are thermal radiation from the medium. What is very surprising is that the $v_2$ of the thermal photons is so large, the same or slightly greater than that of $\pi^0$'s. 

\section{A charming surprise}
\begin{figure}[!ht]
\begin{center} %\vspace*{-1pc}
\begin{tabular}{cc}
\hspace*{-0.02\linewidth}\includegraphics*[width=0.51\linewidth]{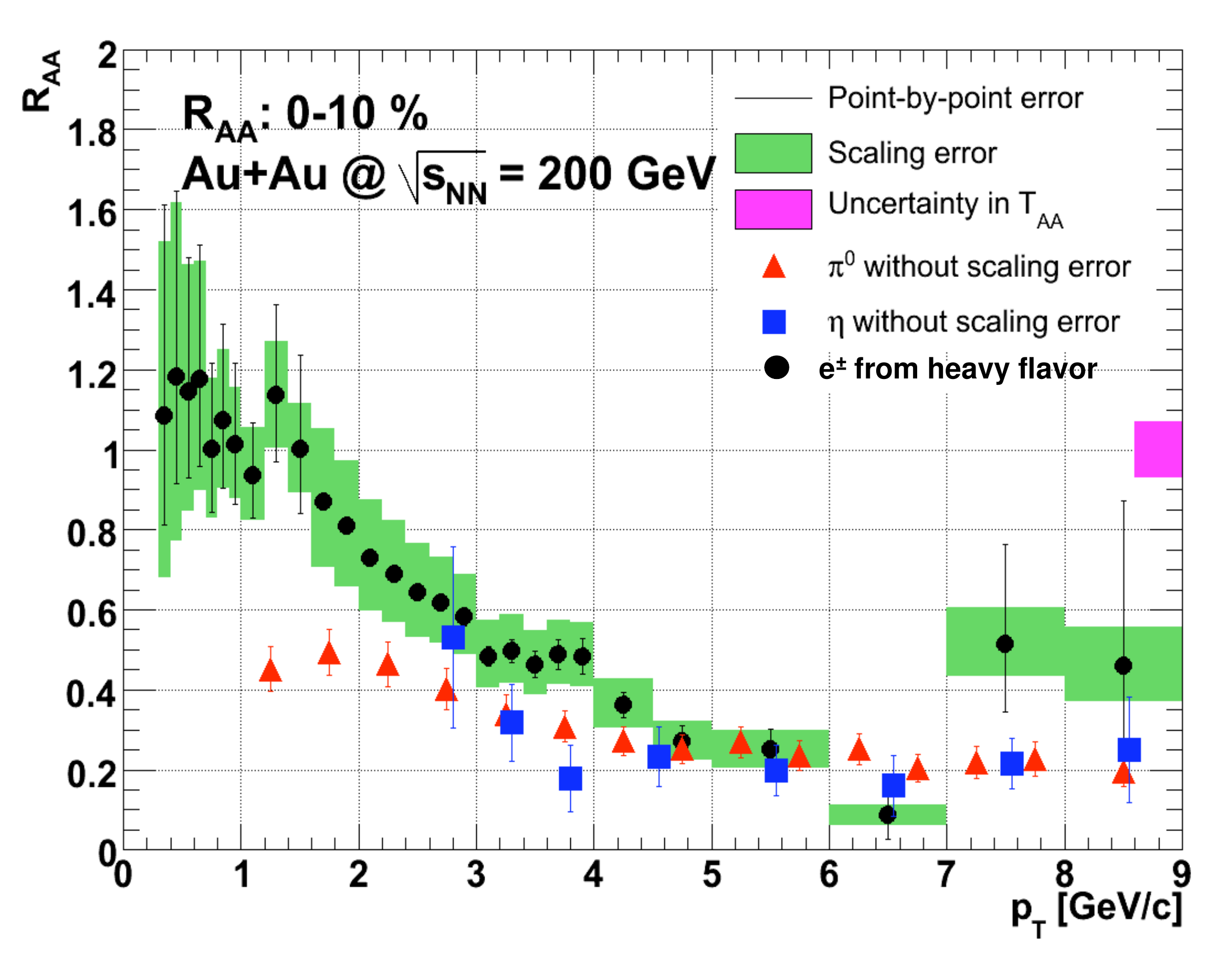} & 
%%\hspace*{-0.08\linewidth}\includegraphics[width=0.51\linewidth]{figs11/fig3se_data_only_logo.eps} 
\hspace*{-0.02\linewidth}\includegraphics[width=0.50\linewidth]{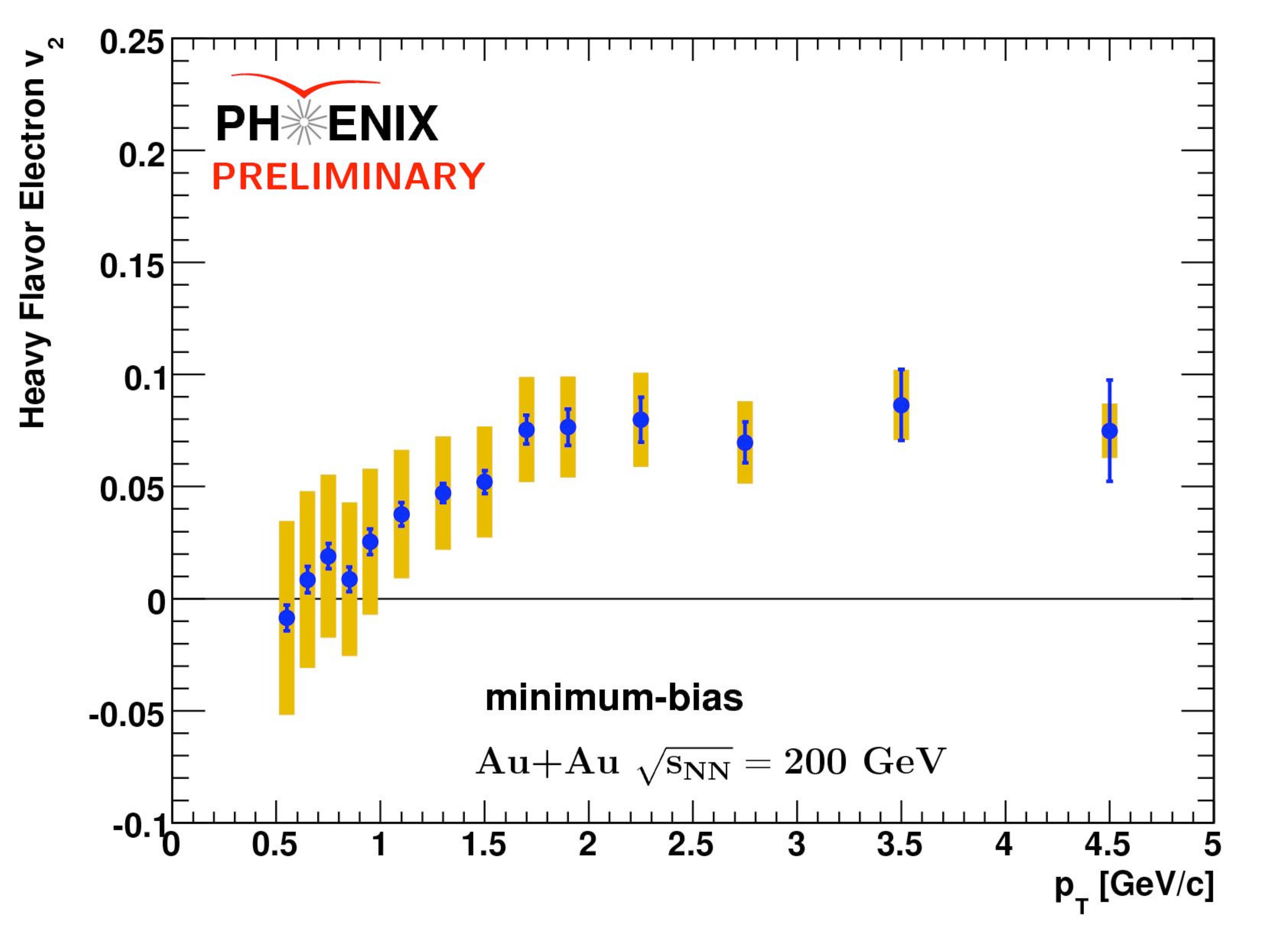} 
\end{tabular}
\end{center}\vspace*{-1.5pc}
\caption[]{PHENIX~\cite{PXcharmAA06}: a) (left) $R_{AA}$ (central Au+Au) b) (right) $v_2$ (minimum bias Au+Au) as a function of $p_T$ for direct-$e^{\pm}$ at $\sqrt{s_{NN}}=200$ GeV. }
\label{fig:fcrisis}
\end{figure}

  One of the most exciting discoveries at RHIC, now confirmed at the LHC, is the suppression of heavy quarks comparable to that of $\pi^0$  from light quarks for $p_T\gsim 4$ GeV/c as observed at RHIC using direct-single-$e^{\pm}$ from heavy quark ($c$, $b$) decay (Fig.~\ref{fig:fcrisis}a)~\cite{PXcharmAA06}. Also seen at RHIC is that heavy quarks exhibit collective flow ($v_2$) (Fig.~\ref{fig:fcrisis}b), another indication of a very strong interaction with the medium. 

\begin{figure}[!h]
\begin{center}
\includegraphics[width=0.45\textwidth]{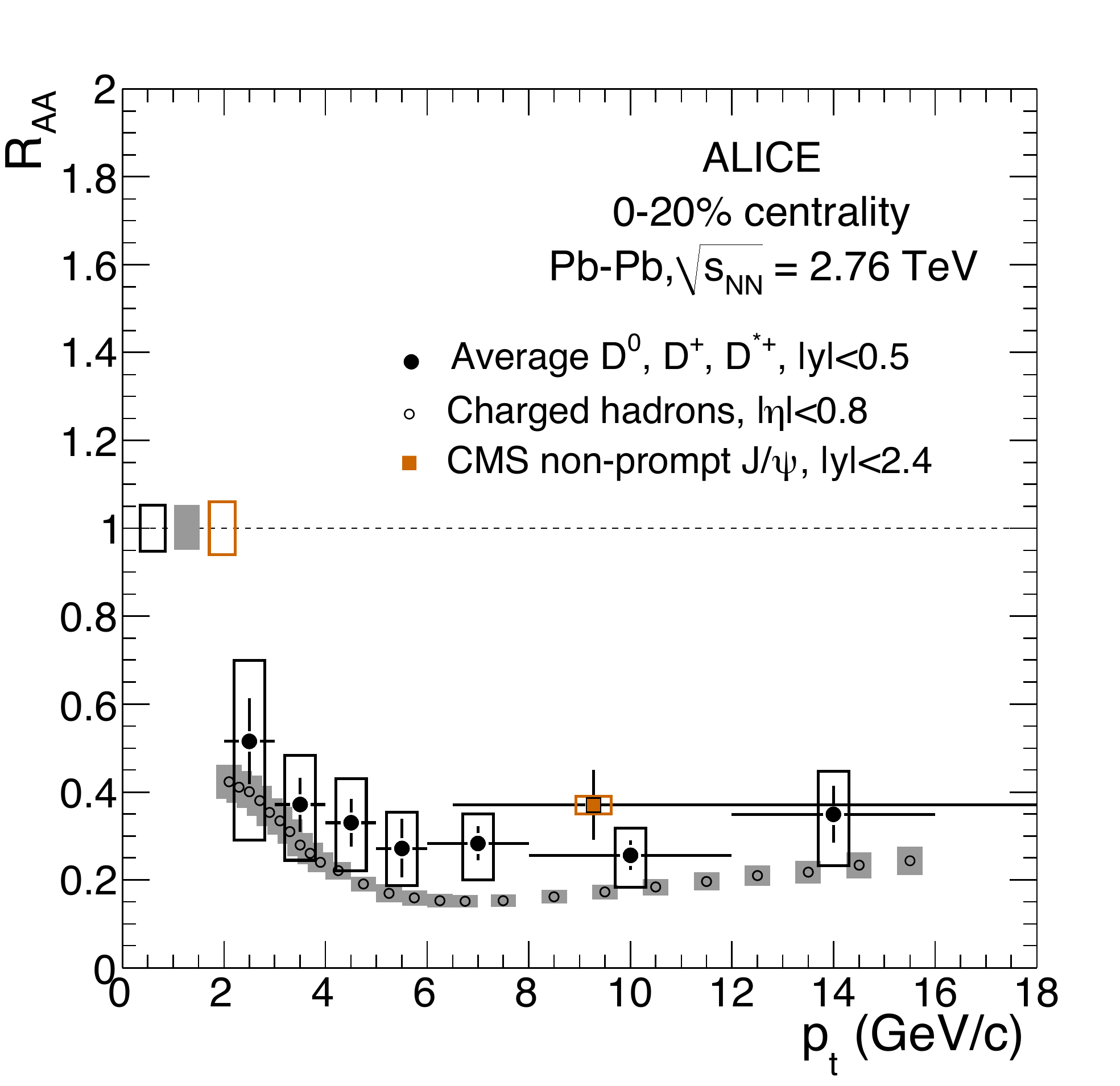}
\includegraphics[width=0.54\textwidth]{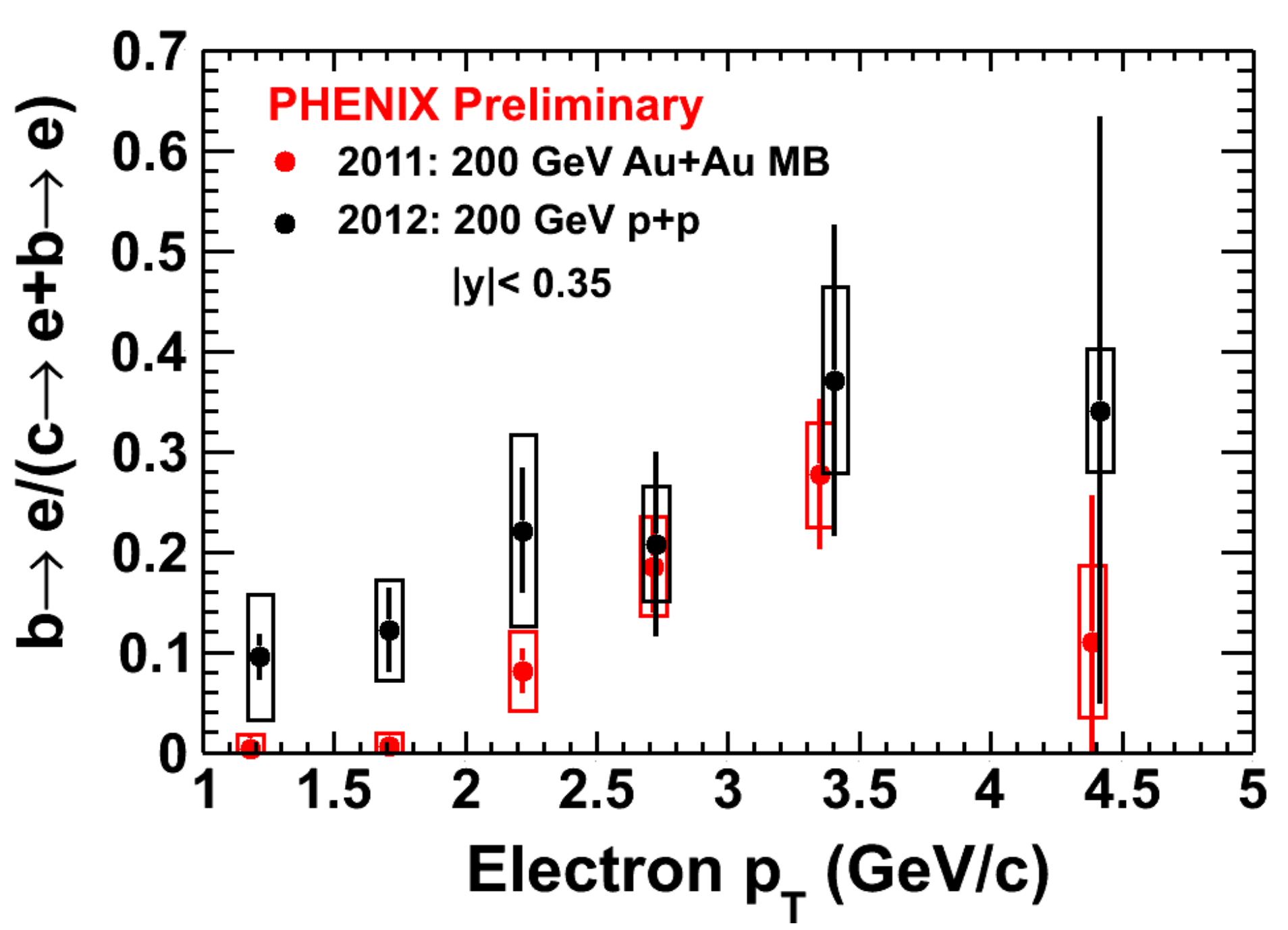}
\end{center}\vspace*{-0.05pc}
\caption[]{a) (left) $R_{AA}$ of ALICE~\cite{ALICEDsuppression} $D$-mesons, charged hadrons, and CMS~\cite{CMSUps1Ssupp} non-prompt $J/\Psi$, in central (0-20\%) Pb+Pb collisions at $\sqrt{s_{NN}}=2.76$ TeV b) (right) $b$-quark fraction ${F}_{b}={b\rightarrow e}/(c\rightarrow e+b\rightarrow e)$  of direct single-$e^{\pm}$ in p-p and Au+Au from PHENIX measurement of the Distance of Closest Approach (DCA) of the displaced vertex. }
\label{fig:heavy}\vspace*{-0.05pc}
\end{figure}

At the LHC, ALICE~\cite{ALICEDsuppression} measured the suppression of reconstructed $D$ mesons containing $c$-quarks, and CMS~\cite{CMSUps1Ssupp} observed suppression of non-prompt $J/\Psi$ from $b$-quark decay (Fig.~\ref{fig:heavy}a). 
The discovery at RHIC in 2007 was a total surprise and a problem since it appears to disfavor the radiative energy loss explanation~\cite{BDMPS} of suppression (also called jet-quenching) because heavy quarks should radiate much less than light quarks or gluons. 

Many explanations have been offered including some from string theory; but the explanation I prefer was by Nino Zichichi~\cite{AZYukawa} who proposed that since the standard model Higgs Boson,  which gives mass to the Electro-Weak vector Bosons, does not necessarily give mass to Fermions, ``it cannot be excluded that in a \QCD\ coloured world (a \QGP), the six quarks are all nearly massless''. If this were true it would certainly explain why light and heavy quarks appear to exhibit the same radiative energy loss in the medium. This idea can, in fact, be tested because the energy loss of one hard-scattered parton relative to its partner, e.g. $g+g\rightarrow b+\bar{b}$ , can be measured by experiments at RHIC and LHC using two-particle correlations in which both the outgoing $b$ and $\bar{b}$ are identified by measurement of the Distance of Closest Approach (DCA) of their displaced decay vertices in silicon vertex detectors. When such results are available, they can be compared to $\pi^0$-charged hadron correlations from light quark and gluon jets, for which measurement of the relative energy loss has been demonstrated at RHIC (recall Fig.~\ref{fig:AuAupp79})~\cite{MJT-Utrecht}. 

Of course, measurement of the Yukawa couplings to Fermions of the 125 GeV Higgs Boson~\cite{Nobel2013} at the LHC may be available in the not too distant future; but, already at Quark Matter 2012, the first direct measurement of $b$-quarks in p-p and Au+Au collisions at RHIC by their displaced vertices in the new PHENIX Silicon VTX detector was presented~\cite{RachidQM2012}.  Figure~\ref{fig:heavy}b shows the measurement of the $b$-quark fraction ${F}_{b}={b\rightarrow e}/(c\rightarrow e+b\rightarrow e)$ of direct single $e^{\pm}$ in p-p and Au+Au collisions at $\sqrt{s_{NN}}=200$ GeV, using the PYTHIA $c$ and $b$ quark $p_T$ distributions in p-p collisions to calculate the DCA distributions of the $e^{\pm}$ in both p-p and Au+Au.  The fact that the Au+Au measurements for all $p_T^e$ are below the p-p measurements indicates clearly that the $b$-quark $p_T$ spectrum is modified in Au+Au compared to p-p. However, the correct conditional DCA distribution requires the actual (modified) $b$-quark $p_T$ spectrum in Au+Au,  which must be obtained by iteration. Once the iteration has converged, the $R^b_{AA}(p_T^e)$ can be calculated from the measured $R_{AA}(p_T^e)$ of the direct-single-$e^{\pm}$ by the relation $R^b_{AA}(p_T^e)=R_{AA}(p_T^e)\times {F}_{b}^{AA}/{F}_{b}^{pp}$. For example, if the final ${F}_{b}^{AA}={F}_{b}^{pp}$ then $R^b_{AA}(p_T^e)=R_{AA}(p_T^e)$. 

\section{Do latest $J/\Psi$ results from ALICE prove deconfinement?}
\begin{figure}[!b]
\begin{center}
\begin{tabular}{cc}
%\hspace*{-4cm}
\includegraphics[width=0.43\linewidth]{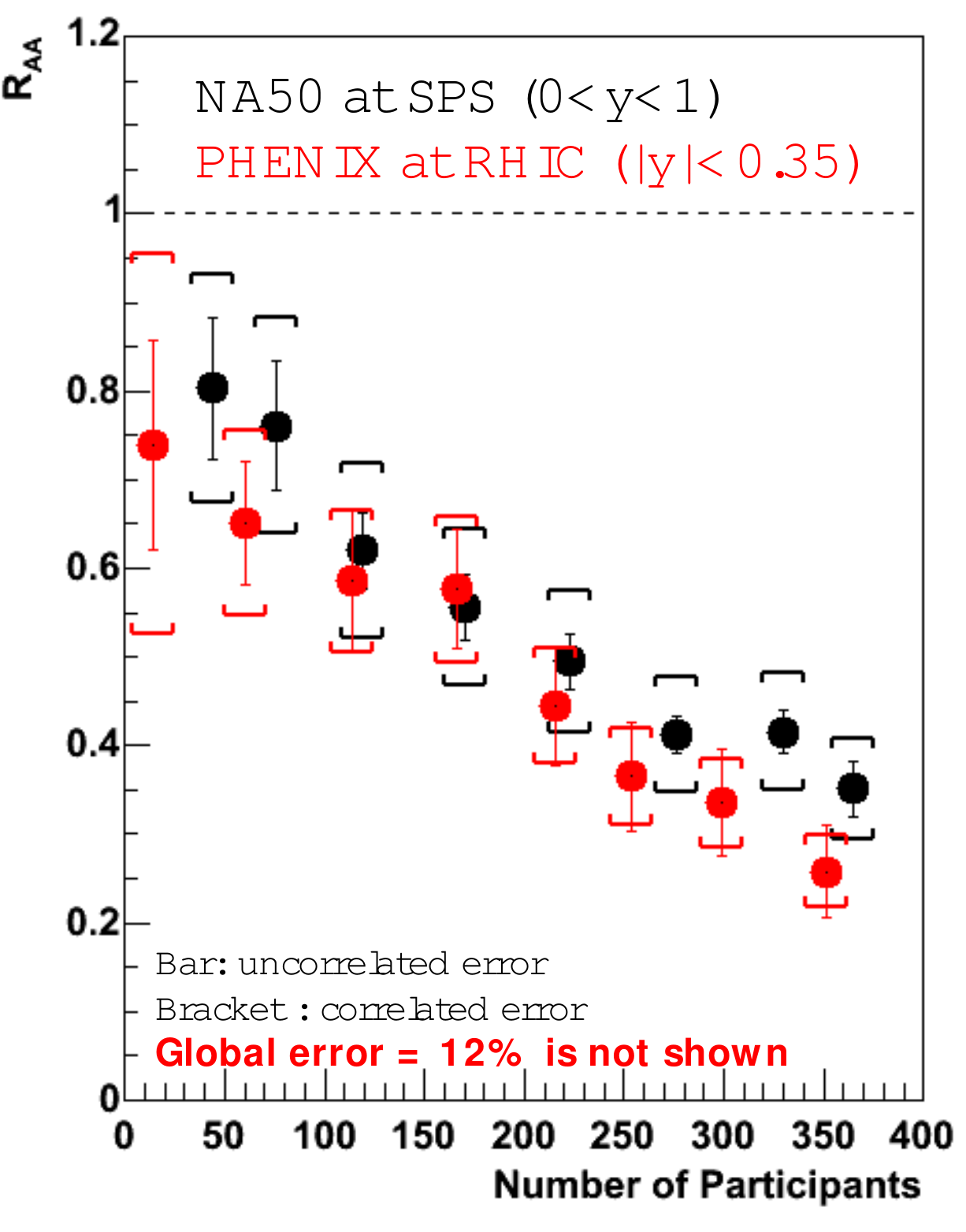}&\hspace*{0.2cm}
\includegraphics[width=0.44\linewidth,height=0.40\linewidth]{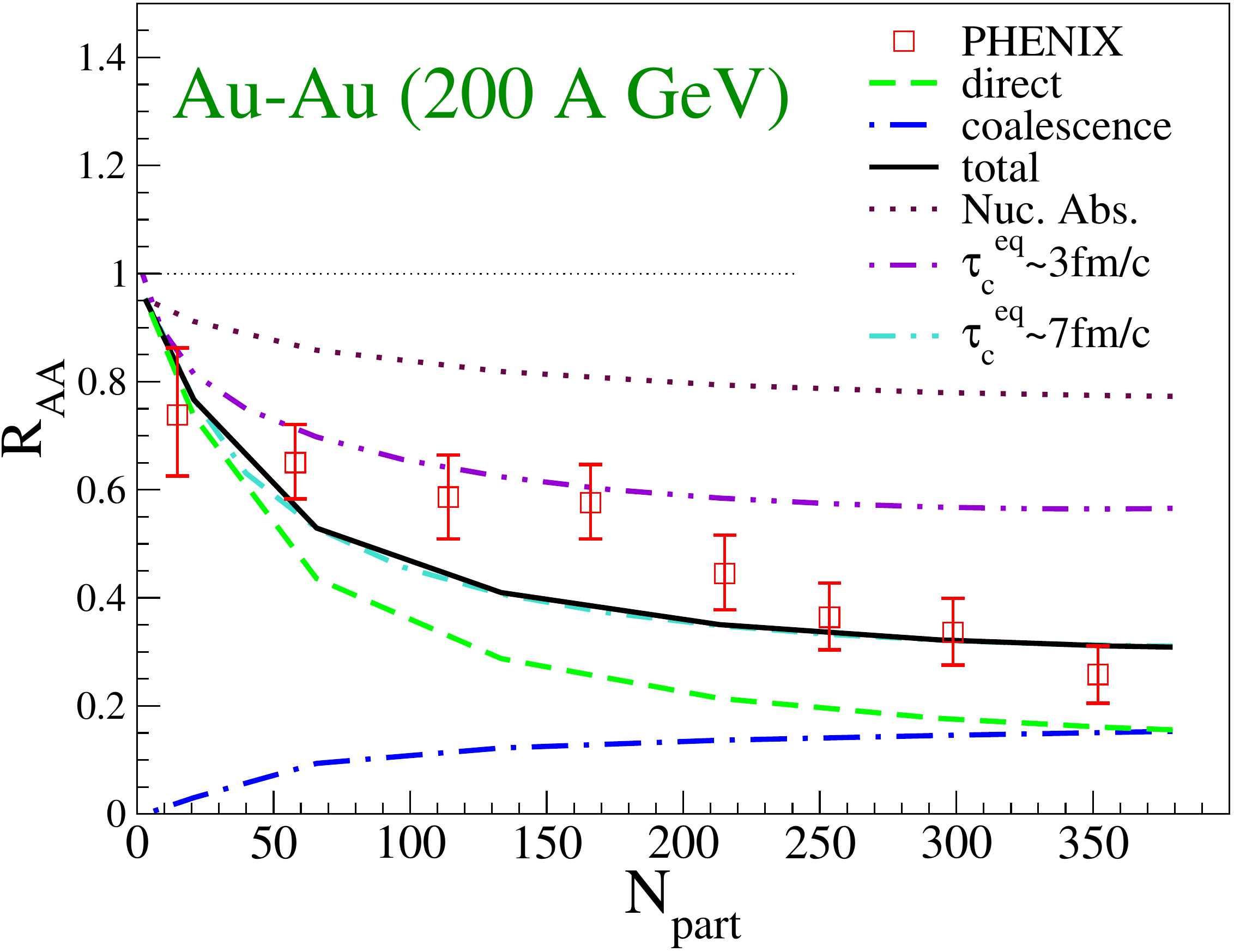}%&\hspace*{0.2cm}
\end{tabular}
\end{center}\vspace*{-1.0pc}
\caption[]{a) (left) $J/\Psi$ suppression relative to p-p collisions ($R_{AA}$) as a function of centrality ($N_{\rm part}$) at RHIC~\cite{PXJPsiAuAu200,GunjiQM06} and at the CERN/SPS~\cite{NA50EPJC39}. b) (right) PHENIX measurement of $R_{AA}$ as a function of centrality from (a) together with prediction from a coalescence model~\cite{RappPLB664} \label{fig:JPsiRHICSpS}}
\end{figure}

   In a previous article in this journal~\cite{MJTIJMPA2011}, I had noted that the dramatic difference in $\pi^0$ and $h^{\pm}$ suppression from SpS to RHIC c.m. energy, $\sqrt{s_{NN}}=17.2$ to 200 GeV (Fig.~\ref{fig:twoRAA}a), is not reflected in $J/\Psi$ suppression, which is nearly identical at mid-rapidity at RHIC~\cite{PXJPsiAuAu200,GunjiQM06} compared to the NA50 measurements at SpS~\cite{NA50EPJC39} (Fig.~\ref{fig:JPsiRHICSpS}a). 
Coupled with the large, well-known, CNM effect on the $J/\Psi$ (recall Fig.~\ref{fig:JPsiAB}), this reinforced my skepticism on the value of $J/\Psi$ suppression as a probe of deconfinement, in spite of the beauty and importance of the Matsui and Satz proposal~\cite{MatsuiSatz86}.

The equality of $J/\Psi$ suppression at $\sqrt{s_{NN}}=17.2$ and 200 GeV was was elegantly explained~\cite{RappPLB664} as recombination or coalescence of $c$ and $\bar{c}$ quarks in the \QGP\ to regenerate $J/\Psi$. Miraculously this made the observed $R_{AA}$ equal at SpS and RHIC c.m. energies (Fig.~\ref{fig:JPsiRHICSpS}b)~\cite{GunjiQM06,RappPLB664}. I called this my ``Nightmare Scenario'' because I thought that nobody would believe it. The good news was that such models are testable because they predict the reduction of $J/\Psi$ suppression or even an enhancement ($R_{AA}> 1$) at LHC energies~\cite{PBMStachelPLB490,ThewsPRC63,AndronicNPA79}, which would be spectacular, if observed. 

  \begin{figure}[!t]
\begin{center}
\begin{tabular}{cc}
%\hspace*{-4cm}
\includegraphics[width=0.49\linewidth]{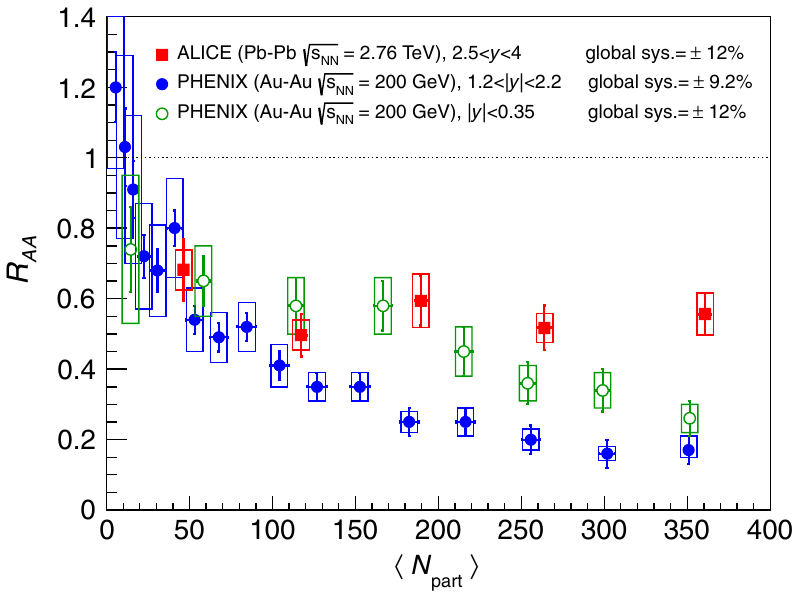}&\hspace*{0.1pc}
\includegraphics[width=0.44\linewidth,height=0.40\linewidth]{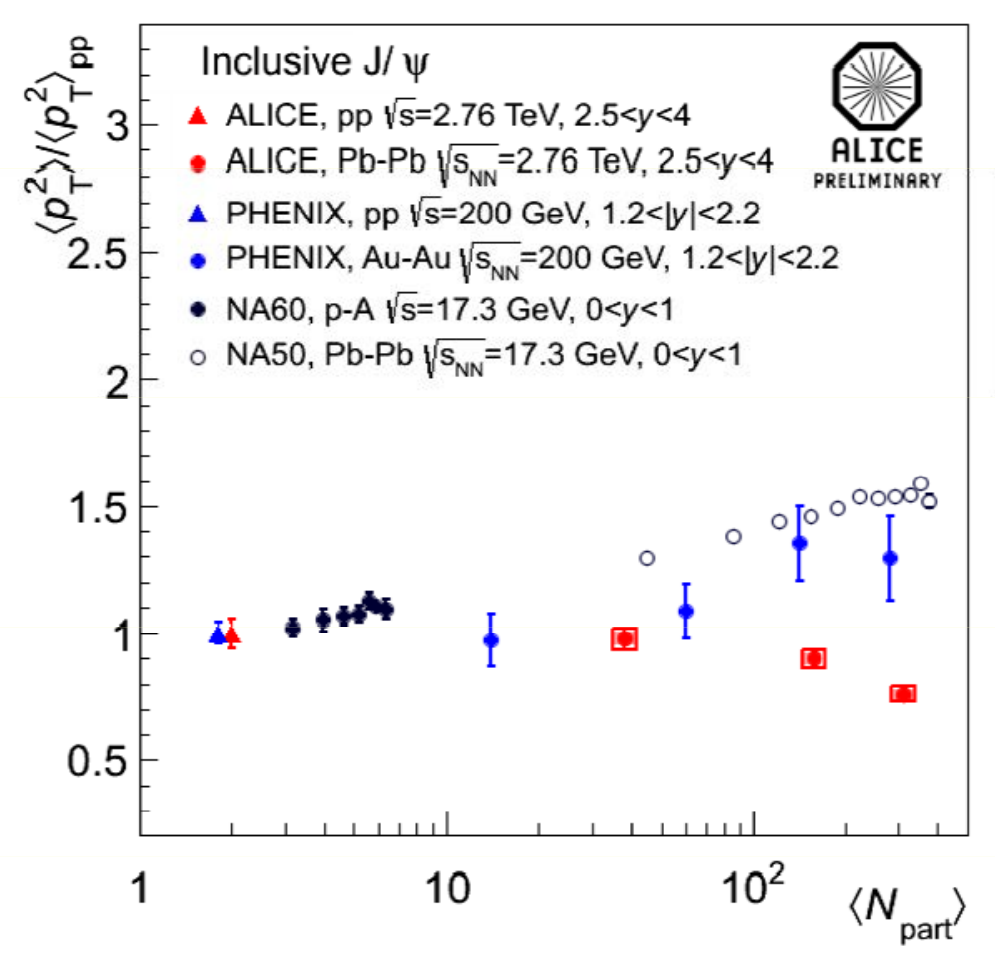}%&\hspace*{0.2cm}
\end{tabular}
\end{center}\vspace*{-1.0pc}
\caption[]{a) (left) $J/\Psi$ suppression relative to p-p collisions ($R_{AA}$) as a function of centrality ($N_{\rm part}$) at RHIC~\cite{PXJPsiAuAu200} and at the CERN/LHC~\cite{ALICEJPsiPRL109}. b) (right) ALICE measurement of $\mean{p_T^2}_{AA}/\mean{p_T^2}_{pp}$ for the $J/\Psi$ as a function of centrality compared to measurements at lower $\sqrt{s_{NN}}$~\cite{PaoloGISSP2013} \label{fig:ALICEPHENIX}}
\end{figure}
In 2012, the ALICE experiment at LHC~\cite{ALICEJPsiPRL109} presented a convincing measurement of reduced $J/\Psi$ suppression at $\sqrt{s_{NN}}=2.76$ TeV at forward rapidity compared to $\sqrt{s_{NN}}=200$ GeV (Fig.~\ref{fig:ALICEPHENIX}a) which, in my opinion,  was clear proof the regeneration prediction.   
However, in 2013, Paolo Giubellino presented ALICE preliminary results~\cite{PaoloGISSP2013} at this meeting for the ratio $\mean{p_T^2}_{AA}/\mean{p_T^2}_{pp}$ as a function of centrality in Pb+Pb (Fig~\ref{fig:ALICEPHENIX}b) which show a decrease from unity in p-p and mid-peripheral  collisions to $\approx 0.7$  for central collisions while both the SpS and PHENIX data continue rising to values $\approx 1.2-1.5$. Paolo claimed that this proves deconfinement in central collisions. 

In the discussion afterwards, I disagreed and claimed that the reduction of $\mean{p_T^2}_{AA}/\mean{p_T^2}_{pp}$ proves regeneration which is more probable at low $p_T$~\cite{ThewsMangano}. Deconfinement would remove $J/\Psi$ at low $p_T$ in central collisions which would increase $\mean{p_T^2}_{AA}/\mean{p_T^2}_{pp}$ as shown by the PHENIX data. This is still good news for CERN because the clear observation of regeneration also proves directly the existence of the \QGP\ at the LHC, since it is evidence that the $c$ and $\bar{c}$ quarks (with their color charge fully exposed) freely traverse the medium (with a large density of similarly exposed color charges) to find each other and form $J/\Psi$. Professor Zichichi uncharacteristically cut off the discussion of deconfinement due to time pressure and said that he agreed with Paolo.

After the session, Professor Tawfik pointed out to me that Helmut Satz had recently discussed a way to distinguish deconfinement in the presence of regeneration~\cite{SatzCalibrating-2013}. The crucial issue is whether the medium modifies the fraction of produced $c-\bar{c}$ pairs which form $J/\Psi$. Dissociation of $J/\Psi$ in the medium would reduce the observed $J/\Psi/(c-\bar{c})$ ratio in A+A compared to p-p collisions, i.e. $R_{AA}^{J/\Psi}/R_{AA}^{c-\bar{c}}\ll1$. 

\begin{figure}[!t]
\begin{center} %\vspace*{-1pc}
\begin{tabular}{cc}
%%\hspace*{-0.08\linewidth}\includegraphics[width=0.51\linewidth]{figs/fig3se_data_only_logo.eps} 
\hspace*{-0.06\linewidth}\includegraphics[width=0.51\linewidth]{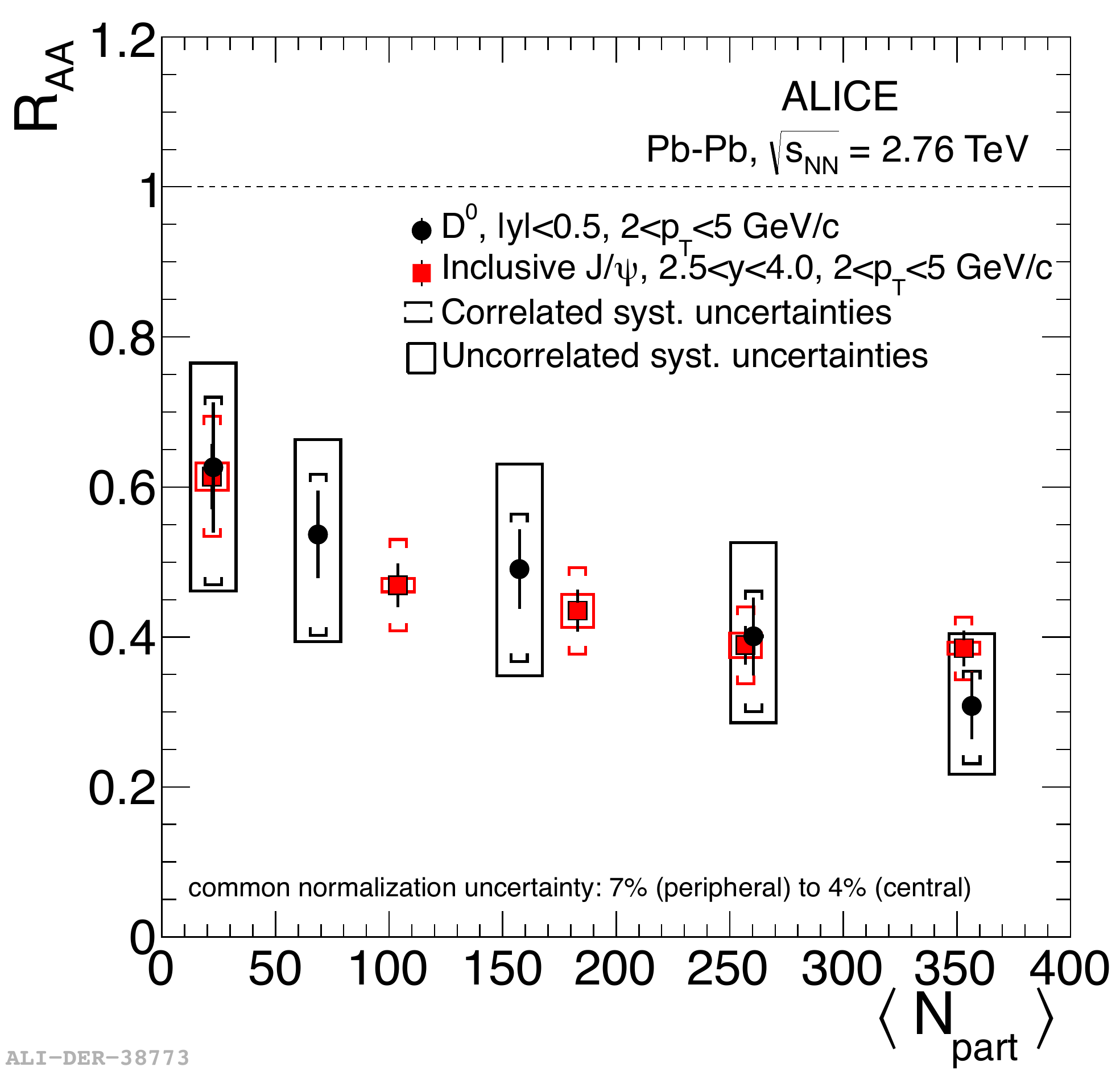} &
\includegraphics[width=0.51\linewidth]{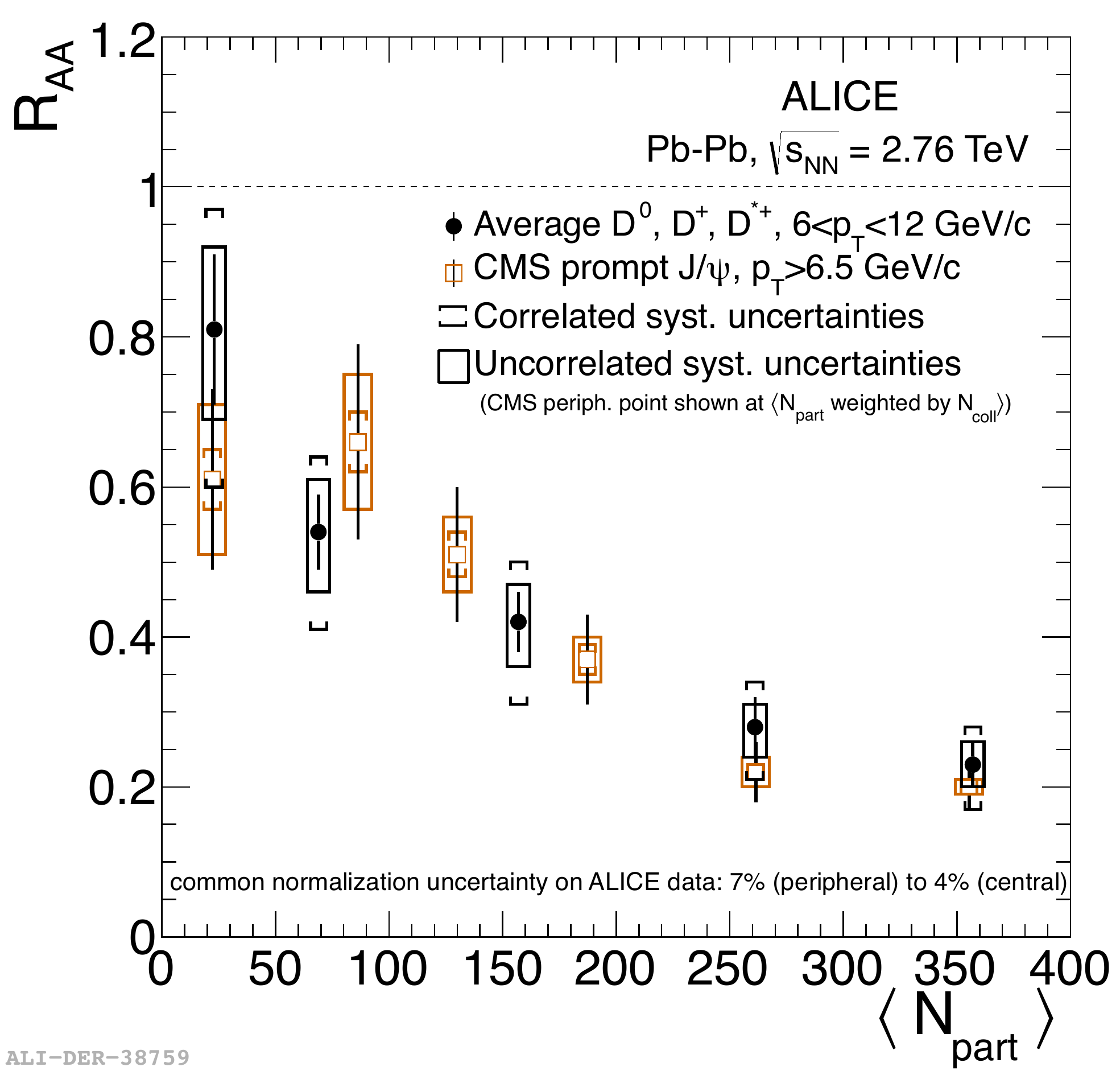} 
\end{tabular}
\end{center}\vspace*{-1.0pc}
\caption[]{LHC data from ALICE~\cite{ALICEDsuppression,ALICEJPsiPRL109} and CMS~\cite{CMSUps1Ssupp}, comparing $J/\Psi$ production to open charm production at intermediate (a) and higher (b) transverse momenta~\cite{SatzCalibrating-2013}.}
\label{fig:SatzLHC}
\end{figure}

Satz~\cite{SatzCalibrating-2013} first compared $R_{AA}^{c-\bar{c}}$ to $R_{AA}^{J/\Psi}$ (Fig.~\ref{fig:SatzLHC}) using LHC data  from \mbox{ALICE~\cite{ALICEDsuppression,ALICEJPsiPRL109}} and CMS~\cite{CMSUps1Ssupp} at intermediate (2.5-4 GeV/c) and higher (6-12 GeV/c) $p_T$. In both cases the $R_{AA}$ for open charm and $J/\Psi$ production show the same behavior within errors. ``In other words, the reduction of the $J/\Psi$ is in complete agreement with that of open charm; there is neither suppression nor enhancement''~\cite{SatzCalibrating-2013}. However the conclusion is premature because the total $c-\bar{c}$ production and $R_{AA}$ at low $p_T$ was not yet available at LHC. 

  \begin{figure}[!t]
\begin{center} %\vspace*{-1pc}
\begin{tabular}{cc}
%%\hspace*{-0.08\linewidth}\includegraphics[width=0.51\linewidth]{figs/fig3se_data_only_logo.eps} 
\hspace*{-0.06\linewidth}\includegraphics[width=0.51\linewidth]{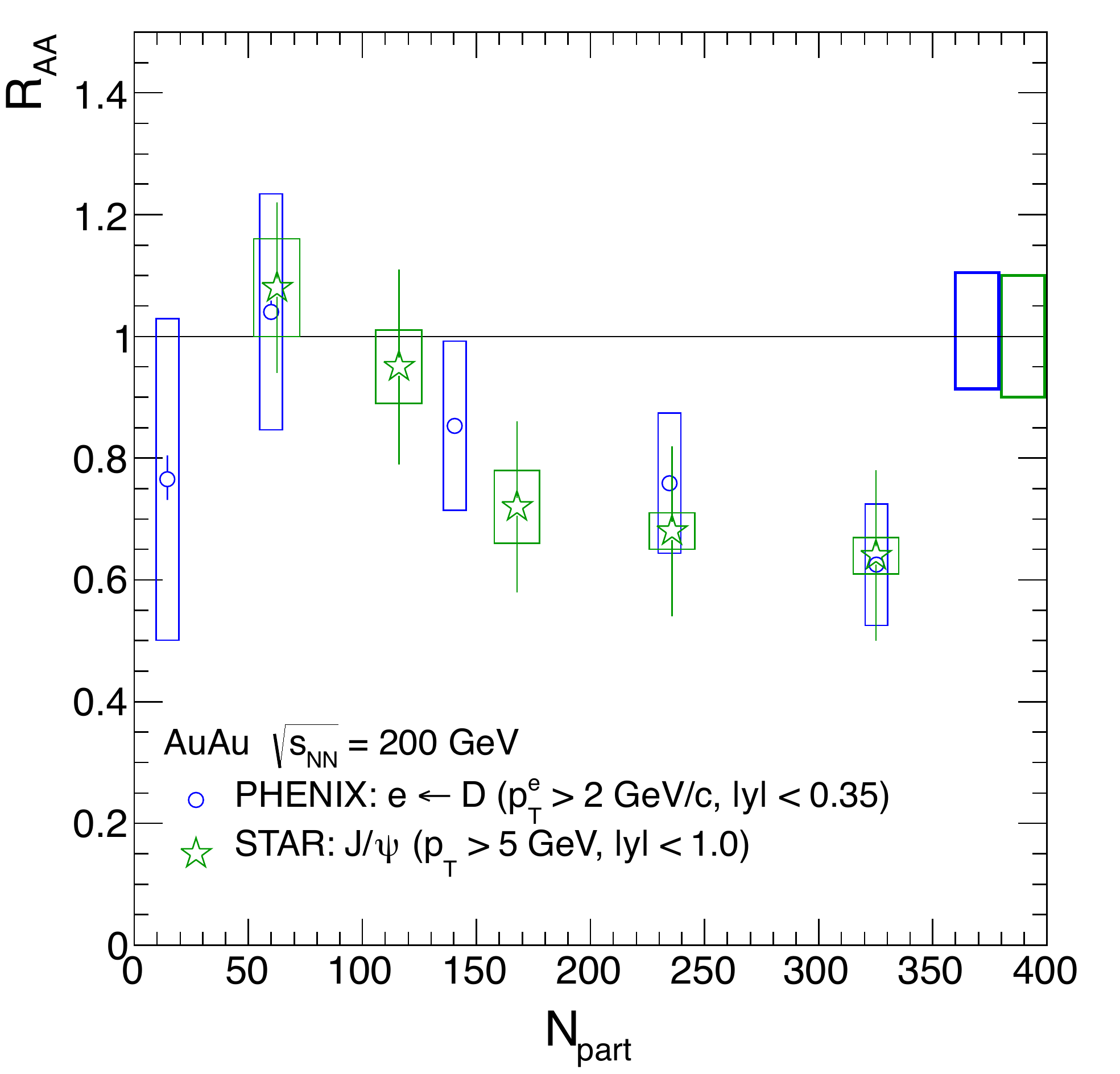} &
\includegraphics[width=0.51\linewidth]{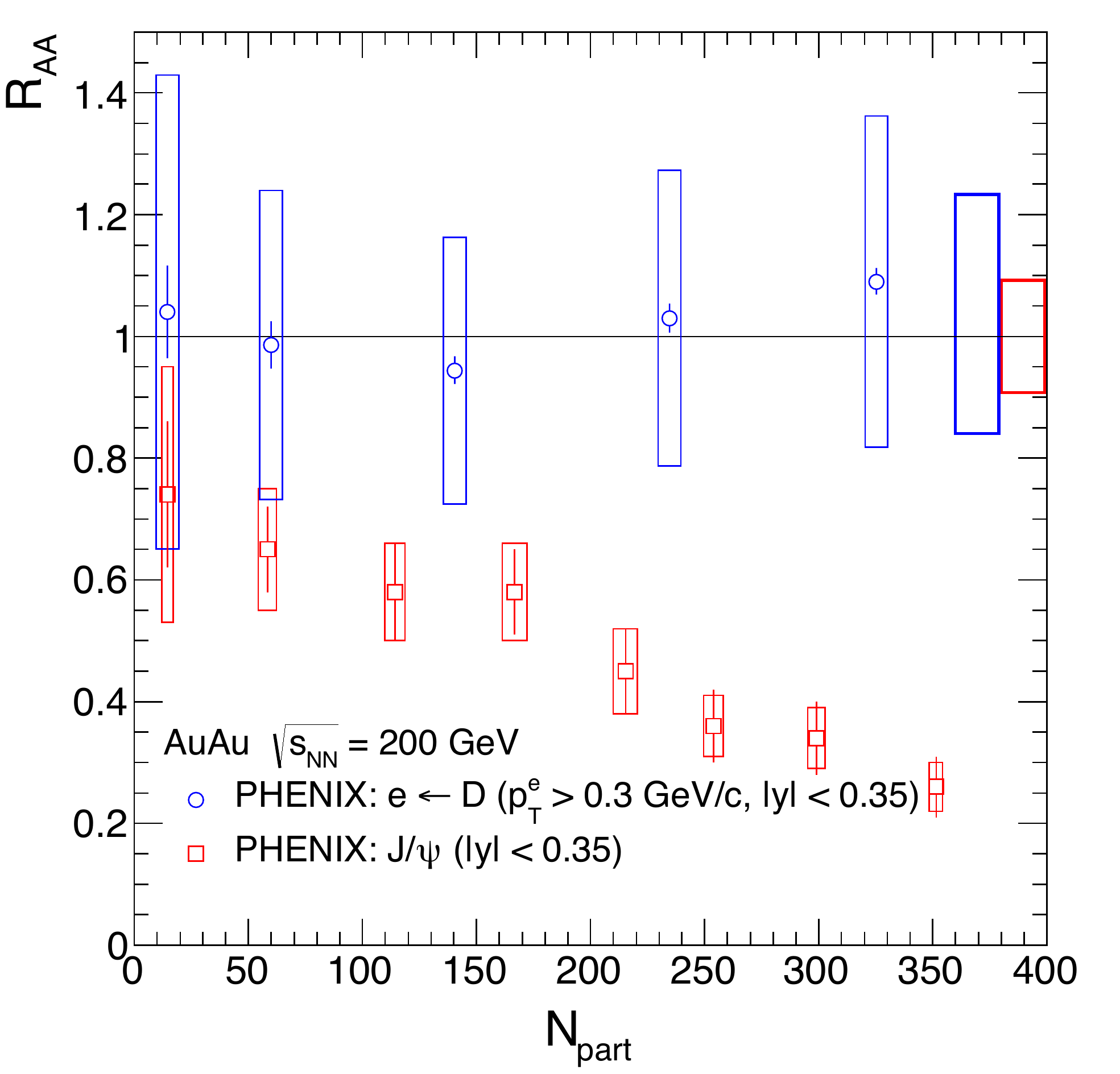} 
\end{tabular}
\end{center}\vspace*{-1.0pc}
\caption[]{RHIC data from PHENIX~\cite{ppg077,PXJPsiAuAu200} and STAR~\cite{STARPLB722}, comparing $J/\Psi$~production to open
charm production at high (a) and low (b) transverse momenta~\cite{SatzCalibrating-2013}.}
\label{fig:f7}
\end{figure}

Satz~\cite{SatzCalibrating-2013} contrasted this with the measurements from RHIC (Fig.~\ref{fig:f7}) where at high $p_T$ the $R_{AA}$ of open charm and $J/\Psi$ track each other, as at LHC. However, at low $p_T$ ``the $R_{AA}^{c-\bar{c}}$ is within errors unity over the entire centrality range; in contrast $R_{AA}^{J/\Psi}$ decreases strongly and thus here gives the correct $J/\Psi$ survival probability \ldots with increasing centrality, a smaller and smaller fraction of $c\bar{c}$ pairs go into $J/\Psi$ production, with a suppression of up to 75\% for the most central collisions''~\cite{SatzCalibrating-2013}.

I suppose this means that at RHIC there is evidence of deconfinement via ``calibrated'' $J/\Psi$ suppression~\cite{SatzCalibrating-2013}, while for LHC the data are inconclusive until $R_{AA}$ of the total (i.e. low $p_T$) $c-\bar{c}$ production is available.  Also, since new measurements of CNM effects at RHIC and LHC have produced surprising results in some cases, measurements of CNM effects in $J/\Psi $ and $c-\bar{c}$ production at both RHIC and LHC are needed for an unambiguous conclusion about whether $J/\Psi$ suppression is evidence for deconfinement via Debye screening in the \QGP.

\section{Surprises in d+Au and p+Pb measurements}
\subsection{Hard-Scattering}

In 2013, the major event was the p+Pb run at LHC which also spurred new or improved d+Au results from RHIC. Apart from one hard-scattering result to be presented first, all the results involve the predominant soft physics of multiplicity and $E_T$ distributions as well as flow. 
A new measurement of identified hadron production in both Au+Au d+Au at $\sqrt{s_{NN}}=200$ GeV~\cite{PXPRC88} gives  some insight into the baryon anomaly. 
         \begin{figure}[!h]
   \begin{center}
\includegraphics[width=0.49\textwidth]{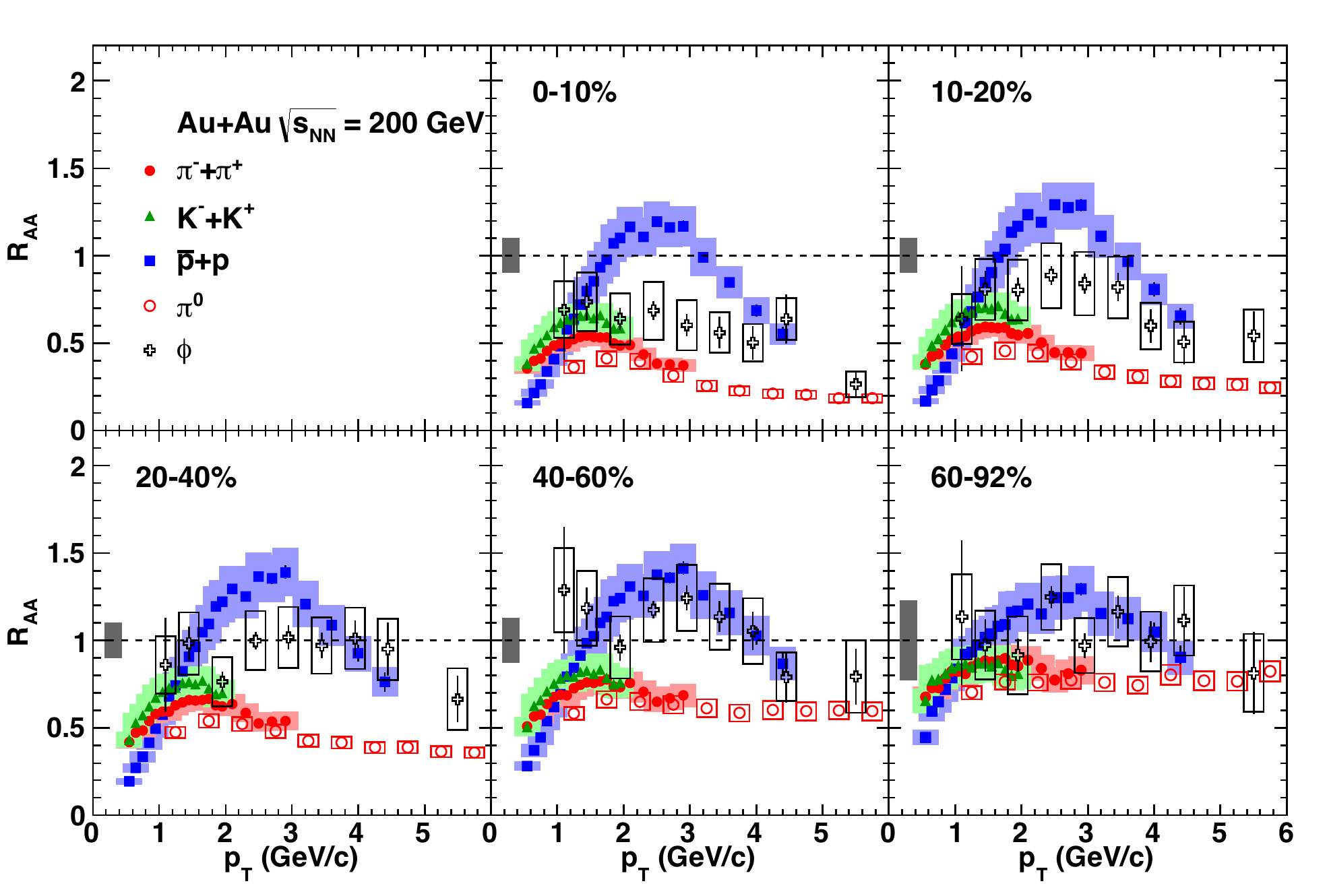}
\includegraphics[width=0.49\textwidth]{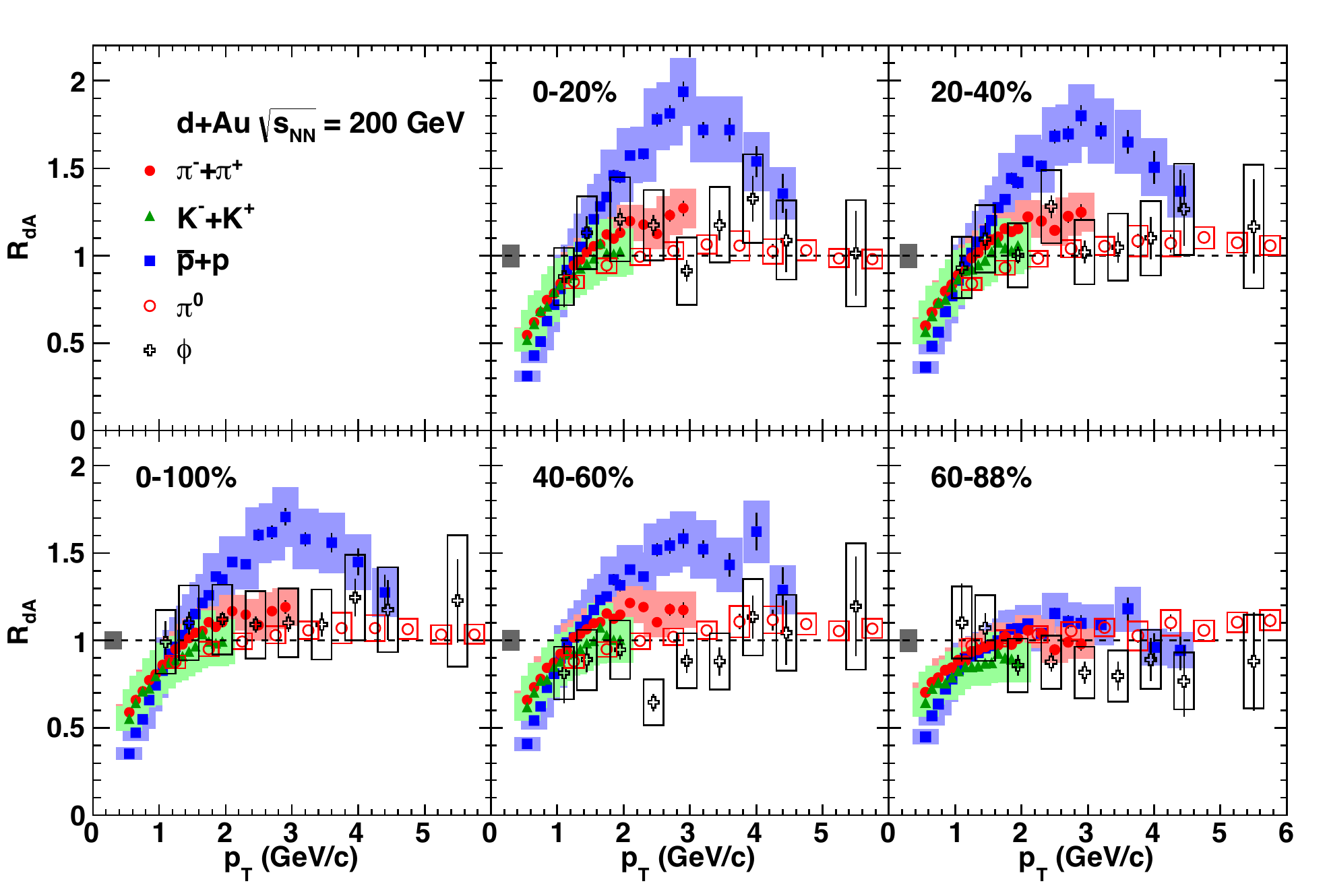}
\end{center}\vspace*{-1.0pc}
\caption[]{Measurements of $R_{AA}$ of identified particles as a function $p_T$ and centrality at $\sqrt{s_{NN}}=200$ GeV~\cite{PXPRC88}: a)(left) Au+Au; b) (right) d+Au. 
\label{fig:RAAdAuAuAu}}\vspace*{-0.1pc}
\end{figure}
Figure~\ref{fig:RAAdAuAuAu}a shows $R_{AA}$ in Au+Au for protons and mesons in the range $0.5<p_T<6.0$ GeV/c, where, in central collisions (0-10\%), all the mesons are suppressed for {$p_T>2$ GeV/c}  while the protons are enhanced for $2<p_T<4$ GeV/c and then become suppressed at larger $p_T$. The d+Au results in Fig.~\ref{fig:RAAdAuAuAu}b show no effect for the mesons, $R_{AA}\approx 1$ out to $p_T=6$ GeV/c; while the protons show a huge enhancement (Cronin effect) in all centralities except for the most peripheral (60-88\%).  This suggests the need for a common explanation of the proton enhancement in both Au+Au and d+Au collisions, which is lacking at present.  

\subsection{Multiplicity and ${E_T}$ Distributions}
\begin{figure}[!h]\vspace*{-1.0pc}
   \begin{center}
\includegraphics[width=0.52\textwidth]{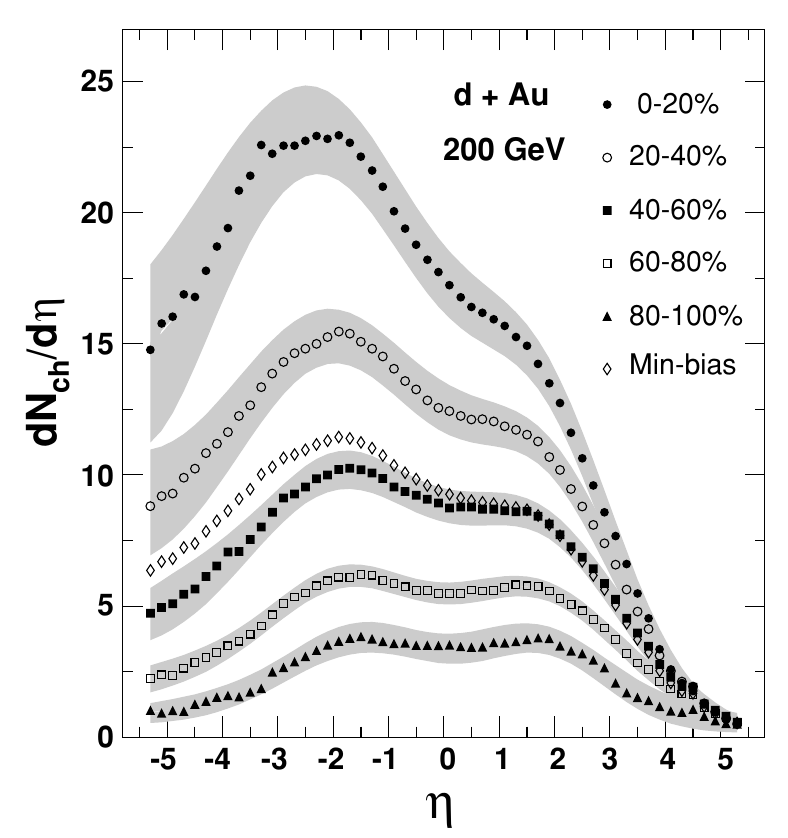}\hspace{0.1pc}
\includegraphics[width=0.46\textwidth]{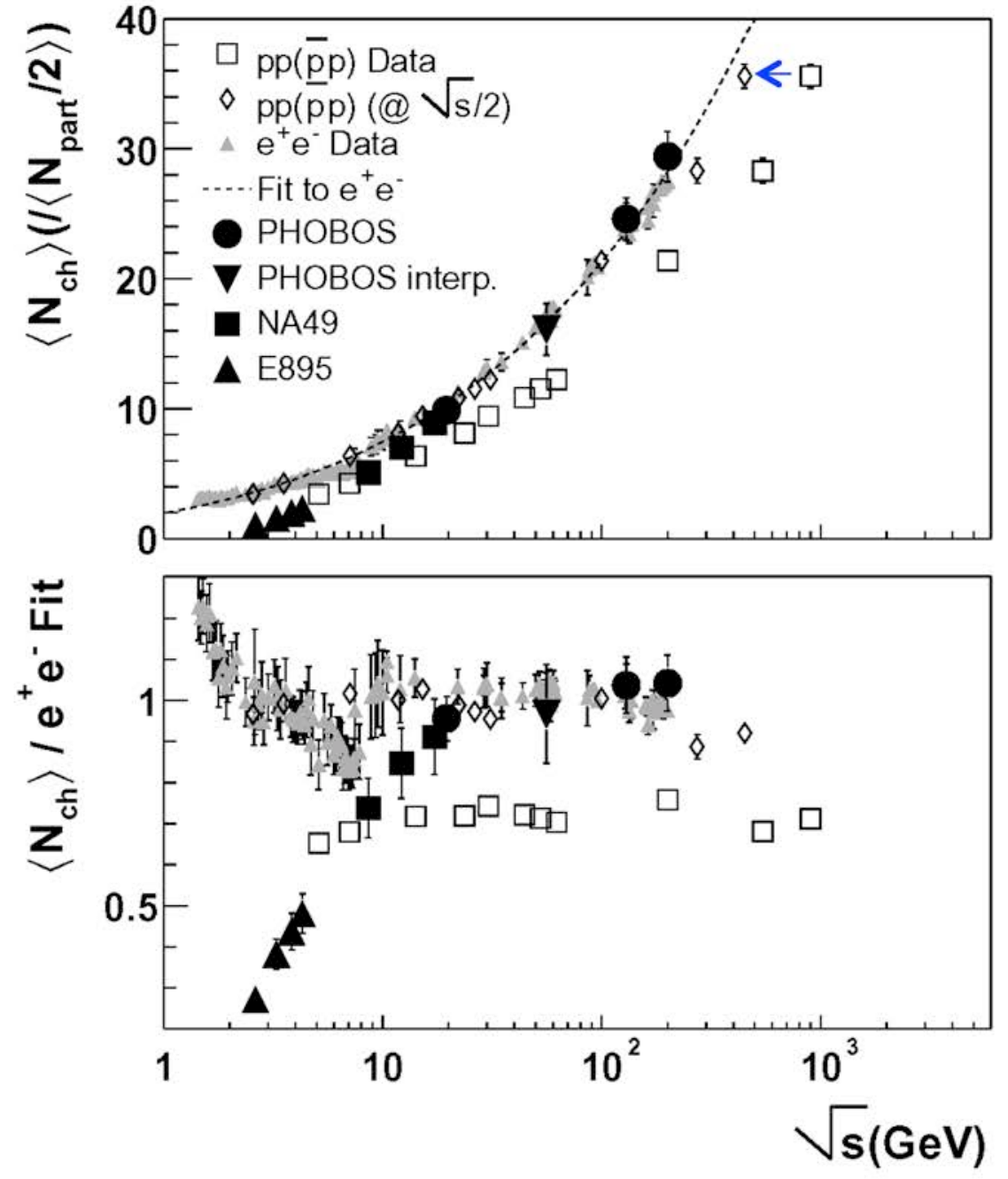}
\end{center}\vspace*{-1.0pc}
\caption[]{a) (left) Charged particle multiplicity density in rapidity, $dN_{\rm ch}/d\eta$, as a function of centrality in d+Au collisions at $\sqrt{s_{NN}}=200$ GeV~\cite{PHOBOSPRC72}. b)~(right) Total charged multiplicity per nucleon pair in p-p and A+A collisions as a function of c.m. energy $\sqrt{s}$ compared to $e^+ + e^-$ collisions~\cite{PHOBOSPRC74}.
\label{fig:PHOBOSdAAA}}\vspace*{-0.5pc}
\end{figure}

Returning to the soft physics of multiplicity and $E_T$ distributions, the PHOBOS experiment at RHIC, with a large pseudo-rapidity acceptance $-5<\eta<+5$ over the full azimuth had presented an instructive measurement of the charged multiplicity density, $dN_{\rm ch}/d\eta$ from the first d+Au run in 2003 (Fig.~\ref{fig:PHOBOSdAAA}a)~\cite{PHOBOSPRC72}.
For peripheral collisions, the distribution is symmetric around mid-rapidity, as in p-p collisions. However, with increasing centrality, the multiplicity increases over the whole $\eta$ range but with a larger increase at negative $\eta$ (the Au rapidity) such that the peak of the distribution steadily shifts in the direction of the Au nucleus. 

These features which are similar to what was first observed in fixed target p+A experiments at $\sqrt{s_{NN}}\sim 19.4$ GeV could be explained (c. 1976) by a simple model, the Wounded Nucleon Model (WNM)~\cite{WNM}. From relativity and quantum mechanics the only thing that can happen to a relativistic nucleon when it interacts with another nucleon in a nucleus is to become an excited nucleon with the same energy but reduced longitudinal momentum (rapidity). It remains in that state inside the nucleus because the uncertainty principle and time dilation prevent it from fragmenting into particles until it is well outside the nucleus. 
If one makes the further assumptions that an excited nucleon interacts with the same 
cross section as an unexcited nucleon and that the successive collisions of the excited nucleon do not affect the excited state or its eventual fragmentation products, this leads to the conclusion that the elementary process for particle 
production in nuclear collisions is the excited nucleon, and to the prediction 
that the multiplicity in nuclear interactions should be proportional to 
the total number of projectile and target participants (Wounded Nucleons)~\cite{WNM}, rather than to the 
total number of collisions. 

Another interesting effect observed by PHOBOS~\cite{PHOBOSPRC74} is that the ``leading particle effect'' in p-p collisions, as discovered by Zichichi and collaborators~\cite{BasilePLB95}, in which the total multiplicity at c.m. energy $\sqrt{s_{\rm pp}}$ is equal to that in $e^+ e^-$ collisions at $\sqrt{s_{\rm ee}}=\sqrt{s_{\rm pp}}/2$ (the ``effective energy'') because the leading protons carry away half the p-p c.m. energy, is absent in A+A collisions (Fig.~\ref{fig:PHOBOSdAAA}b). This observation seems to contradict the WNM, in which the key assumption is that what counts is whether or not a nucleon was struck, not how many times it was struck.
\begin{figure}[!t]
   \begin{center}
\includegraphics[width=0.44\textwidth]{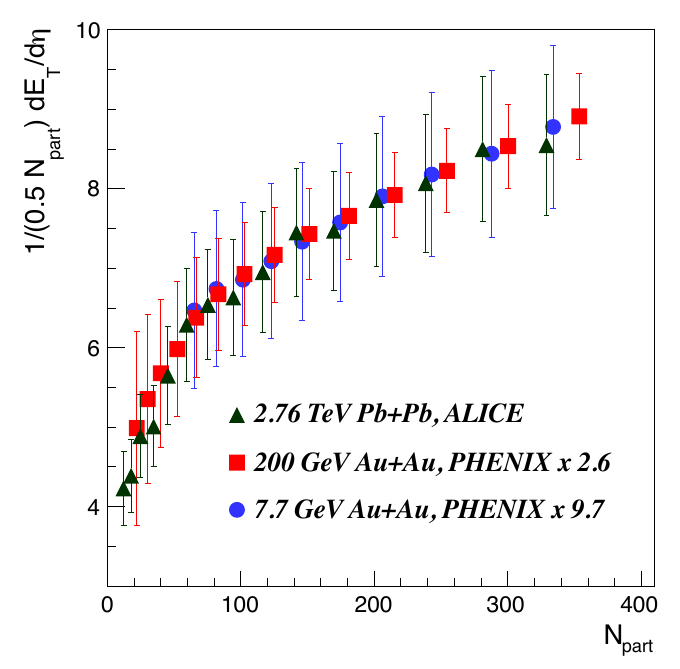}\hspace*{0.4pc} 
\includegraphics[width=0.55\textwidth]{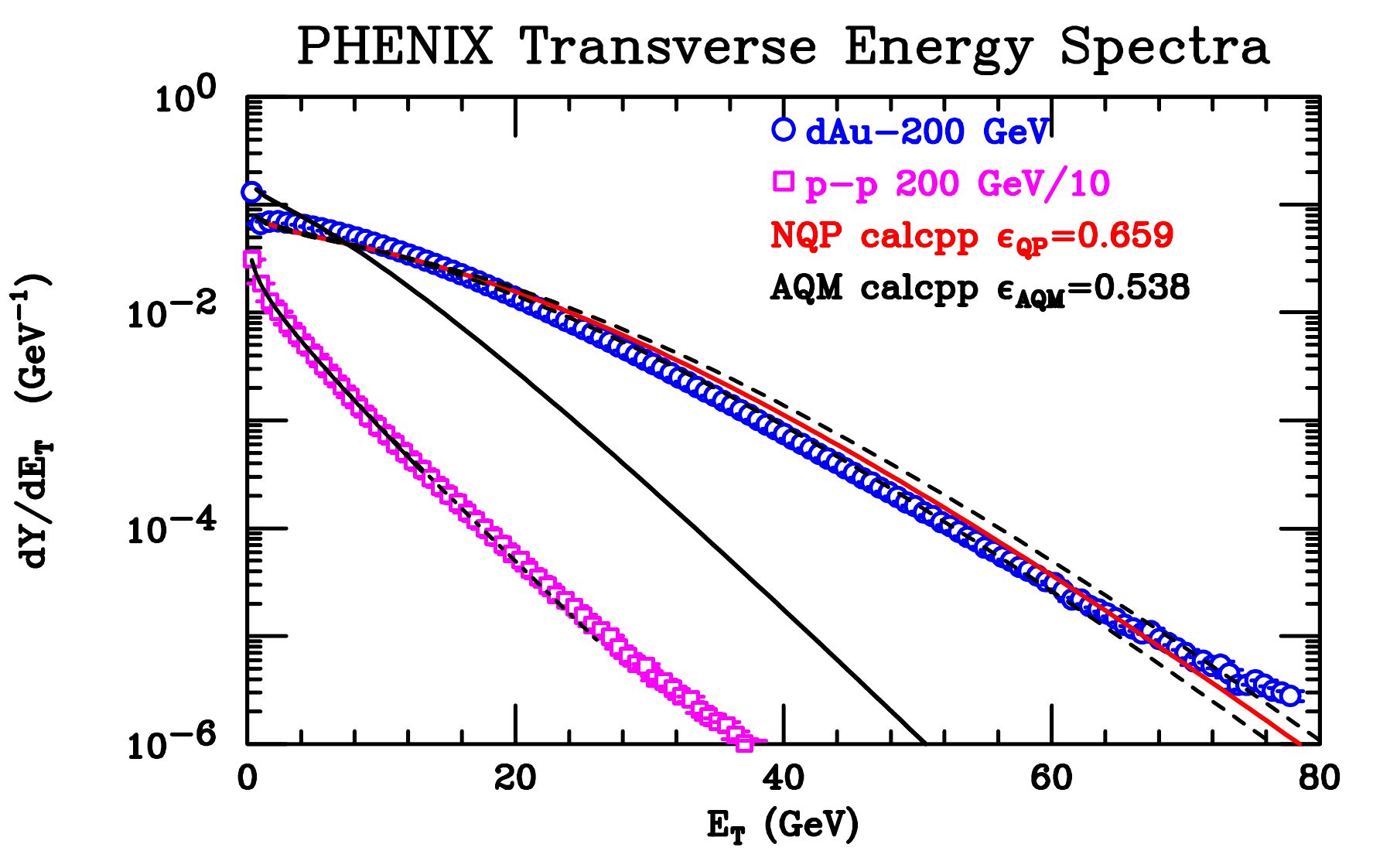}
\end{center}\vspace*{-1.0pc}
\caption[]{a) (left) $dE_T^{AA}/d\eta/(0.5 \mean{N_{\rm part}})$ vs. $\mean{N_{\rm part}}$ in Au+Au and Pb+Pb collisions from $\sqrt{s_{NN}}=0.0077$ to 2.76 TeV. b) (right) PHENIX measurement~\cite{ppg100} of $E_T$ distributions for p+p and d+Au at $\sqrt{s_{NN}}=200$ GeV  with calculations of the d+Au spectrum based on the AQM (color-strings) and the number of constituent-quark participants (NQP).
\label{fig:NQP}}\vspace*{-0.8pc}
\end{figure}

In fact, the WNM fails badly at mid-rapidity for both $dN_{ch}/d\eta$ and $dE_T/d\eta$ as shown by a plot of $dE_T/d\eta/(N_{\rm part}/2)$ vs. $N_{\rm part}$ from PHENIX which should be constant if the WNM were true (Fig.~\ref{fig:NQP}a). 
The fact that the scaled evolution with centrality is the same from $\sqrt{s_{NN}}=7.7$ GeV to 2.76 TeV indicates that the dominant effect is the nuclear geometry of the A+A collision. It has been shown that the evolution in Fig.~\ref{fig:NQP}a can be explained by a nuclear geometry based on the number of constituent-quark participants, the NQP model~\cite{VoloshinNQP,NouicerNQP}.
Thus the shape of the data in Fig.~\ref{fig:NQP}a is simply the number of constituent-quark participants/nucleon participant, ${N_{\rm qp}}/N_{\rm part}$.  

For symmetric systems such as Au+Au, the NQP model is identical to another model from the 1970's, the Additive Quark Model (AQM)~\cite{AQM}.  The AQM is actually a model of particle production by color-strings in which only one color-string can be attached to a constituent-quark participant. Thus, for asymmetric systems such as d+Au, the maximum number of color-strings is limited to the number of constituent-quarks in the lighter nucleus, or six for d+Au, while the NQP allows all the quark participants in both nuclei to emit particles. A new PHENIX measurement (Fig.~\ref{fig:NQP}b)~\cite{ppg100} shows that the NQP model gives the correct $E_T$ distribution in d+Au, while the AQM has a factor of 1.7 less $E_T$ emission due to the restriction on the number of effective constituent-quarks in the larger nucleus.

\begin{figure}[!t]
   \begin{center}
\includegraphics[width=0.49\textwidth]{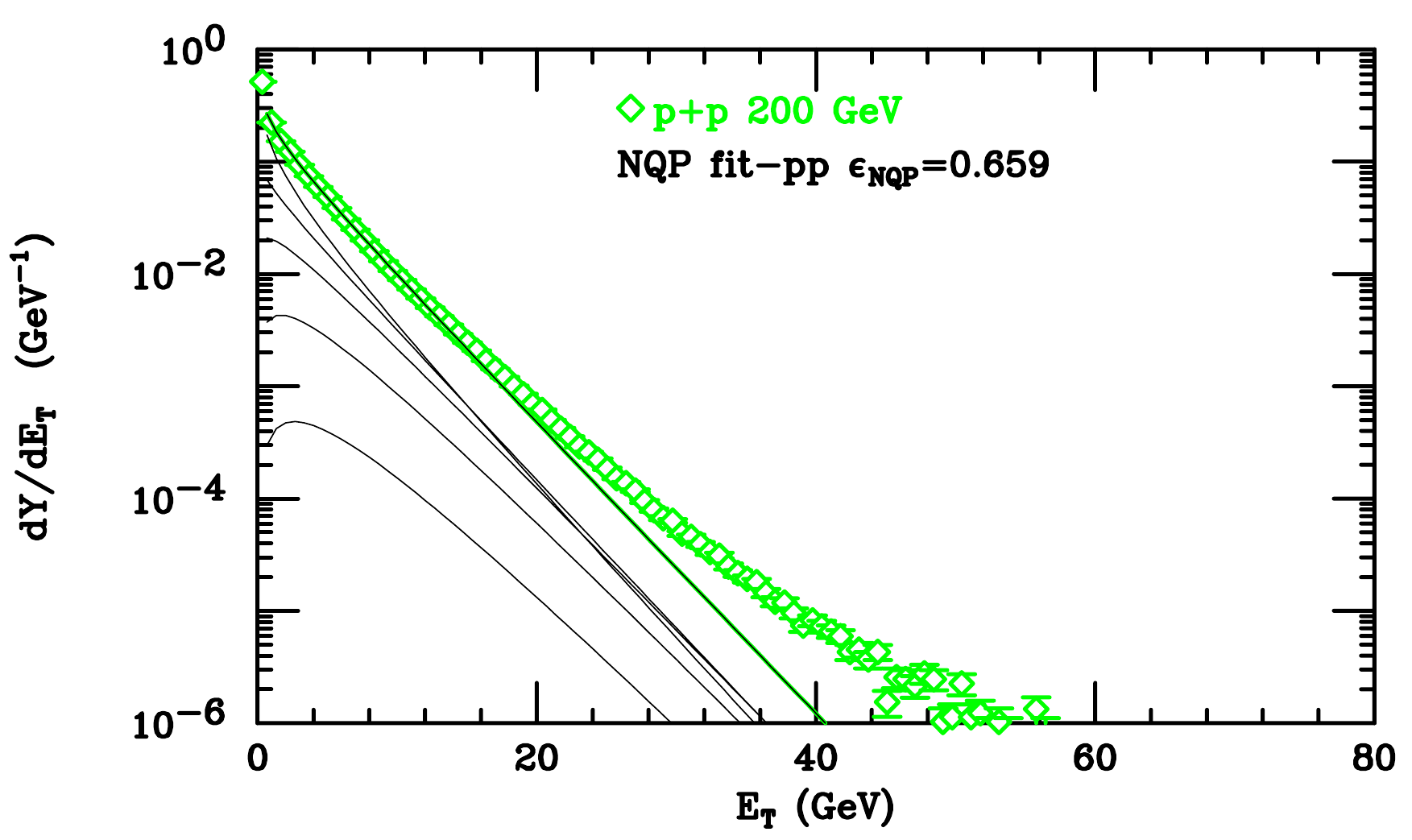}
\includegraphics[width=0.49\textwidth]{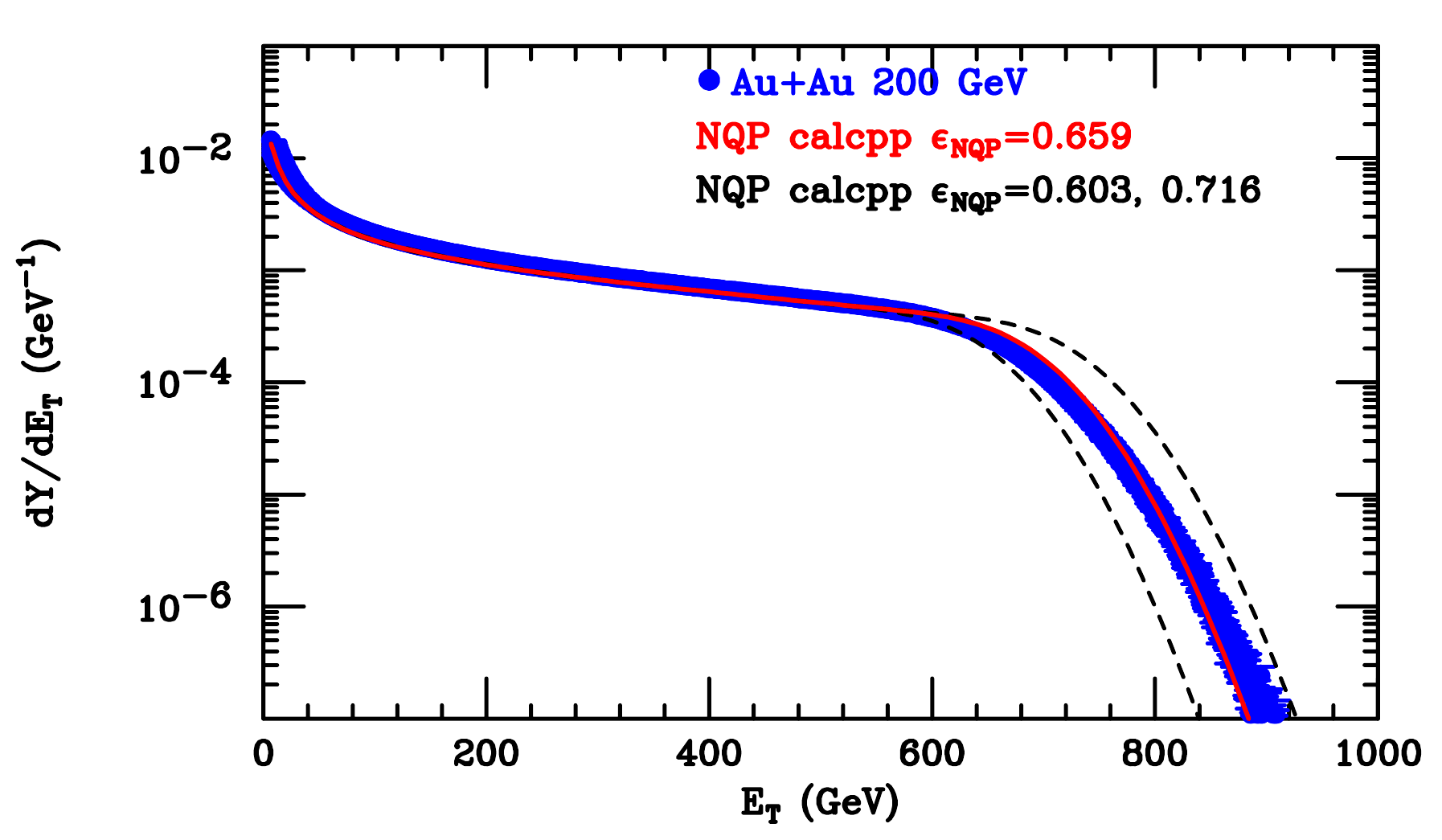}
\end{center}\vspace*{-1.0pc}
\caption[]{PHENIX~\cite{ppg100} $E_T$ distribution in $\sqrt{s_{NN}}=200$ GeV p-p and Au+Au collisions. a) (left) Fit of p-p distribution to a sum of properly weighted 2,3,\ldots 6 constituent-quark-participant $E_T$ distributions. Lines indicate the individual distributions plus the sum. The $q-q$ cross section had been tuned to reproduce the inelastic p-p cross section.  b) (right) Calculation of the Au+Au $E_T$ distribution in the NQP model as the sum of properly weighted convolutions of the constituent-quark-participant $E_T$ distribution derived in (a). Dashed lines indicate the systematic uncertainty. 
\label{fig:PXNQPppAA}}\vspace*{-1.0pc}
\end{figure}

The Wounded Nucleon (WNM), Additive Auark (AQM) and constituent-Quark-Participant (NQP) models are all examples of extreme-independent models in which the nuclear geometry of the interaction can be calculated in independently of the dynamics of particle production, which can be derived from experimental measurements, usually the $p$$+$$p$ (or $p$+$A$) 
measurement in the same detector.
The constituent-quark method for $E_T$ distributions~\cite{ppg100} is illustrated in Fig.~\ref{fig:PXNQPppAA}. 
The \mbox{p-p} $E_T$ distribution (Fig.~\ref{fig:PXNQPppAA}a) is calculated as the properly weighted sum of 2 to 6 convolutions of the fitted $E_T$ distribution of a constituent-quark-participant, which is then applied to d+Au (Fig.~\ref{fig:NQP}b) and Au+Au collisions (Fig.~\ref{fig:PXNQPppAA}b), with excellent agreement in both cases.  In a standard Monte Carlo Glauber calculation of the nuclear geometry,   the positions of the three-constituent quarks are generated about the position of each generated nucleon according to the measured charge distribution of the proton, which gives a physical basis for ``proton size fluctuations'' discussed at LHC~\cite{ATLAS-Gribov}.  

Constituent-quark-participants might also explain the increase of the ``effective energy'' in A+A collisions compared to p-p collisions  discussed previously (Fig.~\ref{fig:PHOBOSdAAA}b). The $\mean{N_{\rm qp}/N_{\rm part}}$ is 1.5 for a p-p collision but rises to 2.3-2.7 for more central (0-50\%) A+A collisions. Thus the ``effective energy'' for particle production increases due to an increase in the number of (constituent-quark) participants, not because of additional collisions of a given participant. This preserves the assumption in these ``extreme-independent'' participant models that successive collisions of a participant do not increase its particle emission.  
\subsection{Collective Flow in d+Au (and p+Pb) collisions?}
The most surprising soft-physics result in p+Pb and d+Au physics in 2013 concerned what looks very much like collective flow observed in these small systems, where no (or negligible) medium or collective effect was expected.  
         \begin{figure}[!h]
   \begin{center}
\includegraphics[width=0.95\textwidth]{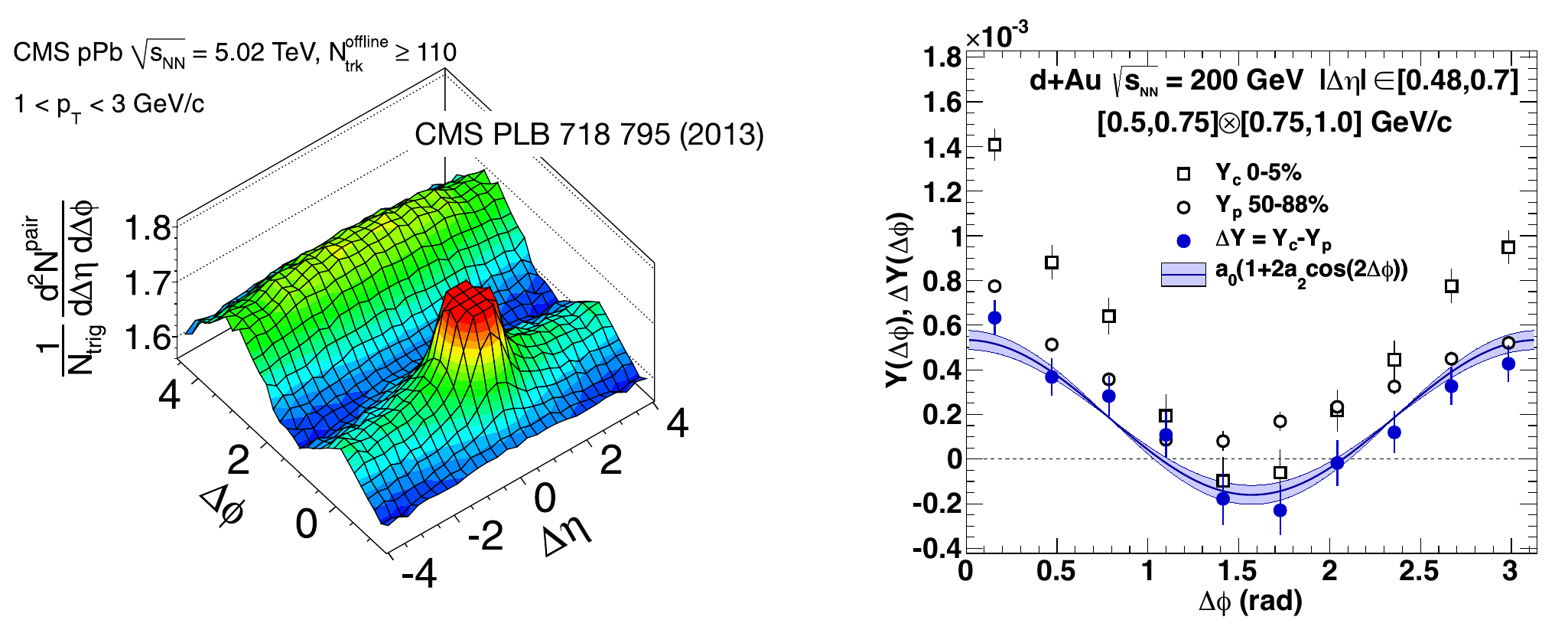}
\end{center}\vspace*{-1.0pc}
\caption[]{a) (left) CMS ridge in two-particle correlations~\cite{CMSPLB718}. b) (right) PHENIX two-particle azimuthal correlations in d+Au at RHIC~\cite{PXv2dAu}.
\label{fig:RidgedAu}}\vspace*{-0.8pc}
\end{figure}
Fig.~\ref{fig:RidgedAu}a shows a LEGO plot of $\Delta\eta$, $\Delta\phi$, the difference in polar and azimuthal angles from correlations of two particles with $1<p_T<3$ GeV/c in p+Pb by CMS at $\sqrt{s_{NN}}=5.02$ TeV~\cite{CMSPLB718}. A clear $1+2v_2\cos 2\Delta\phi$ modulation of the distribution independent of $\Delta\eta$ is observed, called the `ridge' in Au+Au collisions where the modulations $v_2$, $v_3$, \ldots $v_n$ are attributed to collective flow of the \QGP\ medium. %The ALICE~\cite{ALICEPLB719} and ATLAS~\cite{ATLASPRL110} experiments at the LHC and also 
At RHIC, PHENIX confirmed this result in d+Au collisions at $\sqrt{s_{NN}}=200$ GeV (Fig.~\ref{fig:RidgedAu}b)~\cite{PXv2dAu}. In order to remove any $v_2$ effect due to two-particle correlations of hard-scattering, lower $p_T$ triggers were used as well as cuts in $\Delta\eta$ to remove the same-side peak. Also, since there is no suppression of hard-scattering in p+Pb or d+Au collisions (recall Fig.~\ref{fig:RAAdAuAuAu}), the conditional two-particle yield from di-jets is independent of centrality. Thus, any residual hard-scattering effect was removed by subtracting the peripheral (50-88\%) from the central (0-5\%) measurement which revealed the beautiful $\cos 2\Delta\phi$ curve characteristic of elliptical flow shown in Fig.~\ref{fig:RidgedAu}b.  
        \begin{figure}[!h]
   \begin{center}
\raisebox{0.6pc}{\includegraphics[width=0.51\textwidth]{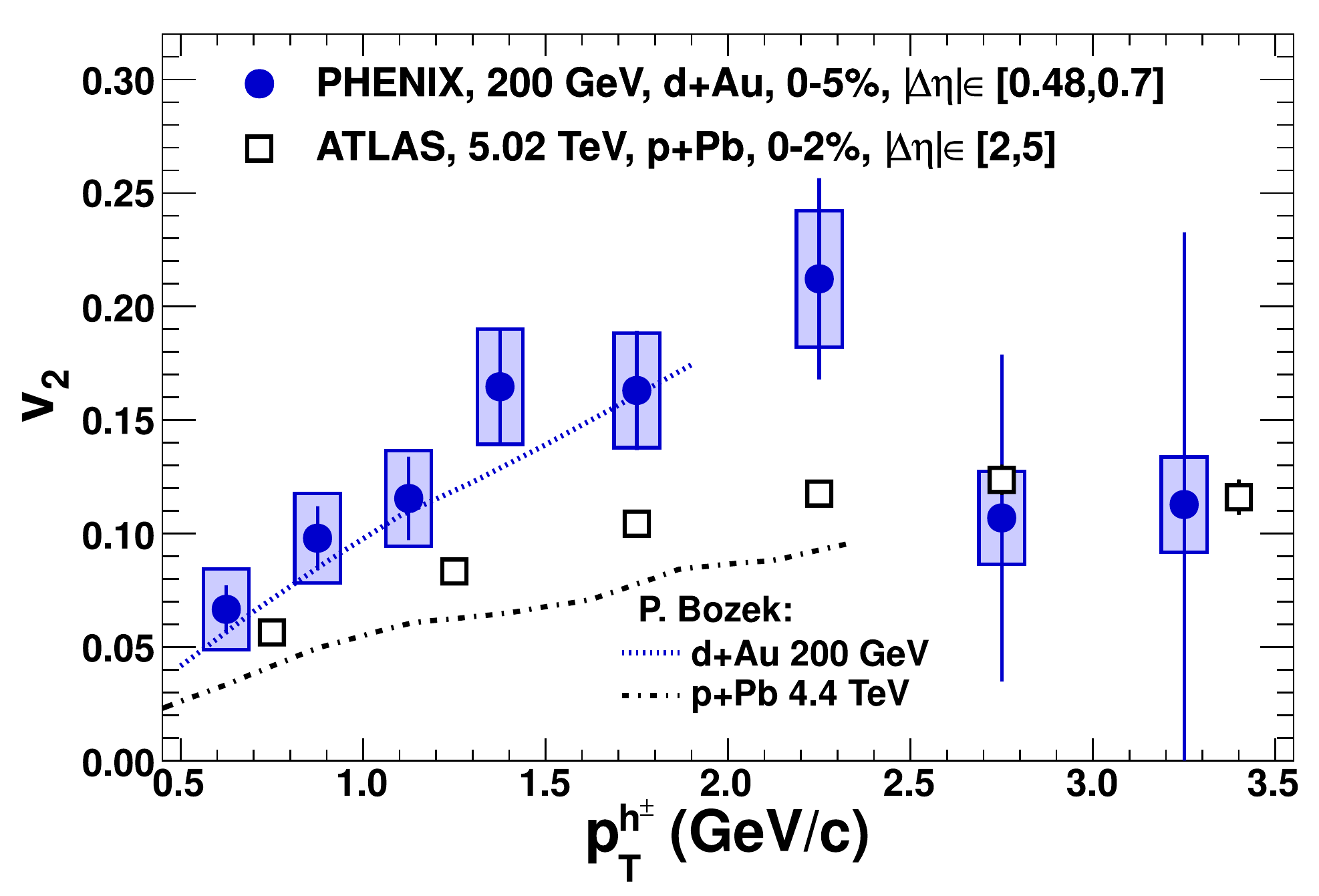}}
{\includegraphics[width=0.47\textwidth]{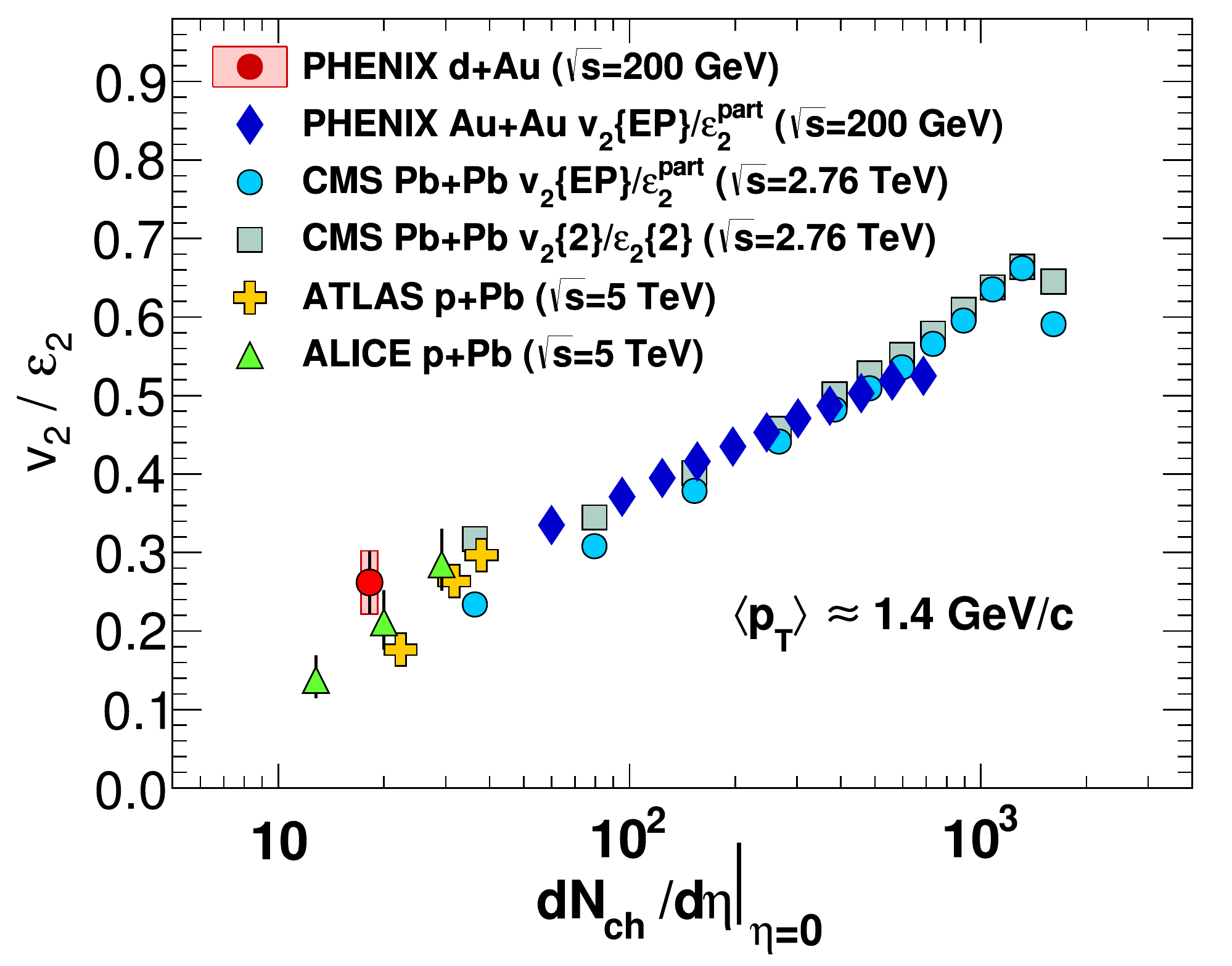}}
\end{center}\vspace*{-1.0pc}
\caption[]{a) (left) Comparison of $v_2$ vs. $p_T$ in d+Au at RHIC to p+Pb at LHC~\cite{PXv2dAu}. b) (right) Compilation of $v_2/\varepsilon$ vs. $dN_{\rm ch}/d\eta$ at $p_T=1.4$ GeV/c in d+Au, p+Pb, Au+Au and Pb+Pb collisions at RHIC and LHC~\cite{PXv2dAu}. 
\label{fig:PXv2dAu}}\vspace*{-1.0pc}
\end{figure}

Figure~\ref{fig:PXv2dAu}a compares the $v_2$ measurements vs $p_T$ from d+Au at $\sqrt{s_{NN}}=200$ GeV and p+Pb at 5.02 TeV.  
The larger values from the d+Au results are thought to be due to the larger eccentricity ($\varepsilon$) of the two-nucleon deuteron compared to the single nucleon proton. In fact, the values of $v_2/\varepsilon$ from d+Au and p+Pb are consistent with the dependence of $v_2/\varepsilon$ on $dN_{\rm ch}/d\eta$ (Fig.~\ref{fig:PXv2dAu}b) observed in Au+Au and Pb+Pb collisions, which was taken as proof of collective flow from hydrodynamics. 
These new results again underscore the importance of p-p and p+A comparison data to understand the observations in A+A collisions, where the detailed physics of the \QGP\ is far from understood. 
\section{The Future}
\begin{figure}[!t]
   \begin{center}
\includegraphics[width=0.95\textwidth]{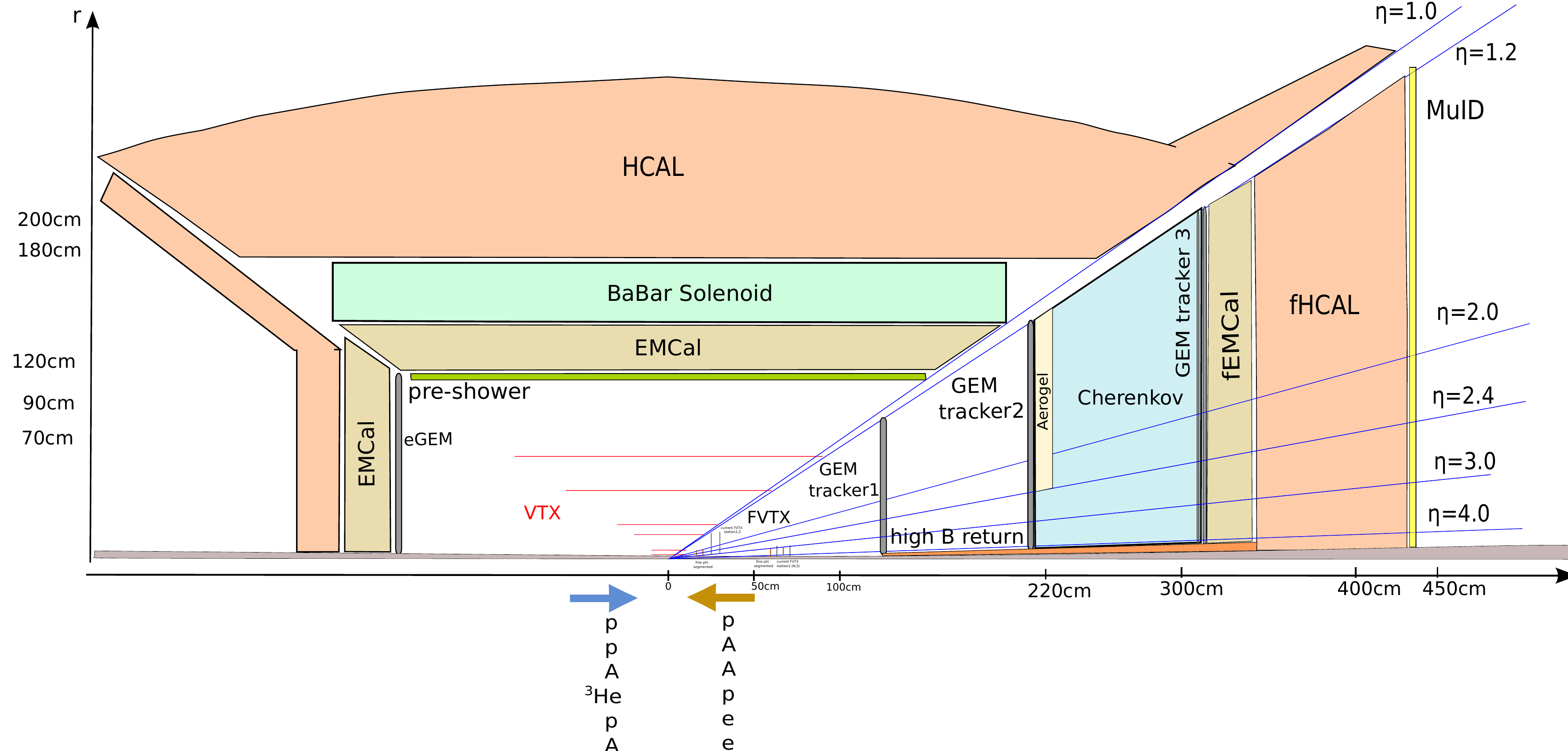}
\end{center}\vspace*{-1.0pc}
\caption[]{a) sPHENIX concept with forward detector~\cite{sPHENIX2013}.
\label{fig:ePHENIX}}\vspace*{-0.8pc}
\end{figure}

Toward this goal, PHENIX has proposed a new more conventional collider detector, s(uper)PHENIX, based on a thin-coil superconducting solenoid, to concentrate on hard-scattering and jets. This would replace the very successful but 15 year old special purpose small aperture two-arm spectrometer designed to measure $J/\Psi\rightarrow e^+ + e^-$ down to zero $p_T$ at mid-rapidity, the original expected signal for deconfinement, as well as identified particles such as single $e^{\pm}$ from heavy quark decay, $\pi^0$, $\eta$ and other hadrons (recall Fig.~\ref{fig:Tshirt}b) that could cause background to the $J/\Psi$ but which turned out to be valuable probes of the \QGP\ . This year sPHENIX got a big boost by acquiring the (made in Italy) BABAR solenoid magnet from SLAC which became available when the B-factory in Italy was unfortunately cancelled. Conceptual design of the new experiment is well underway, with mid-rapidity, forward and eRHIC capability~\cite{sPHENIX2013} (Fig.~\ref{fig:ePHENIX}). New collaborators would be most welcome.   

%\section*{Acknowledgments}
%This research was supported by U.S.Department of Energy, DE-AC02-98CH1088. This manuscript has been authored by employees of Brookhaven Science Associates, LLC under Contract No. DE-AC02-98CH10886 with the U.S. Department of Energy. The publisher by accepting the manuscript for publication acknowledges that the United States Government retains a non-exclusive, paid-up, irrevocable, world-wide license to publish or reproduce the published form of this manuscript, or allow others to do so, for United States Government purposes.

%\section*{References}

%\begin{thebibliography}{000} %for 3 digits


\begin{thebibliography}{00}  %for 2 digits
%\begin{thebibliography}{0}    %for 1 digit
%%%2011
%\begin{thebibliography2011}{99}
\bibitem{BearMountain} {\it Report of the Workshop on BeV/Nucleon Coliisions of Heavy Ions---How and Why}, Bear Mountain, NY, 29 November--1 December 1974. (BNL-50445, Upton NY, 1975). 
\bibitem{seeMJTROP} See Ref.~\cite {MJTROP} for a more extensive list of references. 
\bibitem{MJTROP} M.~J.~Tannenbaum, \Journal{\RPP}{69}{2005--2059}{2006}.
\bibitem{Shuryak80} E.~V.~Shuryak, \Journal{\PLC} {61} {71--158}{1980}. 
\bibitem{BRWP} BRAHMS Collab. (I.~Arsene {\it et al.}),  \Journal{\NPA}{757}{1--27}{2005}.
\bibitem{PHWP} PHOBOS Collab. (B.~B.~Back {\it et al.}), \Journal{\NPA}{757}{28--101}{2005}.
\bibitem{STWP} STAR Collab. (J.~Adams {\it et al.}), \Journal{\NPA}{757}{102--183}{2005}.
\bibitem{PXWP} PHENIX Collab. (K.~Adcox {\it et al.}), \Journal{\NPA}{757}{184--283} {2005}.
\bibitem{THWPS} D.~Rischke and G.~Levin, eds., \Journal{\NPA}{750}{1--171}{2005}. 
\bibitem{CERNBaloney} Press Conference, {\it A New State of Matter Created at CERN} \href{http://newstate-matter.web.cern.ch/newstate-matter/}{http://newstate-matter.web.cern.ch/newstate-matter/Story.html}
\bibitem{NYT02102000} New York Times, February 10, 2000, Page 1, \href{http://www.nytimes.com/2000/02/10/world/particle-physicists-getting-closer-to-the-bang-that-started-it-all.html?scp=4&sq=A%20new%20state%20of%20matter&st=cse}  {\em Particle Physicists Getting Closer To the Bang That Started It All}. 
\bibitem{EriceProcPR} Discoveries and methods at RHIC have also been discussed by the present author in ISSP proceedings for the years 2008 (\href{http://arxiv.org/abs/0906.0745}{arXiv:0906.0745}), 2009 (Ref.~\cite {MJTIJMPA2011}) as well as in a recent textbook (Ref.~\cite {RATCUP}). 
\bibitem{MJTIJMPA2011} M.~J.~Tannenbaum, \Journal{\IJMPA}{26}{5299-5335}{2011}. 
\bibitem{RATCUP} Jan Rak and Michael J. Tannenbaum, {\it High $p_T$ Physics in the Heavy Ion Era}, (Cambridge Univ. Press, Cambridge \& New York, 2013).
\bibitem{ATLASdijet} A.~Aad, {\it et al.} (ATLAS Collab.), \Journal{\PRL}{105}{252303}{2010}.
\bibitem{CMSdijet} S.~ Chatrchyan, {\it et al.} (CMS Collab.), \Journal{\PRC}{84}{024906}{2011}. 
\bibitem{RHICNIM} M.~Harrison, T.~Ludlam and S.~Qzaki (eds.), {\em The Relativistic Heavy Ion Collider Project: RHIC and its Detectors}, \Journal{\NIMA}{499}{235--880}{2003}.
\bibitem{NSRL} NASA/BNL Space Radiation Program, \url{http://www.bnl.gov/medical/NASA/LTSF.asp}
\bibitem{Bunce-ARNPS50} G.~Bunce, {\it et al.}, \Journal{\ARNPS}{50}{525}{2000}. %N.~Saito, J.~Soffer and W.~Vogelsang, 
\bibitem{3DSC-CC1012} M.~Blaskiewicz and W.~Fischer, \href{http://cerncourier.com/cws/article/cern/50795}{\Journal{CERN Courier\ }{52}{17}{Oct. 2012}}.
\bibitem{CBAmagnetsNIMA235} E.~J.~Bleser, {\it et al.}, \Journal{\NIMA}{235}{435}{1985}.
\bibitem{Bozorth} R.~M.~Bozorth, {\it Ferromagnetism} (VanNostrand, New York, 1951)
\bibitem{NaturePhysics7} ALPHA Collab. (G.~B.~Andresen {\it et al.}), {\Journal{Nature Physics\ }{7}{558--564}{2011}}; \Journal{Nature\ }{468}{673--676}{2010}.
\bibitem{T2KPR} Press Release, ``Indication of Electron Neutrino Appearance at the T2K Experiment'' June 15, 2011, \url{http://www.kek.jp/intra-e/press/2011/J-PARC_T2Kneutrino.html}
\bibitem{IEEENS22} R.~Chasman, G.~K.~Green and E.~M.~Rowe, \Journal{IEEE Trans Nucl. Sci.\ }{NS-22}{1765--1767}{1975}.
\bibitem{NSLSIIsource} NSLS-II Source Properties and Floor Layout \url{http://www.bnl.gov/ps/docs/pdf/SourceProperties.pdf}
\bibitem{BNL-g-2} G.~W.~Bennett, {\it et al.}, \Journal{\PRD}{73}{072003}{2006}.
\bibitem{g-2-2-1966} F.~J.~M.~Farley, J.~Bailey, R.~C.~A.~Brown, M.~Giesch. H.~Jostlein, S.~van~der~Meer, E.~Picasso and M.~Tannenbaum, \Journal{\NCA}{45A}{281-286}{1966}.
\bibitem{CERN-first-g-2-final} G.~Charpak, F.~J.~M.~Farley, R.~L.~Garwin, T. Muller, J.~C.~Sens and A.~Zichichi, \Journal{\NCA}{37}{1241-1363}{1965}.
\bibitem{MuP1968} R.~W.~Ellsworth, A.~C.~Melissinos, J.~H.~Tinlot, H.~von~Briesen, Jr., T.~Yamanouchi, L.~M.~Lederman, M.~J.~Tannenbaum, R.~L.~Cool and A.~Maschke, \Journal{\PR}{165}{1449-1465}{1968}. 
\bibitem{Danby1962} G.~Danby, J-M.~Gaillard, K.~Goulianos, L.~M.~Lederman, N.~Mistry, M.~Schwartz and J.~Steinberger, \Journal{\PRL}{9}{36-44}{1962}.
\bibitem{Fitch1963} J.~H.~Christenson, J.~W.~Cronin, V.~L.~Fitch and R.~Turlay, \Journal{\PRL}{13}{138-140}{1964}.
\bibitem{CoolICHEP72} CCR Collab. (R.~L.~Cool {\it et al.}), in {\em Proc. XVI Int. Conf. High Energy Physics}, eds. J.~D.~Jackson and A.~Roberts (NAL, Batavia, IL, 1973), Vol 3, p. 317.
\bibitem{OwensKimel} J.~F.~Owens and J.~D.~Kimel, \Journal{\PRD}{18}{3313-3319}{1978}.
\bibitem{FFF} R.~P.~Feynman, R.~D.~Field, and G.~C.~Fox, \Journal{\PRD}{18}{3320-3343}{1978}.
\bibitem{NA5PLB112} C.~De~Marzo, {\it et al.}, \Journal{\PLB}{112}{173-177}{1982}.
\bibitem{BjPRD8} J.~D.~Bjorken, \Journal{\PRD}{8}{4098-4106}{1973}.
\bibitem{E260NPB134} C.~Bromberg, {\it et al.}, \Journal{\NPB}{134}{189-241}{1978}. 
\bibitem{UA1Paris82} UA1 Collab. (G. Arnison {\it et al.}), Transverse Energy Distributions in the Central Calorimeters (1982), CERN-EP-82-122. 
\bibitem{UA2JetICHEP82} UA2 Collab. (J.-P. Repellin {\it et al.}), in {\em Proc. 21st Int. Conf. High Energy Physics}, eds P. Petiau and M. Porneuf (Journal de Physique Colloques, Paris, 1982), vol. 43, pp. C3-571$-$C3-578.
\bibitem{UA1PLB107} UA1 Collab. (G. Arnison {\it et al.}), \Journal{\PLB}{107}{320-324}{1981}.
\bibitem{PDG} Particle Data Group (C.~Amsler {\it et al.}), \Journal{\PLB}{667}{1}{2008}.
\bibitem{MatsuiSatz86} 
T.~Matsui and H.~Satz, \Journal{\PLB}{178}{416}{1986}. 
\bibitem{NA50EPJC39} NA50 Collab. (B.~Alessandro {\it et al.}),  \Journal{\EPJC}{39}{335--345}{2005}. See also, F.~Prino, Proc. XXX International Symposium on  Multiparticle Dynamics, \href{http://arxiv.org/abs/hep-ex/0101052}{arXiv:hep-ex/0101052v1}.
\bibitem{E772} E772 Collab. (D.~M.~Alde {\it et al.}) , \Journal{\PRL}{66}{2285--2288}{1991}. See, also, M.~J.~Leitch, \Journal{\EPJC}{43}{157--160}{2005} and references therein.  
\bibitem{LHCJINST} A.~Breskin and R.~Voss (eds.), {\it The Large Hadron Collider: Accelerator and Experiments}, \href{http://jinst.sissa.it/LHC/}{\Journal{J.~Instrum.\ }{3}{S08001--S08007}{2008}}.
\bibitem{JTMQM12} PHENIX Collab. (J.~T.~Mitchell {\it et al.}), \Journal{\NPA}{904,905}{903c}{2013}.
\bibitem{ppg019} PHENIX Collab. (S.~S.~Adler {\it et al.}), \Journal{\PRC}{71}{034908}{2005}.
\bibitem{ALICEmult} ALICE Collab. (K.~Aamodt {\it et al.}), \Journal{\PRL}{106}{032301}{2011}.
\bibitem{egseeMJTISSP2009} See, for example,  Refs.~\cite {MJTIJMPA2011,RATCUP}.  
\bibitem{egseePT} The detector is so non-conventional that it made the cover of \href{http://www.phenix.bnl.gov/phenix/WWW/docs/covers/phystoday/2003oct/phystoday03oct.jpg}{Physics Today, October 2003}.
\bibitem{STARNature473} STAR Collab. (H.~Agakishiev {\it et al.}) , \Journal{Nature\ }{473}{353--356}{2011}. 
\bibitem{0909.0566} STAR Collab. (B.~I.~Abelev, {\it et al.}), \href{http://arxiv.org/abs/0909.0566v1}{arXiv:0909.0566v1 [nucl-ex]}.
\bibitem{LaceyQM05} R.~A.~Lacey, \Journal{\NPA}{774}{199--214}{2006}.
\bibitem{KanetaQM04} PHENIX Collab. (M.~Kaneta {\it et al.}),  \Journal{\JPG}{30}{S1217--S1220}{2004}.
\bibitem{PXArkadyQM06} PHENIX Collab. (A.~Adare {\it et al.}),  \Journal{\PRL}{98}{162301}{2007}.
\bibitem{Ollitrault} J.-Y.~Ollitrault, \Journal{\PRD}{46}{229--245}{1992}; \Journal{\NPA}{638}{195c--206c}{1998}.
\bibitem{HeiselbergLevy} H.~Heiselberg and A.-M.~Levy, \Journal{\PRC}{59}{2716--2727}{1999}.
\bibitem{VoloshinQM02} S.~A.~Voloshin, \Journal{\NPA}{715}{379c--388c}{2003}.
\bibitem{TeaneyPRC68} D.~Teaney, \Journal{\PRC}{68}{034913}{2003}.
\bibitem{Kovtun05} P.~K.~Kovtun, D.~T.~Son, and A.~O.~Starinets, \Journal{\PRL}{94}{111601}{2005}. 
\bibitem{LaceyPRL98} R.~A.~Lacey, {\it et al.}, \Journal{\PRL}{98}{092301}{2007}. 
\bibitem{CKMPRL97} L.~P.~Csernai, J.~I.~Kapusta, and L.~D.~McLerran, \Journal{\PRL}{97}{152303}{2006}.
\bibitem{ppg083} PHENIX Collab. (A.~Adare {\it et al.}), \Journal{\PRC}{78}{014901}{2008}; \Journal{\PRC}{77}{011901(R)}{2008}.
\bibitem{ppg067} PHENIX Collab. (A.~Adare {\it et al.}), \Journal{\PRL}{98}{232302}{2007}.
\bibitem{CSST-Coney05} J.~Casalderrey-Solana, E.~V.~Shuryak and D. Teaney, \Journal{\JPCS}{27}{22-31}{2005}.
\bibitem{egseeppg083} For example, see Ref.~\cite {ppg083} for a discussion and list of references. 
\bibitem{JTMQM06} PHENIX Collab. (J.~T.~Mitchell {\it et al.}), \Journal{\JPG}{34}{S911--S914}{2007}. 
\bibitem{STARridgePRC80} STAR Collab. (B.~I.~Abelev {\it et al.}), \Journal{\PRC}{80}{064912}{2009}.
\bibitem{PutschkeHP06} STAR Collab. (J.~Putschke {\it et al.}), \Journal{\NPA}{783}{507c--510c}{2007}.
\bibitem{JacobsHP04} P.~Jacobs, \Journal{\EPJC}{43}{467--473}{2005}.
\bibitem{AlverOllitrault} B.~H.~Alver, C.~Gombeaud, M.~Luzum, and J.~Y.~Ollitrault, \Journal{\PRC}{82}{034913}{2010}.
\bibitem{AlverRoland} B.~Alver and G.~Roland, \Journal{\PRC}{81}{054905}{2010}.
\bibitem{BrazilNuXuv3} J.~Takahashi, {\it et al.}, \Journal{\PRL}{103}{242301}{2009}. Also, see A.~P.~Mishra, {\it et al.}, \Journal{\PRC}{77}{064902}{2008}.  
\bibitem{EsumiQM11} PHENIX Collab. (S.~Esumi, {\it et al.}) , \href{http://arxiv.org/abs/1110.3223v1}{arXiv:1110.3223v1 [nucl-ex]}.
\bibitem{MarekQM11} Presentation by M.~Ga\'zdzicki at \href{http://indico.cern.ch/confSpeakerIndex.py?confId=30248}{Quark Matter 2011},  \Journal{\JPG}{38}{124024}{2011}. Also see NA49 Collab. (M.~Ga\'zdzicki {\it et al.}), \Journal{\JPG}{30}{S701--S708}{2004}.
\bibitem{LKBMQM11} STAR collab. (L.~Kumar {\it et al.}), \Journal{\JPG}{38}{124145}{2011}, (B.~Mohanty, {\it et al.}), \Journal{\JPG}{38}{124023}{2011}.
\bibitem{CleymansOeschlerPLB615} J.~Cleymans, H.~Oeschler, K.~Redlich and S.~Wheaton, \Journal{\PLB}{615}{50--54}{2005}.
\bibitem{LBLJune23} \url{http://newscenter.lbl.gov/news-releases/2011/06/23/when-matter-melts/}.
\bibitem{AsakawaHMPRL85} M.~Asakawa, U.~Heinz, and B.~M\"uller, \Journal{\PRL}{85}{2072}{2000}.
\bibitem{KochCFRNC06} e.g. see V.~Koch, \href{http://pos.sissa.it/archive/conferences/030/008/CFRNC2006_008.pdf}{PoS(CFRNC2006)008}
\bibitem{AMuellerPRD4} A.~H.~Mueller, \Journal{\PRD}{4}{150}{1971}. 
\bibitem{TarnowskyQM2011} Presentation by T.~J.~Tarnowsky at \href{http://indico.cern.ch/confSpeakerIndex.py?confId=30248}{Quark Matter 2011}, STAR Collab. (T.~J.~Tarnowsky {\it et al.}), \Journal{\JPG}{38}{124054}{2011}.
\bibitem{STARnetPPRL105} STAR Collab. (M.~M.~Aggarwal, {\it et al.}), \Journal{\PRL}{105}{022302}{2010}.
\bibitem{Science332} S.~Gupta, X.~Luo, B.~Mohanty, H.~G.~Ritter and N.~Xu, \Journal{Science\ }{332}{1525}{2011}. 
\bibitem{ppg104} PHENIX Collab. (A.~Adare, {\it et al.}), \Journal{\PRD}{85}{092004}{2012}.
\bibitem{STAR14021558} STAR Collaboration (L.~Adamczyk, {\it et al.}), Beam energy dependence of moments of the net-charge multiplicity distributions in Au+Au collisions at RHIC (2014), {\tt arXiv:1402.1558}. 
\bibitem{Westfall2013} T.~J.~Tarnowsky and G.~D.~Westfall, \Journal{\PLB}{724}{51}{2013}.
\bibitem{Barndorff2013} O.~E.~Barndorff-Nielsen, D.~G.~Pollard and N.~Shephard, \Journal{Quant. Finance\ }{12}{587--605}{2012}.
\bibitem{BDMPS} See R.~Baier, D.~Schiff and B.~G.~Zakharov, \Journal{\ARNPS}{50}{37--69}{2000}, and references therein. 
\bibitem{QCDCompton} H.~Fritzsch and P.~Minkowski, \Journal{\PLB}{69}{316}{1977}.
\bibitem{ppg003} PHENIX Collab. (K.~Adcox {\it et al.}), \Journal{\PRL}{88}{022301}{2002}.
\bibitem{ppg054} PHENIX Collab. (S.~S.~Adler {\it et al.}), \Journal{\PRC}{76}{034904}{2007}.
\bibitem{ppg086} PHENIX Collab. (A.~Adare, {\it et al.}), \Journal{\PRL}{104}{132301}{2010}.
\bibitem{ppg133} PHENIX Collab. (A.~Adare, {\it et al.}), \Journal{\PRC}{87}{034911}{2013}.
\bibitem{ALICEPLB696}  ALICE Collab. (K.~Aamodt, {\it et al.}), \Journal{\PLB}{696}{30}{2011}.

\bibitem{ppg138} PHENIX Collab. (A.~Adare, {\it et al.}), \Journal{\PRL}{109}{152301}{2012}.
\bibitem{STARhiQM2012} STAR Collab. (Xin Dong {\it et al.}), STAR Highlights, Quark Matter 2012, \mbox{\url{https://drupal.star.bnl.gov/STAR/files/STARhighlight_QM2012_0810.pdf}}
\bibitem{ppg140} PHENIX Collab. (A.~Adare, {\it et al.}), \Journal{\PRC}{87}{054907}{2013}. 
\bibitem{ppg136} PHENIX Collab. (A.~Adare, {\it et al.}), \Journal{\PRD}{86}{072008}{2012}.
\bibitem{Tingpizff} L3 Collab. (B.~Adeva {\it et al.}), \Journal{\PLB}{259}{199--208}{1991}.
\bibitem{ppg029} PHENIX Collab. (S.~S.~Adler, {\it et al.}), \Journal{\PRD}{74}{072002}{2006}.
\bibitem{BW06} N.~Borghini and U.~A.~Wiedemann, \Journal{\NPA}{774}{549--522}{2006}; see also {arXiv:hep-ph/0506218v1}.
\bibitem{ppg095} PHENIX Collab. (A.~Adare {\it et al.}), \Journal{\PRD}{82}{072001}{2010}. 
\bibitem{TASSOZPC47} TASSO Collab. (W.~Braunschweig {\it et al.}), \Journal{\ZPC}{47}{187}{1990}.
\bibitem{PXPRL111gamh} PHENIX Collab. (A.~Adare {\it et al.}), \Journal{\PRL}{111}{032301}{2013}. 
\bibitem{CMSQM2012} CMS Collab. (G.~Roland {\it et al.}), \Journal{\NPA}{904-905}{43c-50c}{2013}.
\bibitem{ppg106} PHENIX Collab. (A.~Adare {\it et al.}),  \Journal{\PRL}{104}{252301}{2010}.
\bibitem{MJT-Utrecht} PHENIX Collab. (M.~J.~Tannenbaum {\it et al.}), \href{http://arxiv.org/abs/1109.0760v1}{arXiv:1109.0760v1 [nucl-ex]}. 
\bibitem{CMS-dijet-PLB712} CMS Collab. (S.~Chatrchyan {\it et al.}), \Journal{\PLB}{712}{176-197}{2012}
\bibitem{ppg126} PHENIX Collab. (A.~Adare {\it et al.}),  \Journal{\PRL}{109}{122302}{2012}.
\bibitem{ppg042} PHENIX Collab. (S.~S.~Adler, {\it et al.}), \Journal{\PRL}{94}{232301}{2005}.
\bibitem{PXcharmAA06} PHENIX Collab. (A.~Adare {\it et al.}),  \Journal{\PRL}{98}{172301}{2007}.
\bibitem{ALICEDsuppression} ALICE Collab. (B.~Abelev {\it et al.}),  \Journal {JHEP\ }{09}{112}{2012}. 
\bibitem{CMSUps1Ssupp} CMS Collab. (S.~Chatrchyan {\it et al.}), \Journal{JHEP\ }{05}{063}{2012}. 
\bibitem{AZYukawa} A.~Zichichi, \href{http://dx.doi.org/10.1016/j.nuclphysa.2008.02.229}{\Journal{\NPA}{805}{36c}{2008}}.
\bibitem{Nobel2013} {\em The Nobel Prize in Physics 2013}, \href{http://www.nobelprize.org/nobel_prizes/physics/laureates/2013/}{Nobelprize.org}.
\bibitem{RachidQM2012} PHENIX Collab. (R.~Nouicer {\it et al.}),  \Journal{\NPA}{904-905}{647c-652c}{2013}.

\bibitem{PXJPsiAuAu200} PHENIX Collab. (A.~Adare {\it et al.}),  \Journal{\PRL}{98}{232301}{2007}.
\bibitem{GunjiQM06} PHENIX Collab. (T.~Gunji {\it et al.}),  \Journal{\JPG}{34}{S749--S752}{2007},   {\tt http://www.sinap.ac.cn/qm2006/parallel.htm\#p21} .
\bibitem{RappPLB664} X.~Zhao and R.~Rapp, \Journal{\PLB}{664}{253--257}{2008}
\bibitem{PBMStachelPLB490} P.~Braun-Munzinger and J.~Stachel, \Journal{\PLB}{490}{196--202}{2000}.
\bibitem{ThewsPRC63} R.~L.~Thews, M.~Schroedter and J.~Rafelski, \Journal{\PRC}{63}{054905}{2001}.
\bibitem{AndronicNPA79} A.~Andronic, P.~Braun-Munzinger, K.~Redlich and J.~Stachel, \Journal{\NPA}{789}{334--356}{2007}.
\bibitem{ALICEJPsiPRL109} ALICE Collab. (B.~Abelev, {\it et al.}), \Journal{\PRL}{109}{072301}{2012}. 
\bibitem{PaoloGISSP2013} ALICE Collab. (P.~Giubellino, {\it et al.}), Highlights from ALICE, Presented at {\em International School of Subnuclear Physics, 51st Course}, Erice, Sicily, Italy, 24 June-3 July 2013, \url{http://www.ccsem.infn.it/issp2013/docs/Erice_Giubellino_2013.pdf}
\bibitem{ThewsMangano} R.~L.~Thews and M.~L.~Mangano, \Journal{\PRC}{73}{014904}{2006}.
\bibitem{SatzCalibrating-2013} H.~Satz, \href{http://dx.doi.org/10.1155/2013/242918}{\Journal{Advances in H E P\ }{2013}{242918}{2013}}. Also see {\tt arXiv:1303.3493v2}
\bibitem{ppg077} PHENIX Collab. (A.~Adare {\it et al.}),  \Journal{\PRC}{84}{044905}{2011}. 
\bibitem{STARPLB722} STAR Collab. (L.Adamczyk {\it et al.}), \Journal{\PLB}{722}{55-62}{2013}.
\bibitem{PXPRC88} PHENIX Collab. (A.~Adare, {\it et al.}), \Journal{\PRC}{88}{024906}{2013}.
\bibitem{PHOBOSPRC72} PHOBOS Collab. (B.~B.~Back, {\it et al.}), \Journal{\PRC}{72}{031901(R)}{2005}.
\bibitem{WNM} A.~Bia\l as, A.~B\l eszynski and W.~Czy\. z, \Journal{\NPB}{111}{461}{1976}.
\bibitem{PHOBOSPRC74} PHOBOS Collab. (B.~B.~Back, {\it et al.}),  \Journal{\PRC}{74}{021901(R)}{2006}. 
\bibitem{BasilePLB95} M.~Basile, \ldots A.~Zichichi, {\it et al.}, \Journal{\PLB}{95}{311-312}{1980}.
\bibitem{ppg100} PHENIX Collab. (S.~S.~Adler {\it et al.}), \Journal{\PRC}{89}{044905}{2014}.%%{\tt arXiv:1312.6676}, submitted to  \PRC. 
%{Transverse-energy distributions at midrapidity in p+p, d+Au, and Au+Au collisions at $\sqrt{s_{NN}}=62.4-200$ GeV and implications for particle-production models} (2014) 
\bibitem{VoloshinNQP} S.~Eremin and S.~Voloshin, \Journal{\PRC}{67}{064905}{2003}.
\bibitem{NouicerNQP} R.~Nouicer, \Journal{\EPJC}{49}{281-286}{2007}.
\bibitem{AQM} A.~Bia\l as, W.~Czy\. z and L.~Lesniak, \Journal{\PRD}{25}{2328}{1982}.
\bibitem{ATLAS-Gribov} ATLAS Collaboration, ATLAS-CONF-2013-096 (September, 2013).
\bibitem{CMSPLB718} CMS Collab. (S. Chatrchyan, {\it et al.}), \Journal{\PLB}{718}{795}{2013}.
\bibitem{PXv2dAu} PHENIX Collab. (A.~Adare, {\it et al.}), \Journal{\PRL}{111}{212301}{2013}.   
\bibitem{sPHENIX2013} PHENIX Collaboration, The sPHENIX MIE Proposal (2013) \url{http://www.phenix.bnl.gov/plans.html}




\end{thebibliography}
\end{document}